\documentclass[a4paper,12pt,twoside,openright]{book}
\pdfoutput=1
\usepackage{indentfirst}
\usepackage{fancyhdr}
\usepackage{graphicx}
\usepackage{url}
\usepackage[small,bf]{caption}

\usepackage{sidecap}
\usepackage{amssymb}
\usepackage{amsmath}
\usepackage{latexsym}
\usepackage{amsthm}

\usepackage[numbers]{natbib}
\bibpunct{(}{)}{,}{s}{}{;}

 \fancypagestyle{plain}{\fancyhead{}}
\begin{document}
\begin{titlepage}
\begin{center}
\begin{figure}[htbp]
\centering
\includegraphics[width=30mm]{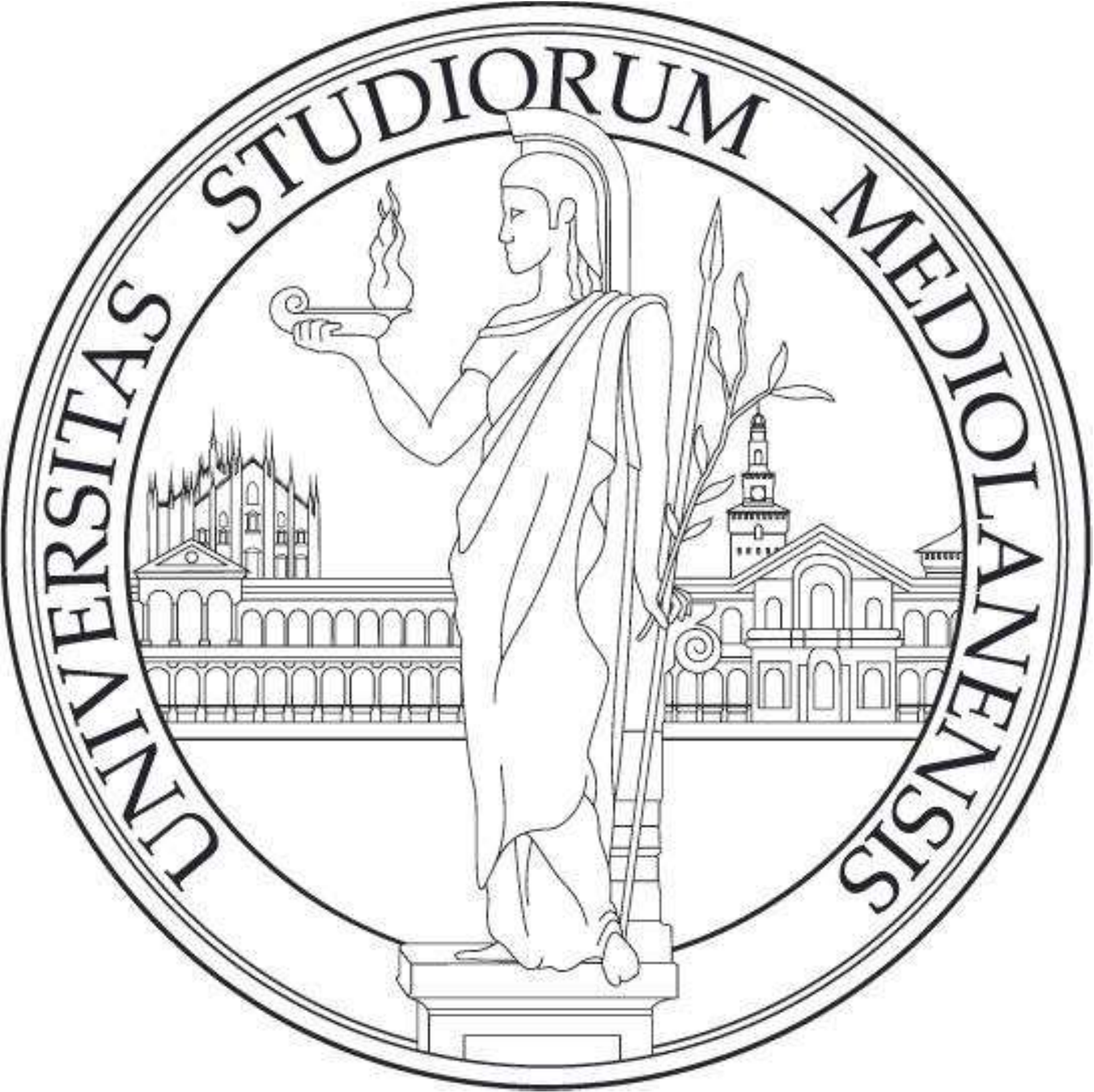}
\qquad
\includegraphics[width=30mm]{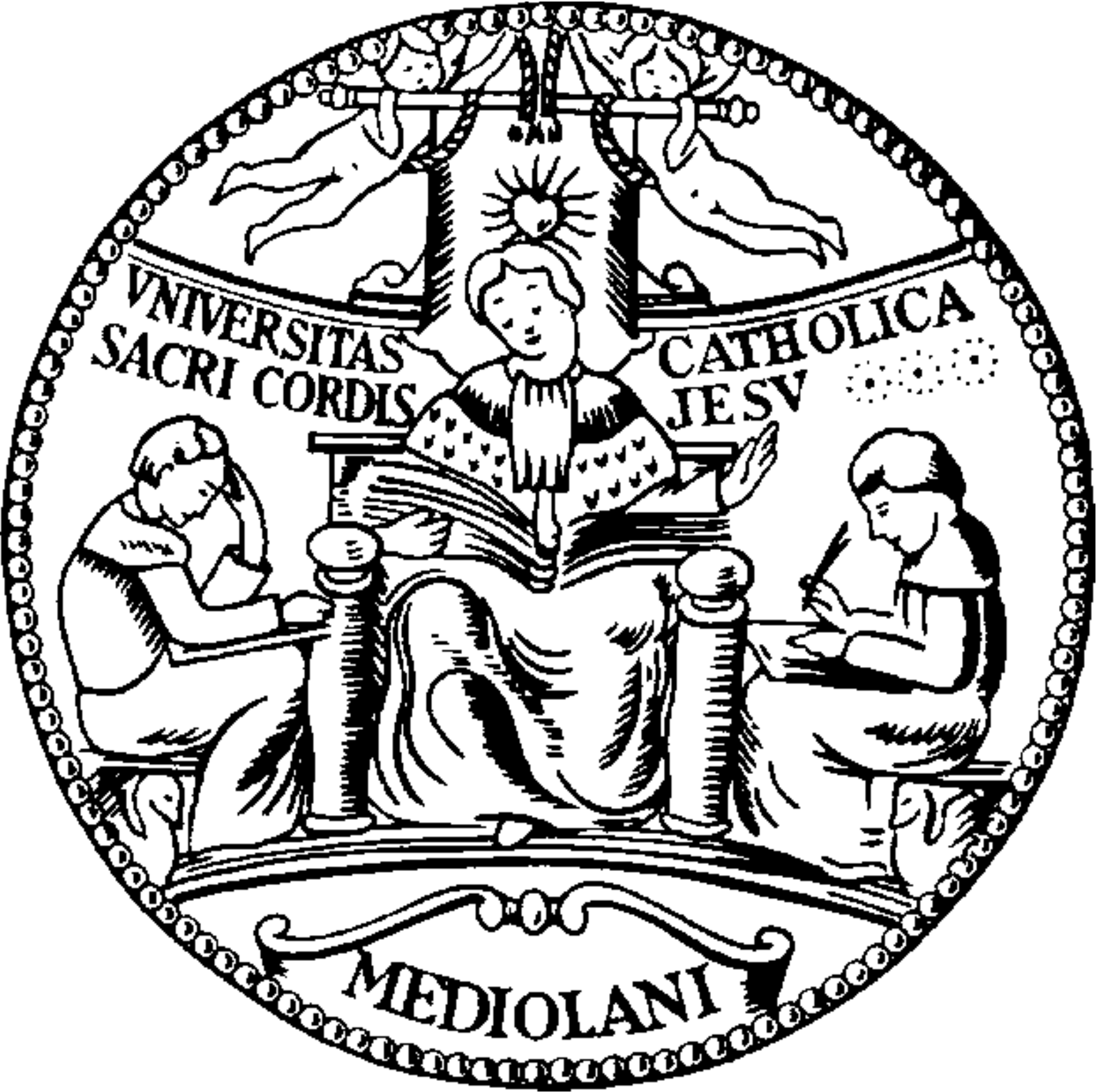}
\end{figure}
\textbf{\normalsize{UNIVERSIT\'A DEGLI STUDI DI MILANO}}\\
\normalsize{Facolt\'a di Scienze Matematiche, Fisiche e Naturali}\\
\normalsize{\&}\\
\textbf{\normalsize{UNIVERSIT\'A CATTOLICA DEL SACRO CUORE}}\\
\normalsize{Facolt\'a di Scienze Matematiche, Fisiche e Naturali}\\[0.7cm]
\textbf{\normalsize{DOTTORATO DI RICERCA IN FISICA, ASTROFISICA E FISICA APPLICATA}}\\[0.7cm]
\Large{\textbf{Electronic structure of TiO$_2$ thin films and LaAlO$_3$-SrTiO$_3$ heterostructures: the role of titanium 3d$^1$ states in magnetic and transport properties}}\\[0.7cm]
\large{Settore scientifico disciplinare FIS/03}\\[1cm]
\end{center}
\begin{flushleft}
\normalsize{\textbf{Coordinatore:} Prof. Marco Bersanelli}\\
\normalsize{\textbf{Tutore:} Prof. Luigi Sangaletti}\\
\end{flushleft}
\begin{flushright}
\normalsize{\textbf{Tesi di Dottorato di:}}\\
\normalsize{Giovanni Drera}\\
\normalsize{\textbf{Ciclo XXIV}}
\end{flushright}
\vfill
\begin{center}
\normalsize{Anno Accademico 2010-2011}
\end{center}
\end{titlepage}


\newpage
\thispagestyle{empty}
\begin{flushright}
\null\vspace{\stretch{1}}
{\emph{to Luisa, Francesco and Martina}\\
\emph{my Best PhD achievements.}}
\vspace{\stretch{2}}\null
\end{flushright}
\pagebreak


\tableofcontents
\chapter{Introduction}
The transition metal oxides (TMO\cite{book_TMO}) play a fundamental role in a wide range of technological applications. They are used as dielectrics, gate insulators and for magnetic applications. The transition metal oxides surfaces\cite{book_surface_TMO} have a primary importance in catalytic processes, in gas-sensing, in fuel-cells and in solar panels. Their conduction properties\cite{book_oxide_conduction} range from insulators to semiconductors and metals, sometimes in the same compounds. This is the case of the so called Metal-to-Insulator transition, in which a material can change from an insulator to a metal state upon suitable doping or under the application of an external pressure.

In each and every of these cases, much of the relevant physics is due to the complexity of the TM electronic structure, which still poses challenging problems to the physics community. In fact, while the electronic structures of elemental TM's have already been fairly described, the same cannot be stated for the TM oxides. The electron-electron interaction in the TM partly filled 3d shell can induce complex correlation effects, hardly described with the simple theories of the early solid state physics. For instance, the best known empirical Hamiltonian that should describe the correlation aspect of these systems, the Hubbard model, cannot still be solved exactly though it has been introduced 50 years ago.

The complexity of the TM-oxides resides also in their bulk and surface crystal structure. Planar dislocations, point defects, cationic or oxygen vacancies are often present in TMO crystals, but their positions in the lattice can be hardly controlled. The theoretical calculations on defect sites in TMO are also rather complex, due to the aforementioned correlation effects. Much of the effort of experimental and theoretical physicists (and chemists) is thus devoted to the characterization and to the control of these defects.

The present thesis is focussed on the Ti-O bond in TiO$_2$ based systems, namely TiO$_{2-\delta}$ thin films and LaAlO$_3$-SrTiO$_3$ heterostructures. The attention on these system has been drawn by the occurrence of unexpected physical properties in formally 3d$^0$ insulators: magnetism and 2D conductivity, respectively. Both properties are thought to be related to the 3d level population of titanium cations. In the following, a brief introduction to these topics is provided.

\section{The TiO$_{2-\delta}$ magnetism}
A peculiar case treated in this Thesis work, is the titanium dioxide (TiO$_2$). This system is a formally simple d$^0$ oxide, a transparent insulator (the gap is nearly 3 eV) widely applied in pigments, cosmetics and coatings because it is cheap, highly refractive and bio-compatible. Due to the vastity of published literature and to the easiness in the growth, its surface is considered as the prototype\cite{Die2003} of the metal-dioxide systems. The various surface terminations of TiO$_2$ single-crystal are very important for catalysis applications and are still subject of an intensive research work.

Titanium dioxide has also been claimed to be a room-temperature ferromagnet upon transition-metal doping\cite{DMO_coey_nat}. Even if at the beginning the dopant ions were considered as the source of the ferromagnetism (since pristine TiO$_2$ is diamagnetic), the researcher shifted their attention to the electronic states generated by the oxygen vacancies. In fact, the same kind of (weak) magnetism have been recently measured in formally undoped samples; this phenomenon is known as d$^0$ magnetism \cite{DMO_TiO2_d0} and has been found also in HfO$_2$\cite{DMO_hfo2_d0}, SnO$_2$, ZnO, MgO and MgAl$_2$O$_4$\cite{DMO_various_d0}. Each of these materials, when grown with state-of-the-art methods, are diamagnetic. The introduction of defects that occurs in non-equilibrium growth techniques seems to induce the observed ferromagnetism.

However, the claim of intrinsic room-temperature ferromagnetism in these closed-shell oxides has not yet been universally accepted by the researcher community\cite{DMO_fact_or_fiction}, with the most important objection being related to sample quality (i.e. to possible external contaminations). A fine control of growth and characterization is then necessary when dealing with the magnetism of these materials.

The defect-related electronic states in these oxides are usually too weak to be probed with standard X-ray photoemission (XPS). One of the experimental techniques that can properly address this problem is Resonant Photoemission spectroscopy (ResPES), in which the valence band spectra are collected by tuning the photon energy close to a selected atomic shell absorption edge. In the case of TiO$_2$, ResPES can enhance the titanium electronic states in the valence band and, if present, also in the band gap.

In the first part of this Thesis, a comprehensive magnetic characterization in a set of TiO$_2$ samples is given, together with the analysis of Ti 3d-related states. A series of TiO$_{2-\delta}$ films have been grown with the RF-sputtering technique on various non-magnetic substrates; subsequently, a protocol of annealing procedures has been applied in order to alter the sample stoichiometry. A full magnetic, crystalline and electronic structure characterization has been carried out at each step of the films treatments. A set of N-doped TiO$_{2-\delta}$ thin films have also been grown, in order to verify the effect of the N doping in the TiO$_2$ magnetism.
The electronic states related to the presence of defects in these films have been further investigated through ResPES. The hypothesis of a clustered V$_O$ origin of the FM is further discussed in the light of the experimental and theoretical results.

\section{The conductivity of the LaAlO$_3$-SrTiO$_3$ interface}
Another interesting oxide system in which the stoichiometry of Ti ions play a fundamental role is the LaAlO$_3$-SrTiO$_3$ interface (LAO-STO in short). LAO and STO, separately, are two band insulators, with an empty shell electronic structure (3d$^0$ for STO, 4f$^0$ for LAO) and a similar perovskite structure; however, the interface created by growing LAO on the top of STO (001) has found to become metallic, hosting a quasi-2D electron gas\cite{LAOSTO_science}. This heterostructure becomes conductive when the STO is terminated with the TiO$_2$ plane while remain insulating with the SrO plane termination; the Ti-related electronic states are thus expected to carry the metallic states.

The transition from the insulating to the metallic state is observed only for a LAO capping larger than 4 unit cells, making the puzzle of the conductivity even more complicated. Furthermore, in some cases a low T$_c$ superconductivity and isolated magnetic moments have also been detected, but not at the same time in a sample. As in the case of the TiO$_2$ magnetism, the growth conditions and the oxygen stoichiometry are crucial in the physics of the LAO-STO system\cite{LAOSTO_rev2}. In fact the interfaces grown in a low partial pressure of oxygen yield a 3D conductivity, induced by the presence of oxygen vacancies. Besides, STO itself is an interesting case of defect-induced physical properties: superconductivity\cite{STO_sc} has been observed in defected SrTiO$_{3-\delta}$, obtained by the annealing of pristine SrTiO$_3$. The possibility of growing high quality samples has been related to the detection of the anomalous Shubnikov de Haas effect\cite{LAOSTO_sub_deHaas} in these interfaces, which is a clear proof of the 2D nature of the conductive layer.

Different mechanisms have been invoked in order to explain the physical phenomena in the LAO-STO heterostructures. By recognizing that the polar nature of LAO could induce a diverging electric potential, an electronic reconstruction mechanism has been proposed, based on a charge-transfer from the surface to the Ti atoms at the interface\cite{LAOSTO_rev1}. However, there are several proof of cationic disorder, i.e. of a non-abrupt LAO-STO interface\cite{LAOSTO_chambers}; the polar discontinuity could then be energetically relaxed by the disorder and not by the electronic reconstruction. Finally, also the V$_O$'s should be taken into account, as well as the possible charge transfer effects from surface defects\cite{bristowe}. Each of these mechanisms don't exclude the others; at the moment there is not a general consensus in the researcher community.

As in the case of TiO$_2$, a direct experimental proof of the conductive electronic states can be hardly obtained with conventional photoemission techniques. The main difficulty is the surface sensitivity of the soft X-ray photoemission, since the relevant electronic states are buried beneath the LAO capping. Moreover, even when hard X-ray photoemission is used, the titanium contribution to valence band spectra is usually overwhelmed by the oxygen 2p states. In a similar way of d$^0$ ferromagnetism, the resonant photoemission taken at Ti edge could be the best technique to directly measure the Ti contribution to the conductivity\cite{apl_LAOSTO}.

The second part of this Thesis work is devoted to the characterization of conductive and insulating LAO-STO interfaces, carried on with X-ray photoemission (XPS), X-ray absorption (XAS) and with ResPES techniques. The stoichiometry of each atomic species has been evaluated by comparison with LAO and STO single crystals. A resonance enhancement of the conductive Ti states, associated to a small fraction of Ti$^{3+}$ ions is reported and compared to theoretical calculations. In addition, a characterization of the intermixing and the disorder at the LAO-STO interface has been done through angle-resolved XPS, which in this case has required a careful control over the theoretical model and over the experimental data handling. The results are then compared to the theoretical models proposed in the literature.

\section{Thesis outline}
In \textbf{Chapter 2} the experimental and the computational details that concern the X-ray spectroscopies of interest are presented. When possible, the XPS, XAS and ResPES spectra have been compared with theoretical calculations; in this Chapter a brief introduction of the multiplet theory and the ab-initio density functional theory (DFT) is thus given. The Hubbard correction to these models, the charge-transfer multiplet and the DFT+U, have been adopted to calculate the spectral weight and to interpret the experimental results. Finally, a section of this Chapter is dedicated to the theory of the AR-XPS spectroscopy, applied to the LAO-STO interface.

A detailed description of the ResPES spectra at Ti L$_{2,3}$-edge in pure TiO$_2$ is given separately in \textbf{Chapter 3}; these data can be used as a reference for the experimental results given in the following chapters. The XAS and ResPES results have been interpreted both with multiplet and with DFT+U calculations.

In \textbf{Chapter 4}, the photoemission and magnetization data for a series of TiO$_2$ and nitrogen doped TiO$_2$ films are shown. The samples have been thoroughly characterized with X-ray diffraction, Raman spectroscopy, SQUID magnetometer and XPS at each step of annealing treatments aimed to control the density of defects in the films. XAS and ResPES have also been performed on the thinnest TiO$_{2-\delta}$ and N-doped TiO$_{2-\delta}$ films. The role of nitrogen in magnetism is further explained with spin-resolved DFT calculations.

Finally in \textbf{Chapter 5} the XPS, XAS and ResPES data taken on the LAO-STO samples are shown, compared to the single-crystal LaAlO$_3$ and SrTiO$_3$. The experimental results of insulating and conductive samples are compared. The angle-resolved X-ray photoelectron spectroscopy (AR-XPS) and X-ray
photoelectron diffraction (XPD) techniques have been used to check the structural properties of the interface and the capping layer.\\
\pagebreak

\section{Publications}
Here is the list of the articles already published ($\dag$) or in preparation ($\ddag$) related to this Thesis work:
\begin{enumerate}
  \item $\ddag$ M. C. Mozzati, G. Drera, L. Malavasi, Y. Diaz-Fernandez, P. Galinetto and L. Sangaletti; \textsl{A systematical study of magnetism in TiO$_{2-\delta}$ thin films grown by RF-sputtering}, in preparation.
  \item $\ddag$ G. Drera, G. Salvinelli, J. Huijben, M. Huijben, G. Rijnders, D. H. A. Blank, H. Hilgenkamp, A. Brinkman and L. Sangaletti; \textsl{Growth condition effects on the basic electronic properties of LaAlO$_3$, SrTiO$_3$ and LaAlO$_3$-SrTiO$_3$ heterostructures}, in preparation.
  \item $\ddag$ G. Drera, G. Salvinelli, J. Huijben, M. Huijben, G. Rijnders, D. H. A. Blank, H. Hilgenkamp, A. Brinkman and L. Sangaletti; \textsl{Origin of titanium 3d$^1$ electronic states in insulating and conductive LaAlO$_3$-SrTiO$_3$ heterostructures}, in preparation.
  \item $\ddag$ G. Drera, G. Salvinelli, J. Huijben, M. Huijben, G. Rijnders, D. H. A. Blank, H. Hilgenkamp, A. Brinkman and L. Sangaletti; \textsl{Unveiling cation intermixing effects in LaAlO$_3$-SrTiO$_3$ interfaces by AR-XPS experiments}, in preparation.
  \item $\ddag$ G. Drera, L. Malavasi, Y. Diaz-Fernandez, F. Bondino, M. Malvestuto, M.C. Mozzati, P. Galinetto and L. Sangaletti; \textsl{Origin, excitation dynamics, and symmetry of the insulating TiO$_{2-\delta}$ electronic states probed by resonant photoelectron spectroscopies}, revised version submitted to \textit{Phys. Rev. Lett.}
  \item $\dag$ G. Drera, F. Banfi, F. Federici Canova, P. Borghetti, L. Sangaletti, F. Bondino, E. Magnano, J. Huijben, M. Huijben, G. Rijnders, D.H.A. Blank, H. Hilgenkamp and A. Brinkman; \textsl{Spectroscopic evidence of in-gap states at the SrTiO$_3$/LaAlO$_3$ ultrathin interfaces}, \textit{Appl. Phys Lett.} \textbf{98} (2011), 052097.
  \item $\dag$ G. Drera, M.C. Mozzati, P. Galinetto , Y. Diaz-Fernandez, L. Malavasi, F. Bondino, M. Malvestuto and L. Sangaletti; \textsl{Response to ``{Comment on `Enhancement of room temperature ferromagnetism in N-doped TiO$_{2-x}$ rutile: Correlation with the local electronic properties' }''}, \textit{Appl. Phys Lett.} \textbf{97} (2010), 186102.
  \item $\dag$ G. Drera, M.C. Mozzati, P. Galinetto , Y. Diaz-Fernandez, L. Malavasi, F. Bondino, M. Malvestuto and L. Sangaletti; \textsl{Enhancement of room temperature ferromagnetism in N-doped TiO$_{2-x}$ rutile: Correlation with the local electronic properties},\textit{Appl. Phys Lett.} \textbf{97} (2010), 012506.
  \item $\dag$ F. Rossella, P. Galinetto, M. C. Mozzati, L. Malavasi, Y. Diaz Fernandez, G. Drera and L. Sangaletti; \textsl{TiO$_2$ thin films for spintronics application:a Raman study}, \textit{J. Raman Spec.} \textbf{41} (2010), 558-565.
\end{enumerate}

Other publications related to magnetism and X-ray spectroscopy techniques applied to diluted magnetic oxides and semiconductors:
\begin{enumerate}
  \item L. Sangaletti, M. C. Mozzati, G. Drera, V. Aguekian, L. Floreano, A. Morgante, A. Goldoni and G. Karczewski; \textsl{Local electronic properties and magnetism of (Cd,Mn)Te quantum wells}, \textit{Appl. Phys. Lett.} \textbf{96} (2010), 142105.
  \item L. Sangaletti, A. Verdini, S. Pagliara, G. Drera, L. Floreano, A. Goldoni and A. Morgante; \textsl{Local order and hybridization effects for Mn ions probed by resonant soft X-ray spectroscopies: the Mn:CdTe(110) surface revisited}, \textit{Phys. Rev. B} \textbf{81} (2010), 245320.
  \item L. Sangaletti, G. Drera, E. Mangano, F. Bondino, C. Cepek, A. Sepe and A. Goldoni; \textsl{Atomic approach to core-level spectroscopy of delocalized system: Case of ferromagnetic metallic Mn$_5$Ge$_3$}, \textit{Phys. Rev. B} \textbf{81} (2010), 085204.
   \item L. Sangaletti, F. Federici Canova, G. Drera, G. Salvinelli, M. C. Mozzati, P. Galinetto, A. Speghini and M. Bettinelli; \textsl{Magnetic polaron percolation on a rutile lattice: A geometrical exploration in the limit of low density of magnetic impurities}, \textit{Phys. Rev. B} \textbf{80} (2009), 033201
\end{enumerate}

\newpage
\section{Acknowledgements}
The author wishes to acknowledge the (numerous) people involved in this Thesis work:
\begin{itemize}
\item Maria Cristina Mozzati, Pietro Galinetto, Yuri Diaz-Fernandez and Lorenzo Malavasi of the Universit\'a di Pavia for their (enormous) work on the growth and characterization of the ferromagnetism in TiO$_2$;
\item Alexander Brinkmann and co-worker, for the growth and characterization of the LAO-STO heterostructures at the University of Twente, Enschede (The Netherlands);
\item Federica Bondino, Elena Magnano, Marco Malvestuto and the staff of the BACH beamline at the ELETTRA synchrotron (Basovizza, Trieste);
\item Aleberto Verdini, Luca Floreano and Alberto Morgante and the staff of the ALOISA beamline at the ELETTRA synchrotron.
\item Luigi Sangaletti, Patrizia Borghetti, Dash Sibashisa, Filippo Federici Canova and Gabriele Salvinelli of the Surface Science and Spectroscopy Lab. of the Universit\'a Cattolica di Brescia.
\end{itemize}
The work on magnetic oxides has been partly funded by the Cariplo Foundation.

Finally, I wish to express my gratitude to professor Ralph Claessen for his refereeing work and to dr. Maddalena Drera for the meticulous proofreading.

\chapter{Experimental and theoretical details}
\section{Introduction}
A relevant part of this thesis work have been devoted to explore in details some core-level based X-ray spectroscopic techniques. The main advantage of using X-rays is the possibility to be chemically selective and thus have access to the electronic structure of a single ionic species. The experimental methods applied in this work are X-ray photoemission (XPS), absorption (XAS) and photoemission taken at resonance condition (ResPES). In the TiO$_{2-\delta}$ case, the magnetic characterization has been done with a SQUID magnetometer at Universit\'a di Pavia, as well as the Raman spectroscopy.

X-ray photoemission measurements have been carried out mostly in the Surface Science and Spectroscopy Lab at Universit\'a Cattolica del Sacro Cuore, while X-ray absorption and resonant photoemission data have been collected at BACH and ALOISA beam lines at the Elettra synchrotron in Trieste (Italy).
Electron spectroscopy spectra have been calculated with various electronic structure calculation methods; in particular, an atomic multiplet approach has been adopted for transition metal core levels and an ab-initio density functional theory approach (DFT) for calculation of valence and conduction bands.
The intensities of the angle-resolved photoemission (AR-XPS) peaks on LAO-STO have been simulated with a theoretical model which accounts for inelastic as well as elastic scattering.

In this Chapter, a description of the experimental techniques and theoretical calculations is given; some XPS and XAS results measured on manganese ions in germanium and cadmium telluride, already published by our research group, are provided as examples.

\section{XPS and AR-XPS}

\begin{figure}[ht]
\begin{center}
\includegraphics[width=0.9\textwidth]{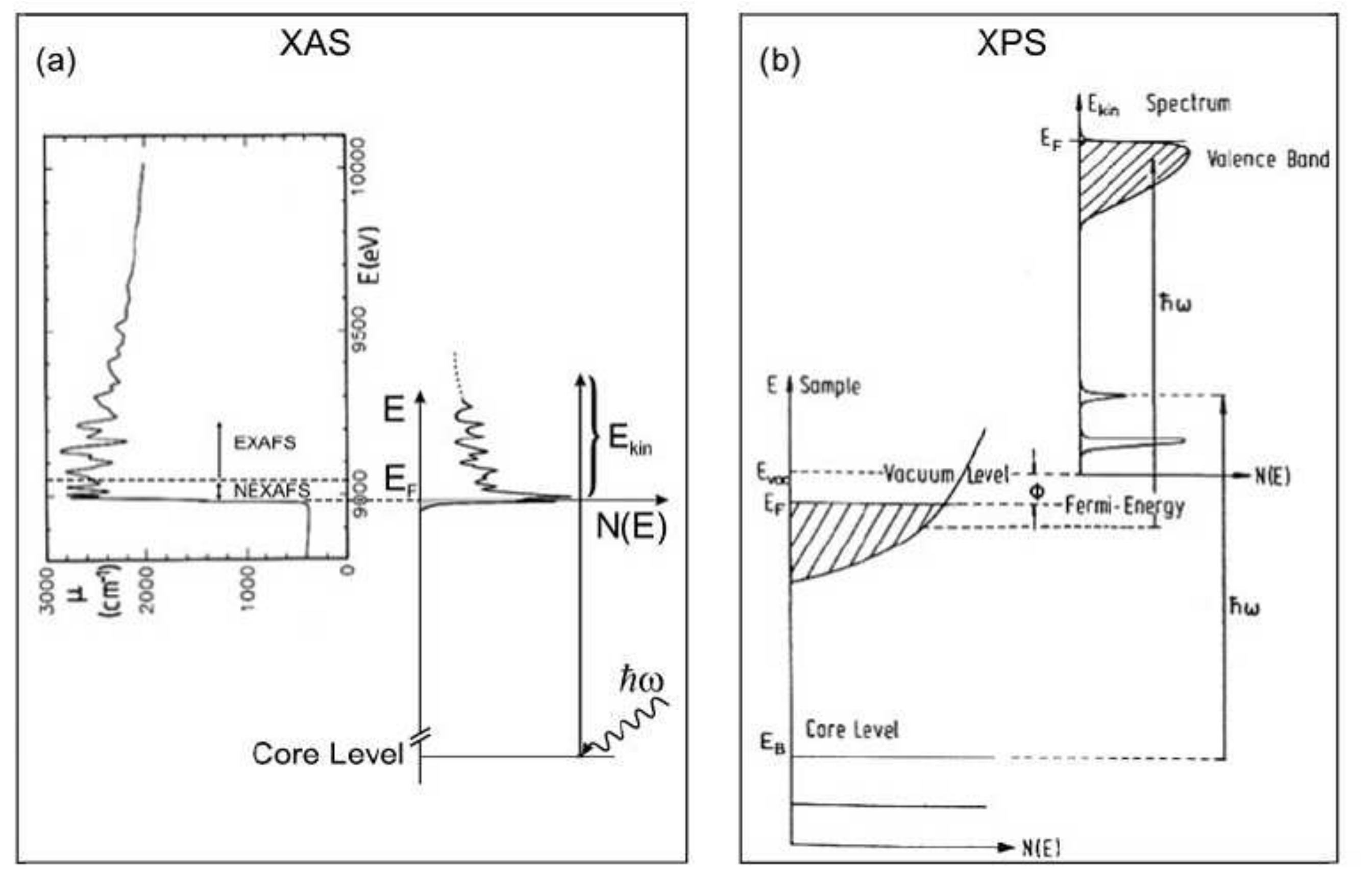}
\caption {Schematic view of X-ray absorption (a) and photoemission (b) spectroscopies.\label{fig_schema_xpsxas}}
\end{center}
\end{figure}

\begin{figure}[ht]
\begin{center}
\includegraphics[width=1\textwidth]{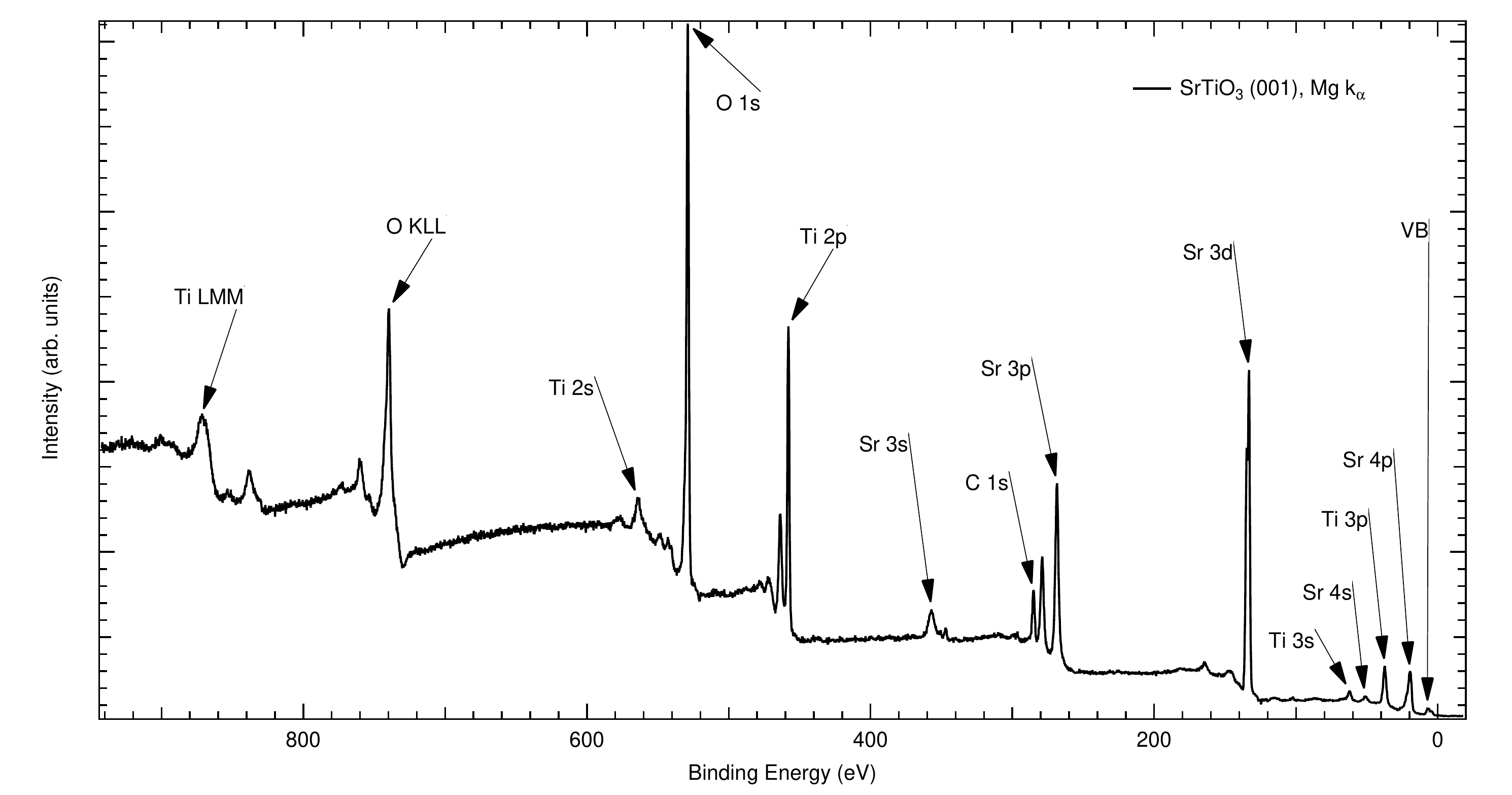}
\caption {XPS survey spectrum on the STO (001) surface, taken with a Mg k$_\alpha$ source.\label{fig_exp_survey}}
\end{center}
\end{figure}

X-ray photoemission spectroscopy is a well established photon-in electron-out technique that gives access to the electronic and chemical properties of selected atomic species on a sample surface. A schematic view of XPS is shown in Fig. \ref{fig_schema_xpsxas}(b). In a XPS experiment the specimen is exposed to a soft X-ray source (usually $h\nu\geq 100$ eV) in order to induce the photoelectric emission from core-levels and valence band. The number of electrons vs the kinetic energy (E$_k$) spectrum is measured by an electron spectrometer. The survey spectra are usually composed by a series of peaks superimposed to a stair-like structure (see for example, the STO (001) survey in Fig. \ref{fig_exp_survey}); the peak energy is related to a specific core level of an atoms or to secondary Auger electrons induced by the core-hole, while the step-like background is generated by the inelastic scattering of photoelectrons. XPS peaks ascribed to a core-level with a symmetry different than spherical s, can be split by spin-orbit interaction, which is inversely proportional to $n$ quantum number; for example the p levels are split in p$_{1/2}$ and p$_{3/2}$ components, with an area ratio given by the electron degeneracy (1:2 for p levels, 2:3 for d levels and so on).

Since the core-level energies are mostly determined by the atomic species, XPS allows the identification of the sample composition (chemical elements and their relative ratio); moreover, the specific valence state of the atom  affects the position of the core-level peaks energies, allowing the identification also of the bond ionicity (or covalency) degree. This is the reason why XPS is also known as ESCA, Electron Spectroscopy for Chemical Analysis. While a complete theoretical description of the XPS technique lies outside the purpose of this Thesis work, some basic details about the angular dependence of the probing depth must be explained in order to properly introduce the AR-XPS method. In the next section, technical details about the experimental instrumentation are given.

\subsection{Experimental set-up}
A standard XPS set-up requires an experimental chamber operated in  UHV conditions (10$^{-10}$ mbar) in order to avoid sample contamination and, to lesser extent, to avoid the photoelectrons scattering through the analyzer. The commercial X-Ray sources are usually based on the emission spectra of aluminum (Al k$_\alpha$ line, $h\nu=1486.7$ eV) and magnesium (Mg k$_\alpha$, $h\nu=1256.6$ eV), but of course XPS can be performed with the synchrotron radiation in dedicated beamlines, in order to have tunable photon energies and a higher photon flux.

In the Universit\'a Cattolica Labs a dual anode (Mg/Al) PSP X-ray source with a Scienta R-3000 AR-XPS analyzer with a 2D phosphor detector have been used. The best resolution was  0.7$\pm$0.1 eV, mainly due to the X-ray source. This analyzer can operate both in a ``transmission'' mode and in an ``angular mode'' that allows the simultaneous collection of the XPS spectra in a $\pm$10$^\circ$ range. When an higher resolution was needed, a parallel XPS setup has been used with a monochromatized Al k$_{\alpha}$ source and a VG MkII system, leading to a 0.5 eV total resolution at the expense of the total electron count rate.

\subsection{XPS theory}
From the conservation of total energy, the kinetic energy of an electron photoemitted from a core-level can be expressed as follows:

\begin{equation}
	E_{Kin}=h\nu-E_{Bin}-\Phi
	\label{eq_exp1}
\end{equation}
\\
where E$_{Bin}$ is the electron binding energy and $\Phi$ is the work function, which is the extra energy needed to transport an electron from the sample to the analyzer. The electrons from Auger process are related only to internal atomic relaxation and thus are photon-energy independent (i.e. fixed kinetic energy), while the kinetic energy of core-level photoelectrons changes with the photon (fixed binding energy). A tunable (multiple anode, or synchrotron based) X-ray source can thus be used to better identify  Auger structures, when superimposed to other core-levels.

The probing depth of this technique is due not to X-ray penetration (in the $\mu$m range) but to the inelastic mean free path (IMFP) of electrons in solids which, at these kinetic energies, is in the order of 1-2 nm: XPS is thus a surface-sensitive technique. In Fig. \ref{fig_EAL_LAO} in section 5.4 are shown some calculated IMFP for the LaAlO$_3$ case. In some synchrotron facilities, the photoemission with hard X-ray photons (HAX-PES, $h\nu$ up to 10 keV) is also possible, which in turn is a bulk-sensitive technique; in HAXPES however one has to deal with a reduced X-ray flux and, most important, with smaller photoelectron cross-sections.

From a quantum-mechanics point of view, the photoemission process can be written as follows with the Fermi Golden rule:

\begin{equation}
	W_{ph}=\frac{2\pi}{\hbar}\left|\left\langle i|T|f \right\rangle\right|^{2}\delta(E_k+E_i-E_f-h\nu)
\label{eq_exp4}
\end{equation}
\\
Where the squared matrix element gives the transition rate and the Dirac delta accounts for energy conservation; in photoemission, the final state is the ground state plus a core hole and a free (photoemitted) electron with E$_{k}$ energy. The matrix operator $T$ for the electron-photon interaction can be taken as the simple dipole operator $T=e\cdot\vec{r}$, since the contribution of the quadrupole operator for the soft X-ray regime is negligible\cite{DEGROOT}. This equation can be solved with the Green functions method, in order to account for scattering process and final state effects due to the relaxation of the electronic levels in the proximity of the core-hole; this approach is called the ``one-step'' photoemission theory.

Another way to describe XPS is the so-called ``three-step'' model, in which the process is described in three different steps:

\begin{itemize}
	\item first step is the photoemission from the atom, described with photoemission cross section;
	\item the second step is the drift of the electron in the solid, described by the electronic IMFP;
	\item the third step is the escape of the photoelectron from the solid, described by the work-function.
\end{itemize}

The three-step model is often assumed in the framework of the \textit{sudden approximation}, in which one suppose that the core-hole final state doesn't influence the XPS spectra, while the one-step model gives e a good description of excitonic effects and plasmon resonances.

In the three-step model, the intensity (i.e. the area) of a photoelectron peaks depends on many parameter; disregarding the diffractive effects (very important in single-crystal, as will be shown in Chapter 5) and X-ray attenuation (important only for grazing photon incidence angles), the contribution to the photoemission intensity of an infinitesimal thick layer at depth z from the surface can be expressed as follows:

\begin{equation}
	dI=\phi D_{0}(E_{K}) \sigma(h\nu) \rho A_{0} P(\lambda,z,\theta) dz\label{eq_exp2}
\end{equation}\\
where
\begin{equation}
	P(\lambda,z,\theta)= e^{-\frac{z}{\lambda cos(\theta)}}\label{eq_exp3}	
\end{equation}
\\
and:
\begin{itemize}
	\item $\phi$ is the photon flux;
	\item $D_{0}$ is the analyzer transmission function that depends on the kinetic energy (old electron analyzer could have different sensitivity at different E$_k$);
	\item $\sigma$ is the photoelectron cross section that depends on the atomic species, the electronic level (i.e. 1s,2s,2p...) and the photon energy;
	\item $\rho A_{0}$ is the number of atoms per unit volume per sampling area, that can depend on the X-ray focalization on the sample and/or on the analyzer focus;
	\item $P(\lambda,z,\theta)$ is the probability of an electron at depth z to escape from the specimen and to reach the analyzer placed at an angle $\theta$ relative to the sample normal; $\lambda$ is the IMFP and depends on the sample composition and on the E$_k$.
\end{itemize}

When dealing with the elements ratios only, the photon flux and sampling area can be discarded since they are contributing in the same way to each photoelectron peak. When measuring XPS peaks with close kinetic energy (or with a well calibrated analyzer) also the $D_0$ term can be neglected. The other parameters have to be taken into account to carefully quantify the elements concentration; the typical accuracy of this method in a homogeneous sample is around 1-5\%. The typical XPS low boundary of sensitivity is about 0.1-1\%.

\subsection{Angle resolved XPS}
The parameter $P$ in Eq. \ref{eq_exp3} is also dependent on the relative orientation of specimen with respect to the analyzer; in fact, at grazing emission, photoelectrons of inner layer have to travel for a longer path through the sample in order to reach the vacuum, resulting in a reduced photoelectron intensity from the bulk. The surface sensitivity of XPS is thus improved at grazing emission angles; this can be useful for instance for enhancing the contribution of absorbed molecule respect to the host substrate.

This effect can be also used in a reverse way: for example, in the case of a thin film coverage of a bulk material, one can use AR-XPS data to estimate the overlayer thickness. This is only possible when the composition of each layer is known, in order to fix all the parameter described in Eq. \ref{eq_exp2}; in some cases just a single measure (i.e. taken at only one emission angle) could be sufficient to estimate thickness, but of course a set of data taken at different angles can improve the results reliability.

\begin{SCfigure}
\label{fig_arxps_multi}
\includegraphics[width=0.6\textwidth]{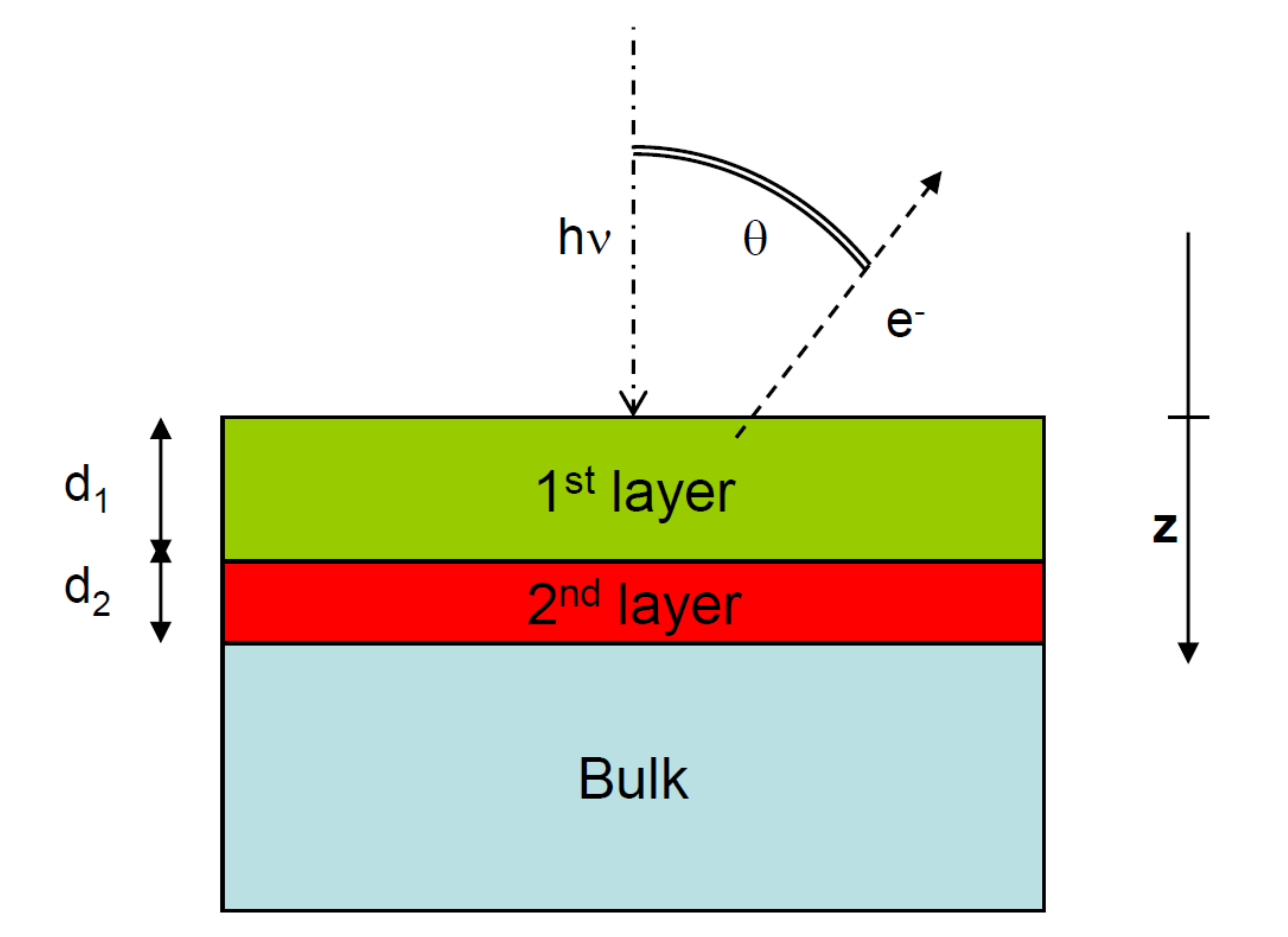}
\caption{Schematic configuration of an AR-XPS model for a multilayer sample. In this Figure the parameters for Eqs. \ref{eq_exp6},\ref{eq_exp7},\ref{eq_exp8} are defined}
\end{SCfigure}

Eq. \ref{eq_exp2} has to be integrated on the z axis to obtain the total XPS intensity; for instance, in the case of a multilayer sample on a substrate (see Fig. \ref{fig_arxps_multi}) with thickness $d_1$ and $d_2$, the contribution on the photoelectron intensity of each layer can be written as follows:

\begin{equation}
I^{1}_{a}\propto\sigma_{a} \rho_{a} \int^{d_1}_{0}e^{-\frac{\lambda_{a}}{z cos(\theta)}} dz =\sigma_{a} \rho_{a} \frac{cos(\theta)}{\lambda_{a}} (1-e^{-\frac{\lambda_{a}}{d_1 cos(\theta)}})
\label{eq_exp6}
\end{equation}
\\
\begin{equation}
I^{2}_{b}\propto\sigma_{b} \rho_{b} \int^{d_1+d_2}_{d_1}e^{-\frac{\lambda_{b}}{z Cos(\theta)}} dz =\sigma_{b} \rho_{b} \frac{Cos(\theta)}{\lambda_{b}} e^{-\frac{\lambda_{b}}{d_1 Cos(\theta)}} (1-e^{-\frac{\lambda_{b}}{d_2 Cos(\theta)}})
\label{eq_exp7}
\end{equation}
\\
\begin{equation}
I^{bulk}_{c}\propto\sigma_{c} \rho_{c} \int^{+\infty}_{d_1+d_2}e^{-\frac{\lambda_{c}}{z Cos(\theta)}} dz =\sigma_{c} \rho_{c} \frac{Cos(\theta)}{\lambda_{c}} e^{-\frac{\lambda_{c}}{d_1 Cos(\theta)}}e^{-\frac{\lambda_{c}}{d_2 Cos(\theta)}}
\label{eq_exp8}
\end{equation}
\\
where $a$, $b$ and $c$ are index for atomic levels; the photon flux has been omitted, since only the relative ratios are usually evaluated. These equations can be used as a model to fit AR-XPS data in order to obtain the $d_1$ and $d_2$. With simple modification of the element profile along the z axis one can deal also with other cases, such as 3D islands on a surface.
The photoelectron cross-sections in the soft X-ray regimes for each element of the periodic table have been calculated by Yeh and Lindau\cite{YehLindau}; the IMFP data can be evaluated through the Tanuma-Powell-Penn\cite{TPP2M} (TPP-2M) formula, that requires the knowledge of material density, bandgap and the number of valence electrons, as well as the electron kinetic energy.

\begin{SCfigure}
\label{fig_tougaard}
\includegraphics[width=0.6\textwidth]{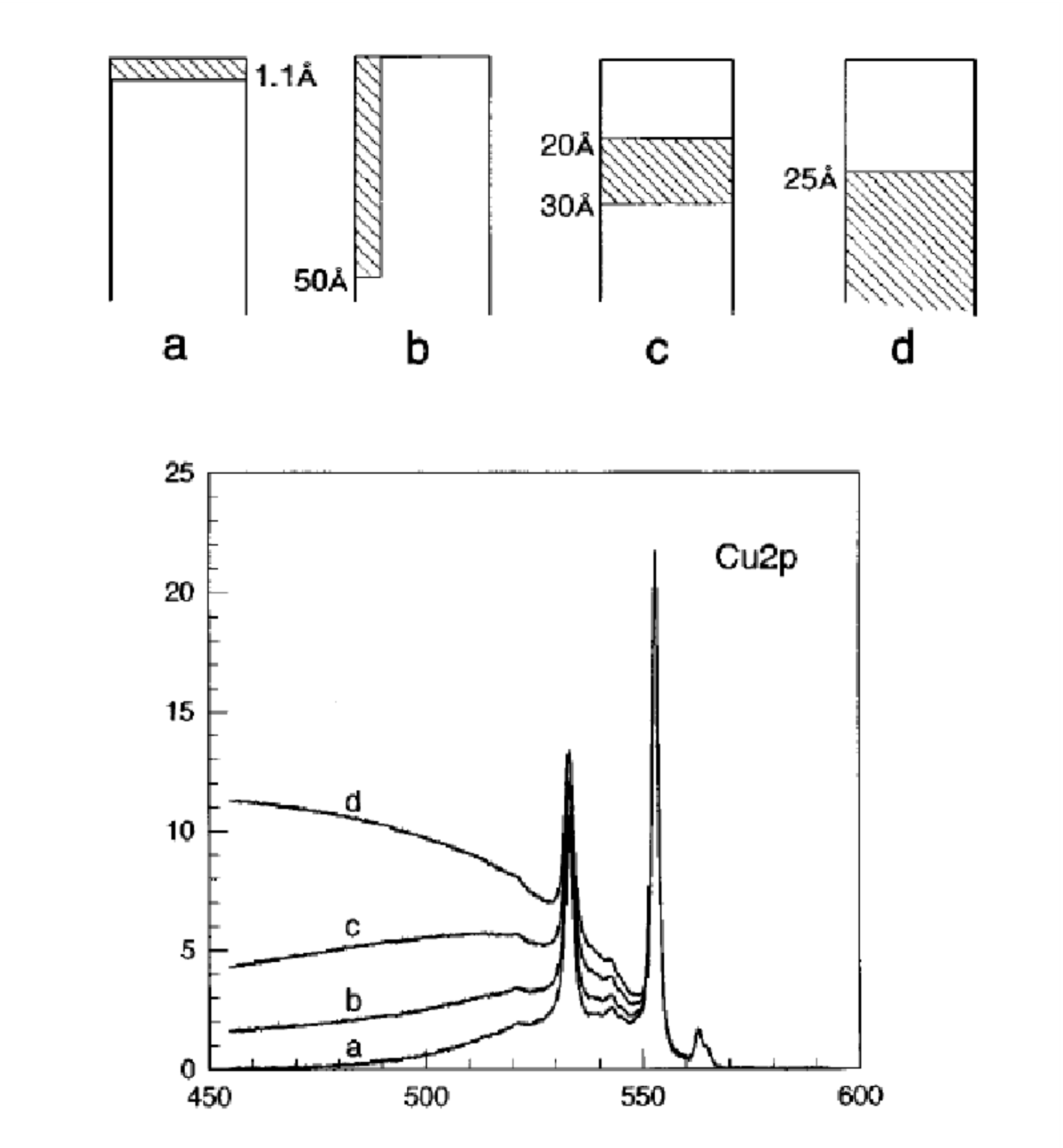}
\caption{Different surface structure of copper in gold which give exactly the same XPS Cu 2p peak intensity. The different configurations can be inferred by inelastic background at higher BE than the photoemission peak. From Ref.\cite{tougaard}}
\end{SCfigure}

Unfortunately, this model suffers from two drawbacks. First, different topological arrangements can lead to exactly the same peak ratios, as shown in Fig. \ref{fig_tougaard}: a comprehensive approach should consider also the slope of the stair-like background, which increases with the number of inelastic scattering events\cite{tougaard}.

Besides, even if the structure composition (overlayer, multilayer, 3d island etc.) were known, the inelastic mean free path alone would not be sufficient to correctly estimate the thickness of the multilayer structure. A soultion to this problem is given by the replacement of IMFP with the Effective Attenuation Length\cite{EAL_theory} (EAL), which is a more complete estimation of the effective opacity of a solid for a given electron energy. The EAL is calculated by introducing the contribution of the elastic scattering length, which can change significantly both the exponential and the angular attenuation of the XPS signal. As a matter of fact, the real attenuation length is dependent to the specific geometry of the system and thus can be only evaluated through computational demanding Monte-Carlo calculations. However, it is possible to obtain EAL values for specific cases (like thin overlayer or thin marker-layer at a specific depth) through simple formulas. EAL values adopted in this Thesis have been calculated with the \texttt{EAL} program of NIST\cite{NIST}.
The expression for the local EAL are:

\begin{equation}
EAL=-[cos(\alpha)\frac{d ln \phi(z,\alpha)}{dz}]^{-1}
\label{eq_exp9}
\end{equation}
\\
for measurement of marker-layer depth and

\begin{equation}
EAL=-[cos(\alpha)\frac{d}{dt} ln (\int^{\infty}_{t}{\phi(z,\alpha)dz})]^{-1}
\label{eq_exp10}
\end{equation}
\\
for measurement of an overlayer-film thickness $t$, where $\phi$ is the emission depth distribution function that depends on depth $z$ and on the analyzer angle $\alpha$. The EAL from Eq. \ref{eq_exp10} can be used instead of $\lambda$ in formulas such as Eq. \ref{eq_exp8} \textit{before} the integration, which thus can't be done analytically (the EAL depends on z). An average value of the EAL, designed to be used for convenience in the XPS intensity formulas after the integration, can also be obtained through the \texttt{EAL} code; however, this ``practical EAL'' is suitable for a thick (more than 3 nm) capping layer and thus has not been adopted in this Thesis work.

\begin{figure}[ht]
\begin{center}
\includegraphics[width=1\textwidth]{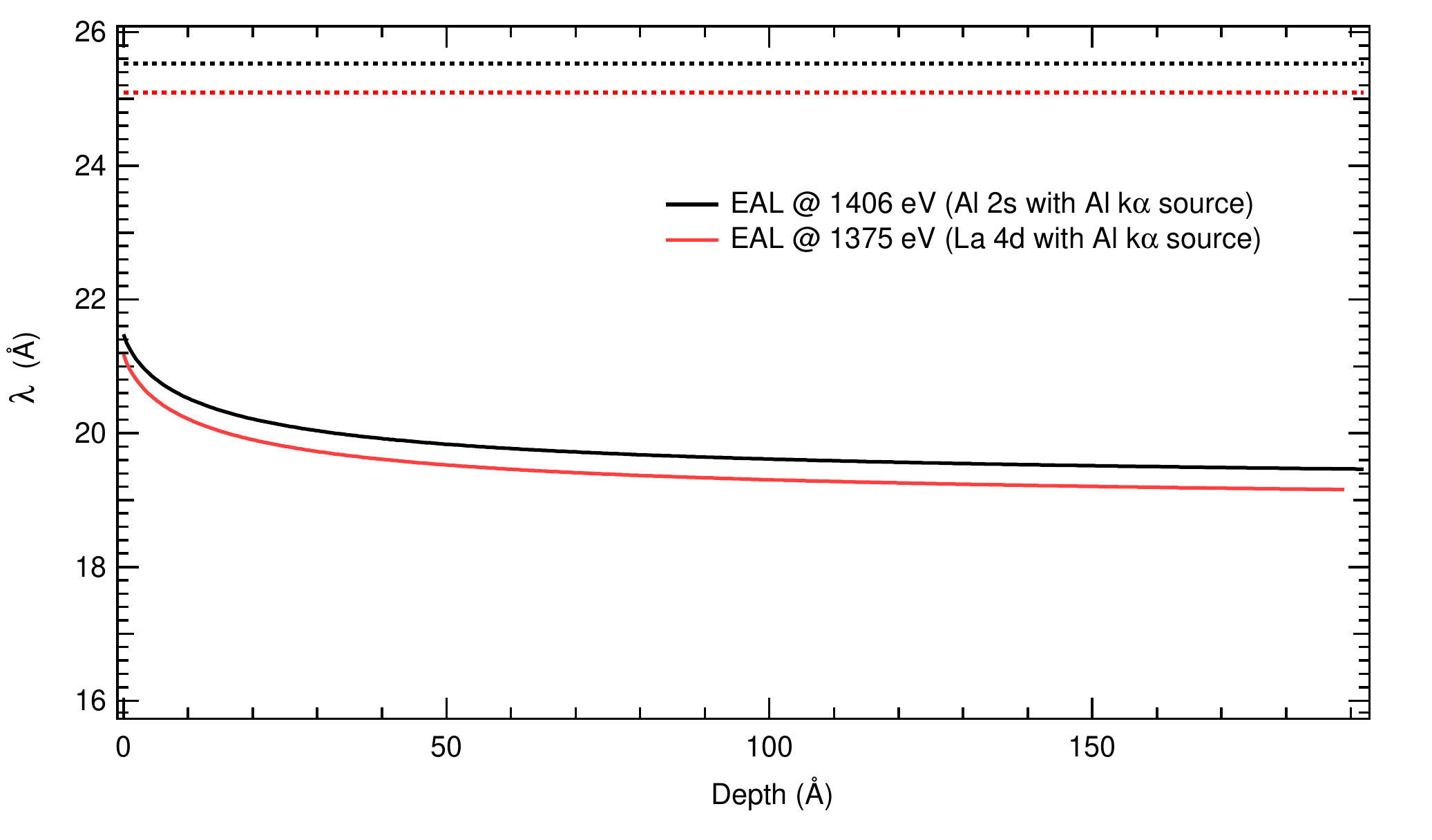}
\caption{Effective attenuation length\cite{EAL_theory,NIST} vs depth (i.e. the distance from the surface) calculated for La 4d and Al 2s core levels photoelectrons on LaAlO$_3$; the dotted lines indicate the corresponding inelastic mean free path, calculated with TPP-2M\cite{TPP2M} formula. The calculations have been done for an Al k$_\alpha$ X-ray source.\label{fig_EAL_LAO}}
\end{center}
\end{figure}

As an example, in Fig. \ref{fig_EAL_LAO} the graph of EAL vs depth for electrons photoemitted from La 4d and Al 2s core levels in LaAlO$_3$ is shown. EAL is usually lower than the corresponding IMFP and rapidly decreases in a short distance from the surface. As it is shown in Chapter 5, the introduction of the EAL is thus fundamental to obtain the correct peak intensity or, on the contrary, the thickness of a very thin overlayer.

\section{XAS}
In a XAS experiment the X-ray total absorption cross-section of the specimen is measured. Therefore, XAS requires the possibility to scan the photon energy and thus it is usually performed on synchrotron facilities, even if similar spectra can be obtained through the electron energy loss spectroscopy (EELS) which formally share the same theoretical formalism. While XAS is the most generic name, many different notation are given according to the required experimental information or to the specific application fields; here is a list of the different XAS definitions:

\begin{itemize}
	\item NEXAFS (Near-Edge X-ray Absorption Fine Structure) or XANES (X-ray Absorption Near-Edge Structure): in both cases the absorption cross section is measured ``near'' the absorption edges of a specific elements. Although from an experimental point of view are both similar, in general NEXAFS is used in surface or molecular studies and XANES in crystal or bulk studies. The NEXAFS experimental information covers the empty states, the adsorbate geometry (NEXAFS) and the symmetry of a specific atomic species (XANES).
	\item EXAFS (Extended X-ray Absorption Fine Structure): in this case, the absorption cross section is collected over a wider energy range. In ordered structures, these spectra show oscillations due to multiple scattering of excited electrons: by analyzing the Fourier transformation of the EXAFS spectra it is possible to evaluate the nearest-neighbors bonding lengths.
	\item SEXAFS (Surface Extended X-ray Absorption Fine Structure): same as EXAFS, but tuned to give more information on the surface bond lengths.
	\item XMCD and XLD (X-ray Magnetic Circular Dichroism and X-ray Linear Dichroism): XAS spectra taken with different X-ray polarization. XMCD requires a magnetic field (both of specimen or external) and gives direct information on the local magnetism; XLD can give further information on the symmetry distortion of the ionic environment. These techniques can be done also in an imaging-like fashion, both scanning the sample position or using electronic lenses, giving the possibility to map the specimen different magnetic or structural domains.
\end{itemize}

In this Thesis XANES, XMCD and XLD measurement are reported. Further details about experimental set-up and theory will be given in the next paragraphs.

\subsection{Experimental set-up}
\begin{figure}
\begin{center}
\includegraphics[width=0.9\textwidth]{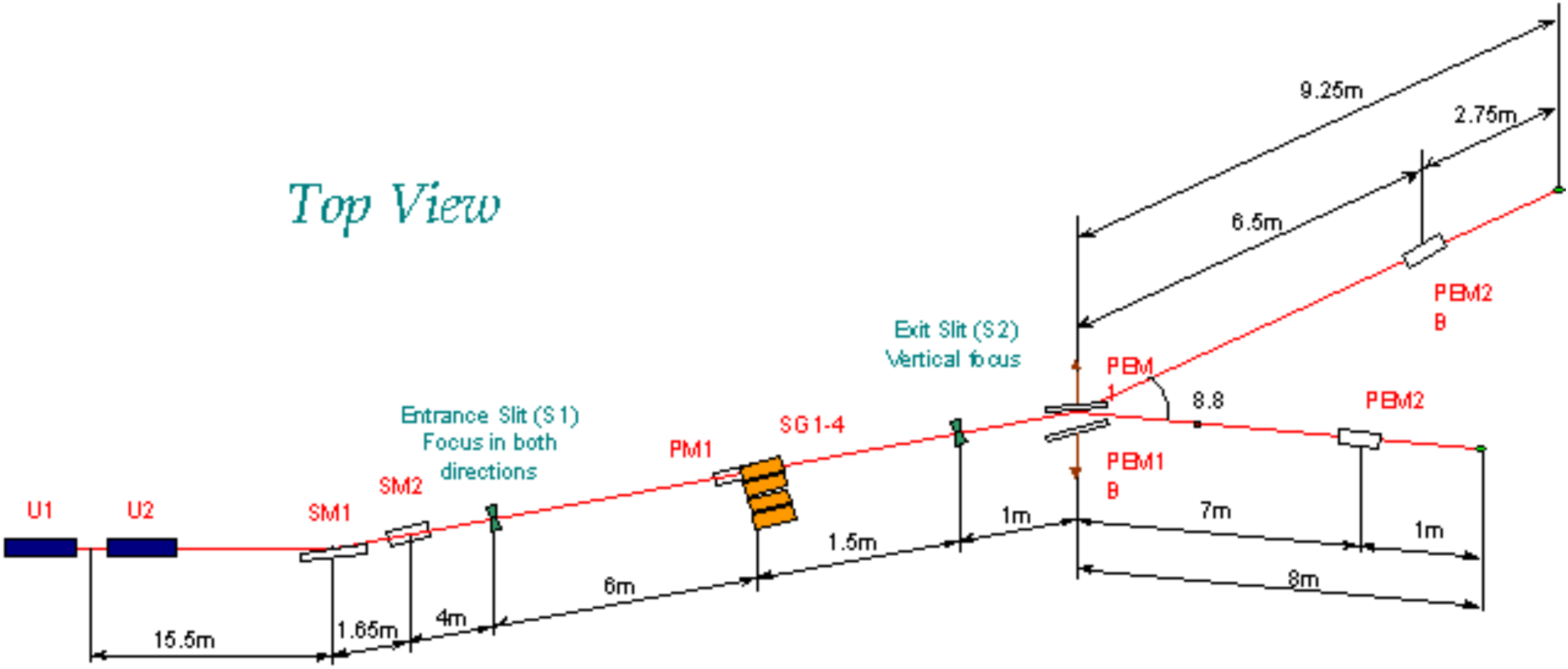}
\caption{Top view of BACH beamline layout at ELETTRA synchrotron. From left to right are depicted the undulators (U$_1$ and U$_2$) on the storage rings, then the mirrors, the slits, the monochromator and the user end-stations.\label{fig_Bach_Layout}}
\end{center}
\end{figure}

Most of XAS data shown on this thesis have been taken on BACH beamline at ELETTRA synchrotron in Trieste (Italy). The reference spectra taken on Mn:CdTe and Mn:Ge have been measured at the ALOISA beamlines, again at ELETTRA. A synchrotron X-ray source combines an high brilliance, a small on-target X-ray focus as well as the photon-energy and polarization tunability. In Fig. \ref{fig_Bach_Layout} a schematic view of the BACH (Beamline for Advanced diCHroism) beamline set-upis given. The photon energy is tuned by changing the distance (usually referred as ``gap'') of undulators magnets and by means of a two mirror monochromator. The resolving power ($\Delta E/E$) at BACH beamline is 20000-6000, 20000-6000, and 15000-5000 in the energy ranges 40-200 eV, 200-500 eV, and 500-1600 eV, respectively. As in XPS, XAS requires UHV conditions and, because of synchrotron high photon-flux, can suffer of charging effects. The X-ray polarization can be tuned with dedicated insertion devices (helical undulators) or by exploiting the natural polarization characteristic of synchrotron light.

The absorption cross section could be in principle detected by measuring the photon flux before and after the sample; in practice, this method can be applied only to very thin (in the range on $\mu m$) samples. Usually XAS is performed by measuring secondary de-excitation process caused by the absorption of X-rays. Here is a summary of detection technique:

\begin{itemize}
	\item Fluorescence: XAS can be measured by detecting the rate of fluorescence given by the recombination of the electrons with the core-hole created by the excitation. This detection technique requires the presence of a silicon-based photon detector in the measurement chamber and its sensitivity is related to the X-rays penetration depth in the sample (in the \textnormal{$\mu m$} range). Fluorescence detection is less effective in light materials, since the Auger decay is the most probable de-excitation process, and can be quenched in very dense material, because of the re-absorption of the emitted photon. It doesn't suffer of charging effects, even in insulating samples.

	\item Total yeld: XAS can be detected by measuring the electrical current (``drain current'') generated by the X-ray absorption. A picoammeter is needed, since this current ranges typically on the 10$^{-10}$- 10$^{-7}$ A scale. The drain current is generated by a cascade of Auger process that are also related to the electrons inelastic scattering. Since only the electrons that reach the surface contribute to this current, the probing depth is lower than in fluorescence detection. The typical total yield MPD (maximum probing depth) is in the 4-10 nm range, the Ti L-edge being at the lower MPD side\cite{drain_current}.

	\item Partial yeld: the photoemission intensity is usually proportional to the absorption cross section, thus XAS can be measured by integrating the photoelectron emission in a defined energy range, through a channeltron detector or an electron analyzer. This detection technique is the most surface-sensitive and, like XPS, can be affected by charging effects in insulating samples.
\end{itemize}

Spectra should be normalized with the incoming photon flux, which is usually measured on the monochromator last mirror through the drain current method.

\subsection{XAS theory}
\begin{figure}
\begin{center}
\includegraphics[width=0.9\textwidth]{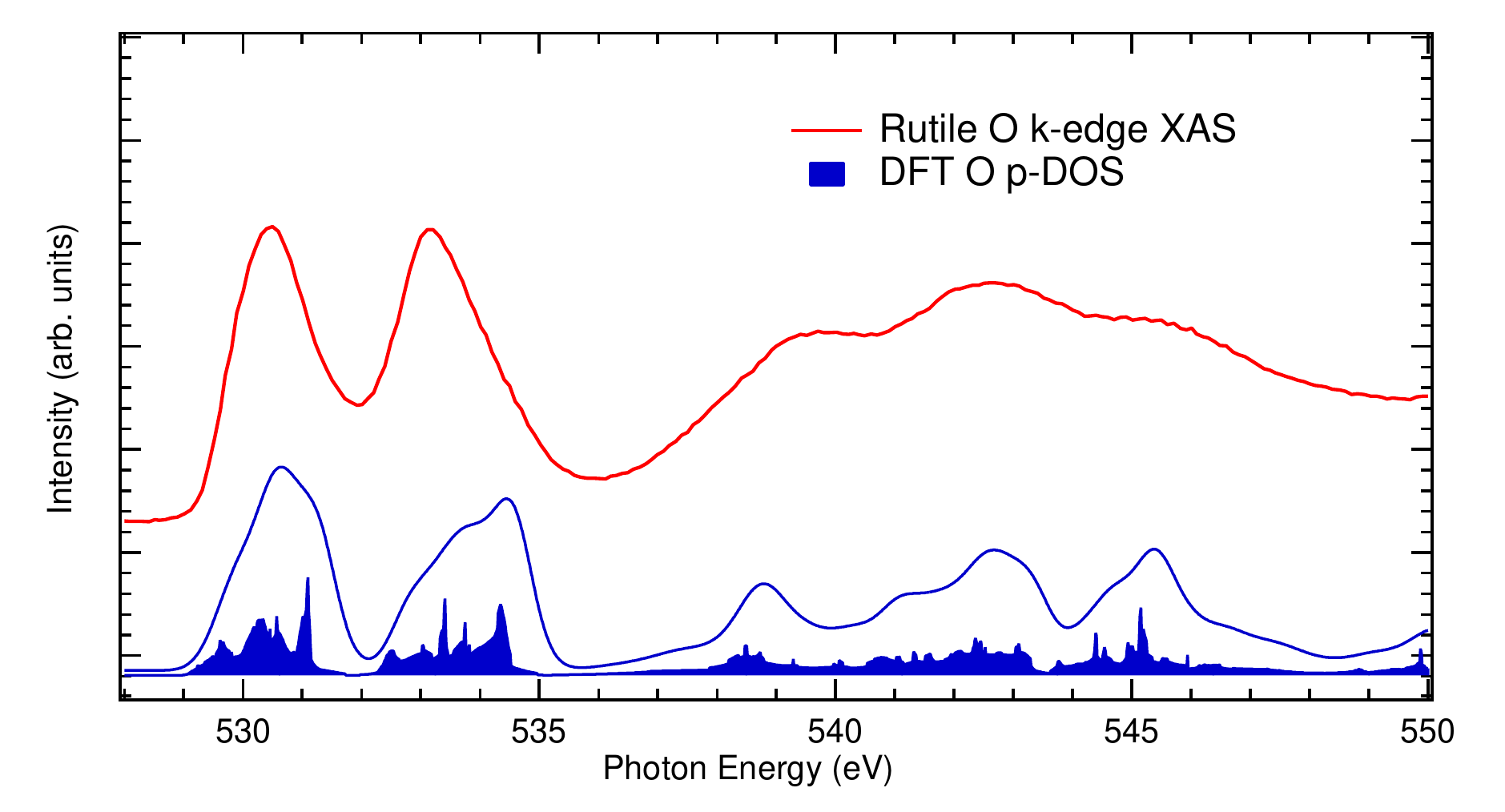}
\caption {Rutile TiO$_2$ K-edge XAS spectrum, as compared to ab-initio DFT calculations. In blue, the calculated p-projected density of states is shown. Taken from Ref.\cite{APL_ntio2}.\label{fig_XAS_rutile}}
\end{center}
\end{figure}

A schematic view of XAS is shown in Fig. \ref{fig_schema_xpsxas}(a). In short, when the photon energy is higher than a core-level binding energy, an electron from that core level could be excited into an empty states below the Fermi edge. The transition rate for this process can be described with Fermi Golden rule, similarly to Eq. \ref{eq_exp4}:

\begin{equation}
W_{ph}=\frac{2\pi}{\hbar}\left|\left\langle i|T|f \right\rangle\right|^{2}\rho(E_f-E_i-h\nu)
\label{eq_FGR_XAS}
\end{equation}
\\
where $\rho$ is the empty level density of states (DOS). As in XPS, operator $T$ can be taken as the usual dipole operator $e\cdot\vec{r}$. In the case of most core-level edges, except in transition metals and rare earths, with photon energy between 100-2000 eV the dipole operator is slowly varying with $h\nu$; in this situation, the XAS spectra becomes a direct measurement of the empty DOS, plus a contribution from electrons multiple scattering\cite{XAS_feff}. A proof of this can be seen in Fig. \ref{fig_XAS_rutile}, where an oxygen K-edge XAS spectrum on TiO$_2$ is compared to the DOS results of a DFT ab-initio calculation. The dipole operator is still important, since it gives the selection rules and the polarization dependence.

In the case of open-shell system, such as transition-metals (TM) or rare-earths (RE), the electrons excited in the empty states can strongly interact together and with the core-hole. In such system, the matrix element of Eq. \ref{eq_FGR_XAS} becomes the leading term and multiplet features appears in the spectrum. Since many experimental data of this Thesis have been taken at the titanium 2p threshold, the multiplet effects are explained in more details in the next section.

\section{Multiplet effects on core-levels spectroscopies}
\begin{figure}
\begin{center}
\includegraphics[width=0.9\textwidth]{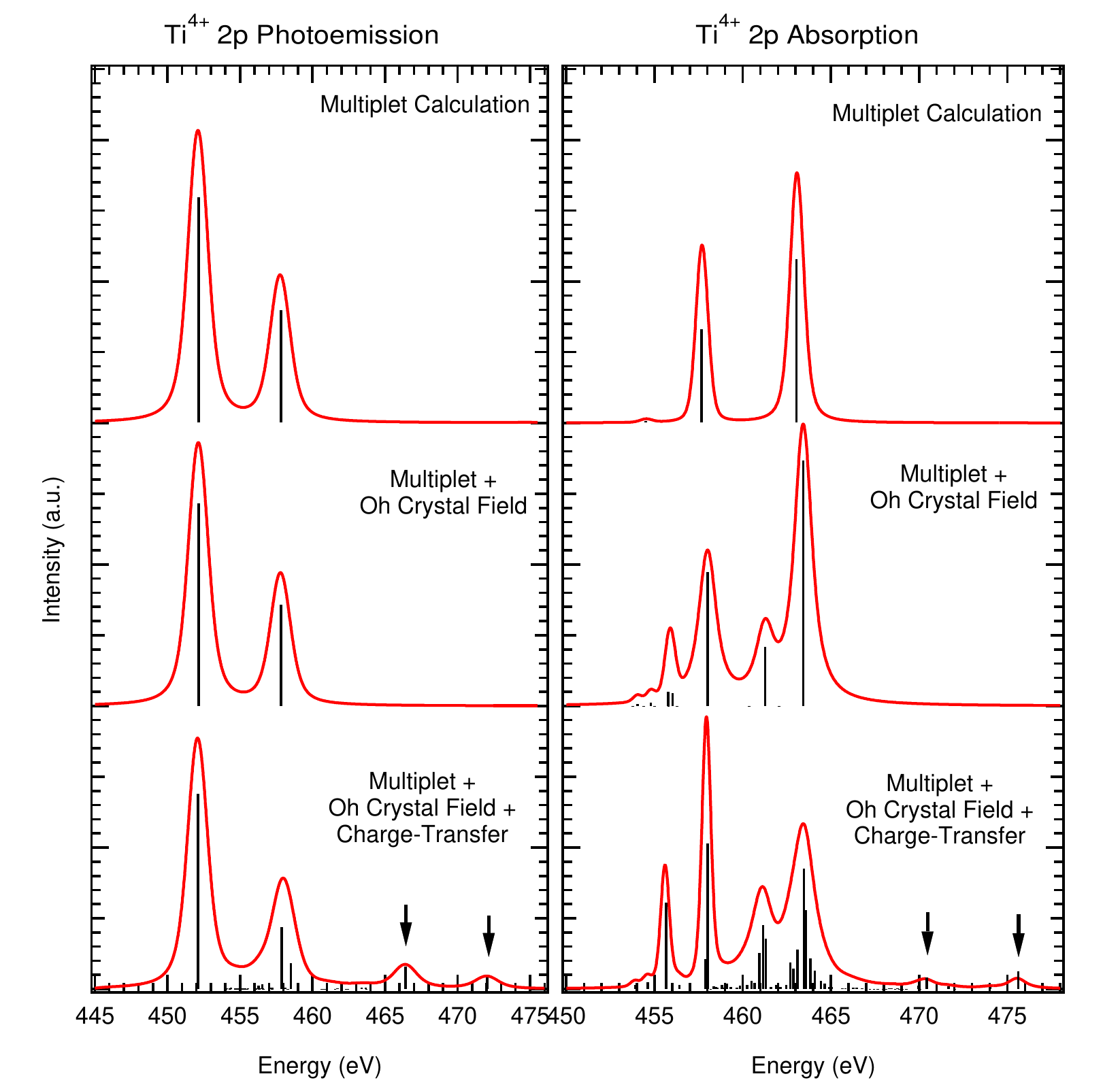}
\caption {Step in charge-transfer multiplet calculation of 2p XPS (left) and XAS (right) spectra for a Ti$^{4+}$ ion. The upper graphs are the atomic multiplet calculation, the middle ones are the results after the introduction of an octahedral crystal field and the lower ones are results with the charge-transfer (CT). Black arrows indicate the CT satellites. CT parameters are 10Dq=1.7 eV, $\Delta$=3.0 eV, Q=5.0 eV, U=9 eV and the mixing parameters are 3.0 and 1.0 eV for T$_{2g}$ and E$_{g}$ symmetries, respectively.\label{Image_multiplet}}
\end{center}
\end{figure}

There are certain cases in which the XPS and XAS matrix elements of Eqs. \ref{eq_exp4} and \ref{eq_FGR_XAS} have a complex photon energy dependence; in fact, in the case of TM and RE multiple final states are allowed, because of the electron-electron interactions in the empty d shell (for TM) or f shell (for RE). In the TM cases the 2p, 3p and even the 3s level XPS and XAS spectra are thus a superposition of many electronic transition between the core-hole level and the 3d shell (see Fig. \ref{Image_multiplet}).

Simple ground-state DFT calculations cannot account for these effects because they are deeply related to the excited states; on the contrary, Hartree-Fock (HF) calculations can be done also for excited states, even if the electronic correlation is only partly described. The latter approach (HF), mixed with a proper electronic angular momenta algebra\cite{COWAN}, has proved to be an effective tool in predicting core-level spectroscopic data, under the name of \textit{atomic multiplet} theory. This model can be expanded with subsequent steps in order to better match the experimental results on crystals.

The starting point is an HF calculation for a single ion, carried out with the \texttt{COWAN}\cite{COWAN} code, which gives the energy levels and the transition matrix from the ground to the final states. The results is an atomic multiplet spectrum that describes the XAS (or XPS) process in a spherical symmetry (SO$_3$). The eigenvalues and the matrix elements are then modified according to crystal field (CF) theory and an finally to an Hubbard model to mimic the ionic environment. Atomic multiplet plus CF calculations have been carried out in this Thesis with the \texttt{MISSING}\cite{MISSING} package and the full calculations, which also include the ligand-metal charge-transfer (CT), have been performed with the CTM4XAS\cite{DEGROOT} code. The complete process is shown for a 3d$^0$ atom in Fig. \ref{Image_multiplet}.
To account for both inter-atomic charge-transfer and intra-atomic multiplet splitting effects, a configuration interaction (CI) description of the wavefunctions is usually adopted. Several configurations, denoted as $d\,^{n}$, $d\,^{n+1}\, \underline{L}$, $d\,^{n+2}\, \underline{L^2}$ and so on  ($\underline{L}$ denotes a ligand hole) are used to describe the open shell of the 3d transition metal ion during the photoemission process. Accordingly, the initial state wavefunction is written as:

\begin{equation}
\Psi_{g.s.}=\alpha_1|3d^{n}\rangle+\alpha_2|3d^{n+1}\underline{L}^1\rangle+\alpha_3|3d^{n+2}\underline{L}^2\rangle
\label{eq_exp11}
\end{equation}
\\
where $\underline{L}$ denotes a configuration with a p-hole in the anion states. In the present case, the p-hole represents the CT from the O 2p levels to TM 3d levels. The CT energy is defined as $\Delta=E(d\,^{n+1}\underline{L}\,)-E(d\,^{n})$, whereas the Coulomb d-d interaction is represented by $U=E(d\,^{n-1})+E(d\,^{n+1})-2E(d\,^{n})$, where $E(d\,^{n}\underline{L}\,^{m})$ is the center of mass of the $d\,^{n}\underline{L}\,^{m}$ multiplet. The Coulomb and CT energies are usually regarded as model parameters. Interaction between $d\,^{n+m}\underline{L}\,^{m}$ and $d\,^{n+m+1}\underline{L}\,^{m+1}$ configurations is accounted for by the T$_{pd}$ off-diagonal term in the Hamiltonian matrix.

In the case of core level photoemission the final state interaction between the core hole and the 3d electrons in the outer shell is explicitly accounted for by an energy parameter Q.
Therefore the CT term is corrected as $\Delta-Q$ upon creation of the core hole. The final state wavefunction is given by:

\begin{equation}
\Psi_{f.s}=\beta_1|2\underline{p}3d^n\rangle + \beta_2|2\underline{p}
3d^{n+1}\underline{L}^1\rangle+\beta_3|2\underline{p}3d^{n+2}\underline{L}^2\rangle + e^{-}
\label{eq_exp12}
\end{equation}
\\

where $\underline{c}$ represents the core hole. The spectral weight in a photoemission experiment is calculated, in the sudden approximation, by projecting the final state configurations on the
the ground state, i.e.

\begin{equation}
I_{XPS}(BE)\propto\sum_{i}\left|\langle\Psi_{GS}|\Psi_{i,fs}\rangle
\right|^{2}\delta(BE-\varepsilon_{i})
\label{eq_sudden_CT1}
\end{equation}\\
where
\begin{equation}
\left|\langle\Psi_{GS}|\Psi_{i,fs}\rangle \right|^{2}= \left|
\alpha_{1}\beta_{1,i} + \alpha_{2}\beta_{2,i} + \ldots \right|^{2}
\label{eq_sudden_CT2}
\end{equation}\\

and the sum is run over all final state configurations $|\Psi_{i,fs}\rangle$ with energy $\varepsilon_{i}$. Each of the d$^n$ configurations can be further described with the crystal field theory, with the multiplet calculations or with both approaches. The CF is introduced by reducing the symmetry from the spherical SO$_3$ group to the desired symmetry; in this thesis, an octahedral (O$_h$) and tetragonal (D$_{4h}$) space group are used. The adequate level of the theory approximations is fixed by the experimental technique. For example, in XAS a multiplet-CF approach is usually sufficient to describe most of the experimental features, while in XPS the CT effects cannot be neglected.

An example of these calculations is given in Fig. \ref{Image_multiplet} for Ti$^{4+}$ (3d$^0$) 2p XAS and XPS in an octahedral symmetry (O$_h$). Without multiplet, only two peaks separated by the spin-orbit interaction are predicted. The interactions in the 3d levels (and with the 2p core-hole) induce a complex multiplet structure which is further split by the introduction of the CF. Eventually, the addition of charge-transfer configurations adds small (strong) satellites to XPS (XAS) spectrum.
As can be noted in Fig. \ref{Image_multiplet}, the shape of the Ti L$_{2,3}$ edge XAS spectra is mostly determined by CF effect, which are less important in XPS; on the contrary, CT satellites are best observed in XPS. The small CT satellites in XAS are due to the conservation of the total charge in the absorption process.

\begin{figure}[ht]
\begin{center}
\includegraphics[width=0.8\textwidth]{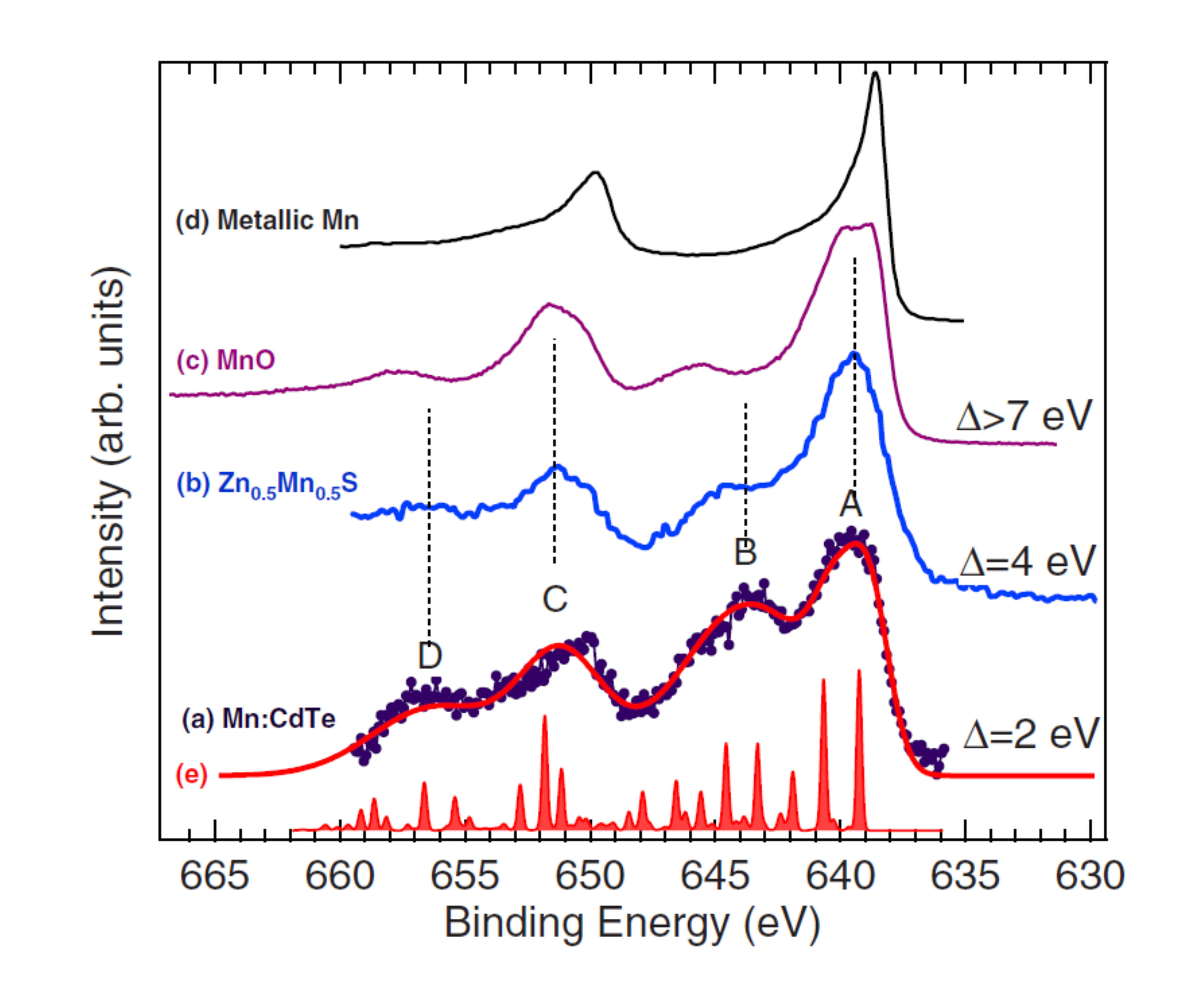}
\caption {Comparison of Mn 2p XPS spectra from Mn:CdTe (100), Zn$_{1-x}$Mn$_{x}$Te, MnO and a reference metallic Mn thick film deposited on a silicon wafer. Calculated Mn 2p XPS spectrum for Mn$^{2+}$ ion with charge-transfer derived configurations ($\Delta$=2.1 eV, T=2.2 eV, Q=6.0 eV, and U=5.1 eV). Taken from Ref.\cite{PRB81MnCdTe}.\label{graph_CdMnTe}}
\end{center}
\end{figure}

The relative intensities of the XPS satellites with respect to main peaks are related to the type of  chemical bond. In more ionic compounds (such as MnO) weak satellites are observed, therefore the spectra can be often described with multiplet calculations only; in samples with covalent bonds, strong XPS satellites are present and thus in order to evaluate the experimental spectra the charge-transfer model is required. As an example, in Fig. \ref{graph_CdMnTe} multiplet CT calculations (without CF) and experimental data\cite{PRB81MnCdTe} for manganese ions in different host matrices are shown. The shift from an atomic-like spectrum (small satellites, in MnO) to a covalent bond (large satellites, Mn:CdTe) is clearly visible. As shown in Fig. \ref{graph_CdMnTe}, the charge-transfer parameter ($\Delta$) determines the strength of the satellites peaks. Insulating samples where $\Delta<U$ are referred as charge-transfer insulators\cite{DEGROOT}; in that case, which includes TiO$_2$, $\Delta$ is close to the band-gap (see the $\Delta$ parameter in Fig. \ref{Image_multiplet}). In the case of a metallic material the XPS peaks assume a typical asymmetric shape (Doniach-Sunjic\cite{doniach}), which cannot be described with atomic calculations.

The ligand-field charge-transfer multiplet calculations provide an effective tool for calculating XAS and XPS spectra in TM; however, this theory requires numerous empirical parameters: a reduction factor for the Slater integrals (obtained by HF), the crystal field parameters related to the crystal point group (i.e. 10Dq, ds,dt and so on) and the U, $\Delta$, Q an T parameters set. Even if most of these parameters can be obtained from various experimental techniques, a complete ab-initio approach would be highly desirable.

\section{Resonant Photoemission}
In a ResPES experiment, the valence band photoemission spectra are collected by changing the photon energy across an absorption edge. In a single-particle approach (i.e. multiplet calculations and following approximations) the direct valence band photoemission channel ($c^{m}v^{n}\rightarrow c^{m} v^{n-1}+e^{-}$) can interfere with the autoionization channel caused by the presence of a core-hole ($c^{m-1} v^{n+1}\rightarrow c^{m} 3d^{n-1} + e^{-}$). As a results of this, the part of the valence band related to the specific atomic species can be greatly enhanced (or suppressed). This effect is known to occur for most of the elements, both organic and inorganic.

\begin{figure}
\begin{center}
\includegraphics[width=1\textwidth]{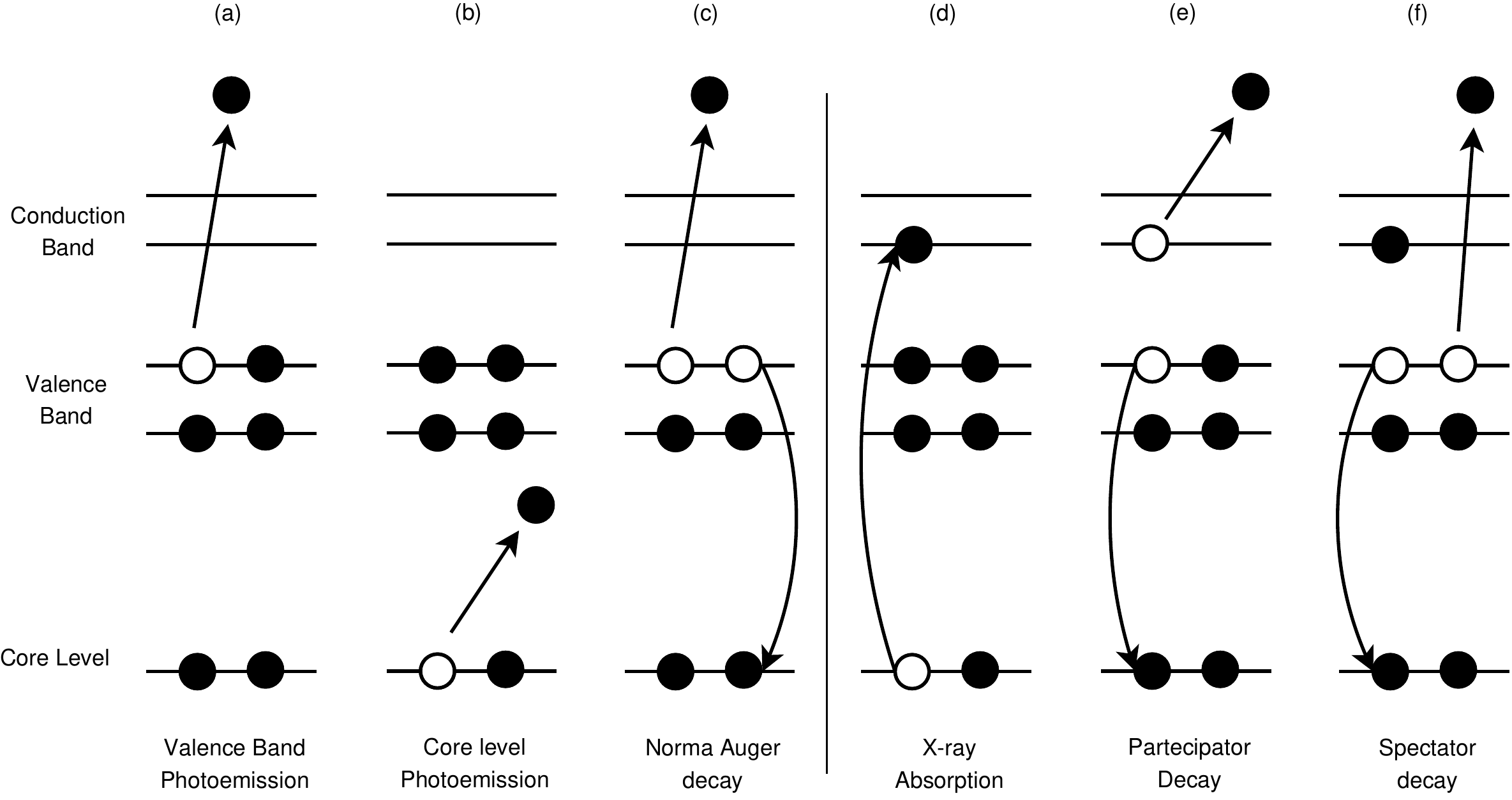}
\caption{Schematic of the possible excitation and de-excitation channels in a ResPES experiment.\label{graph_Respes_channel}}
\end{center}
\end{figure}

Following the description of Br\"{u}hwiler et al.\cite{Bruwiler}, a pictorial representation of the process is given in Fig. \ref{graph_Respes_channel}. In this figure the most important excitation-deexcitation processes related to the ResPES technique are shown.

On the left part the photoemission-related channels are shown: Fig. \ref{graph_Respes_channel}(a) shows the normal valence band photoemission (VPES) and Fig. \ref{graph_Respes_channel}(b) the core-level photoemission. In the latter case, the system can fill the core-hole through a normal Auger decay (Fig. \ref{graph_Respes_channel}(d)) or through fluorescence, which is not usually measured during a conventional ResPES experiment. In Fig. \ref{graph_Respes_channel}(d) the resonant transition from the core level to the empty state is depicted (i.e.,the X-ray absorption). In this case the system can relax by filling the core-hole with a valence band electron, opening two different autoionization channels: Fig. \ref{graph_Respes_channel}(e) shows the emission of the excited electron from the empty states, labeled as \emph{partecipant decay} and Fig. \ref{graph_Respes_channel}(f) shows the emission of another electron from the valence band, labeled \emph{spectator decay}. The final state of the participant decay is equivalent to the valence band photoemission one, leading to the (constructive or destructive) interference effect that is called Resonant PES (RPES or ResPES).

\begin{figure}
\begin{center}
\includegraphics[width=0.5\textwidth]{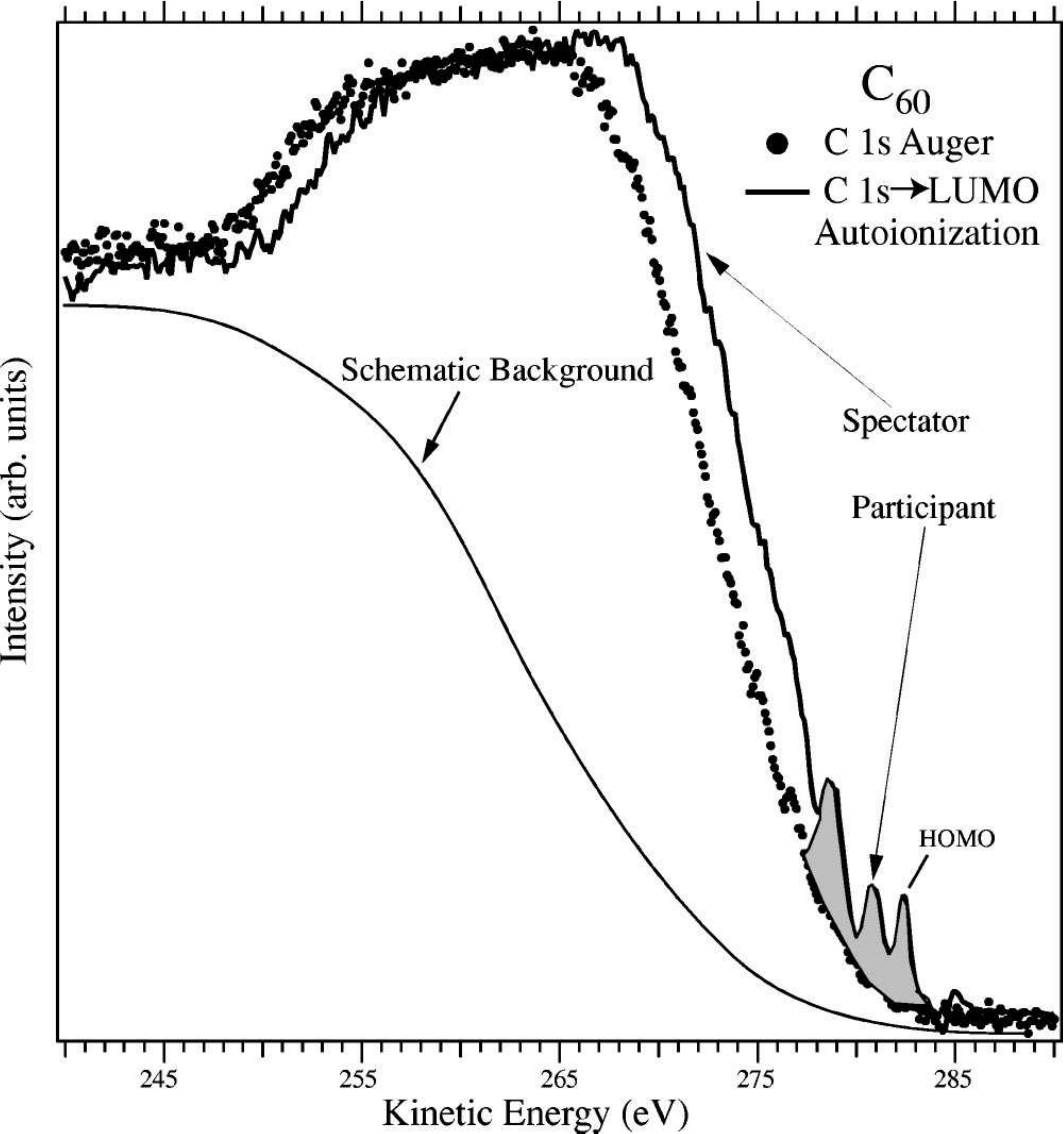}
\caption {The Auger (dots) and resonant valence band photoemission spectra (thick line) on fullerene C$_60$. The energy difference between Auger and spectator decay, called \emph{spectator shift} is clearly visible (adapted from Ref.\cite{Bruwiler}).\label{Spec_part}}
\end{center}
\end{figure}

The spectral weight related to the spectator decay and to the normal Auger channels are usually similar in shape, with a shift in energy due to the lower screening effects induced by the extra electron in the conduction band (see Fig. \ref{Spec_part}).

\begin{figure}
\begin{center}
\includegraphics[width=0.6\textwidth]{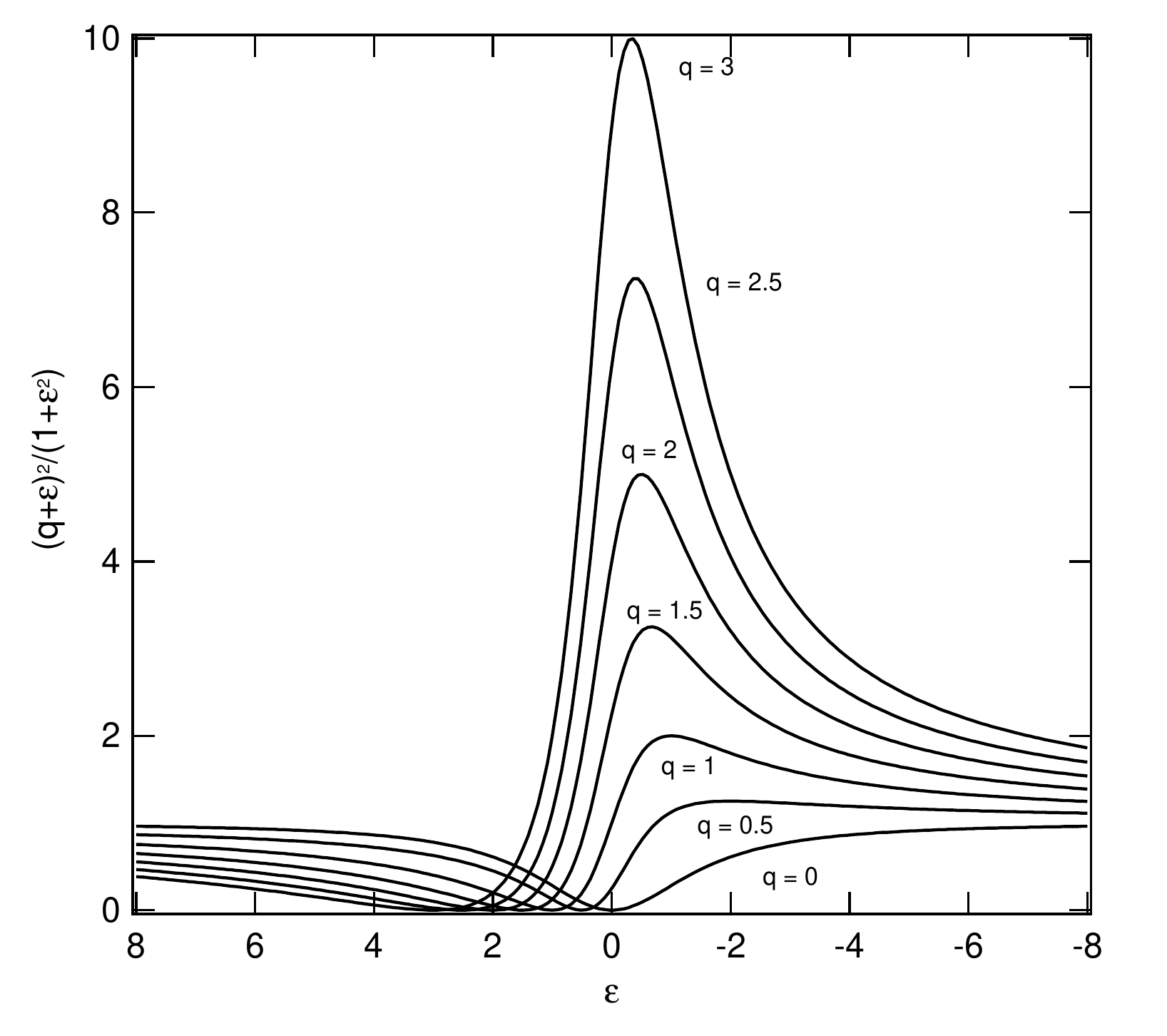}
\caption {Effect of Fano q-parameter on ResPES intensity.\label{fano_fig}}
\end{center}
\end{figure}

The ResPES technique can give information also on the excited electron dynamics, because the intensity of the interference effects between VPES and partecipator channels is related to the excited electron lifetime.
In fact, in an highly delocalized (metallic) band the extra electron can be quickly removed, quenching the ResPES channel to the normal Auger. ResPES is thus a useful experimental tool to evaluate the charge-transfer process, especially between a substrate and deposited molecules or atoms. When applied for this purpose, the ResPES technique is usually referred to as Core-Hole-Clock (CHC) spectroscopy. The interference intensity can be calculated in the Fermi golden rule approach as follows\cite{Martennson}:

\begin{equation}
\label{eq:rpes} \omega=2\pi\sum\limits_{f}\left| \langle
f|V_r|g\rangle + \sum\limits_{m}\frac{\langle f|V_A|m\rangle \langle
m|V_r|g\rangle}{E_g-E_m-i \Gamma_{m}/2} \right|^2 \delta(E_f-E_g)
\end{equation}
\\

where the left part of the matrix element is the transition rate for normal photoemission from ground (g) to final state (f), the right part is the summation over the possible intermediate state (m) and $\Gamma_m$ is the lifetime of excited state. V$_r$ and V$_A$ are the radiative (dipole) and the Coulomb (Auger) operators. Under simple assumptions (only one core-hole excitation and a continuum of empty states) this formula can be simplified in the Fano formalism\cite{Fano}, obtaining:

\begin{equation}
\label{eq:Fano} \omega=\sum\limits_{m}\left| \langle f|V_r|g\rangle
\right|^2
\frac{(q+\varepsilon)^2}{1+\varepsilon^2}\frac{E_m/\pi}{(E_f+E_g-E_m)^2+\Gamma_m}
\end{equation}
\\
where the ResPES intensity is the normal VPES one multiplied by the factor $(q+\varepsilon)^2/(1+\varepsilon^2)$ with $\varepsilon=(E-E_m)/\Gamma$. The parameter \emph{q} is called \emph{Fano factor} and describes an antiresonant effect for $q=0$ and an high resonant effect for high $q>2$, as can be seen in Fig. \ref{fano_fig}. In the case of transition metals (TM) and rare earths (RE) the q factor can be rather high ($q>3$), leading to a pronounced resonance effect called ``giant resonance''.

\begin{SCfigure}
\label{graph_CdMnTe_RPES}
\includegraphics[width=0.5\textwidth]{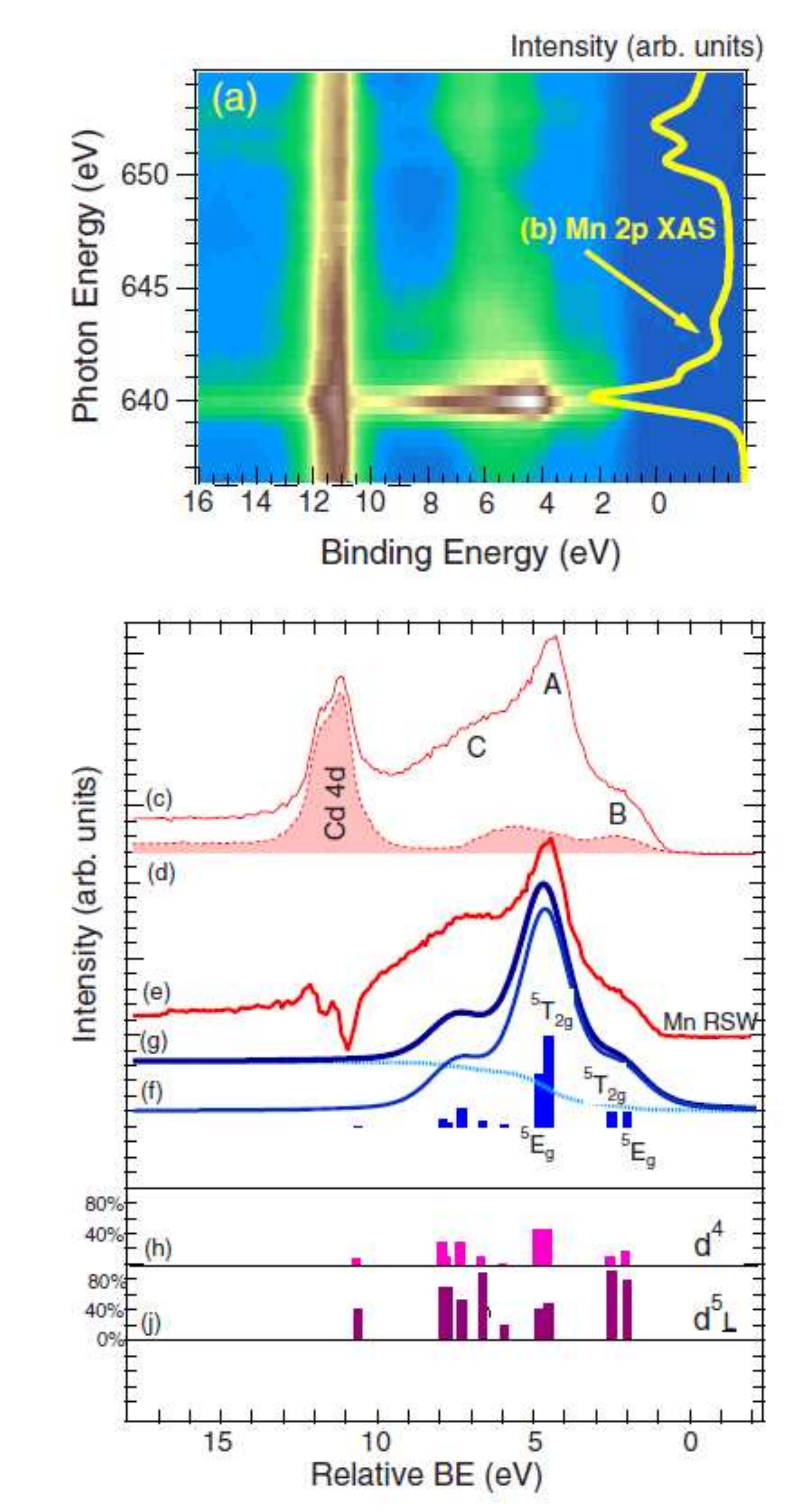}
\caption {Resonant photoemission data of Mn:CdTe taken at Mn L$_{2,3}$ edge. (a) Two dimensional plot of ResPES data for photon energies ranging from 636 to 655 eV; (b) XAS spectra at Mn L$_{2,3}$; (c) and (d) are the on-resonance and off resonance valence band spectra, and their difference (e) is fitted with multiplet calculation (f) composed by d$^4$ and d$^5$\underline{L} configurations. Adapted from Ref.\cite{PRB81MnCdTe}}
\end{SCfigure}

An example of ResPES, in the case of Mn over a CdTe (110) surface\cite{PRB81MnCdTe}, is given in Fig. \ref{graph_CdMnTe_RPES}. A clear resonance effect is seen in Fig. \ref{graph_CdMnTe_RPES}: the on-resonance spectra is different in shape and overall intensity from the off-resonance spectra. The difference between the two (spectrum (e) in Fig. \ref{graph_CdMnTe_RPES}) is related to the partial DOS of manganese 3d electrons in the valence band that, in this case, can be calculated with a multiplet approach. At least, two final state configurations have to be considered in order to reproduce the experimental data: the expected 3d$^4$, that results from a photoemission process $3d^5 \rightarrow 3d^4 + e^-$ resonating with the autoionization process $2p^5 3d^6 \rightarrow 3d^4 + e^-$, and a charge-transfer 3d$^5$\underline{L} configuration due to the Mn-Te hybridization. In order to evaluate the non-resonant valence band contribution a more detailed method should be used, such as ab-initio DFT.

\begin{figure}
\begin{center}
\includegraphics[width=0.5\textwidth]{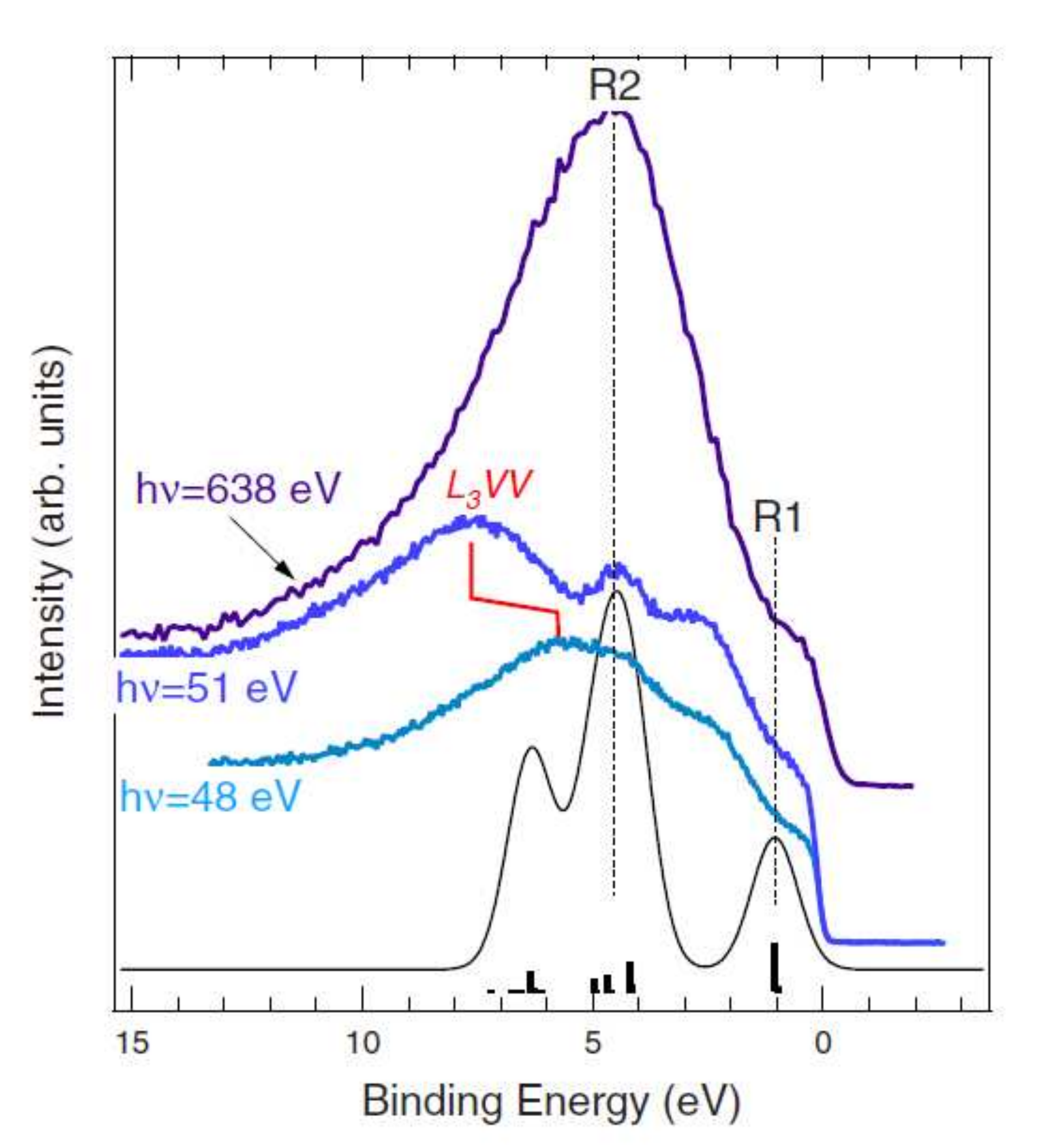}
\caption {Valence band spectra on Mn$_5$Ge$_3$ taken at Mn L$_3$ edge, as well as with h$\nu$=48 and 51 eV (i.e., across the Mn 3p-3d resonance), and calculated photoemission spectrum for the Mn$^+$ 3d$^6$ ion in the atomic-multiplet approach. Adapted from Ref.\cite{PRB81MnGe}.\label{mn5ge3}}
\end{center}
\end{figure}

Another interesting case where ``valence band'' (i.e. 3d electrons) multiplet calculations can reproduce the average energy position of the resonating features in the VB is the Mn$_5$Ge$_3$ alloy. In Fig. \ref{mn5ge3} are reported the resonant spectra\cite{PRB81MnGe} taken at Mn L$_{2,3}$-edge, with theoretical calculation; the metallic character of this material can be simulated by considering, instead of a Mn$^{2+}$ ($3d^5$) ion, a Mn$^{1+}$ ($3d^6$) ion without the spin-orbit interaction on the d-level. The same approximation has proved to be effective also in calculating core-level photoemission and absorption.

Since many experimental results of this Thesis work have been achieved through the ResPES technique at Ti L-edge, the Chapter 2 is dedicated entirely to a detailed description of the ResPES process in the TiO$_2$ d$^0$ system.

\section{Ab-initio electronic structure calculations}
As already pointed out, ``simple'' multiplet calculations can deal very well with transition matrix to excited states in open-shell system, but fail in reproducing completely the valence band spectra and the DOS in general. For this reason, ab-initio\cite{DFT_martins} calculations have to be resorted, with a complete description of the wavefunctions that participate to the chemical bond. The problem addressed in ab-initio calculations can be described with the Hamiltonian for N interacting electrons (in \textbf{r} position) in a lattice of M ions (in \textbf{R} position):

\begin{equation}
\label{eq_Ham}
\hat{H}=-\frac{\hbar}{2m}\sum^N_i \nabla^2_i - \sum^{N,M}_{i,j}\frac{Z_je}{\left|\boldsymbol{r_{i}}-\boldsymbol{R_j}\right|}+\sum^{N}_{i<j}\frac{e^2}{\left|\boldsymbol{r_i}-\boldsymbol{r_j}\right|}
\end{equation}
\\
\begin{equation}
\label{eq_Shro}
\hat{H}\Psi=E\Psi
\end{equation}
\\
This is already an approximation (Born-Oppenheimer), since it lacks of the kinetic term for ions and the ion-ion repulsion (the latter can be taken into account with empirical formulas). However, even with these approximations, the solution of this Hamiltonian is a formidable task that requires intensive numerical calculations. Many advances in the field of quantum chemistry, such as DFT, the LDA approximation, pseudopotentials and so on, allow now to perform complex (and realistic) ab-initio calculation even on a single desktop/workstation machine. In this Thesis, various ab-initio calculations results on simple and doped crystals are shown, mostly based on the Density Functional Theory (DFT). A brief description of this method is given in this section, starting from the Hartree-Fock equations.

\subsection{Hartree-Fock method}
The first successful attempt to solve the Eqs. \ref{eq_Shro} and \ref{eq_Ham} was the Hartree-Fock (HF) method\cite{hfbook}, in which the total wave function $\Psi$ is expressed as a Slater determinant of single particle functions:

\begin{equation}
\label{eq_slater}
\Psi=Det\left|\psi_i(\boldsymbol{r_j})\right|
\end{equation}
\\
This formulation gives a correct antisymmetric description of the total wave function. It is possible then to define the total energy of the system and to minimize it through Lagrange multipliers in order to obtain the correct single-particle wavefunctions $\psi(\boldsymbol{r})$ and the corresponding energies. By knowing the total energy of the system, it is possible to find the lattice (or molecular) equilibrium position, the binding energies and even the phonon frequencies; by knowing the single-particle wavefunctions, it is possible to compute also the band structure, the density of states and evaluate the stoichiometry of each atom.

This method, at least in the initial formulations, suffers from some drawbacks: the exchange interaction energy has to be computed with a two particle integral (since the exchange interaction is non-local), which has to be often approximated to a single particle one to speed up calculations; moreover, the HF wavefunction cannot properly describe the electronic correlation because of the oversimplification of a single-particle approach. In order to include the correlation to HF, Configuration Interaction (CI) methods have been introduced\cite{hfbook}, in which the total wave function is described as a sum of many Slater determinant. The expansion of the ground and excited state in charge-transfer multiplet calculations is reminiscent of this approach. The full-CI is probably the most accurate electronic structure method but it is also very computational demanding, and can be applied only to small molecules (around 10 electrons).

\subsection{Density Functional Theory}
Another answer to the problem was given by the Density Functional Theory (see Ref.\cite{DFT_Gross} and Refs. therein), which is probably the most popular electronic-structure ab-initio method at the moment. The theoretical foundation of DFT are given by the Hoenberg-Kohn (H-K) theorem, which states that in a system of interacting electrons in an external potential (in this case, the potential generated by the ion lattice) there's a bijective correspondence between the electronic density and the external potential.
An equivalent statement is that the ground-state energy of such systems, apart for an uninteresting additional constant, is a functional of the total electron density \textit{n} solely:

\begin{equation}
\label{eq_DFT}
E[n]=T[n]+U[n]+V[n]=T[n]+U[n]+\int d^3\boldsymbol{r} n(r) v_{ext}
\end{equation}
\\
where T is the energy associated to kinetic term, U the contribution of the electron-electron interaction (both Coulomb and exchange) and v$_{ext}$ is the external potential.

By means of the variational theory it should be possible to find the ground-state electronic density of the system by minimizing the total energy. The power of the H-K theorem resides in the fact that the total energy can be computed out of a 3 variable function (the total density) instead of the complex 3N $\Psi(\boldsymbol{r_1},\boldsymbol{r_2},...,\boldsymbol{r_n})$ wave function.
However, even if V[n] is a simple term, it is rather difficult to express T[n] and U[n] as analytical density functionals, therefore the direct minimization of Eq. \ref{eq_DFT} cannot be easily carried out without further approximations.

This equation can be solved instead in the so called Kohn-Sham (KS) scheme, in which the total density is expressed through a fictitious single-particle system in the form of Slater determinant of $\psi_1,...,\psi_N$ functions. The T[n] is thus broken in two parts:

\begin{equation}
\label{eq_KS}
T[n]=T_h[n]+T_c[n]=-\frac{\hbar}{2m}\sum_i^N\int d^3\boldsymbol{r} \nabla^2 \psi^2_i(\boldsymbol{r})+T_c[n]
\end{equation}
\\
where T$_h$ is the ``Hartree'' part of kinetic term, while T$_c$ is the extra correlation energy which is unknown but, due to H-K theorem, is still a functional of the density; the same can be done with the U[n] term, which is splitted in a Coulomb part (U$_h$) and an exchange part (U$_e$). The total energy is then rearranged in this form:

\begin{equation}
\label{eq_DFT_KS}
E[n]=T[n]+U[n]+V[n]=T_h[n]+U_h[n]+\int d^3\boldsymbol{r} n(\boldsymbol{r})v_{ext}+E_{xc}[n]
\end{equation}
\\
where E$_{xc}$[n]=T$_c$[n]+U$_e$[n]. The Eq. \ref{eq_DFT_KS} is formally correct, except for the fact that exchange-correlation energy (E$_{xc}$) is not known. The real success of DFT came with a suitable (and effective) approximation the exchange-correlation potential V$_{xc}$ (i.e. $\delta E_{xc}/\delta n$): the local density approximation (LDA). In this approximation, the V$_{xc}$[n(r)], functional of the position and of the density at that position, is equal to the exchange-correlation potential of an homogeneous electron gas (HEG) with the same electron density.

Data of V$^{HEG}_{xc}$ as function of the density have already been calculated with numerical Monte-Carlo methods by Ceperley and Alder\cite{LDA}, and thus can be easily interpolated with a parametric function and implemented in the calculation software.
Eq. \ref{eq_DFT_KS} thus can be solved through self-consistent methods like HF but in a faster way because in DFT the exchange potential is fully local. Moreover, the results contain (in an approximate way) also the correlation energy, introduced by the LDA formalism. However, when dealing with TM or RE, the correlation effects became very strong and further refinements have to be included in the theory.

Even if LDA should be a rough approximation, since electron density is not a slowly varying quantity in  molecules or crystals, the results in terms of bonding length, binding energies and other properties are usually rather good. DFT can be also applied to magnetic systems, by considering two separate spin-up and spin-down electronic densities ($n^\uparrow$ and $n^\downarrow$) within the Local Spin Density Approximation (LSDA). Ground state DFT calculation are now considered as the starting point for more accurate theoretical models.

The improvements of DFT have gone in two directions. On the first side, the DFT has been extended to excited states (since KS-DFT is a ground-state theory) and to highly correlated materials. In fact, with a correct description of excitations it is possible to properly evaluate the band-gap, the optical spectra and other relevant physical properties. A well known DFT formulation for excited states is the Time Dipendent DFT (TD-DFT). In highly correlated materials, a better agreement with the experiments can be achieved by adding extra-energy terms to the Eq. \ref{eq_DFT_KS}, such as in the GW approximation (a self energy correction) or in the DFT+U (an extra Hubbard term in selected energy levels).

On the other side, many attempts have been done to improve the quality of the exchange-correlation potentials. The first correction has been the Generalized Gradient Approximation (GGA), where a V$_{xc}$ is a function of the position, of the electronic density and of the density gradient. A proper GGA formulation, which respects the same internal sum-rules that are the key of the LDA effectiveness, was given by Perdew, Burke and Ernzerhof\cite{PBE} (GGA-PBE). In the so called meta-GGA exchange-correlation functionals also the density laplacian is included and in hybrid functionals an exact Hartree exchange term is considered. The quality of the improvements is not systematical, though: in some cases the LDA performs better than the GGA\cite{DFT_Gross}; for this reason the research in this field is still very active.

Of course, a DFT calculation involves a lot of technical details, related to the basis set, the pseudopotentials (when needed), the periodic conditions and so on. The specifics for the calculation of this Thesis work are given in the next paragraph.

\subsection{Computing details}
The calculations of this Thesis have been done with the \texttt{ABINIT} package\cite{ABINIT}, based on a plane-wave basis set. Each KS single particle $\psi$ wavefunction is thus described with a sum of plane-waves (PWs); this basis-set requires 3D periodic conditions and thus is best suitable for crystals, where the electronic states can be expressed with the Block theorem. The main advantages of this method are the generality, (since the PW are unbiased) and the speed; in fact, with plane-waves the KS equations can be solved in the reciprocal space with the aid of the Fast Fourier Transform algorithm. The accuracy level of the results can be tuned by the energy cut-off parameter (E$_{cut}$) that fixes the total number of PWs. The complexity of the calculation scales with the real-space cell size. Molecules and surfaces can be simulated within larger supercells with a suitable region of empty space at the border.

\begin{SCfigure}
\label{fig_pseudo}
\includegraphics[width=0.4\textwidth]{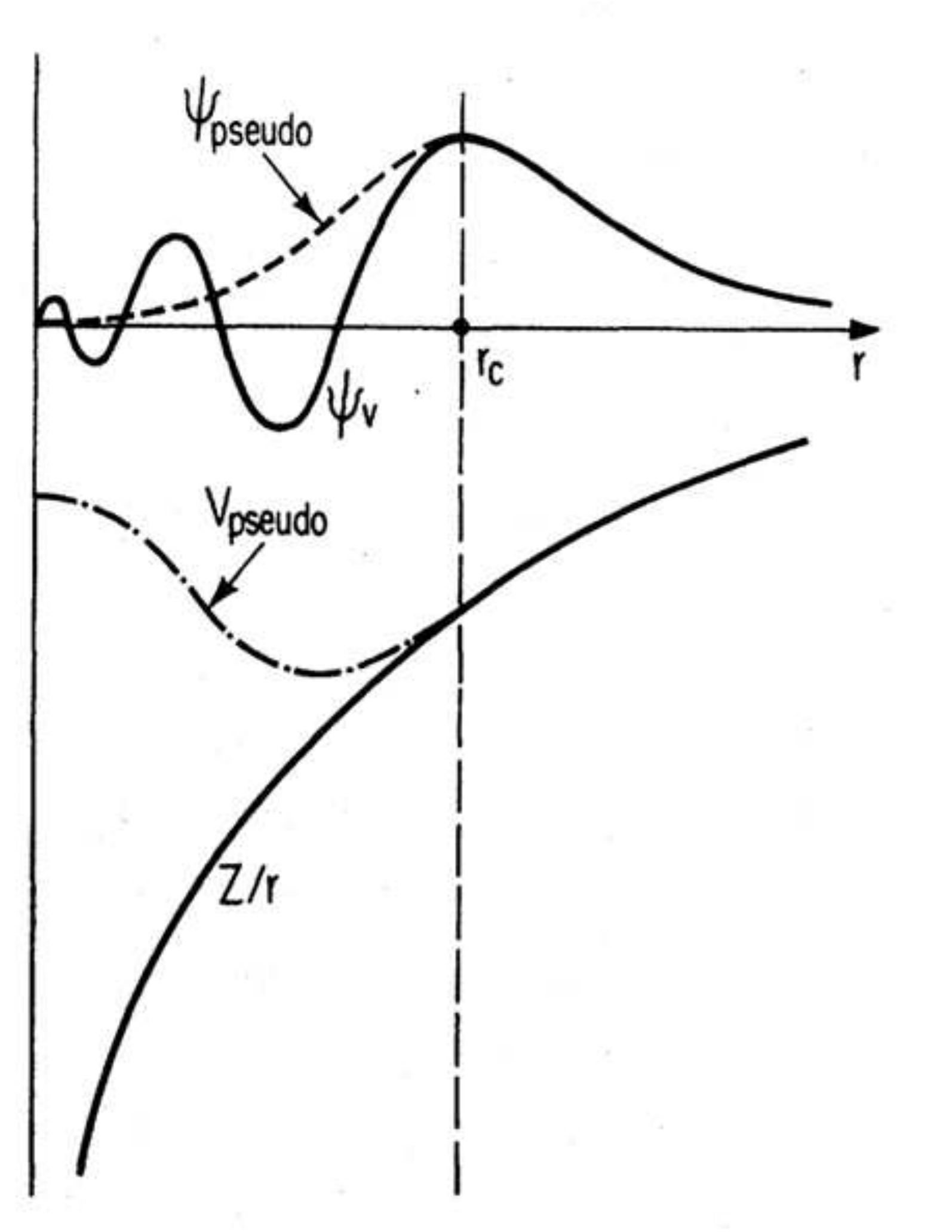}
  \caption{Schematic view of a ionic (solid line) and a pseudized (dashed line) potentials and wave functions, designed to match at a radius r$_c$. Adapted from Ref.\cite{Rev_DFT}.}
\end{SCfigure}

However, because of the diverging shape of the Coulomb potential, the core-level wavefunctions are highly oscillating around the nucleus and thus require a very large number of Fourier components to be properly described. For this reason, in a PW-DFT calculation the real ionic potential is usually substituted with the so called pseudopotentials (see Fig. \ref{fig_pseudo}). A Pseudopotential (PSP) is specifically designed in order to produce wavefunctions that match the exact ones outside a certain distance from the nucleus, while a smooth non-physical solution is left inside. The pseudopotentials are non-local, since are different for each $l$ quantum number. The diffusion of DFT calculation is related to the computational speed-up obtained by the introduction of PSP.
The chemical properties are more dependent to valence electrons than core ones; therefore, a good description of the properties of molecules and crystals can be achieved by discarding a defined set of core electronic levels and by working with the few remaining electrons in a ``pseudoatom'' with a lower ionic charge.

The other possibility in the basis-set choice is to work with a localized set of functions; the most common are Slater-type Orbitals described with a linear combination of Gaussian peaks (STO-NG basis, where N is the number of gaussian peaks). In transition metals it is possible to add a specific additional basis subset aimed to mimic the d orbitals. This basis set has the advantage of being localized around an atom and to have a relatively small number of parameters. Moreover, because of the gaussian peak properties, many of the electronic integrals can be calculated analytically. The disadvantage is the relatively fixed shape of the functions, which in some cases can poorly describe the long-range electronic states. The typical example are the carbon-nanotubes with small diameter, where the conductive states in the tube axis cannot be predicted by means of DFT with a localized basis set\cite{DFT_martins}. The STO-NG and the numerical equivalent(see for instance the \texttt{SIESTA} code\cite{SIESTA}) basis sets are thus often used in quantum chemistry for molecules, since they don't require periodic conditions.

Another opportunity is to mix both worlds by separating the basis set in two part: a localized (for core-levels) function set inside a critical radius (r$_c$), and plane waves in the remaining space in order to describe the valence electrons. This field includes the muffin-tin (LMTO) based approach, the so called all-electron calculations (LAPW methods) and, to lesser extent, the Projector Augmented-Wave (PAW) method used in this Thesis work. In particular, in the PAW method a linear transformation that connects the all-electron wave function $\Psi_n$ with a ``soft'' one $\tilde{\Psi_n}$ is defined through a set of projectors $\tilde{p_i}$:

\begin{equation}
\left| \Psi_n\right\rangle =\left| \tilde{\Psi_n}\right\rangle + \sum_i(\phi_i-\tilde{\phi_i})\left\langle \tilde{p_i}| \tilde{\Psi_n} \right\rangle
\label{eq_PAW}
\end{equation}
\\
where the all-electron $\phi_i$ and the pseudized $\tilde{\phi_i}$ partial waves are equal outside the PAW sphere with radius r$_c$. The wavefunctions of Eq. \ref{eq_PAW} are given by the superposition of pla\-ne\-wa\-ves and a localized set of spherical harmonics functions. The PAW method is usually faster than the norm-conserving pseudopotential method, because the loss of speed due to the higher complexity of the basis set is counterbalanced by a much lower E$_{cut}$.
In this Thesis, the PAW formalism has been used because it also allows the introduction of localized Hubbard terms in specific orbitals. This is the formulation of the total energy in the LDA+U approach\cite{DFT_U}:

\begin{equation}
E_{LDA+U}[n(\boldsymbol{r})]=E_{LDA}[n(\boldsymbol{r})]+E_{Hub}[{n^{I\sigma}_m}]-E_{dc}[{n^{I\sigma}}]
\label{eq_DFT_U}
\end{equation}
\\
where E$_{Hub}$ is the Hubbard correction added to the specific $\sigma$ orbital of atom $I$ and E$_{dc}$ is an additional term that removes a double counting error on orbitals. When applied to TM oxides, the DFT+U theory allows a correct description of band-gap and magnetic ordering.

Calculations in this Thesis have been performed on a quad-core desktop machine. Both norm-conserving pseudopotential and PAW basis have been used. For most of the elements some shallow core levels have been included in the basis set, in order to obtain results consistent with the experiments; for example, titanium ions have been described in a 3s$^2$ 3p$^6$ 4s$^2$ 4d$^2$ configuration, instead of considering only the 4s and 4d electrons. The convergence on total energy and k-grid density has been checked carefully, and every crystal cell has been relaxed to an equilibrium position. The Hubbard U values have been found in the literature; these are not completely empirical parameters, since they can be evaluated theoretically with other DFT ab-initio calculations\cite{DFT_U}.

\chapter{TiO$_{2-\delta}$ Resonant photoemission}
\section{Introduction}
\begin{figure}
\begin{center}
\includegraphics[width=0.7\textwidth]{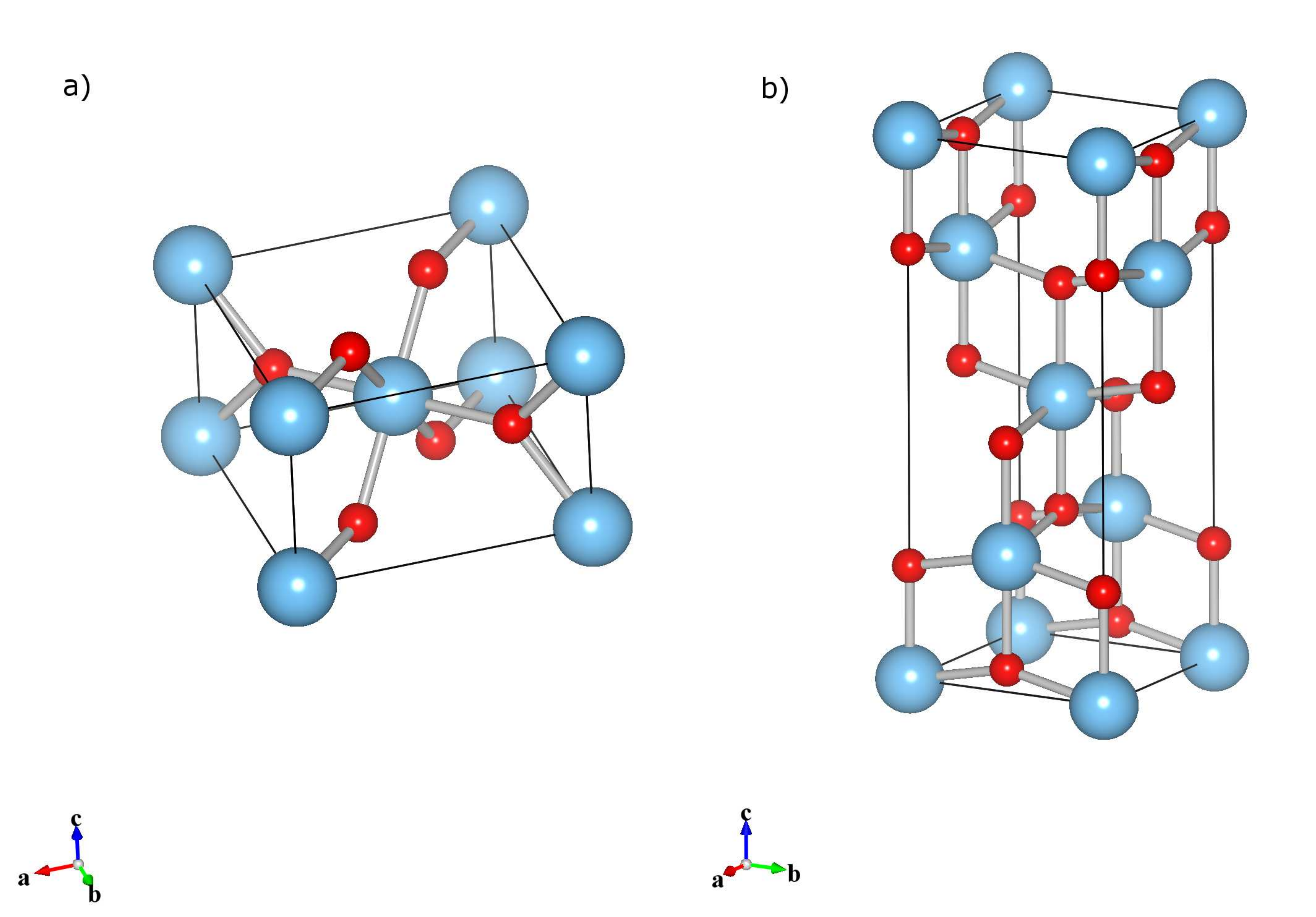}
\caption{Crystal structure of rutile (a) and anatase (b) unit cell. Oxygen atoms are depicted in red, titanium ones in cyan.\label{fig_rutilo_anatase}}
\end{center}
\end{figure}

Titanium dioxide (TiO$_2$) is one of the most studied metal-oxides. The constant interest on the TiO$_2$ surface is due to the applicative properties (especially as a single-crystal surface), to its simplicity in growth and in its electronic structure. The two most common TiO$_2$ minerals are rutile and anatase, the former being the most abundant (stable) form. The rutile single cell is composed of six atoms (two titanium and four oxygen atoms) and belong to the tetragonal space group (P$_{4_2/mnm}$, n$^\circ$136); each titanium atom is at the center of a distorted octahedral cage of oxygen, with a D$_{4h}$ symmetry (see Fig. \ref{fig_rutilo_anatase}).

The valence electrons are localized in the oxygen 2p levels, while the empty states are composed by Ti d levels. TiO$_2$ should be a (formally simple) closed-shell d$^0$ band insulator; however, the hybridization of titanium 3d and oxygen 2p states can induce a partial occupation of Ti d-level. In fact, TiO$_2$ is a charge-transfer insulator, like other TM oxides\cite{early_TM_xps}. The (direct) band-gap of rutile and anatase is of nearly 3.2 eV.

The main applications of TiO$_{2}$, such as photocatalysis, photodegradation of carbon compounds, solar cells and high-k gate for transistors\cite{Die2003}, require a fine tuning of the electronic structure, especially of titanium 3d states. In particular, the TiO$_2$ based photocatalytic process require to shine the surfaces with UV light, in order to pump electrons from the top of the valence band to the empty states. A reduction of the TiO$_2$ band-gap, achieved by selective doping, could in principle extend the catalytic properties up to the visible (solar) light range, with an obvious impact on the applications.

The understanding of the relationship between the defects electronic structures and the so called ``d$^0$ magnetism''\cite{DMO_TiO2_d0} in TiO$_2$ poses an even more challenging task, from both the theoretical and the experimental point of view. A detailed analysis of the magnetism in RF-sputtered anatase and rutile thin films is given in the next Chapter.

The Resonant photoemission spectroscopy (ResPES) on transition metals (TM) provides the opportunity to emphasize elemental-specific electronic structures in the gap and in the valence band. Theoretical prediction of resonant behavior at 2p edge of early TM oxides have been done by Tanaka and Jo\cite{tanaka}, followed by the experimental work of Prince et al.\cite{prince} on TiO$_2$ rutile. Resonant photoemission on different TiO$_2$ phases have already been measured\cite{thomas} on 3p-3d (or M$_{3,4}$) edge, which gives higher sensitivity to surface states than the 2p-3d (L$_{2,3}$) edge. While highly defective TiO$_2$ films have already been reported\cite{aloisa}, data on non-conductive, slightly defected specimens are still missing, although recent studies on Ti-based oxide have shown the importance of the in-gap defect states in slightly substoichiometric systems\cite{apl_LAOSTO}. In this chapter, a ResPES study of transparent, non-conducting rutile TiO$_{2-\delta}$ film is presented.

\section{Resonant Photoemission on TiO$_2$}
As already explained in Chap. 2, in the case of TM compounds, the L-edges ResPES effects are expected to be rather pronounced. The Fano factor in TM-based systems is usually large ($q>3$) and it will be shown, is even larger in TiO$_2$. In our case study, photon energies have been scanned through the L$_{2,3}$ edge; in a configuration-interaction approach the initial, the final and intermediate states in a general TM can be labeled as follow:

\begin{equation}
\label{eq:pes_TM}
2p^6 3d^{n}\rightarrow 2p^6 3d^{n-1}+e^{-}
\end{equation}
\\
\begin{equation}
\label{eq:intermediate_TM}
2p^6 3d^{n}\rightarrow 2p^5 3d^{n+1}
\end{equation}
\\
\begin{equation}
\label{eq:auger_TM}
2p^5 3d^{n+1}\rightarrow 2p^6 3d^{n-1}+ e^{-}
\end{equation}
\\
Eq. \ref{eq:pes_TM} defines the photoemission process, Eq. \ref{eq:auger_TM} defines the autoionization channel and Eq. \ref{eq:intermediate_TM} shows the 2p-3d transition. In the empty shell (3d$^{0}$) TiO$_{2}$ film, the VB resonances shouldn't be observed, because at least two d-electron are needed in the intermediate state in order to yeld the autoionization process. However, in this case ResPES occurs because of the aforementioned hybridization of Ti 3d level with valence-band O 2p level. As shown in Chapter 2, this issue can be addressed with the charge-transfer multiplet theory and it is partly considered in DFT calculation (which already include some electron correlation, as defined by the exchange-correlation potential).

\section{Theoretical calculations for ResPES}
In order to theoretically calculate a ResPES spectrum, at least two different models have to be taken into account: an atomic charge-transfer model for the absorption cross section and an ab-initio density of states calculation, usually ground-state DFT. The first model generally includes several parameters (charge-transfer energies, Hubbard U d-d repulsion, inter-configuration mixing, crystal field) but can successfully describe matrix elements for excited states, which determine the absorption edge on TM L-edges; the latter (DFT) doesn't need many empirical parameters (except U for DFT+U, which can be calculated from similar ab-initio approaches\cite{DFT_U}) but it is difficult to extend to excited states. Usually\cite{Respes_okada}, in a configuration-interaction approach, ground (g.s.) and final (f.s.) states of TiO$_2$ resonant photoemission can be described as:

\begin{equation}
\label{eq:GroundState}
\Psi_{g.s.}=\alpha_0|3d^0\rangle+\alpha_1|3d^1\underline{L}^1\rangle+\alpha_2|3d^2\underline{L}^2\rangle
\end{equation}
\\
\begin{equation}
\label{eq:ExcitedState}
\Psi_{f.s}=\beta_0|2\underline{p}3d^1\rangle+\beta_0|2\underline{p} 3d^2\underline{L}^1\rangle+\beta_0|2\underline{p}3d^3\underline{L}^2\rangle
\end{equation}
\\
where \underline{L} and \underline{p} stands for a hole in the ligand (O) and in the 2p (Ti) level. In rutile TiO$_2$, an average occupation of d-levels of 0.7 can be estimated from the $\alpha$ coefficient, using literature\cite{Respes_okada} data ($\alpha_0$=39.5\%, $\alpha_1$=48.2\%, $\alpha_2$=12.3\%). The charge-transfer multiplet model has proved to be really effective for calculating XAS spectra at L$_{2,3}$ edges of TM, because of the possibility to evaluate the transition matrix elements to the excited states; however, atomic approaches fail in modelling VB spectra\cite{DEGROOT}, which must often be calculated using ab-initio ground-state calculation, like density functional theory (DFT). An highly desiderable complete ab-initio approach is still a matter of research; good results have been achieved in modeling Ti L-edge XAS using multiple scattering theory\cite{kruger}.

\begin{figure}
\begin{center}
\includegraphics[width=1.0\textwidth]{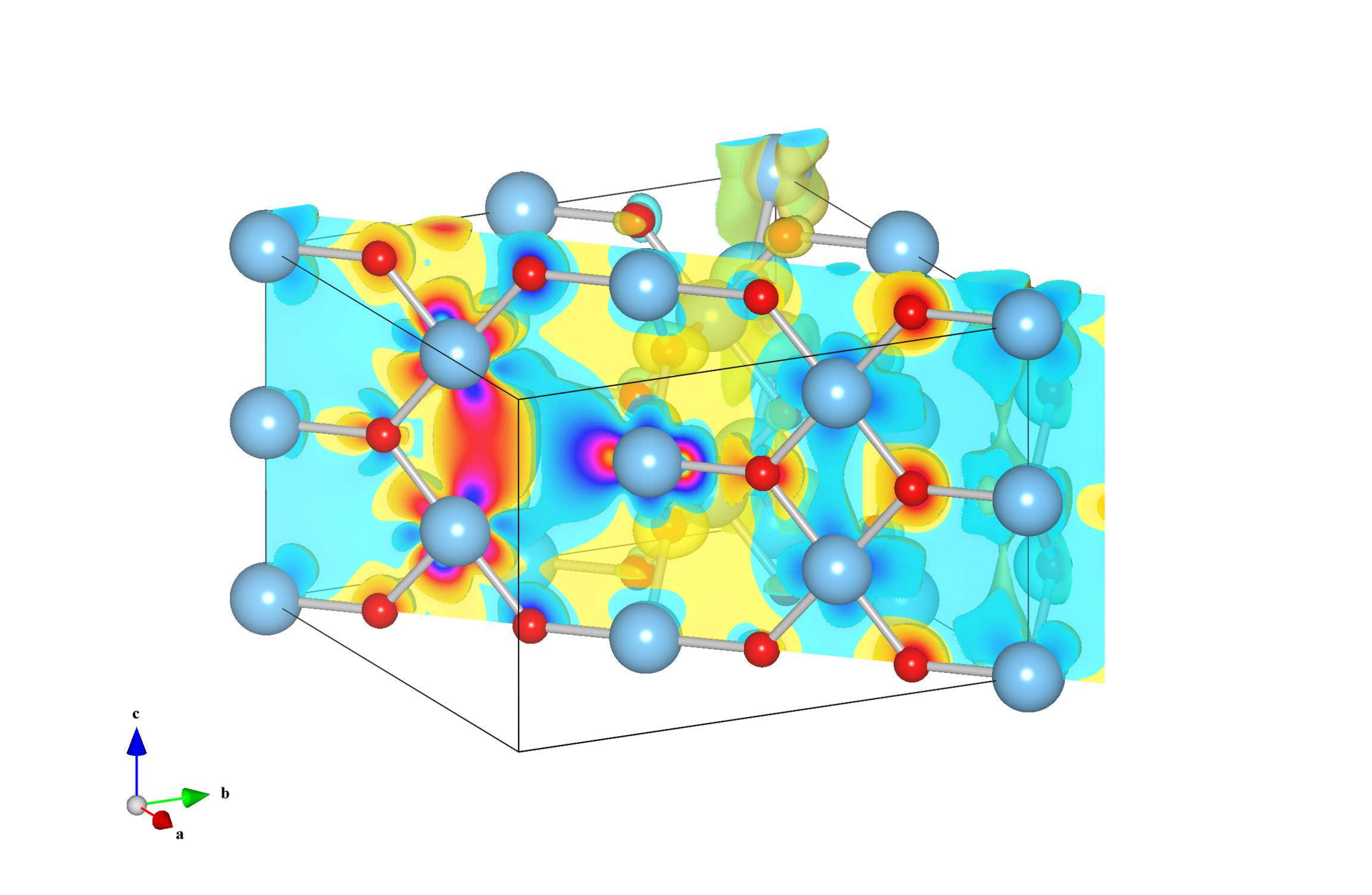}
\caption{LDA+U DFT Relaxed rutile 2$\times$2$\times$2 supercell, cutted on the plane containing the oxygen vacancy (titanium atoms are depicted in cyan, oxygen atoms in red); superimposed to the structure, a slice of the spin-up/spin-down difference electronic density is shown. In this calculation, U=8.0 eV (from Ref.\cite{TIO2_ldau}).\label{supercell}}
\end{center}
\end{figure}

In order to reproduce the non-resonant valence band spectra of TiO$_2$, a DFT calculation have been carried out, both for a single rutile cell and a 2$\times$2$\times$2 48 atoms supercell (with an oxygen vacancy, shown in Fig. \ref{supercell}); the calculations have been performed in the LSDA+U approximation with the ABINIT\cite{ABINIT} code, using a projector augmented waves (PAW) basis set\cite{PAW}. The initial spin has been chosen to be up (down) for the first (second) Ti atom in the single rutile cell. The Hubbard U correction has been necessary to recover the correct band-gap (3.2 eV) and to obtain an insulating ground-state. In fact, an LDA (without U) calculation on this defected supercell usually gives a metallic behavior, as already reported in literature\cite{tio2_lda}.

The LSDA+U calculation shows a net difference in spin-up/spin-down electronic densities, located at the three Ti atoms surrounding the oxygen vacancies. Every oxygen atom is bonded to three titanium atoms: the two Ti atoms belonging to the edge-sharing octahedra take a relative ``plus'' sign (yellow/hot colors in Fig. \ref{supercell}) while the apical octahedra has a ``minus'' sign (blue/cold colors in Fig. \ref{supercell}). These results, that concern the magnetic properties of TiO$_2$, will be extensively discussed discussed in Chapter 3.

From a technical point of view, energy cut off was 25 Hartree, and full cell relaxations have been performed with a maximum residual forces on atoms at 10$^{-3}$ Hartree/\AA.; results are consistent with other all-electron calculations\cite{TIO2_ldau}.
The X-ray absorption spectrum and core-level photoemission on Ti edge have been calculated with the \texttt{MISSING} package\cite{MISSING} by considering full multiplet and crystal field effects, in D$_{4h}$ symmetry; crystal field parameters for this structure have been set at 10Dq=1.7 eV, ds=dt=-0.12 eV.

\section{Experimental details and results}
The sample has been grown with RF sputtering technique from a TiO$_2$ target. The film thickness was thin enough (10 nm) to avoid charging effect. Resonant photoemission and x-ray absorption data have been collected at the BACH beamline of the ELETTRA synchrotron radiation source in Trieste (Italy). Usually, TiO$_2$ surfaces for photoemission experiment are prepared by an heavy annealing in vacuum (600$^\circ$C or higher), followed by an annealing in oxygen atmosphere to recover surface oxidation; this procedure usually generates a very ordered surface reconstruction and also make the TiO$_2$ film conductive, allowing synchrotron measurement. Rutile crystals, after this procedure, usually change the color from transparent to black/blue, due to the annealing-induced defects. Since this work aims to measure the pristine crystal defect electronic structures, the TiO$_2$ film has been left untreated (transparent and insulating); a layer of atmospheric contamination is thus expected on the sample surface. Even if this hydrocarbon contamination layer can attenuate the photoemission intensity, it can help to fix the binding energy scale (through XPS on C 1s peak) and to saturate the dangling bond at TiO$_2$ surface, so that only ``bulk'' Ti$^{4-\delta+}$ can be detected.

The resolution of the monochromator was set to 0.24 eV at the Ti L$_{2,3}$-edge photon energy. X-ray absorption spectroscopy (XAS) and photoemission measurements have been carried out by using a modified 150 mm VSW hemispherical electron analyzer with a 16-channel detector. The total resolution for ResPES was 0.3 eV. XAS spectra have been taken in total yield mode, with circularly polarized light perpendicular to sample.
During photoemission measurement, the sample has been kept normal to the analyzer.

\begin{figure}
\begin{center}
\includegraphics[width=0.9\textwidth]{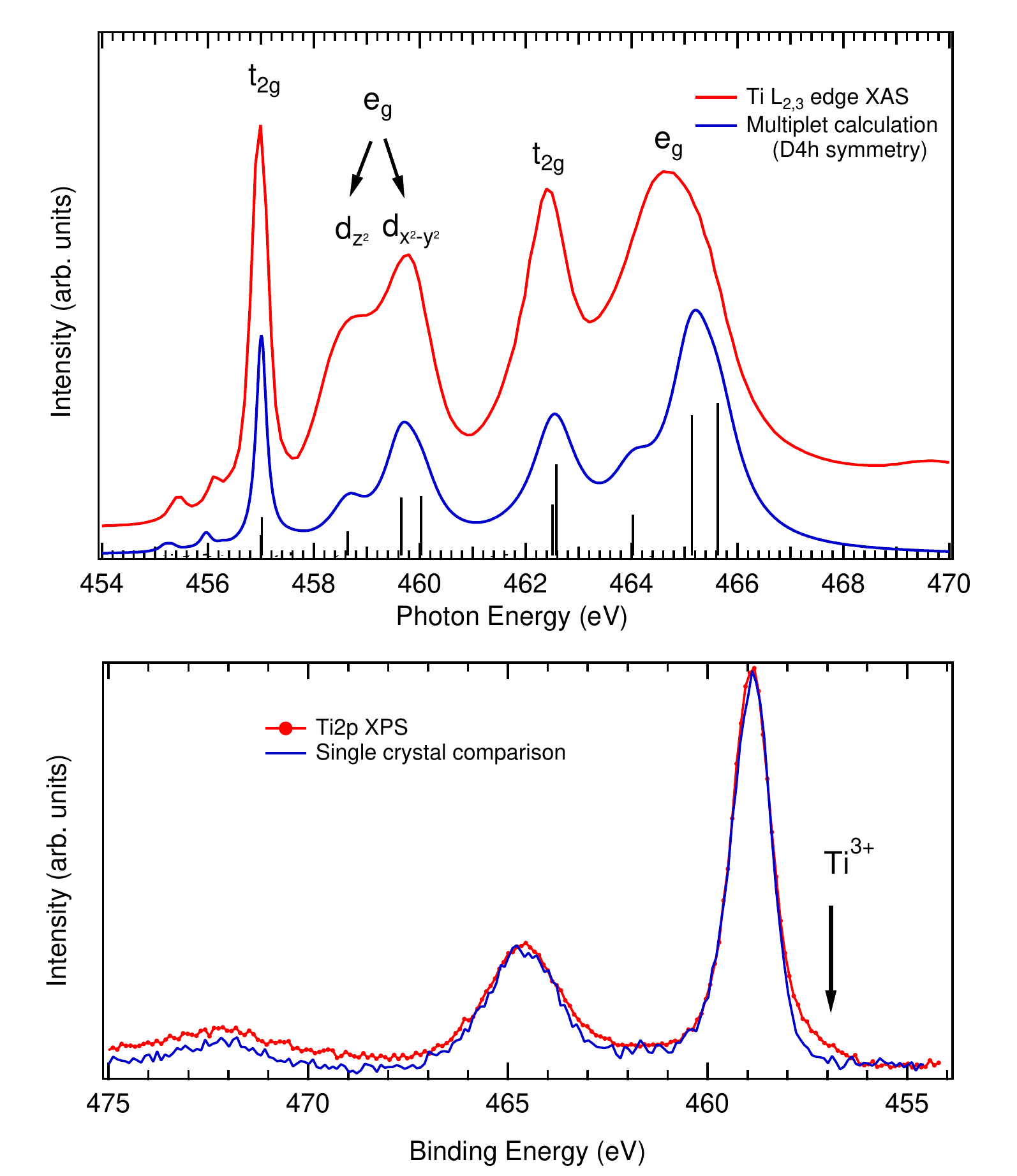}
\caption {Upper panel: Ti 2p-3d X-ray absorption spectra, with multiplet calculation in D$_{4h}$ symmetry; lower panel: Ti 2p XPS spectra from the polycristalline film and from a reconstructed (100) rutile single-crystal surface.\label{fig_XasXps}}
\end{center}
\end{figure}

In Fig. \ref{fig_XasXps} the measured and calculated XAS at Ti L-edge are shown, along with the Ti 2p XPS core levels on the thin film (Fig. \ref{fig_XasXps}, lower panel). Ti 2p core levels are compared to the results obtained from a high-quality single crystal rutile (110) surface. The XAS data compare well to the multiplet calculations and are in good agreement with the results already reported in the literature for high quality samples\cite{Die2003}. The XAS spectrum shows a sequence of four main bands, with alternating t$_{2g}$ and e$_g$ symmetries (labeled in the figure). The first couple of bands at low photon energies is separated from the second at higher photon energies by the spin-orbit interaction of the Ti 2p core hole. The e$_g$ band at about 459 eV is actually split into two features, due to the crystal field effect that reduces the octahedral to a tetragonal symmetry; therefore, the D$_{4h}$ symmetry must be taken into account in order to reproduce the experimental data. The XAS data apparently do not suggest the presence of oxygen vacancies or Ti$^{3+}$ ions, which, in turn, can be related to the shoulder of Ti 2p$_{3/2}$ XPS peak (black arrow in Fig. \ref{fig_XasXps}, lower panel). However, this finding does not allow us to draw further conclusion on the role of oxygen vacancies on the electronic properties of insulating rutile and therefore we resorted to ResPES techniques to enhance the spectral weight of Ti-related states in the valence band region, and to probe the relationship of these states with oxygen vacancies.

\begin{figure}
\begin{center}
\includegraphics[width=0.9\textwidth]{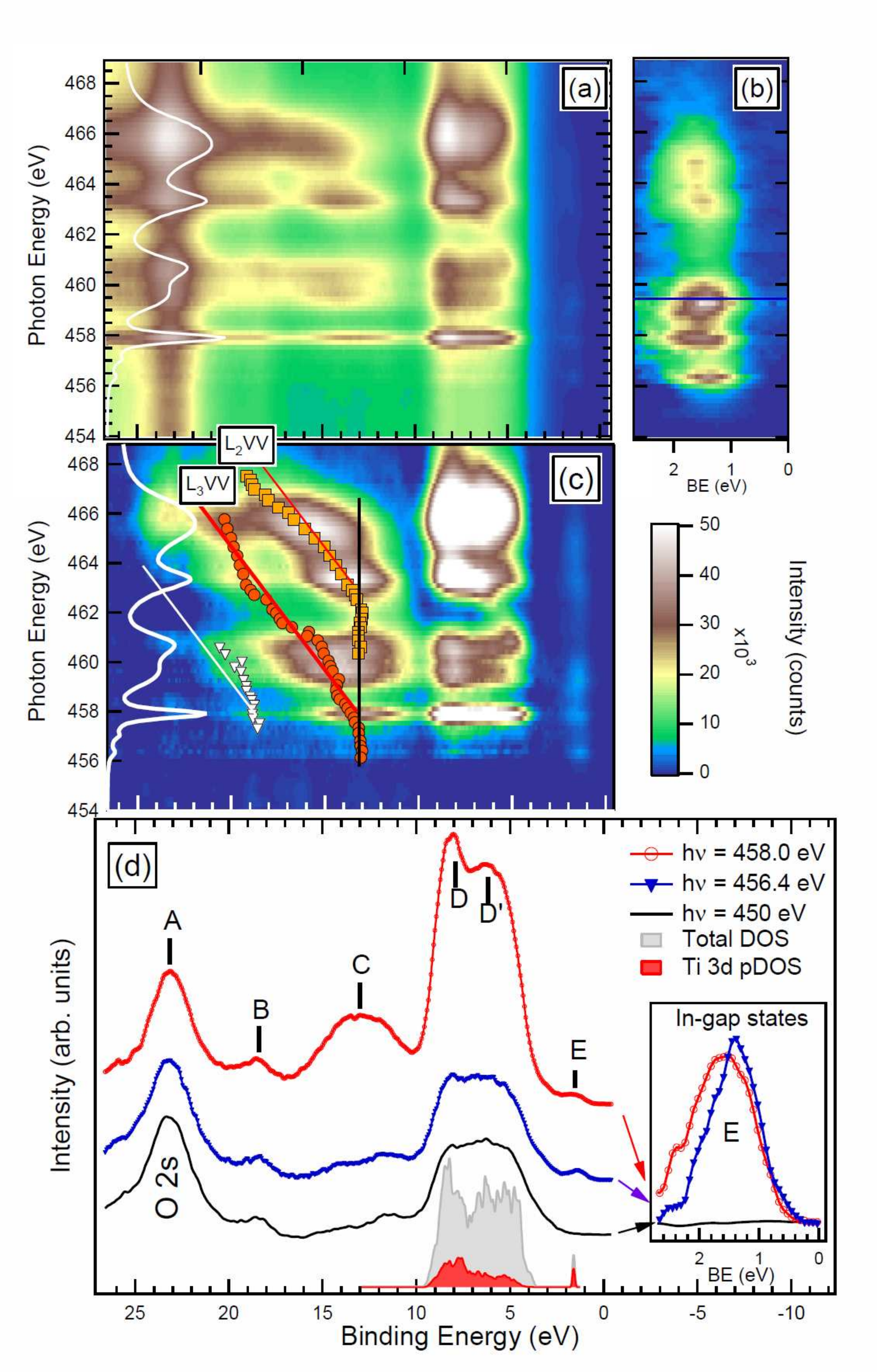}
\caption{RePES data: (a) contour plot of the raw data, along with the Ti 2p-3d absorption
spectrum (left axis); (b) enlarged view of the defect electronic states states nearby the Fermi level; (c) ResPES data after the subtraction of the off-resonance ($h\nu$=450 eV) spectrum; (d) selected photoemission spectra spectra (collected at $h\nu$=450, 456.3 and 458.1 eV) compared to LDA+U calculations of the total and Ti 3d projected DOS. Inset: details of
the in-gap electronic states measured at $h\nu$=450, 456.3 and
458.1 eV. Taken from Ref.\cite{Respes_Article}.}
\label{resLarge}
\end{center}
\end{figure}

In Fig. \ref{resLarge} the ResPES data on the Ti L-edge are shown. Each line of panel (a) represents a single photoemission spectrum taken at the corresponding photon energy, labeled on the vertical axis. The XAS profile across the Ti L-edge is also shown on the left axis. Fig. \ref{resLarge}(c) shows the ResPES data obtained from panel (a) after the subtraction of the off-resonance spectrum collected at $h\nu$ = 450.0 eV; this procedure has been done in order to enhance the contribution of resonating spectral weight over the non-resonating valence band states.

Selected valence band spectra are shown in Fig. \ref{resLarge}(d), while in Fig. \ref{resLarge}(b) and in the inset of panel (d), enlarged views of the in-gap states, with BE ranging from 0 to 2.5 eV, are shown. For a description of the overall features of the valence band photoemission region, we refer to Fig. \ref{resLarge}(d). Peak A is assigned to the oxygen 2s shallow core level. The valence band shows two features, D and D', the former resonating at $h\nu$=458.1 eV.  Therefore, the high energy states C clearly resonates at $h\nu$=458.1 eV, along with the in-gap state E, that resonates also at $h\nu$=456.3 eV. A weaker resonance is detected for peak B.

The analysis of the DFT calculations for ground state of the 2$\times$2$\times$2 supercell provides useful indications for the identification of some of the above-mentioned features. As it can be seen in Fig. \ref{resLarge}(d), the LDA+U calculated density of states (DOS) can reproduce the shape of the valence band (VB), dominated by the oxygen 2p band, and provide information about the Ti-3d projected DOS in the valence band that, because of the O 2p - Ti 3d hybridization, results to be non negligible (especially in the high BE part of the VB, i.e. below the experimental D band). Therefore, peak D (BE=8.2 eV) is the resonating structure related to the Ti-O bonding part of the VB, while the lower BE region corresponding the peak D' is usually referred to as the non-bonding part of the VB\cite{thomas}.

The low intensity peak (E) found at BE=1.5 eV, can be seen in detail in Fig. \ref{resLarge}(b) and in the inset of Fig. \ref{resLarge}(d) for selected photon energies. This peak, that is not detectable off-resonance, appears to resonate at different photon energies with respect to main resonance (maximum at $h\nu$=456.3 eV rather than $h\nu$=458.1 eV) and it is ascribed to oxygen vacancy states, yielding formally Ti$^{3+}$ ions nearby the vacancy itself. Indeed, the DOS calculation on the defected supercell (Fig. \ref{resLarge}(d)) well reproduces the position of these in-gap states. Notably, the position of E peak seems to slightly shift upon changing photon energy from the pre-edge ($h\nu$=456.3 eV) to the main absorption edge ($h\nu$=458.1 eV), revealing a not completely resolved manifold of states. In fact, the slightly asymmetrical lineshape of E peak is similar to the one observed in other Ti-based 3d$^1$ highly correlated systems (see, e.g., Ref.\cite{LAOSTO_ticlo} and Refs. therein).

We now address the origin of peak B and C by referring to the available results of configuration interaction calculations of the photoemission process as well to the tracking of B and C peaks BEs upon photon energy increase. The $|3d^n\underline{L}^m\rangle$-like configurations of Eq. \ref{eq:GroundState} are those accounting for charge transfer (CT) effects from the anion 2p to the TM 3d orbitals, providing a straightforward tool to consider the effect of 3d-2p hybridization in CI calculations. Therefore, the possibility to observe ResPES effects in the formally empty 3d-shell of the Ti$^{4+}$ cation is due to the presence of CT-like configurations both in the initial and final states. Furthermore, the presence of CT-like configurations in the initial state should agree with the finding that the Ti 3d projected DOS (Fig. \ref{resLarge}(d)) in the valence band is not negligible.

As a first step of our analysis, we focus on the most intense peak C of Fig. \ref{resLarge}(c). As it can be observed, the BE of peak C is constant for low photon energies, but it shifts to higher values as the photon energy increases above 458 eV. Therefore, peak C shows a radiationless Raman-to-normal Auger transition\cite{Respes_hufner} and above $h\nu$=458 eV, peak C ends up to be the normal LVV Auger. Hence peak C is assigned to $|3d^2\underline{L}^2\rangle$ initial state configuration, as at least two d electrons are required to yield a normal L$_3$VV Auger emission. Apparently, the dispersion on the BE scale of peak C is not strictly linear, since a sequence of absorption thresholds is accessed in a relatively narrow energy range, introducing new resonances that make the BE tracking of this peak rather difficult. However, it is also clear that above about $h\nu$=463 eV, a new normal-Auger feature (L$_2$VV) becomes detectable, as the photon energy crosses the Ti L$_2$ threshold. Finally, peak B also shows a dispersion above the resonance on the BE scale, though weaker than that observed for peak C.

It is important to note that, unlike peak C and, to a minor extent, peak B, peak D does not show any detectable dispersion on the BE scale. Therefore, in the CI approach, for peak D a major contribution of the $|3d^1\underline{L}^1\rangle$. configuration is expected, as for this configuration the autoionization channel is allowed, but not the normal Auger emission. According to Ruus et al.\cite{Respes_ruus}, peak D is fit by the most intense feature of the CI calculations\cite{tanaka}, and starting from this peak all the higher-BE peaks (B and C) are well fit by the CI model. On the other hand, the energy of peak D corresponds to the region where the Ti-projected DOS has the largest intensity, therefore providing an alignment for the two computational results. However, while the single-configuration, ground-state, LDA+U calculations fail to reproduce the high BE satellites B and C, this indicates that the Ti 3d states are not only localized below the peak D, but they are spread, through band structure effects, to lower BEs, where CI calculations do not predict any Ti contribution, as the band dispersion of oxygen states is usually not included in the CI model.

\begin{figure}
 \includegraphics[width=1.0\textwidth]{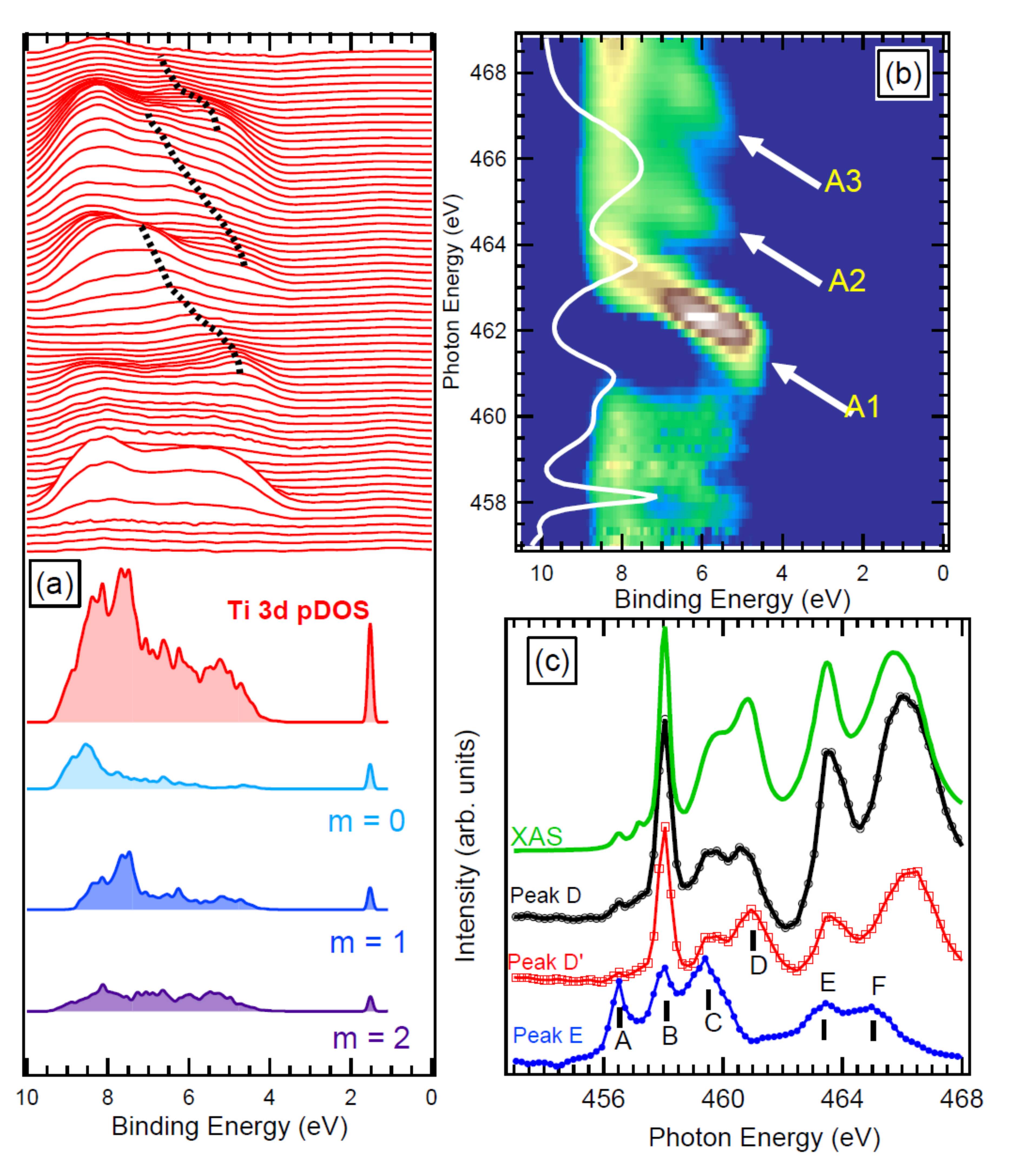}
    \caption { Detail on valence band part of ResPES: in
panel (a), ResPES on VB without the out-of-resonance contribution;in panel (b), spectra taken from image of panel (a) together with DFT m-resolved d-DOS calculations; in panel (c), ResPES normalized to the absorption edge.}\label{img:res_Slice}
\end{figure}

In Fig. \ref{img:res_Slice}(a)-(c) an enlarged view of the ResPES data collected in the low BE range (0-10 eV) is shown, while Fig. \ref{img:res_Slice}(d) shows the constant initial state (CIS) profiles of peak D, (BE=8.2 eV), peak D' (BE=6.0 eV) and defect states (peak E at BE=1.5 eV). These intensity profiles are obtained by vertically slicing Fig. \ref{img:res_Slice}(a), along the three dashed lines at constant BE. Due to a very high Fano q-factor (q$>$10), at the TM L-edge, the CIS profiles of selected valence band states usually closely follow the shape of the absorption edge. However, the CIS spectrum for the defect structure (BE=1.5 eV) is clearly different from the Ti L-edge XAS spectrum, especially in the case of peaks A, C, and F of Fig. \ref{img:res_Slice}(c). This CIS spectrum could be regarded as an effective on-site defect XAS and it is compatible to that found in heavily sputtered rutile Ti L-edge XAS data\cite{aloisa}. The narrow width of the present CIS peaks with respect to those observed in heavily defective samples (see e.g.\cite{aloisa}) could be related to the reduced lattice disorder effects with respect to the heavily sputtered surfaces.

Finally, the ResPES spectrum collected with a $h\nu$ = 461.5 eV photon energy shows a clear resonant enhancement of peak D', that seems to shift to higher BEs while increasing the photon energy. In fact, a careful tracking of peaks BE reveals the presence of three Auger-like features that start to resonate in the non-bonding part of VB at photon energies 461.5 eV, 463.5 eV and 466.2 eV (dotted lines in Fig. \ref{img:res_Slice}(a)).

To further explore this effect, in Fig.\ref{img:res_Slice}(b) the VB ResPES data normalized to XAS spectrum intensity are shown: by doing so we can quench the XAS matrix contribution and enhance the spectral weight due to interference effects in the ResPES process. Three spectral features, A1, A2 and A3, displaying a BE shift appear, A1 being the most intense (marked by arrows). These features are too prominent to be assigned to pure 3d$^1$ defect states, and resonate mainly around the e$_g$ symmetry edges of the absorption, while the more intense LVV Auger peaks are generated on t$_{2g}$ edges. Therefore, this state (peak D') presents a normal Auger (i.e. a two hole correlated satellite) behavior similar to the one observed for peak C, indicating that in peak D' a large contribution of the $|3d^2\underline{L}^2\rangle$ configuration in the initial state is expected. Incidentally, we recall that the e$_g$ XAS bands are those mostly affected by the CF effects arising from the D$_{4h}$ distortion of the TiO$_6$ cluster cage. The symmetry dependence, which is ultimately a dependence on the orbital degrees of freedom, can also be discussed by considering the projected DOS in the VB region as in Fig. \ref{img:res_Slice}(b), where the m-resolved d-DOS (m=0,$\pm1$,$\pm2$) are shown. While the strongest contribution to the bonding part of VB comes from the the m=0 and m=$\pm1$, related to d$_{z^2}$, d$_{xz}$ and d$_{yz}$ orbitals, the m=$\pm2$ projected DOS, which is related to d$_{xy}$, d$_{x_2-y_2}$ orbitals, provides a relevant contribution to the energy region below peak D'.

In conclusion, a ResPES experiment at the Ti L-edge has allowed us to probe the in-gap states of an insulating TiO$_{2-\delta}$ thin film, otherwise missed by off-resonance measurements. The corresponding CIS profile indicates that these states should be ascribed to Ti$^{3+}$ ions. In spite of the rather complex sequence of thresholds, determined both by crystal-field and spin-orbit interactions, by tracking the dependance of the BE of each resonating peak on photon energy we have been able to identify the main configurations contributing to these electronic states. Distinct two-hole correlated satellites in the final state are related to the spectral features at BE=6, 13 and 19 eV. In particular, we provide evidence of a $|3d^2\underline{L}^2\rangle$ character for the peak below the overwhelming non-bonding states of the VB (D' at BE=6 eV), so far virtually neglected by CI calculations, as well as by the analysis of experimental data. Our finding asks for more refined CI calculations aimed to draw a consistent picture of the N-1 electron final state of the photoemission process. It is likely that this split-off peak could be properly described by (i) explicitly accounting for the oxygen 2p band width in CI calculations or (ii) by resorting to a TiO$_6$ cluster symmetry lower than the one (O$_h$) so far assumed in CI calculations for the photoemission process\cite{Respes_okada,tanaka}, in agreement with the symmetry (D$_{4h}$) already adopted for the calculation of the Ti L-edge XAS spectrum.

\chapter{Magnetism in TiO$_{2-\delta}$ and N:TiO$_{2-\delta}$ films}
\section{Introduction}
\begin{figure}
\begin{center}
\includegraphics[width=0.6\textwidth]{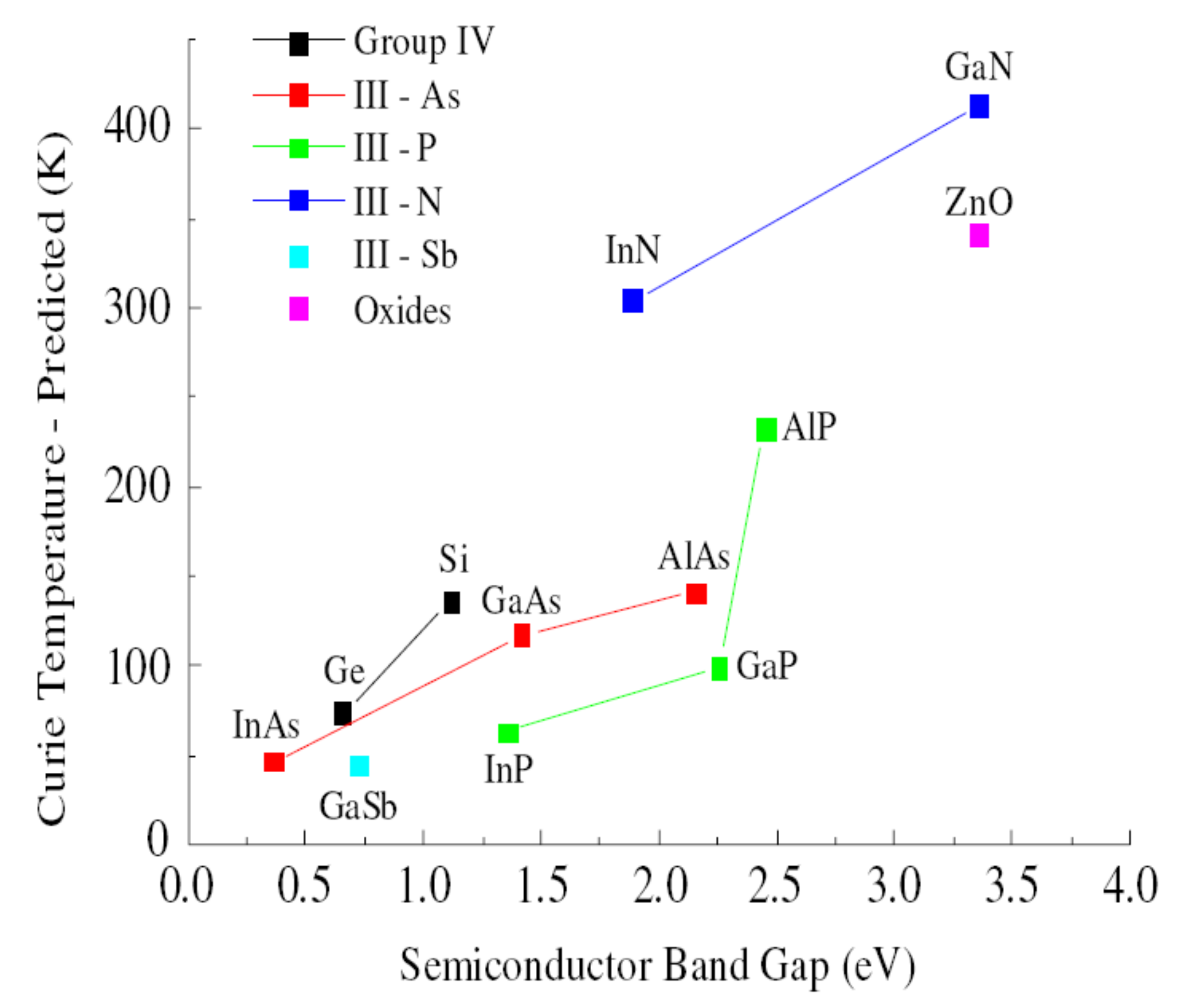}
\caption{Theoretical prediction of ferromagnetic T$_c$ in different materials, upon doping with Mn ions. The wide-gap material are predicted to give higher T$_c$. Taken from Ref.\cite{DMO_dietl}.\label{fig_DMO_dietl}}
\end{center}
\end{figure}

The history of diluted magnetic oxides (DMO) started in 2000 with a work of Dietl\cite{DMO_dietl}, in which the authors theoretically predicted an high-T$_c$ ferromagnetism (FM) in Co-doped ZnO (see Fig. \ref{fig_DMO_dietl}). Such kind of material would display a simultaneous presence of magnetism and optical transparency, due to the high band-gap, and would be a very interesting system for spintronic applications. Subsequent experimental work didn't find FM in ZnO-based compound\cite{DMO_zno_nofm}, but many research group reported the presence of room-temperature FM in Co-doped TiO$_2$ film\cite{DMO_tio2_1,DMO_tio2_2}. The Curie temperature in this cases is expected to be higher than 500 K; however, this kind of FM is usually rather weak (in term of coercive field) and thus it is not suitable for typical applications.

The growth process has proved to be crucial for the magnetism in DMO. For instance, PLD (Pulsed Laser Deposition) grown Ti$_{1-x}$Co$_x$O$_2$ samples display\cite{DMO_tio2_pld} a very high magnetic moment per Co atom (1.5-1.7 $\mu_B$) which is not expected in a diluted sample and is caused by metallic cluster precipitation. Films of the same material, grown by reactive sputtering\cite{DMO_tio2_re_sputt}, display a much lower value (0.96 $\mu_B$), that could be consistent with a non-clustered random Co distribution or with a mixture of metallic cluster and isolated Co atoms.

The difference between an ``intrinsic'' or ``extrinsic'' origin of the magnetism is fundamental for the theoretical models and for real applications. Only an intrinsec magnetism would be useful for the spintronc field, since the charge carriers have to lie in a spin-polarized band. It is generally believed that it is possible to grow diluted Ti$_{1-x}$Co$_x$O$_2$, with a suitable control of the growth parameters; unfortunately, in the attempt to achieve the highest T$_c$, many research groups focussed their work only on cobalt doping, which can easily generate secondary magnetic phases (and thus extrinsic FM).

Subsequently, FM has been reported also in single crystal\cite{DMO_TMtio2} TM-doped TiO$_2$ (TM = V, Cr, Mn, Fe, Co, Ni, Cu); the magnetization measurements appear to be similar for each dopant, with a different saturation magnetization. Other research group also reported FM on TM-SnO$_2$\cite{DMO_TMsno2} and TM-ZnO\cite{DMO_TMzno} systems. Therefore the driving mechanism of such kind of FM, observed for every kind of TM doping (some of which usually don't develop FM phases with oxygen, like Mn), has to be related to a general mechanism instead of a specific dopant. The magnetism is then likely to be mediated by intrinsic defects, among which the most common in these systems are oxygen vacancies (V$_O$'s).

In fact, one side effect of the doping is the replacing of Ti$^{4+}$ ions with stoichiometric non-equivalent ions that can induce structural defects. In this description, the presence of magnetism should be triggered by interacting Ti$^{3+}$ ions, induced by V$_O$'s; the dopant role should be both to create V$_O$ and to increase the saturation moment.

These hypothesis has been further confirmed by the finding of ferromagnetism in undoped TiO$_{2-\delta}$\cite{DMO_TiO2_d0}, HfO$_2$\cite{DMO_hfo2_d0} and other\cite{DMO_various_d0} closed-shell oxides (CaO, ZnO, SrTiO$_3$, MgO, LaAlO$_3$, LSAT, Al$_2$O$_3$) and non oxide materials(AlN, GaN, proton-irradiated graphite); these materials are classified as ``d$^0$ magnets''. In each of these compounds, a sort of structural disorder is needed to observe the FM\cite{cao_theory} and the required density of defects is usually achieved with non-equilibrium growth methods. Moreover, the measured FM in d$^0$ thin films is usually highly anisotropic, being much stronger with an external field parallel than perpendicular to the surface.

From a theoretical point of view, DMO's and d$^0$ FM poses several problems; first of all, magnetic interaction appears at doping-level far beyond the percolation threshold in the lattice: in other words, assuming an homogeneous distribution of magnetic centers, dopant ions (or V$_O$) should not interact, because their average distance is far too high for super-exchange interaction to take place. The RKKY magnetic coupling mechanism is expected to work only in the presence of a metallic conduction. A ``localized'' mediator to magnetic interaction could be given by trapped electronic states at V$_O$ sites, while cationic vacancies are known to carry a finite magnetic moment due to the open shell structure\cite{cao_theory}. The kind of defect that is responsible for the magnetism is still unknown.

From an experimental point of view, dealing with crystalline defects in oxides can be rather challenging. The usual way\cite{tio2_reoxidation} to characterize the defects in TiO$_2$ is to perform measurements (in particular, transport, XPS and EPR) on highly annealed samples (up to 1300 $^\circ$C) that cause the oxygen desorption. Upon annealing, transparent insulating TiO$_2$ crystals usually undergo a pronounced color change (up to dark blue) and to an increased conductivity. Annealed and re-oxidized films are often used in TiO$_2$ photoemission surface studies in order to avoid charging effects, assuming that the annealing in oxygen could restore the full stoichiometry at the surface (for a thickness larger than the probing depth of the selected technique). However, it has been pointed out that bulk defects can play an important role in the surface structure\cite{Die2003} and, for example, that the sample preparation is thus crucial when performing resonant photoemission experiments\cite{thomas}.

\begin{figure}
\begin{center}
\includegraphics[width=0.6\textwidth]{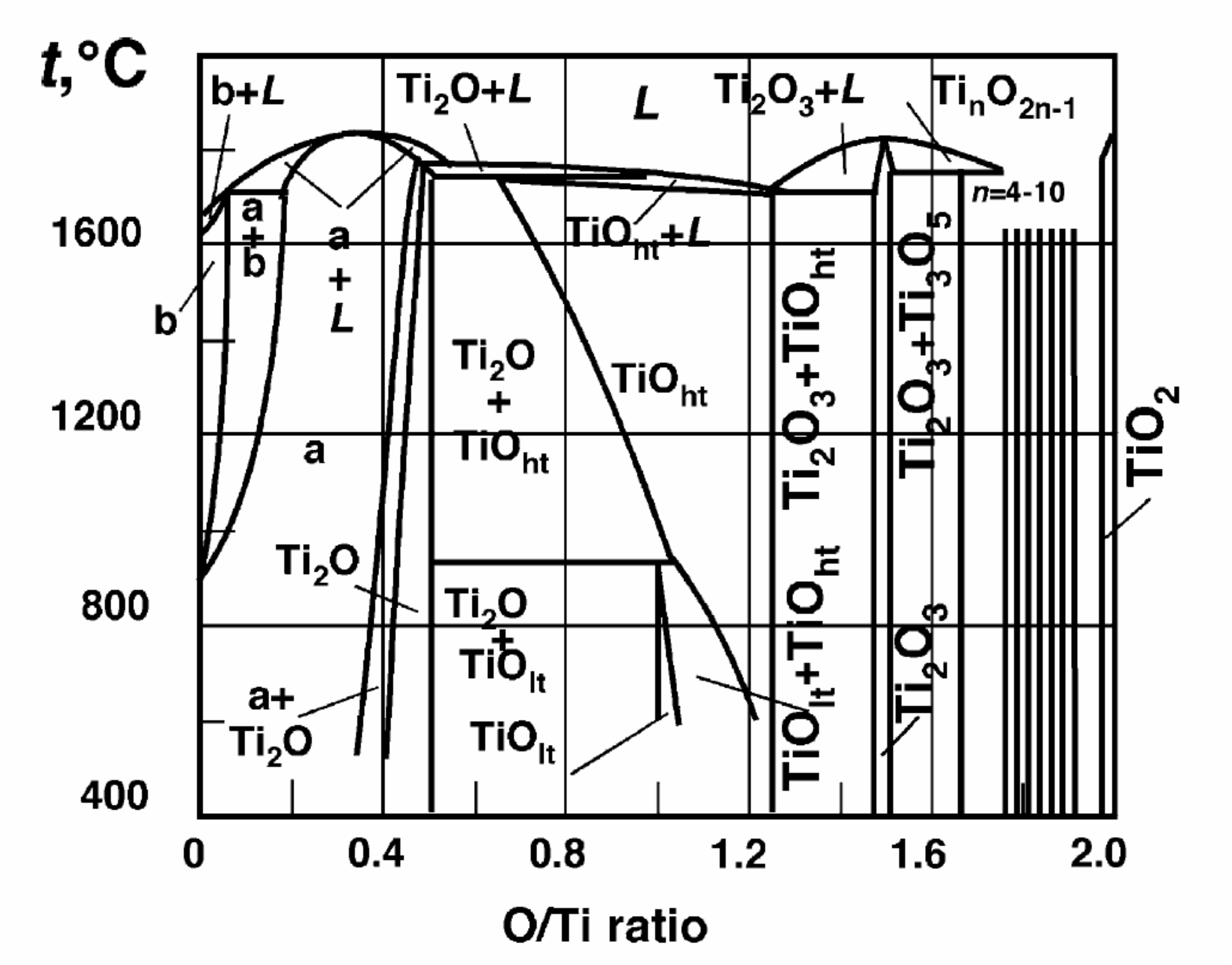}
\caption{The TiO$_x$ phase diagram as a function of x and of the annealing temperature. The vertical lines in the $1.5<x<2.0$ range correspond to the Magn\'e li phases (Ti$_n$O$_{2n-1}$). Taken from Ref.\cite{OxideHandbook}.\label{fig_tio_phases}}
\end{center}
\end{figure}

The presence of oxygen vacancies, interstitial Ti, clustered defects and planar dislocations have been observed\cite{OxideHandbook}, but the question of which kind of defect is prevalent for a defined stoichiometry is still under debate. For example, it has been proven that interstitials Ti (and not the oxygen vacancies) migrate on the surface during the re-oxidization process\cite{tio2_reoxidation}. The Ti-O phase diagram, as a function of the Ti-O ratio, is rich of secondary phases (see Fig. \ref{fig_tio_phases}); in particular, some elements of the homologous Ti$_n$O$_{2n-1}$ series of the Magn\'eli phases (indicated by the vertical lines on the right part of Fig. \ref{fig_tio_phases}), ranging from TiO$_2$ to Ti$_2$O$_3$, can display magnetic properties.

Furthermore, these magnetic effects are sometimes so weak (in terms of saturation magnetization) that a lot of effort has to be made in order to control the possible external contamination on the samples properly. For instance, a clean ceramic sample heater has to be used during the growth and post-annealing treatment, since a metallic one could cause metallic ion to diffuse into the sample. The samples have to be handled with (clean) ceramic or titanium tweezers, and can't be mounted on sample-holder with adhesive tape in order to avoid external contaminations. The XRD spectra have to be collected carefully in order to evidence even small peaks due to metallic inclusions. When working with powders or deposition targets purchased by chemical industries, an evaluation of the impurities at the ppm range should carried out, in order to cross-checking the declared composition.

It is generally believed that, taking the necessary caution, a ferromagnetic phase is really present in these oxides. Even if the possible application of these materials seems unlikely (despite the high T$_c$) the puzzle of this kind of magnetism is challenging from both the experimental and theoretical point of view and deserves further investigations.

In this chapter, a series of magnetic measurements on several TiO$_{2-\delta}$ thin films is given. Hysteresis loops have been recorded with SQUID on ``as deposited'' sample, after an annealing in vacuum (in order to induce more V$_O$), and finally after an annealing in a O$_2$ atmosphere (in order to recover the stoichiometry). Samples have then been characterized with XPS on Ti 2p peaks but, as will be shown in the next sections, the effect of such small quantities of V$_O$'s can't be detected with normal photoemission and ResPES seems to be the proper technique to address this problem.Thus a comparison of ResPES data on rutile TiO$_{2-\delta}$ and N-doped TiO$_{2-\delta}$ thin film is given. Experimental results are compared to DFT+U calculation on TiO$_2$ and N-TiO$_2$ supercells in the last chapter's section.

\section{Growth method and characterization}
The titanium oxide films have been grown with RF-sputtering from a TiO$_2$ target on quartz substrate and, in order to improve the sample crystallinity, on oriented LSAT(001) and Al$_2$O$_3$(1102) surfaces. On the quartz, the control over the TiO$_2$ crystal phase (anatase or rutile, see Fig. \ref{fig_rutilo_anatase}) was achieved by checking the temperature and the Ar/O$_2$ ratio in the gas mixture. The growth on perovskite LSAT ((LaAlO$_3$)$_{0.3}$(Sr$_2$AlTaO$_6$)$_{0.7}$) results in fully Anatase phase samples and the growth on sapphire in the rutile phase, without the need of additional O$_2$ gas during the growth. While rutile is the most thermodynamical stable crystal structure, often thin films are grown in the anatase form; both are expected to display ferromagnetism. In order to evidence interface or surface effects, a ``thick'' and a ``thin'' (less than 60 nm) film have been grown for each substrate.  A list of the samples is given in Table \ref{tab_rf_sputtering}.

\begin{table}
\begin{center}
\begin{tabular}{ccccccc}
  \hline
    Substrate&Temp.&Pressure&O$_2$\%&Phase&Dep. Time&Thickness\\
     &$^\circ$C&mBar&&&min&nm\\\hline
  Quartz & 650 & 5$\cdot$10$^{-3}$ & 9 & Anatase & 45 & 27.4\\
  Quartz & 650 & 3.7$\cdot$10$^{-3}$ & 9 & Anatase & 165 & 83\\
  Quartz & 750 & 4$\cdot$10$^{-3}$ & 0 & Rutile & 30 & 57.7\\
  Quartz & 750 & 4$\cdot$10$^{-3}$ & 0 & Rutile & 120 & 226\\
  LSAT & 750 & 5$\cdot$10$^{-3}$ & 0 & Anatase & 135 & 317\\
  LSAT & 750 & 5$\cdot$10$^{-3}$ & 0 & Anatase & 20 & 47\\
  Al$_2$O$_3$ & 750 & 5$\cdot$10$^{-3}$ & 0 & Rutile & 135 & 220\\
  Al$_2$O$_3$ & 750 & 5$\cdot$10$^{-3}$ & 0 & Rutile & 20 & 33\\
  \hline
\end{tabular}
\caption{List of TiO$_2$ samples grown with RF-sputtering for magnetic measurement.\label{tab_rf_sputtering}}
\end{center}
\end{table}

The samples grown in an higher Ar pressure ($<$ 2$\cdot$10$^{-3}$) or in a lower pressure \@ 650$^\circ$C substrate temperature resulted in a mixture of rutile and anatase phases. The magnetization of each sample has been measured with a SQUID magnetometer after the growth process (labeled as ``as-deposited'' films), after a 4 hours annealing in vacuum (10$^{-5}$ mBar) at 400 $^\circ$C (``reduced'' films) and after the same annealing treatment in a O$_2$ atmosphere (``oxidized'' films). It has to be noted that the annealing temperature was too low to induce structural modifications: the deposited films of table \ref{tab_rf_sputtering} were perfectly transparent and insulating, without any observable color change.

Since photocatalysis is the most important TiO$_2$ application, a nitrogen-doped rutile TiO$_2$ sample has also been grown. Following the suggestion of Elfimov et al.\cite{ndoped_oxide}, an increased magnetism could also be expected in oxides with substitutional nitrogen ions; a structural and magnetic characterization has then been performed also in this case. The details of the growth parameter and the SQUID measurement are given in the dedicated section of this Chapter.

The crystalline order has been checked with XRD using a Bruker D8 Advance diffractometer and with Micro-Raman using a Labram-Dilor spectrometer. The sample growth, the magnetic and the structural characterization have been done at the Universit\'a degli Studi di Pavia. A Raman study on several TiO$_2$ samples grown in the same laboratory has already been published (see Ref.\cite{tio2_raman}).

\section{Magnetic characterization}
\begin{figure}[h!]
\begin{center}
\includegraphics[width=0.6\textwidth]{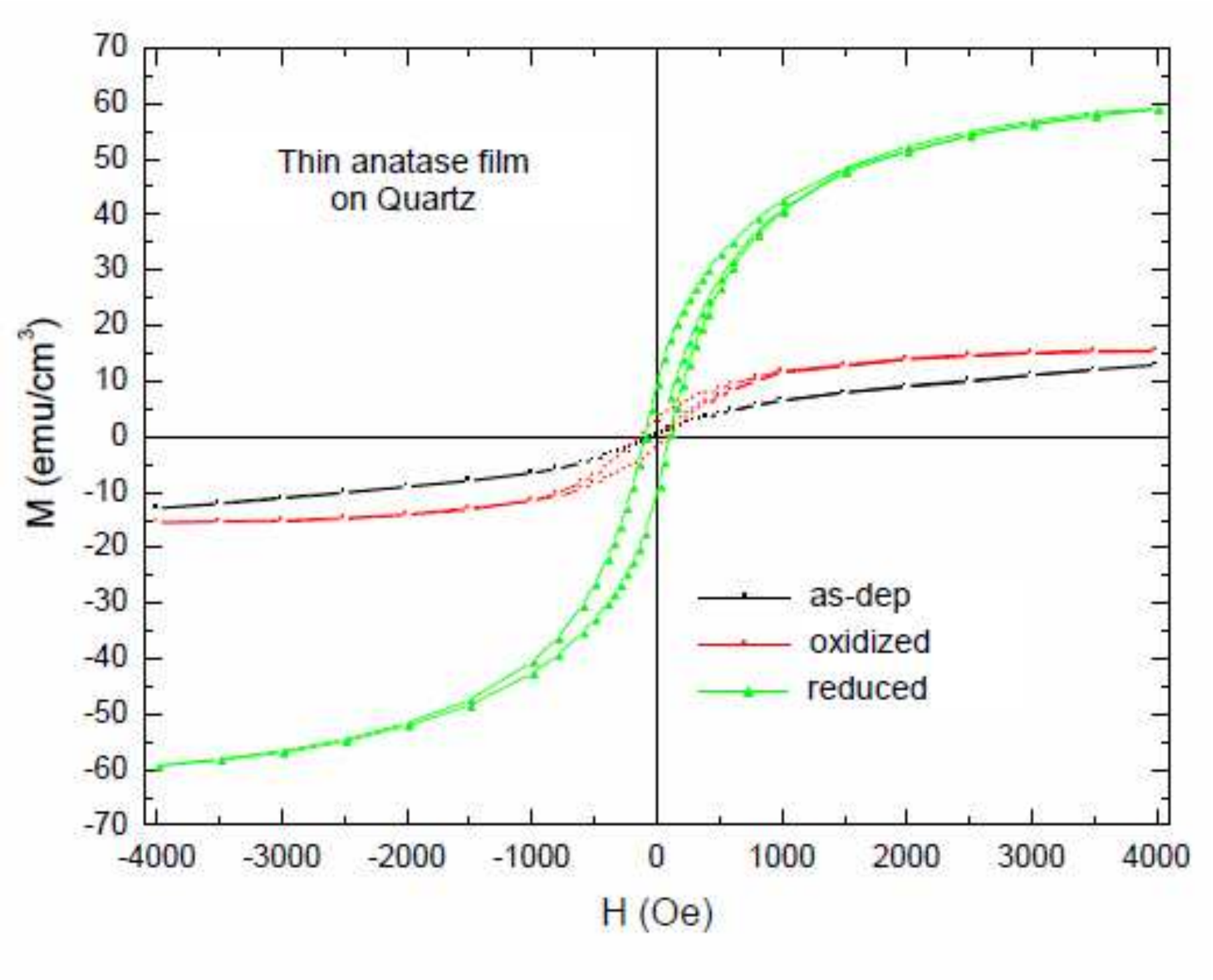}
\caption{Magnetic moment vs external field measured on the thin anatase film deposited on quartz, taken at T=300 K. This graph shows the ``raw'' SQUID data, with the ferromagnetic and paramagnetic contribution.\label{fig_quartz_anatase_MvsH}}
\end{center}
\end{figure}

Static molar magnetization (M$_{mol}$) loops were collected at room temperature for magnetic fields (H) ranging between 0 and $\pm$4000 Oe with a superconducting quantum interference device (SQUID). Reference measurements have been done also on the SQUID sample holder and on the substrates, in order to evaluate only the contribution of the deposited films. A typical plot of the magnetic moment versus the external magnetic field is shown in Fig. \ref{fig_quartz_anatase_MvsH}, which represents the raw data as measured by SQUID on the thin anatase sample on quartz.

A hysteresis cycle is detectable both in the as-deposited and treated films and it is superimposed to a paramagnetic (PM) (or diamagnetic) linear background. In fact, high-quality single-crystal TiO$_2$ should display diamagnetism (DM), which is indicated by a negative slope of M(H); the presence of oxygen vacancies due to the growth or induced by the vacuum annealing may yield a paramagnetic contribution that reduces or even eliminates the diamagnetism. The hysteresis loop of each data-set was extracted by removing the background obtained trough a linear fitting on the high magnetic field part of spectra.

\begin{figure}
\begin{center}
\includegraphics[width=0.9\textwidth]{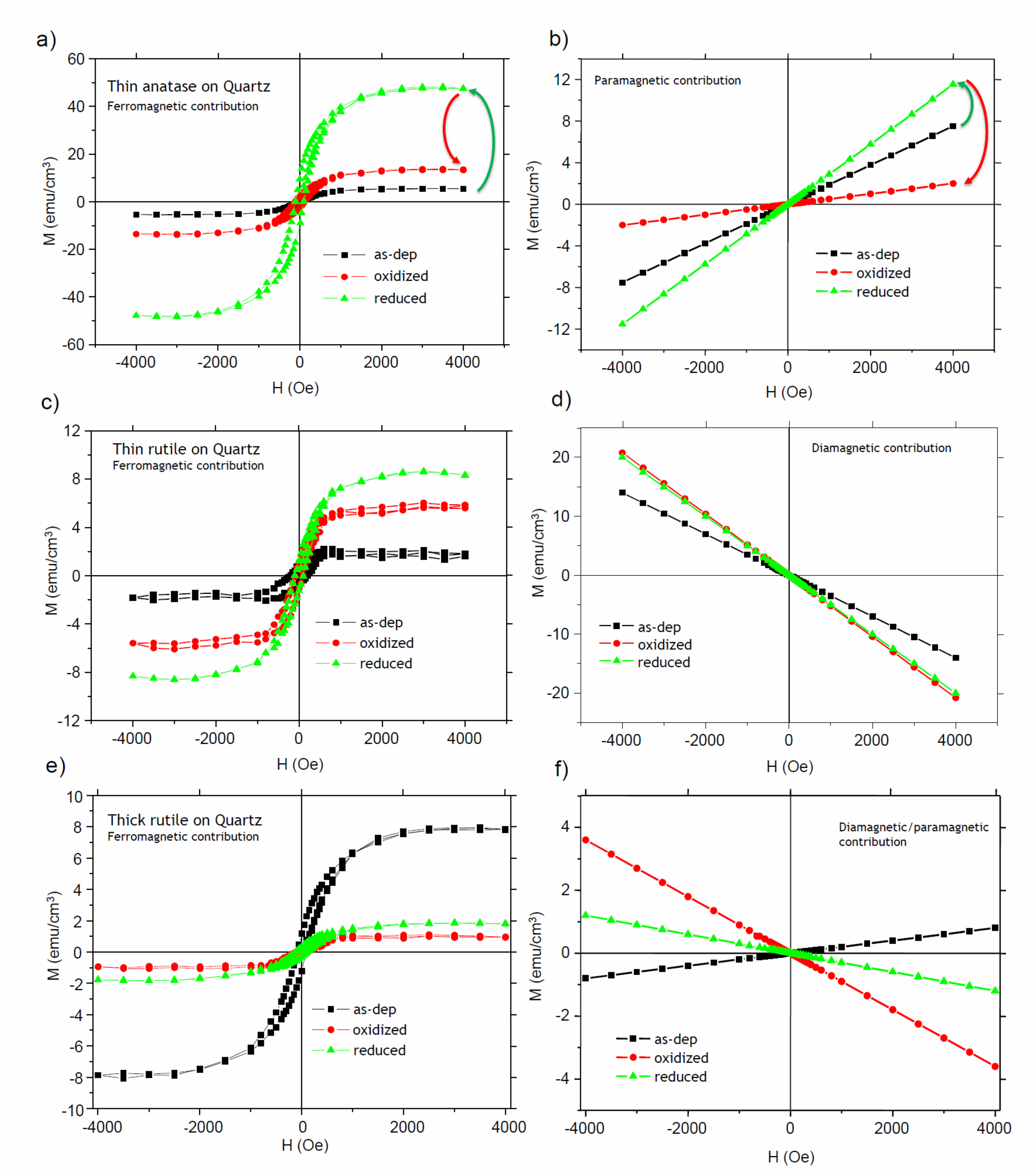}
\caption{Magnetic moment vs external field measured on the TiO$_2$ films deposited on the quartz substrate; in (a)-(b), the thin (27.4 nm of thickness) anatase sample; in (c)-(d), the thin (57.7 nm) rutile; in (e)-(f) the thick (226 nm) rutile. On the left panel are shown the ferromagnetic contribution, on the right panel the paramagnetic or diamagnetic one. The effect of the first (second) annealing is marked by the green (red) arrows on panels (a)-(b).\label{fig_quartz_MvsH}}
\end{center}
\end{figure}

A summary of magnetic measurement of TiO$_2$ film grown on quartz substrates is shown in Fig. \ref{fig_quartz_MvsH}. The data are already separated in the ferromagnetic and paramagnetic (or diamagnetic) contribution. The as-deposited thin rutile film is diamagnetic (as expected for TiO$_2$), while the thin anatase and thick rutile display paramagnetism, which indicates the presence of defects due to growth. The green arrows shown in Fig. \ref{fig_quartz_MvsH}(a)-(b) underline the effect of the first in-vacuum annealing, which increases both the saturation magnetization and the paramagnetic contribution, the latter identified by a counter-clockwise rotation of the linear background. In all cases, the subsequent annealing in oxygen, marked by the red arrows, restores (or improves) the crystalline order of the all samples, quenching the ferromagnetism.

The samples have been carefully manipulated with ceramic tweezers, and the annealing has been done with a proper ceramic heating stage. In fact, assuming that the magnetism were due to the diffusion of FM particles from the heater, the magnetism should have increased also after the second annealing (the one in oxygen atmosphere), which is not the case. The data shown in \ref{fig_quartz_MvsH} thus clearly display the importance of oxygen vacancies in the magnetism of TiO$_2$.

It is possible to evaluate the number of magnetic ions per unit formula from the saturation magnetization. The total molar magnetic moment is given by the following expression:

\begin{equation}
\label{eq:ferromagnetic_M}
M=n\mu_B g S
\end{equation}
\\
where n is the molar density of magnetic ions, g the Land\'e factor and S the spin. In the case of a Ti$^{3+}$ ion, the expected spin is 1/2. In table \ref{tab_quartz_msat} the evaluated film stoichiometry in term of TiO$_{2-\delta}$ is shown, extracted from the data of Fig. \ref{fig_quartz_MvsH}.

\begin{table}
\begin{center}
\begin{tabular}{lccc}
  \hline
    Sample&$\delta_{as-dep}$&$\delta_{reduced}$&$\delta_{oxidized}$\\\hline
  Thin anatase & 0.037 & 0.312 & 0.074\\
  Thin rutile & 0.015 & 0.059 & 0.044\\
  Thick rutile & 0.052& 0.014 & 0.007 \\
  \hline
\end{tabular}
\caption{Value of $\delta$ in the TiO$_{2-\delta}$ films grown on quartz, estimated from the value of the saturation magnetization, using the Eq. \ref{eq:ferromagnetic_M}.\label{tab_quartz_msat}}
\end{center}
\end{table}

Apart from the reduced thin anatase (with a theoretical TiO$_{1.7}$ composition, in which a presence of Magn\'eli secondary phases is expected), the values displayed in table \ref{tab_quartz_msat} are exceedingly low respect to the usual density threshold associated to magnetic properties in diluted samples.

In fact, for a random distribution of dopants, a long-range order (like FM) appears when the size of the larger cluster of interacting atoms is comparable to the size of the material. At fixed interaction radius, the density for which the long-range interaction appears is given by the solution of the so-called percolative problem. The percolative threshold depends on the system geometry (lattice type and dimensionality) and represents a way to simulate the effect of disorder in diluted system. The the magnetism of a Mn doped CdTe can be fairly understood by the percolation of interacting Mn ions in the CdTe lattice (see Ref.\cite{MnCdTe_Theory} and Ref.\cite{aplMnCdTeQW} for a 2D case). Values for the percolative threshold in various 3D lattice (including anatase) are given in the appendix of Ref.\cite{hfo2_theory}.

Indeed, the densities of magnetic centers listed in Table \ref{tab_quartz_msat} are much lower than the percolation threshold in rutile or anatase lattice. This is a common result of d$^0$ ferromagnetic material, and justifies the cautious behavior of the research community about claims of FM ordering. Surprisingly, the thick rutile sample seems to display (in average) a smaller ferromagnetism respect to the thin one. As will be shown, this behavior can be found at higher extent in the oriented substrate films.

In principle, it should be possible to evaluate the number of magnetic centers that contribute to paramagnetism (or diamagnetism), by fitting the slope of the linear contribution with a Curie-type law; however, as paramagnetic effects can also be found in the substrates after annealing treatments, this evaluation is not reliable for the present systems.

\begin{figure}
\begin{center}
\includegraphics[width=0.9\textwidth]{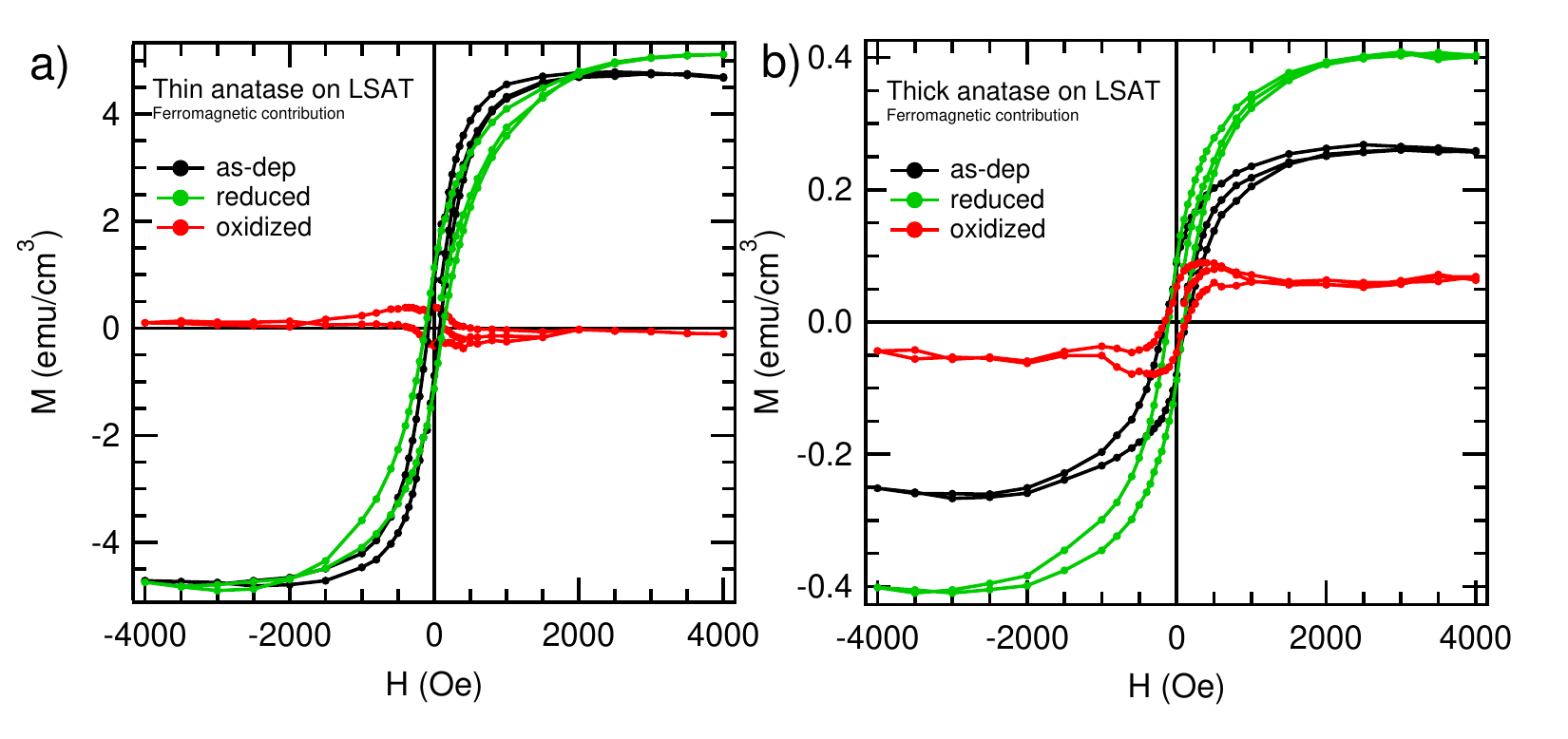}
\caption{Magnetic moment vs external field measured on the thin(a) and thick(b) film grown on LSAT. In black the as-deposited samples, in green the reduced one and in red the oxidized films.\label{fig_anatase_MvsH}}
\end{center}
\end{figure}

\begin{figure}
\begin{center}
\includegraphics[width=1.0\textwidth]{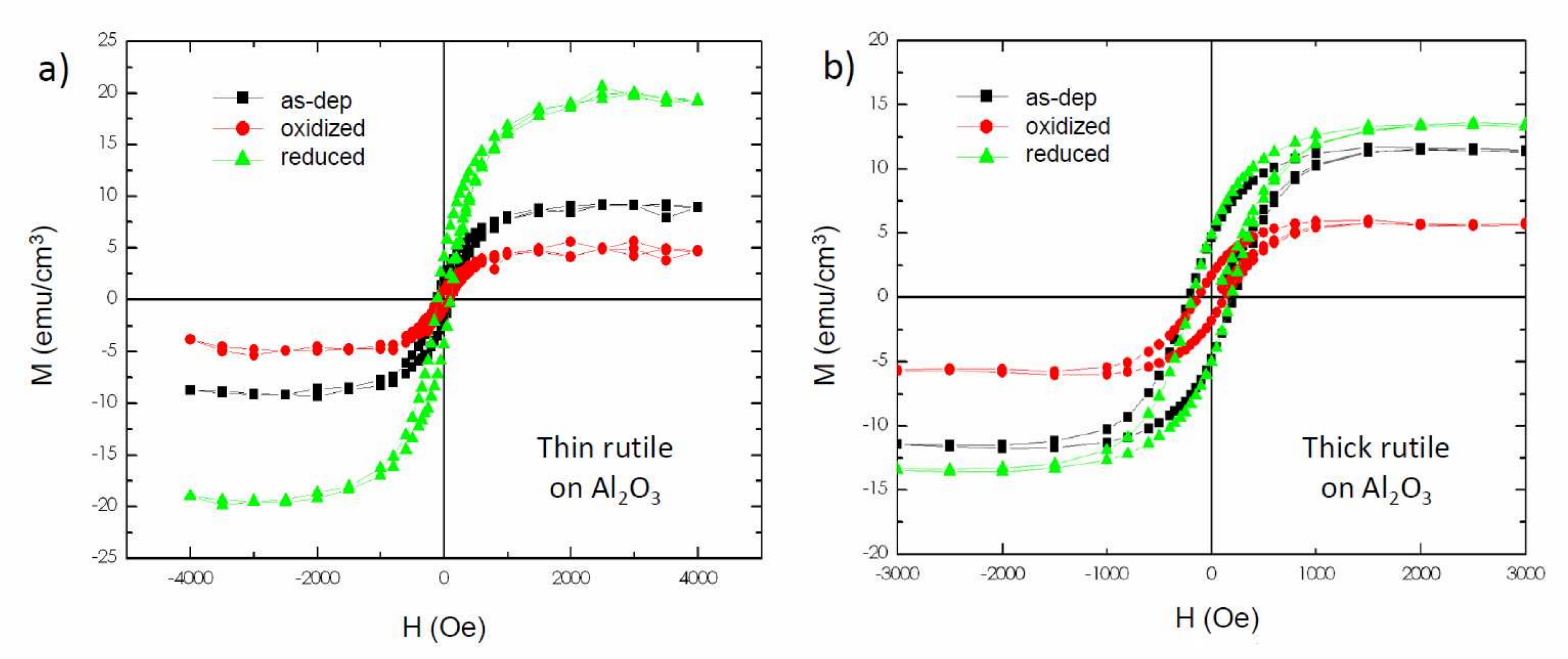}
\caption{Magnetic moment vs external field measured on the thin(a) and thick(b) rutile film grown on Al$_2$O$_3$(1102). In black the as-deposited samples, in green the reduced one and in red the oxidized films.\label{fig_rutile_MvsH}}
\end{center}
\end{figure}

The SQUID data for the anatase films grown on LSAT, already separated from the linear background, are shown in Fig. \ref{fig_anatase_MvsH}, while the data for rutile films on alumina are shown in Fig. \ref{fig_rutile_MvsH}. The results are comparable with the ones obtained on the samples grown on quartz. The annealing procedures results in the expected behavior: the in-vacuum annealing increased the FM by increasing the number of V$_O$'s, while the oxidation resulted in the opposite effect. The absolute saturation magnetization and the relative stoichiometry in terms of TiO$_{2-\delta}$ are listed in Table \ref{tab_ordered_msat}.

\begin{table}
\begin{center}
\begin{tabular}{lccc}
  \hline
    Sample&$\delta_{as-dep}$&$\delta_{reduced}$&$\delta_{oxidized}$\\\hline
  Thin anatase & 0.031 & 0.035 & 0.003\\
  Thick anatase & 0.002 & 0.003 & 0.0003\\
  Thin rutile & 0.051 & 0.136 & 0.034\\
  Thick rutile & 0.078 & 0.088 & 0.040 \\
  \hline
\end{tabular}
\caption{Value of $\delta$ in the TiO$_{2-\delta}$ films grown on LSAT and alumina, estimated from the value of the saturation magnetization, using the eq. \ref{eq:ferromagnetic_M}.\label{tab_ordered_msat}}
\end{center}
\end{table}

The extrapolated density of magnetic ions is of the same order of the one measured in the TiO$_2$ grown on quartz substrate and thus is again too low to justify a long-range magnetic ordering. In particular, in an anatase-type lattice, for a cation-cation interaction the percolation occurs when  27.9\% of Ti atoms have a magnetic moment, by considering only the nearest-neighbor interaction\cite{hfo2_theory}. According to Coey et al.\cite{DMO_TiO2_d0}, the high dielectric constant of these materials can increase the range of magnetic interaction between ions; in the case of HfO$_2$, the cation-cation interaction can be effective up to the fifth-neighbor, which correspond to a doping concentrations of 13.7\%, still higher than the values shown on Table \ref{tab_ordered_msat}.

\begin{SCfigure}
\includegraphics[width=0.55\textwidth]{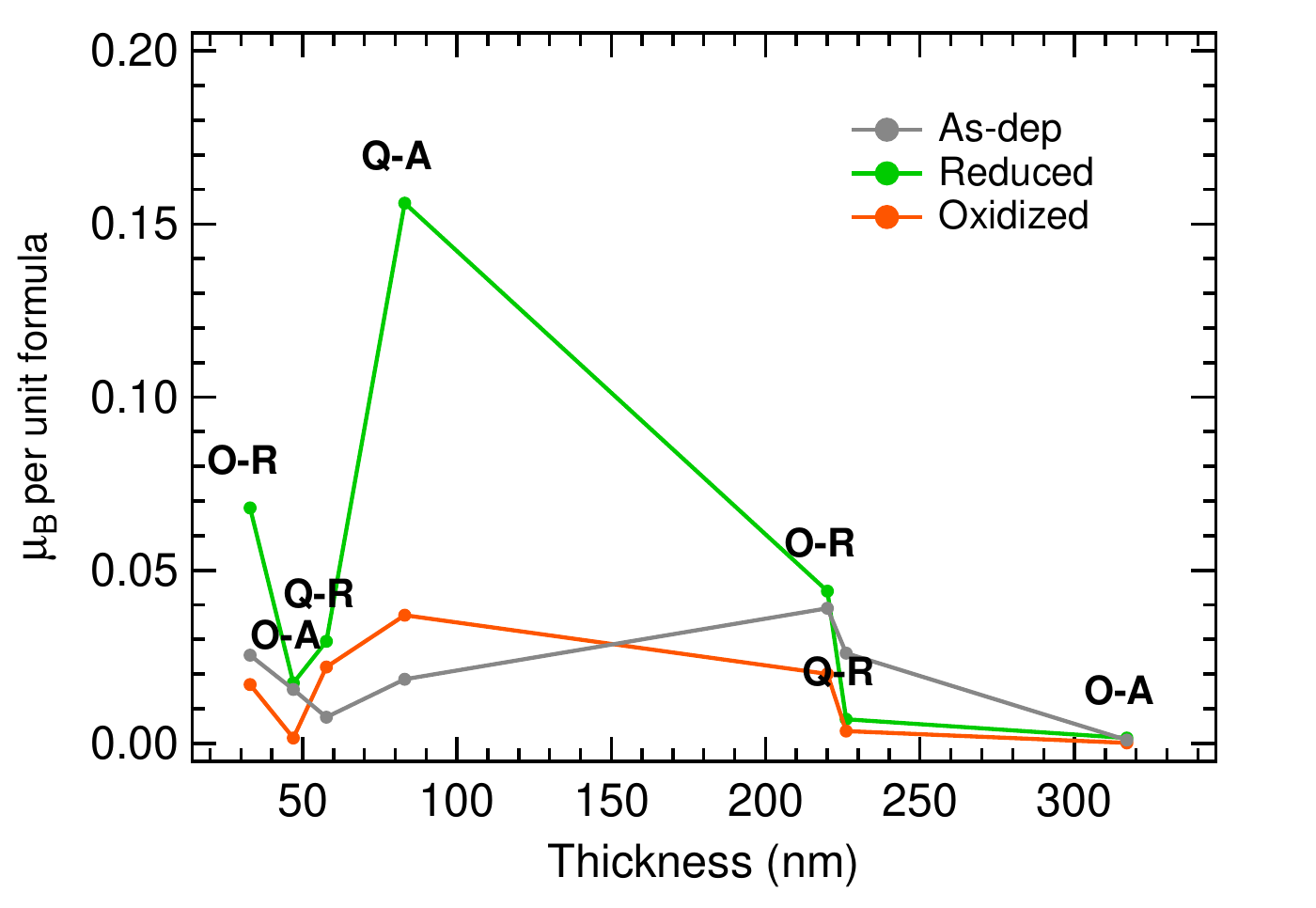}
\caption{Number of Bohr magneton per unit formula versus the film thickness for each sample. The marker label correspond to the sample: Q(O) is for sample grown on Quartz (oriented surface); A(R) is for Anatase (Rutile) film. \label{fig_thick_dep}}
\end{SCfigure}

Nevertheless, the absolute value of the saturation magnetization of our film becomes more significative when compared to a surface or to a 2D-like distribution of spins instead of a 3D one. A proof of a not-homogeneous defects distribution can found by looking at the saturation magnetization in the anatase thick sample of table \ref{tab_ordered_msat}, which is surprisingly low (0.25 emu/cm$^3$) as compared to the thin one. A plot of the estimated number of Bohr magneton per unit formula versus the film thickness for each sample is given in Fig. \ref{fig_thick_dep}. The scattering of the data reflects the disordered nature of the phenomenon. In general, the thin films (especially after the annealing in vacuum) display a higher magnetization than the corresponding thick one. This should not happen in homogeneous diluted ferromagnet, in which the magnetization per unit mass (or volume) is constant.

It should be pointed out that the magnetization has been calculated assuming a uniform distribution of magnetic moments. The difference of m$_{sat}$ of the thin and the thick film can be partly reduced when the films thickness listed in table \ref{tab_rf_sputtering} is considered. An accumulation of defects (V$_O$, V$_{Ti}$ or interstitial O and Ti ions) near the surface or the interface can justify the presence of FM interaction, since the higher local magnetic moment density could reach the percolation threshold. For example, according to the percolation threshold (27.9\%) estimated for nearest-neighbor interaction in the anatase lattice\cite{hfo2_theory}, in the reduced thick anatase film grown on quartz this threshold can be achieved if the total number of magnetic ions were confined in a 1.7 nm thick layer. Furthermore, the theoretical explanations in literature (see Ref.\cite{DMO_TiO2_d0}), discussed at the end of this chapter, suggest that FM is caused by clustered defects. This model could also explain the anisotropic magnetism of d$^0$ thin films.

The location of this ``magnetic'' layer should be near the boundaries of the film, i.e. at the surface or the interface. The reduction of the samples, induced by the annealing, occurs at the surface; however, the magnetism appears also in as-deposited samples, and the lattice mismatch and roughness could induce an higher quantity of structural defect at the beginning of the sputtering process, i.e. at the interface. Moreover, as pointed out in Ref.\cite{tio2_reoxidation}, during the annealing process the reduction of TiO$_{2-x}$ is accompanied by a migration of Ti atoms in the bulk that can increase the density of magnetic moment carriers at the interface.

\section{XPS measurements}
\begin{figure}
\begin{center}
\includegraphics[width=0.7\textwidth]{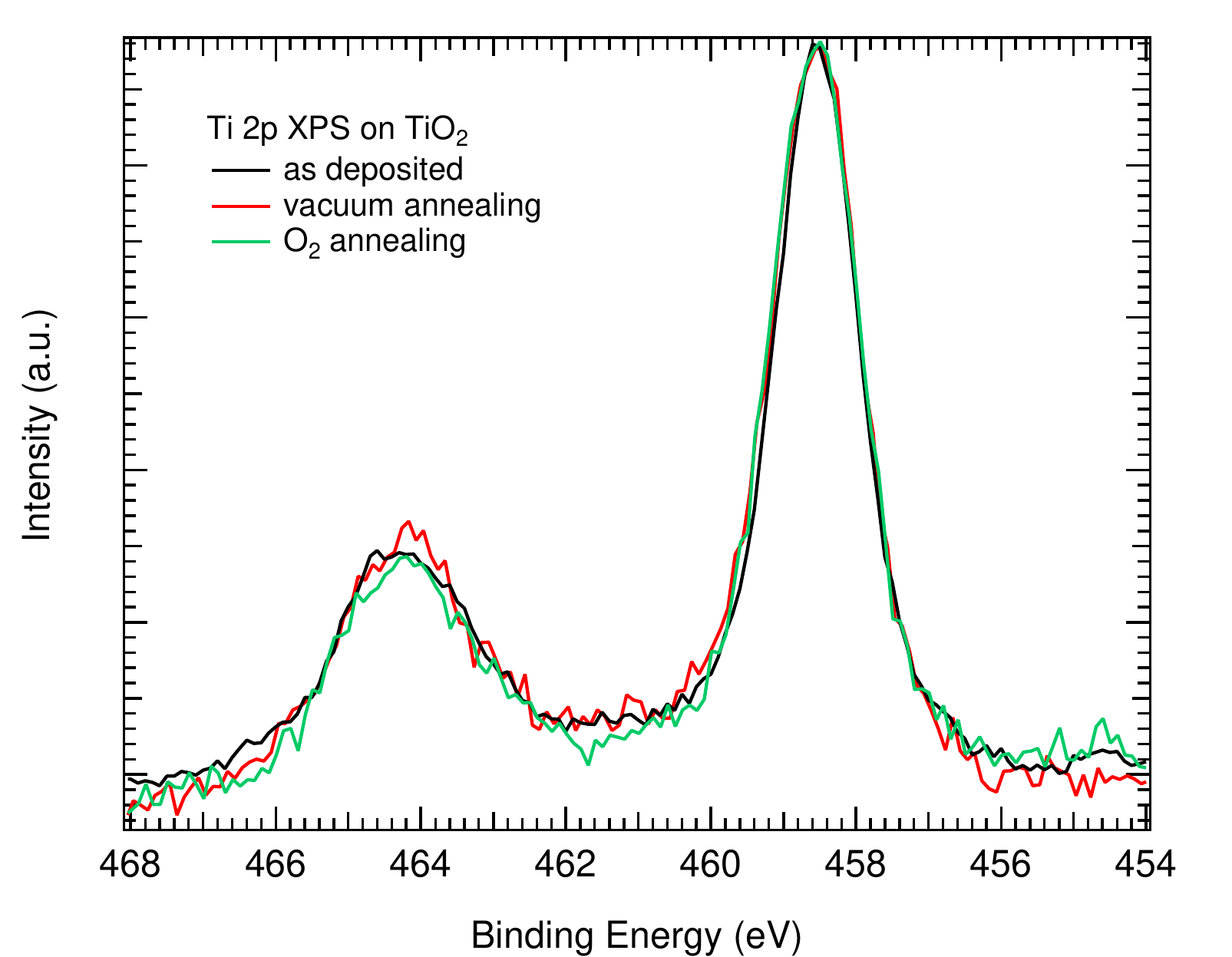}
\caption{XPS spectra of Ti 2p core levels on the thick Rutile film grown on quartz.\label{fig_xps_fqyu8}}
\end{center}
\end{figure}

An XPS analysis has been carried out in order to investigate the possible presence of defects at the surface. The thin and thick rutile sample grown on Quartz and on Sapphire substrate have been measured, both as-deposited, oxidized and reduced. The XPS measurements have been taken with a VG Escalab Mk II apparatus with a modified 32-channel detector and a monochromatized Al $k_\alpha$ x-ray source in the Surface Science lab at the Universit\'a Cattolica of Brescia. The charging effects induced by x-ray exposure have been compensated with an e$^-$ flood-gun. Since the purpose of this study was the characterization of the defects electronic structure, we didn't perform further annealing and sputtering treatment. For transparent and insulating of TiO$_2$ samples, i.e. for such a low amount of oxygen vacancies, it is not possible to directly detect the defect-induced modification in the valence band. However, it should be possible to evaluate the presence of Ti$^{3+}$ states by analyzing the photoemission structure of the Ti 2p core-levels. The results for the thick rutile sample grown on quartz is given in Fig. \ref{fig_xps_fqyu8}.

The three spectra of Fig. \ref{fig_xps_fqyu8} are almost identical; it is not possible to clearly detect the presence of Ti$^{3+}$ related structures, which should appear at lower BE respect to the main peak (2p$_{3/2}$, at BE=458.5 eV). The same results have been obtained for the other samples of Table \ref{tab_rf_sputtering}. Assuming an homogeneous distribution of defects, with the densities listed in Table \ref{tab_quartz_msat} and \ref{tab_ordered_msat}, a difference in spectra should be easily detectable. The typical probing depth of XPS at Ti 2p core-level, with a photon energy of $h\nu=$1486.6 eV is given by the electron inelastic mean free path $\lambda=$2.212 nm, obtained from the TPP-2M formula\cite{TPP2M}. Therefore, a difference in the V$_O$ density at a distance $\lambda$ from the surface should be excluded in these TiO$_2$ film.

\begin{SCfigure}
\includegraphics[width=0.55\textwidth]{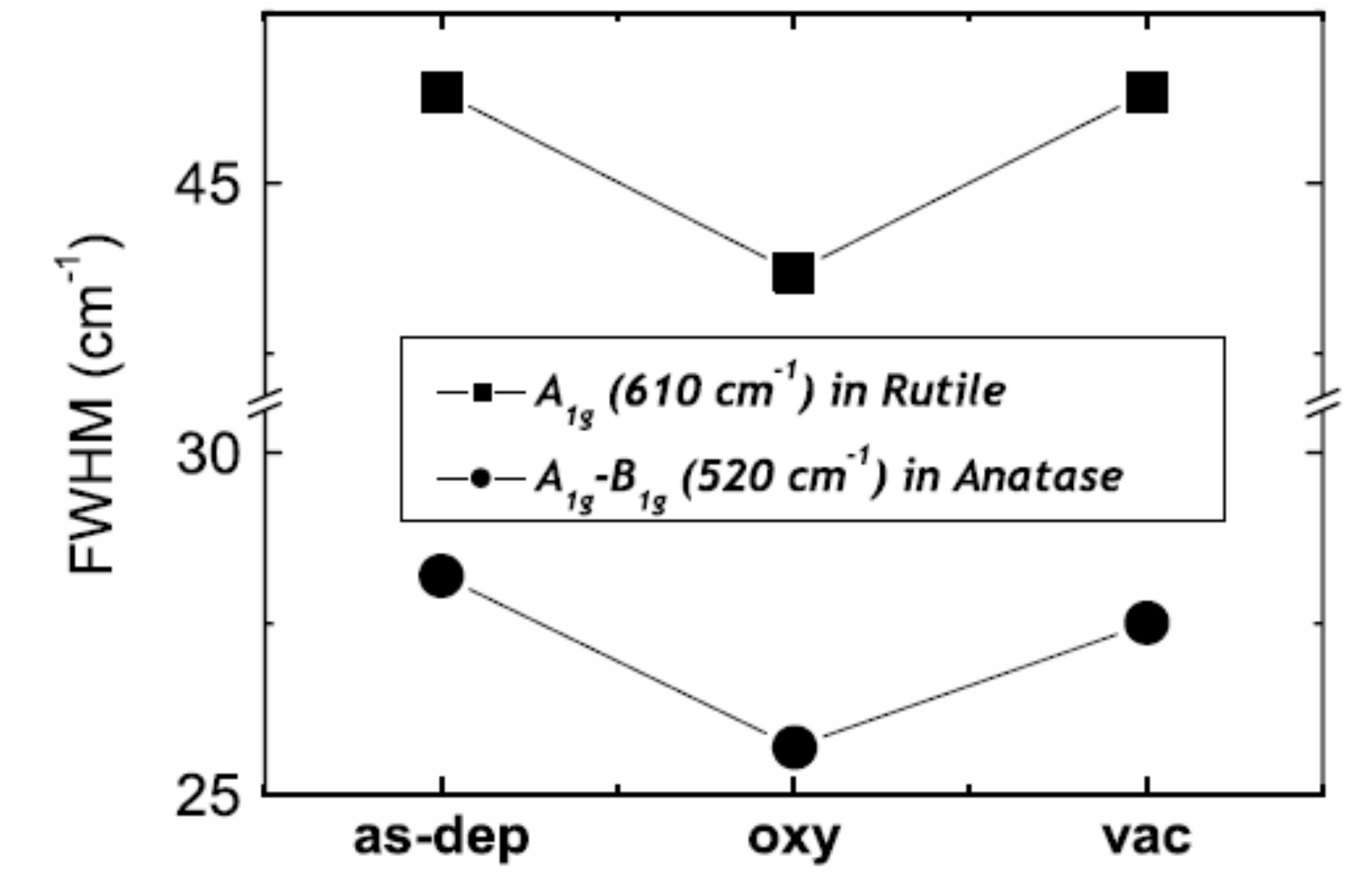}
\caption{Peak width of selected Raman mode for anatase and rutile sample grown on quartz. The in-vacuum annealed samples are labeled as ``vac'', the O$_2$ annealed as ``oxy''. Taken from Ref.\cite{tio2_raman}.\label{fig_raman_width}}
\end{SCfigure}

A more bulk-like technique that can prove the relative disorder of a crystal structure is the Raman spectroscopy. Fig. \ref{fig_raman_width} (Ref.\cite{tio2_raman}) shows the peak width of selected Raman peaks taken on films grown on quartz, as a function of the sample treatment. The Raman modes of fully oxidized samples display a peak width that is considerably lower than the as-deposited or annealed films, as a consequence of the improved crystalline order. These data confirm the effectiveness of the annealing procedure in the creation of defects, even if the annealing temperature (400 $^\circ$C) is considerably lower than the one typically used to induce oxygen vacancies in literature. Together with the XPS data, Raman spectroscopy could support the hypothesis of a localized layer of defects near the interface. The experimental noise and the difficulty of distinguish a small amount of oxygen vacancies in the XPS should also be taken into consideration. We thus resort to Resonant Photoemission (ResPES), in order to probe the electronic structure of even a small concentration of defects near the surface.

Since the ResPES technique requires a synchrotron facility to be performed, and since a typical synchrotron X-ray beam causes much stronger charging effects than normal XPS, only the thinnest rutile TiO$_2$ sample has been probed. The results of this analysis is given in the next sections and have been published in Ref.\cite{APL_ntio2}.

\section{ResPES on TiO$_{2-\delta}$ and N-doped TiO$_{2-\delta}$}

As already pointed out, the typical densities of defects measured in ferromagnetic TiO$_2$ film is usually too low to explain a long-range magnetic interaction. This is particularly true for TiO$_2$ where highly localized deep-gap defect orbitals lead to only a very-short-range defect-defect interaction. For very low densities of defects, these vacancies would fail to percolate through the sample, resulting in a vanishing Curie temperature.
On this basis, a possible explanation for the origin of FM coupling in insulating DMO can be the inhomogeneity of V$_O$ distribution that can be high enough in limited regions to give FM coupling. The point is now to establish whether doping of rutile TiO$_2$ can create better conditions for the insurgence of FM ordering.

\begin{figure}
\begin{center}
\includegraphics[width=0.9\textwidth]{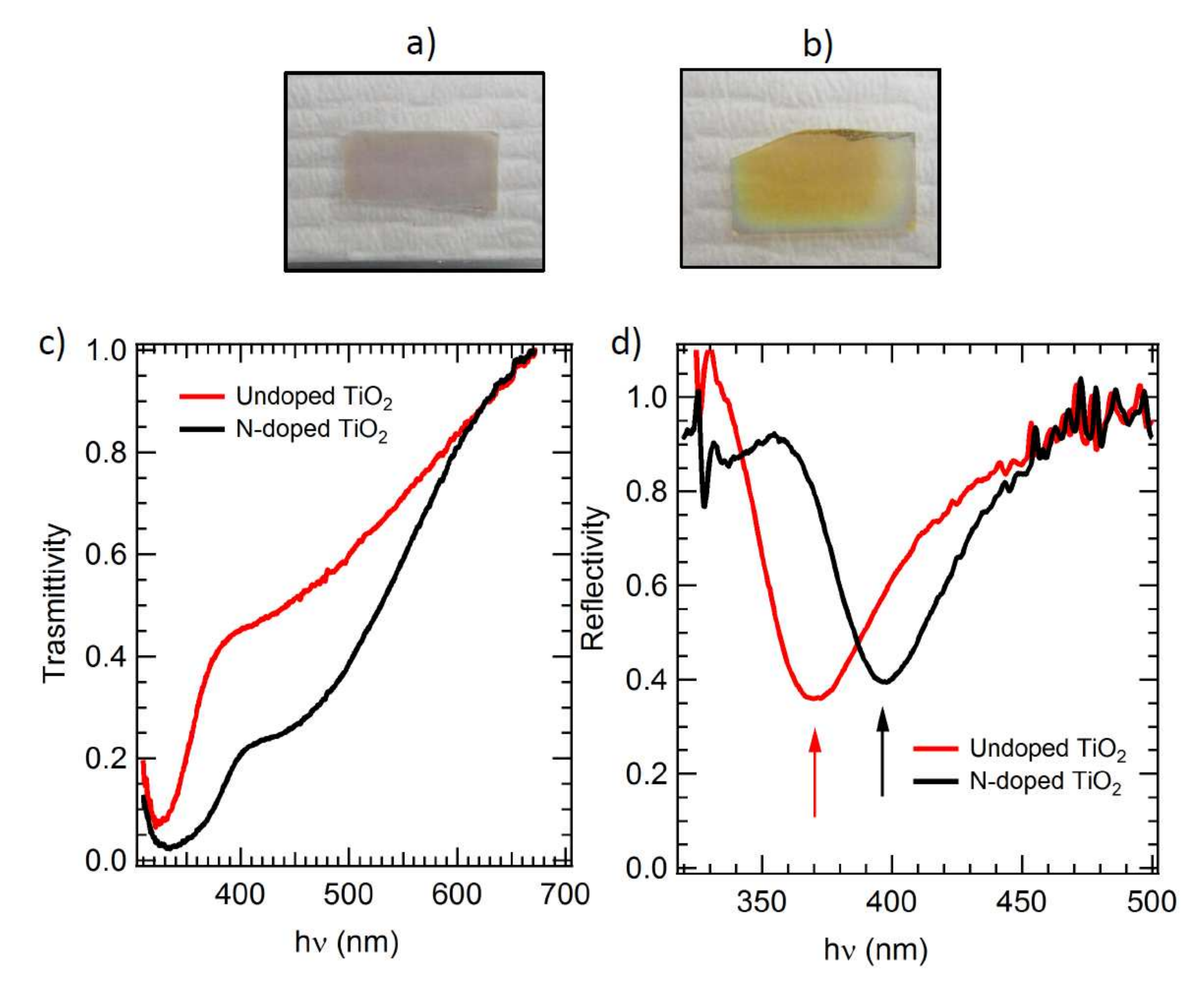}
\caption{Picture of the rutile TiO$_2$ (a) and N-doped TiO$_2$ (b) rutile film; in panel (c), the optical transmittivity; in panel (d), the optical reflectivity.\label{fig_ntio2_trans}}
\end{center}
\end{figure}

In this section, we show the effect of N-doping on the magnetic and electronic properties of TiO$_{2-\delta}$ rutile thin films. The thin films (about 40 nm thick) have been grown by RF-sputtering a TiO$_2$ target on oriented Alumina (0001) substrates at 750 $^\circ$C under a flux of pure Ar (undoped film) or an Ar–N mixture (95\%–5\%, N-doped film), at a pressure of 5$\times$10$^{-3}$ mbar. The crystalline phase of each sample was checked by X-ray diffraction and micro-Raman spectroscopy. While the undoped film is perfectly transparent, the N-doped film display a yellow-gold color (see Fig. \ref{fig_ntio2_trans}(a)-(b)), that can be probed with simple optical transmittivity and reflectivity experiments, shown in Fig. \ref{fig_ntio2_trans}(c)-(d).

The color change reflects the band-gap reduction, from nearly 3.2 eV to 3.0 eV, marked by the arrows in Fig. \ref{fig_ntio2_trans}(d). Many experimental research groups already reported an increased photocatalytic\cite{ntio2_catalysis} activity of TiO$_2$ under sunlight, achieved trough a band-gap reduction upon nitrogen or vanadium doping.

Upon N-doping an enhancement of ferromagnetic ordering in terms of remanent and saturation magnetization is also achieved. By combining soft X-ray spectroscopy and magnetization measurements with ground state density functional calculations, we are able to show that, with respect to the undoped system, N doping does not add electronic states in the region where V$_O$ states are usually found, but introduces additional N-related low-lying acceptor and donor levels through O substitution with N. This ultimately reduces the band gap of rutile providing more favorable conditions for the onset of robust FM ordering in these compounds.

Resonant photoemission spectroscopy (ResPES) and X-ray absorption spectroscopy (XAS) spectra have been collected at the BACH beamline of the Elettra synchrotron in Trieste (Italy). Ground-state density functional theory (DFT) calculations have been carried out, based on the Perdew-Burke-Ernzerhof (PBE\cite{PBE}) parametrization of the generalized gradient approximation (GGA) for the exchange-correlation functional. The projector augmented-wave\cite{PAW} atom description is used, as implemented in the \texttt{ABINIT} (Ref.\cite{ABINIT}) code, adding semicore level to valence states. Calculations have been carried out on a rutile cell (6 atoms/cell) and a 2$\times$2$\times$2 (48 atoms) supercell with one nitrogen atom substitutional to oxygen, corresponding to a x=3.2\% doping level. The k-space grid was a 4$\times$4$\times$4 Monkhorst–Pack grid, and the plane-wave energy cutoff was 25 hartree. We performed a full cell relaxation with a maximum residual force on atoms fixed at 10$^{-4}$ hartree/{\AA}.

\begin{figure}
\begin{center}
\includegraphics[width=0.7\textwidth]{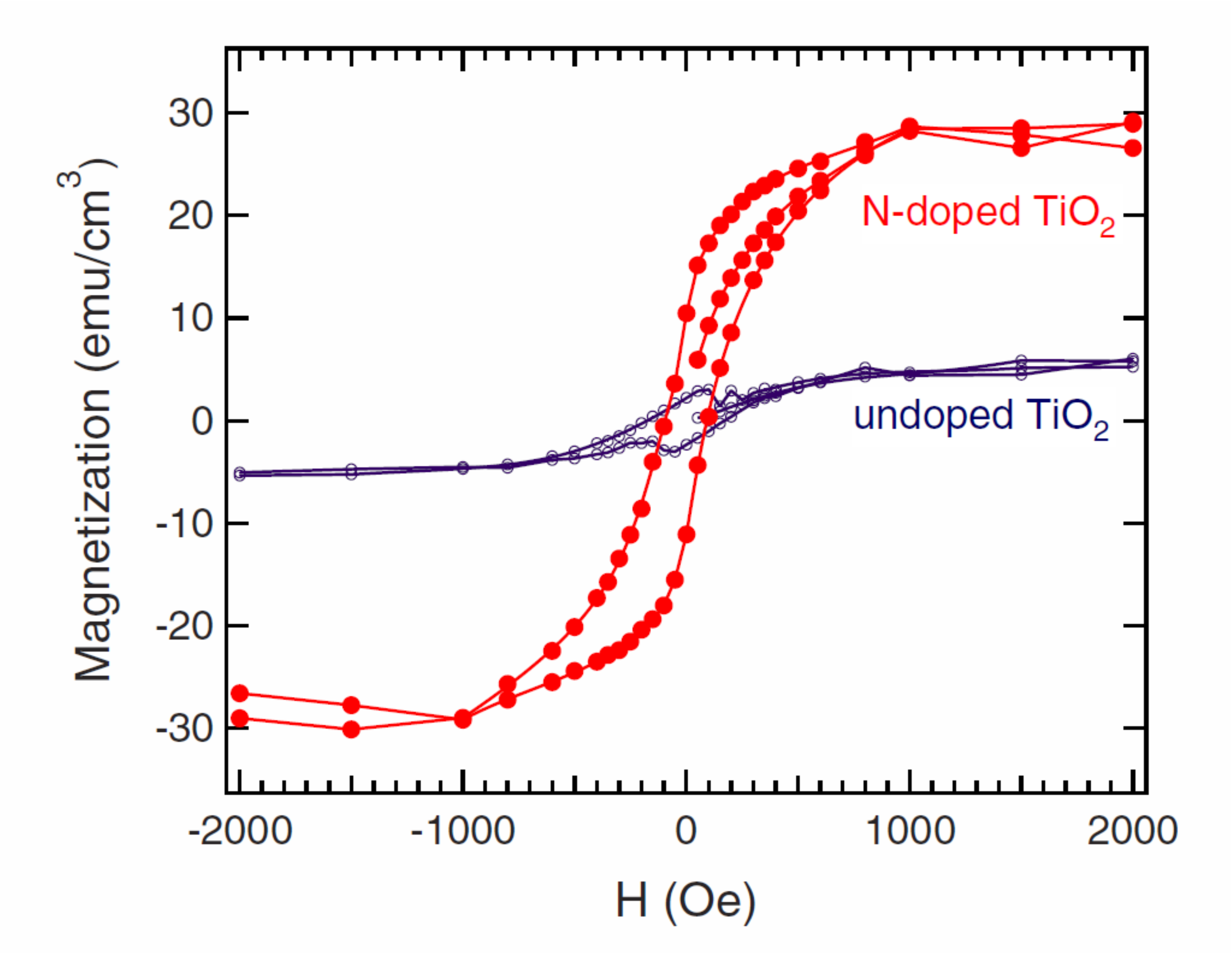}
\caption{Hysteresis loop measured on the undoped TiO$_2$ and on the N-doped TiO$_2$ thin films. The diamagnetic contribution of the substrate, as well as that of the core electrons have been subtracted from both cycles.\label{fig_ntio2_squid}}
\end{center}
\end{figure}

M$_{mol}$ versus H curves collected at RT with the external field H applied parallel to the thin film surface (Fig. \ref{fig_ntio2_squid}) show an hysteresis cycle that mark the presence of a ferromagnetic-like ordering already at RT. For the N-doped sample, saturation is reached at about 30 emu/cm$^3$, a value much higher that the case of the undoped sample (about 5 emu/cm$^3$).

\begin{figure}
\begin{center}
\includegraphics[width=0.9\textwidth]{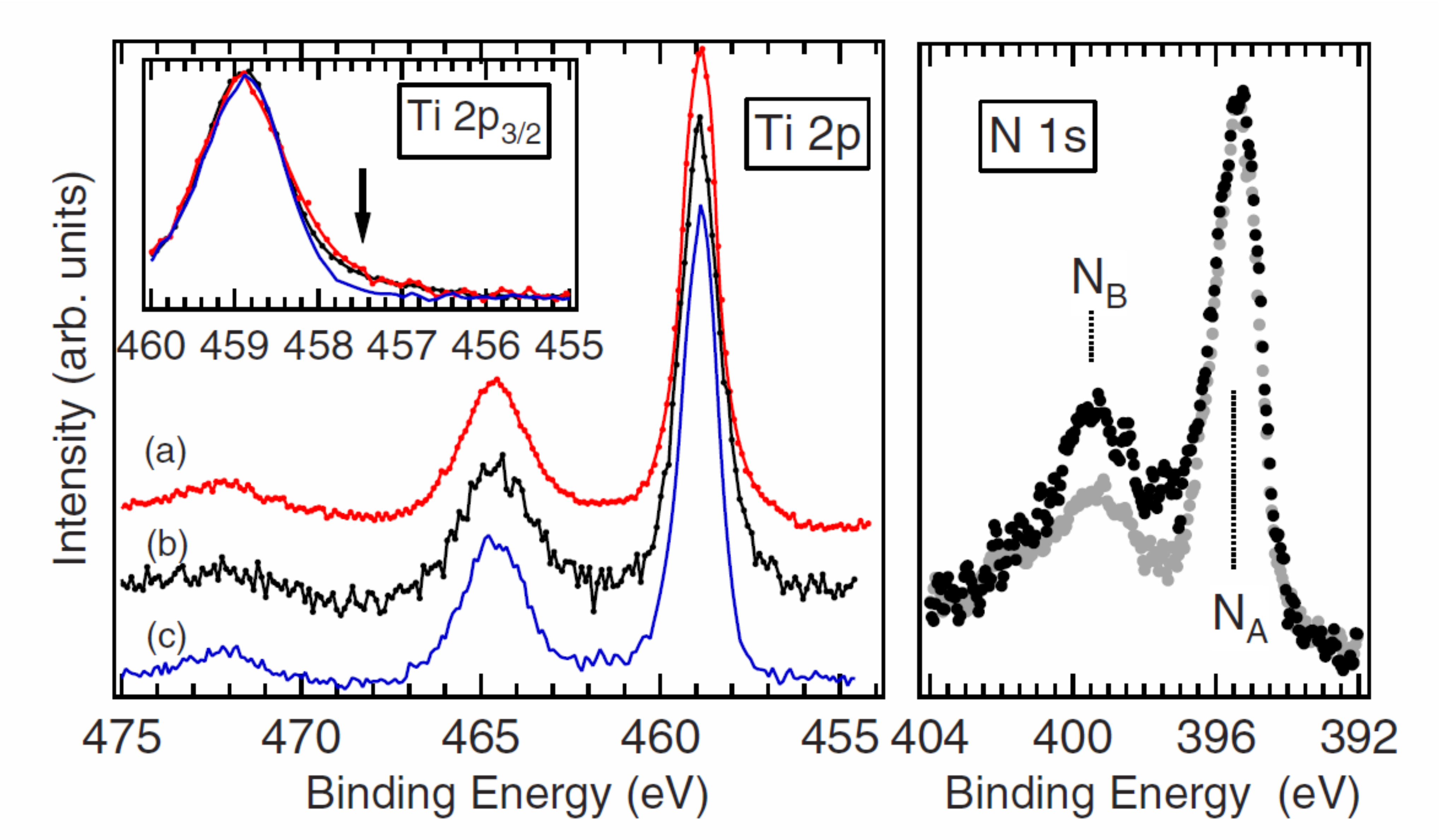}
\caption{Left panel: Ti 2p XPS spectrum from the undoped (a) and the N-doped (b) TiO$_2$ thin film. These spectra are compared with that collected from a stoichiometric TiO$_2$ (110) single crystal surface (c). The inset shows the Ti 2p$_{3/2}$ core line region and the vertical arrow indicates electronic states that can be ascribed to Ti$^{3+}$. Right panel: N 1s photoemission spectrum at normal emission (gray dots) and at 50$^\circ$ off-normal emission (black dots).\label{fig_ntio2_xps}}
\end{center}
\end{figure}

The effect of N on the electronic properties of the doped sample is discussed by examining the Ti 2p and N 1s core level x-ray photoelectron spectroscopy spectra of the undoped and N-doped thin films. Figure \ref{fig_ntio2_xps}, left panel shows the Ti 2p XPS spectrum obtained from the undoped TiO$_2$ thin film and from the N-doped thin film. These spectra are compared with the Ti 2p XPS core lines of a TiO$_2$ rutile single crystal. The Ti 2p spectra show the typical spin-orbit split doublet of the Ti$^{4+}$ ion at a binding energy BE =458.8 eV and BE=464.8 eV. The Ti 2p$_{3/2}$ line shows a shoulder at low BE (inset of Fig. \ref{fig_ntio2_xps}) that is formally assigned to a Ti$^{3+}$ oxidation state (3d$^1$ configuration). This shoulder is absent in the stoichiometric TiO$_2$ single crystal surface. In the present case, it is rather important to observe that, in spite of the N-doping, the Ti$^{3+}$ component, has virtually the same weight in both samples, and therefore, we can assume that the content of 3d$^1$ magnetic ions is the same in both samples.

In the N-doped sample, two components (N$_A$ and N$_B$) are present in the N 1s core level region (Fig. \ref{fig_ntio2_xps}, right panel). The N$_B$ peak at about BE=400 eV can be assigned either to molecular nitrogen(N$_2$) bonded to surface defects or to N bonded to surface O sites\cite{ntio2_chambers}. The intensity of N$_B$ component increases by collecting the spectra at grazing angles, i.e., in the more surface sensitive conditions for the XPS probe, and therefore this peak is ascribed to a surface contamination. In turn, the N$_A$ peak is ascribed to bulk N atoms incorporated during the thin film growth. An estimation of the N content was carried out by considering the area of the Ti 2p, N 1s, and O 1s photoemission peaks, suitably weighted by the photoemission cross section of each atomic level and the analyzer transmission. It is found that the N content in the N-doped film is about 4\%.

\begin{figure}
\begin{center}
\includegraphics[width=0.7\textwidth]{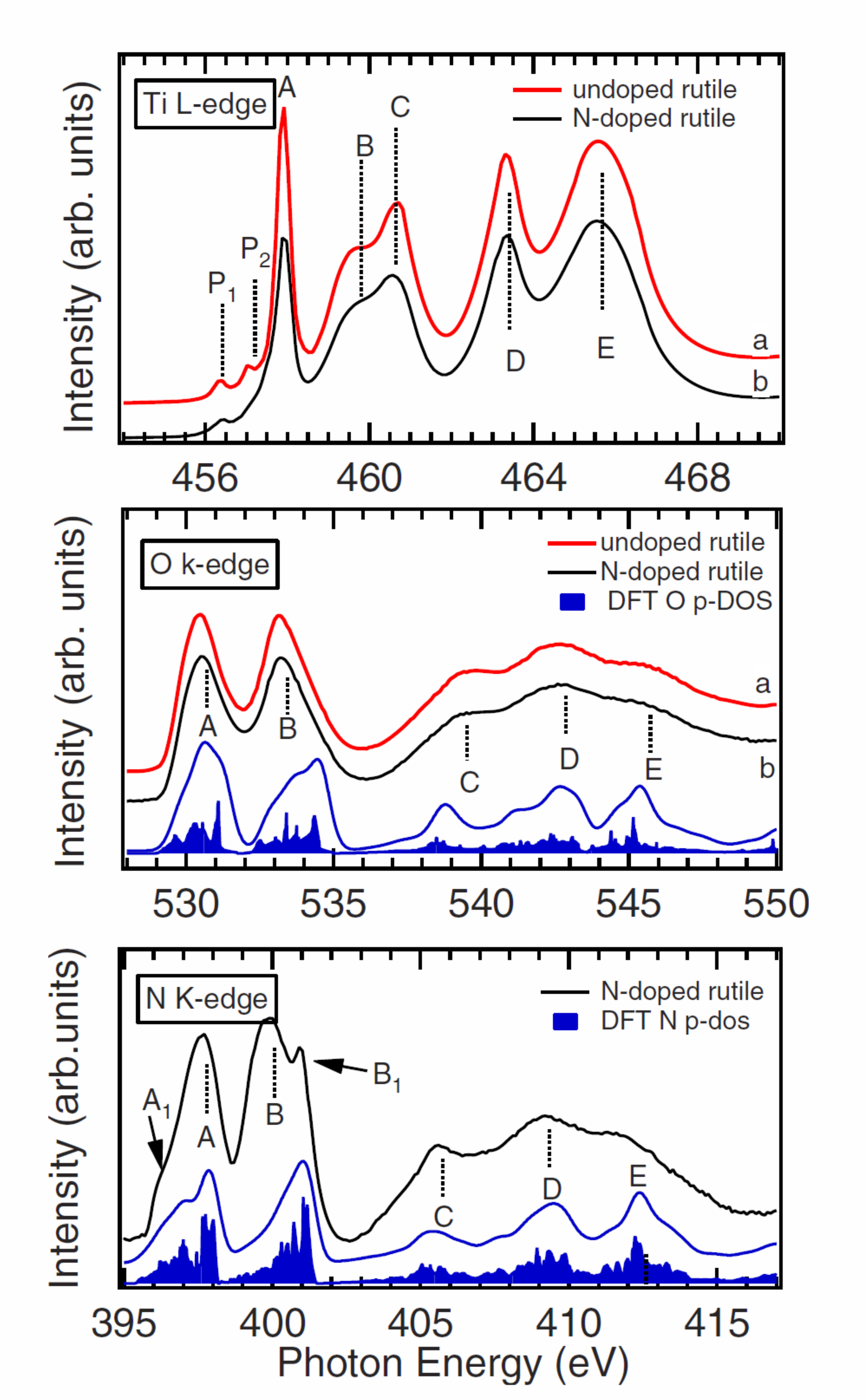}
\caption{Top panel: Ti L$_{2,3}$-edge XAS spectra collected from
the undoped (a) and N-doped (b) rutile thin films. Middle panel: O K-edge
XAS spectra collected from the undoped (a) and N-doped (b) rutile thin
films, compared with the undoped cell DFT calculation. Bottom panel:
N K-edge XAS spectrum collected from the N-doped rutile thin film, compared
with the supercell DFT calculations. The resolution of the monochromator
was set to 0.24 eV at the Ti L$_{2,3}$-edge photon energy.\label{fig_ntio2_xas}}
\end{center}
\end{figure}

The Ti L$_{2,3}$-edge XAS spectrum is shown in Fig. \ref{fig_ntio2_xas}, top panel. This spectrum is in agreement with those already reported for rutile with a set of bands that are related to the L$_3$ edge (A, B, and C) and to the L$_2$ edge (D and E). The A and D bands are ascribed to t$_{2g}$ states arising from crystal field splitting, while the B, C, and E bands are ascribed to e$_g$ states. The P$_1$ and P$_2$ prepeaks originate from multiplet splitting for the 2p$^5$3d$^1$ final state configuration. With respect to the undoped sample, the N-doped sample shows the same energy and intensity for all peaks but a larger width. The same remarks hold also for the O K-edge XAS spectra (Fig. \ref{fig_ntio2_xas}, middle panel). Both spectra are well reproduced by DFT calculation of the undoped cell. The two spectra are identical but the peaks of the N-doped sample are slightly larger than those of the undoped sample, suggesting an higher degree of structural disorder. The fact that this extra broadening isn't observed in XPS spectra can be related to the different probing depth of XAS and XPS.

Finally the N K-edge XAS spectrum is shown (Fig. \ref{fig_ntio2_xas}, bottom panel). This spectrum is rather similar to that measured for oxygen, presenting a sequence of five bands (A to E). However a closer inspection shows that two additional features (A$_1$ and B$_1$) can be detected and the C, D, and E peaks appear to be shifted with respect to the O 1s XAS spectrum. DFT supercell calculations are able to consistently reproduce most of these features. In particular, the introduction of substitutional N determines a transfer of the spectral weight toward the low photon energies in the calculated A band and toward the high photon energies in the calculated B band, consistently with the observed A$_1$ and B$_1$ experimental features. Therefore, the A$_1$ states can be regarded as N-related, donor, levels. For a very low concentration of V$_O$s the spectroscopic signature of the electronic states can be rather elusive and therefore one has to consider also ResPES techniques to probe their existence and determine the BE. Indeed, for both samples, the study of the occupied electronic states in the valence band has been carried out with ResPES with photon energies across the Ti L$_{2,3}$-edge.

\begin{figure}
\begin{center}
\includegraphics[width=0.8\textwidth]{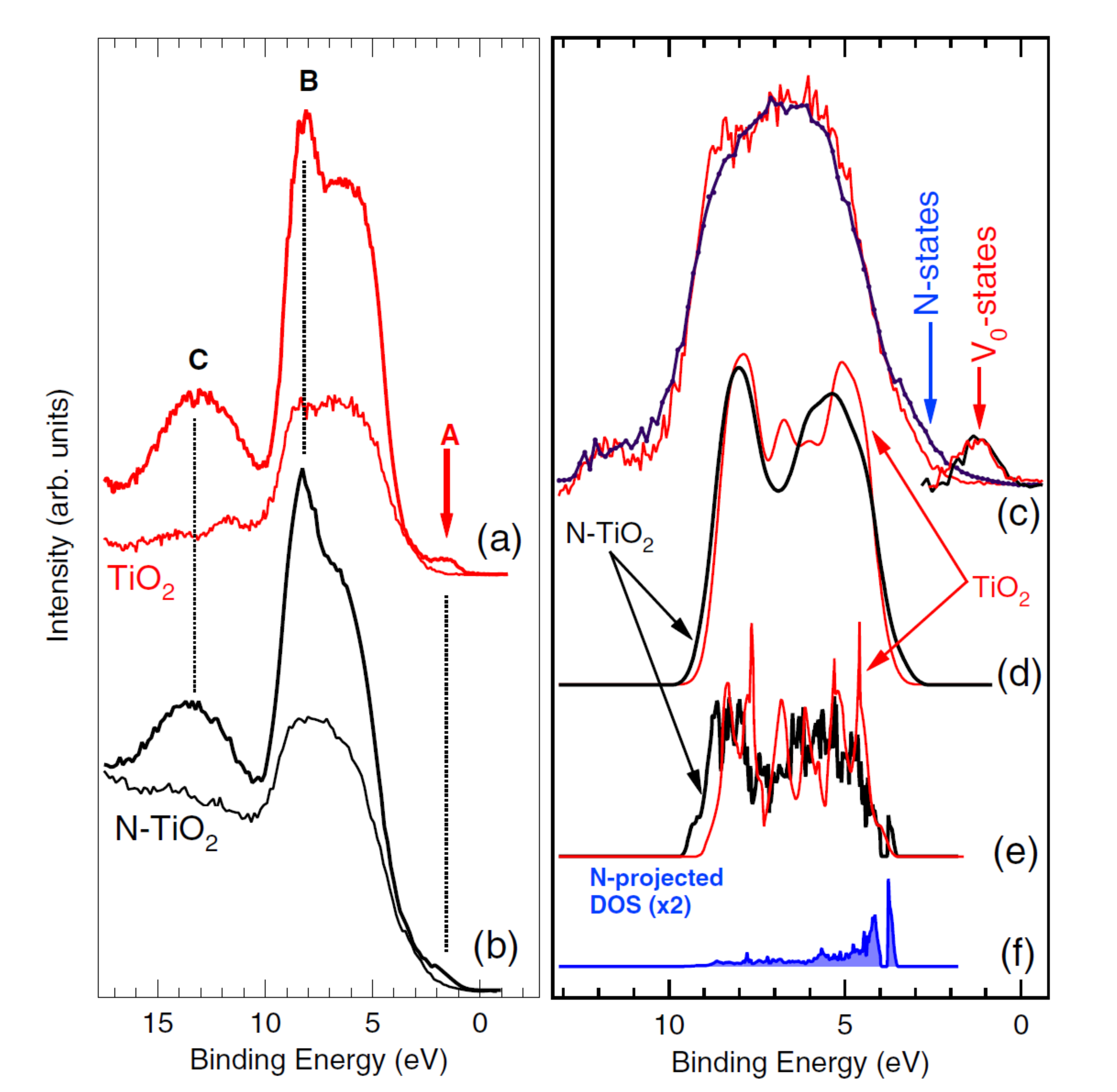}
\caption{Photoemission spectra collected from the undoped (a) and N-doped (b) samples in the valence band region. Spectra drawn with a thin line have been collected with photons below the Ti 2p-3d absorption threshold ($h\nu$ =455 eV), whereas spectra drawn with a thick line have been collected in resonance conditions ($h\nu$ =458.2 eV). (c) Enlarged view of the
off-resonance valence band spectra of the undoped (thin line) and N-doped (thick line) TiO$_2$ rutile. (d) Calculated DOS for the undoped (thin line) and doped samples (thick line) samples. (e) Convolution of the calculated DOS with a Gaussian curve (FWHM=0.6 eV). (f) N-projected calculated DOS for N-TiO$_2$. The calculated DOS curves have been shifted by about -3.5 eV to match the measured DOS.\label{fig_ntio2_respes}}
\end{center}
\end{figure}

In Figs. \ref{fig_ntio2_respes}(a) and \ref{fig_ntio2_respes}(b), for each sample, we show the valence band photoemission spectra collected with photons below the Ti L$_{2,3}$-edge ($h\nu$=450 eV, off-resonance) and with photons corresponding to the A peak ($h\nu$ =458.2 eV, on-resonance) of the Ti XAS spectrum of (Fig. \ref{fig_ntio2_xas}, top panel). As can be observed, in the resonance conditions three photoemission peaks (A, B, and C) results to be enhanced. Peak C is ascribed to the Auger emission, peak B to the resonant enhancement of Ti electronic states hybridized with oxygen, and peak A to the contribution of Ti$^{3+}$ induced by V$_O$. The evidence of these in-gap states further supports the assignment of the tail on the low-BE side of the Ti 2p$_{3/2}$ photoemission peak to Ti$^{3+}$ contribution. The N-doped sample shows a similar peak enhancement.

The A features of both samples have been extracted by subtracting from the ResPES spectrum the off-resonance spectrum. The results are shown in Fig. \ref{fig_ntio2_respes}(c) (V$_O$-states). The two peaks present identical peak energies and intensities, providing an additional proof that the vacancy-induced states contribute to both samples with equal weight. The comparison of the off-resonance valence band spectra (Fig. \ref{fig_ntio2_respes}(c)) also allows to highlight small differences in the spectral weight that are well reproduced by DFT calculations (Figs.  \ref{fig_ntio2_respes}(d) and \ref{fig_ntio2_respes}(e)). In particular, the doped sample displays a higher DOS with respect to the undoped sample for 2$\leq$BE$\leq$4 eV, and below 10 eV. On the basis of the analysis of the DOS projected over the N-derived orbitals (Fig. \ref{fig_ntio2_respes}(f)), the contribution at the top of the valence band, that adds electronic states (labeled as N-states) at low BE as compared to the pure rutile, is ascribed mainly to N.

In summary, we have shown that N-doping of TiO$_2$ rutile yields (i) an enhancement of ferromagnetic behavior, (ii) a replacement of about 4\% O atoms in the rutile lattice with (substitutional) N atoms, and (iii) acceptor and donor levels, that determine a reduction in the rutile energy gap. This finding may have consequences on the enhancement of FM properties. In fact, the band gap reduction may favor the overlap of V$_O$ states with the empty conduction band, rendering the latter magnetic. Moreover, a highly localized deep-gap defect orbital could lead to only a very-short-range defect-defect magnetic interaction that would fail to percolate through the sample. In turn, because of the band gap reduction, the V$_O$ states can loose part of their deep-gap character and display longer-range interactions, creating more favorable conditions for the onset of magnetism.

The different broadening of XAS spectra, as compared to the XPS one, suggest also that the defects could be localized away from the surface (or at least, away from the probing depth of photoemission technique). Moreover, as already shown in Chapter 2, the V$_O$ states are deeply localized inside the bad-gap, while some theories for the d$^0$ ferromagnetism requires the creation of a impurity metallic band\cite{DMO_TiO2_d0} that should be detected at the Fermi level. More details about the theoretical models are given in the next section.

\section{Theoretical models}
The first attempt to explain the magnetism in DMO were based on qualitative considerations about the oxides electronic structure. Coey e al.\cite{DMO_coey_nat} originally proposed a model in which the magnetic impurities interact together with the help of an impurity band created by V$_O$s. This model is based on the assumption that the extra charge given by the oxygen vacancies is localized at the defect site and can be represented with an hydrogen-like atom. The radius of the V$_O$ ``wave-functions'' then should be greatly enhanced by the effect of the local dielectric constant, which is usually high in oxides. According to the authors, the FM should be present also when electrons remain localized in a narrow band because of correlation and potential fluctuations, thus justifying the magnetic interaction in insulating material (like most DMO's).

Recently, the same authors\cite{DMO_TiO2_d0} extended the model also for the (undoped) d$^0$ system, starting from considerations on the Fe doped TiO$_2$ system. The magnetism should be of Stoner-type and takes place in the spin-split impurity band of percolating defects structure, with an half-metallic behavior. In this picture, only a 1\% fraction of the volume should be involved in the FM. The Curie temperature value should be related to the splitting energy of the impurity band (in order of few mEv) and should be consistent with an high T$_C$ ferromagnetism.
Such kind of description fits well with our experimental findings, although is rather qualitative and not based on strict ab-initio calculations.

A more systematical approach has been developed by J. Osorio-Guill\'en et al. for the undoped HfO$_2$\cite{hfo2_theory} and CaO\cite{cao_theory} cases. The authors followed a stepped procedure: the first step is the individuation of the defects that can carry a magnetic moment; the second is the evaluation of the possible (both AF and FM) magnetic coupling through DFT simulations; the third is the evaluation of the percolation threshold for the desired lattice; the last is the evaluation of the lattice stability with the resulting density of defects. The authors identify the \textit{cationic} vacancies as the only source of magnetic moments and estimate that the minimum density of defects needed for the percolation is around 5\% for CaO and 13.5\% for HfO$_2$, thus much higher than what is allowed in an equilibrium-growth process. The latter conclusion is of course due to the choice of the cationic vacancies as the only magnetic moment carriers, while the most common defect in d$^0$ systems is the oxygen vacancy.

Summarizing, the theoretical models supports a defect-based origin of the FM. Because of the weak experimental saturation magnetization of typical d$^0$ systems, the oxygen (or cationic) vacancies should be spatially confined into a small region of the film volume.

\subsection{DFT on TiO$_2$ supercells}
\begin{figure}[htbp!]
\begin{center}
\includegraphics[width=0.75\textwidth]{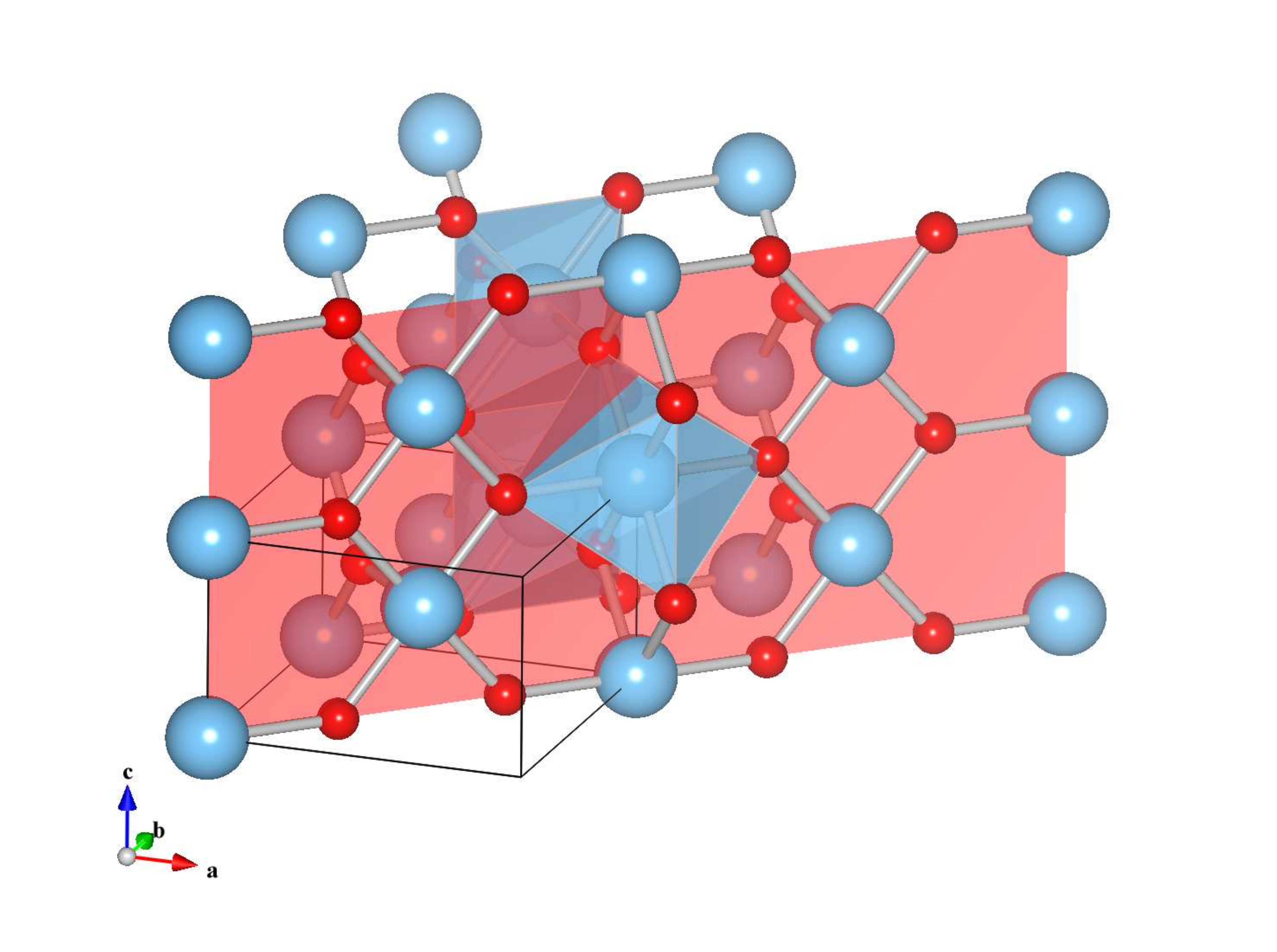}
\caption{Supercell (2$\times$2$\times$2) geometry used for LSDA+U calculations. The single unit cell is shown in the lower left part. The supercell is divided in half by the (1\underline{1}0) plane (in red) which contains the charge density slice shown in Fig. \ref{fig_dft_charge}.\label{fig_dft_cell}}
\end{center}
\end{figure}

To better investigate the origin of magnetism in the rutile N-doped TiO$_2$, another set of DFT calculations has been carried out in the spin-resolved LSDA+U\cite{LDA} approximation, in the same supercell (see Fig. \ref{fig_dft_cell}) described in Chapter 2. The purpose of these calculations is to further explore the contribution of nitrogen-doping to magnetism, apart from the band-gap reduction. Four different supercell geometries have been considered:

\begin{itemize}
  \item an undoped cell;
  \item a 2$\times$2$\times$2 cell with a substitutional N atom (labeled N$_{sub}$);
  \item a 2$\times$2$\times$2 cell with an oxygen vacancy (V$_O$);
  \item a 2$\times$2$\times$2 cell with a substitutional N atom and an oxygen vacancy (V$_O$+N$_{sub}$).
\end{itemize}

\begin{figure}[htbp!]
\begin{center}
\includegraphics[width=0.8\textwidth]{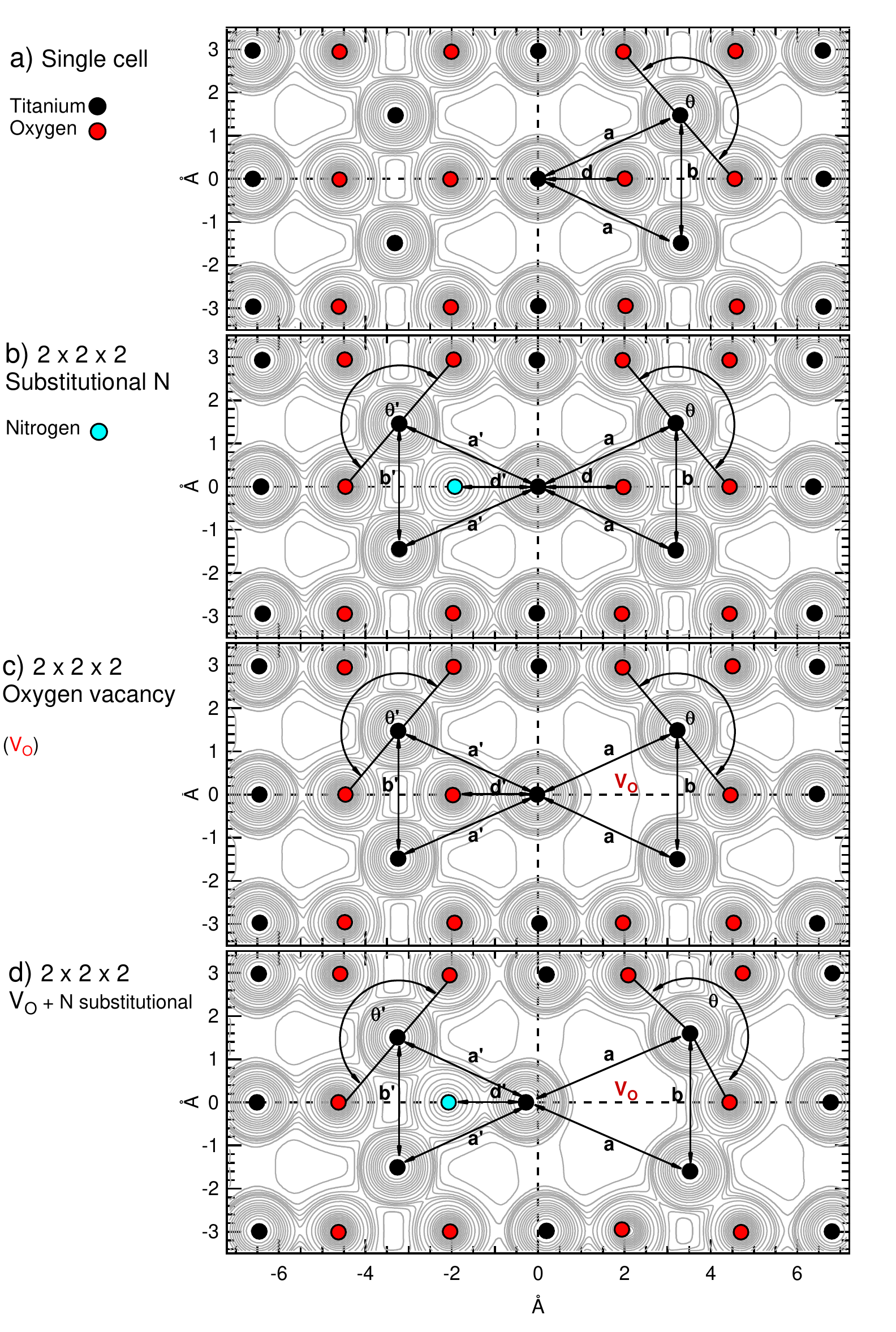}
\caption{Total electronic charge density along the (1\underline{1}0) plane (see Fig. \ref{fig_dft_cell}) for the various LSDA+U calculations. The data on the undoped rutile(a) cell have been extended to match the 2$\times$2$\times$2 range for comparison. The values for the inter-atomic lengths labeled with the reference vectors are given in Table \ref{tab_dft_relax}.\label{fig_dft_charge}}
\end{center}
\end{figure}

The same computational approach (PAW atomic description, with \texttt{ABINIT}\cite{ABINIT} code) has been adopted; the cell geometry has been relaxed to the equilibrium position in each case. A strong distortion of the lattice positions is expected, especially in the V$_O$ case. The results for the electronic charge densities are shown in Fig. \ref{fig_dft_charge}, with reference vectors that point out the relaxation effects induced by V$_O$ and substitutional N. The triangle depicted in Fig. \ref{fig_dft_charge}(a) marks the three titanium atoms that surround an oxygen one in the rutile structure; the removal of this oxygen atom or its substitution with a nitrogen one directly influences the charge at the vertex of the triangle and ultimately the electronic structure.

\begin{table}
\begin{center}
\begin{tabular}{lcccccccccc}
  \hline
    Cell & a & b & d & $\theta$ & a' & b' & d' & $\theta$'\\\hline
  Undoped & 3.622 & 2.959 & 2.018 & 180$^\circ$ & - & - & - & - \\
  N$_{sub}$& 3.514 & 2.947 & 1.985 & 180.2 $^\circ$ & 3.527 & 2.904 & 1.936 & 180.3 $^\circ$\\
  V$_O$& 3.592 & 2.985 & - & 181.3 $^\circ$ & 3.533 & 2.969 & 1.935 & 180.8$^\circ$ \\
  N$_{sub}$+V$_O$& 4.132 & 3.202 & - & 195.7 $^\circ$ & 3.330 & 3.002 & 1.798 & 183.3$^\circ$   \\
  \hline
\end{tabular}
\caption{Structural parameters after the lattice relaxation. These parameters are labeled in Fig.\ref{fig_dft_charge}; the lengths are in {\AA} units.\label{tab_dft_relax}}
\end{center}
\end{table}

The relaxed geometric parameters of the different calculations are listed in Table \ref{tab_dft_relax} and are consistent with other results in literature\cite{TIO2_ldau}.
The N$_{sub}$ supercell (Fig. \ref{fig_dft_charge}(b)) displays only a minor distortion, while the introduction of the oxygen vacancy (Fig. \ref{fig_dft_charge}(c)) clearly elongates the shape of the triangle, pushing the central atom away from the V$_O$. The maximum distortion is found in the case of the V$_O$ plus the substitutional N: the central Ti atom is clearly shifted towards the nitrogen atom, which can accommodate the extra charge given by the V$_O$. This structural modification can be seen, for example, by looking at the $a$ and $d'$ values of the N$_{sub}$+V$_O$ case in Table \ref{tab_dft_relax}, as compared to the other supercells.

The total ionic charges of the Ti and N atoms near to the V$_O$ have been roughly estimated with the Hirshfield method, as implemented in the \texttt{CUT3D} tool included in the \texttt{ABINIT} package. As expected, in the undoped cell the O and Ti total charges are in a -1:2 ratio (O$^{2-}$ , Ti$^{4+}$). In the N$_{sub}$ supercell, the nitrogen atom keeps the same valence of the substituted oxygen (N$^{2-}$). In the V$_O$ supercell, the two electrons left from the V$_O$ are distributed mostly in the surrounding three Ti atoms; the central Ti switches to an approximate Ti$^{3+}$ ionic state, while the other two to a Ti$^{3.5+}$ state. Finally, in the N$_{sub}$+V$_O$ case the central Ti atom reverts to the Ti$^{4+}$ state, while its V$_O$ electronic charge in excess is taken instead by the nitrogen, whose ionic state, evaluated by the Hirshfield method, is N$^{3-}$ (i.e. with a filled 2p shell).

\begin{figure}[htbp!]
\begin{center}
\includegraphics[width=0.8\textwidth]{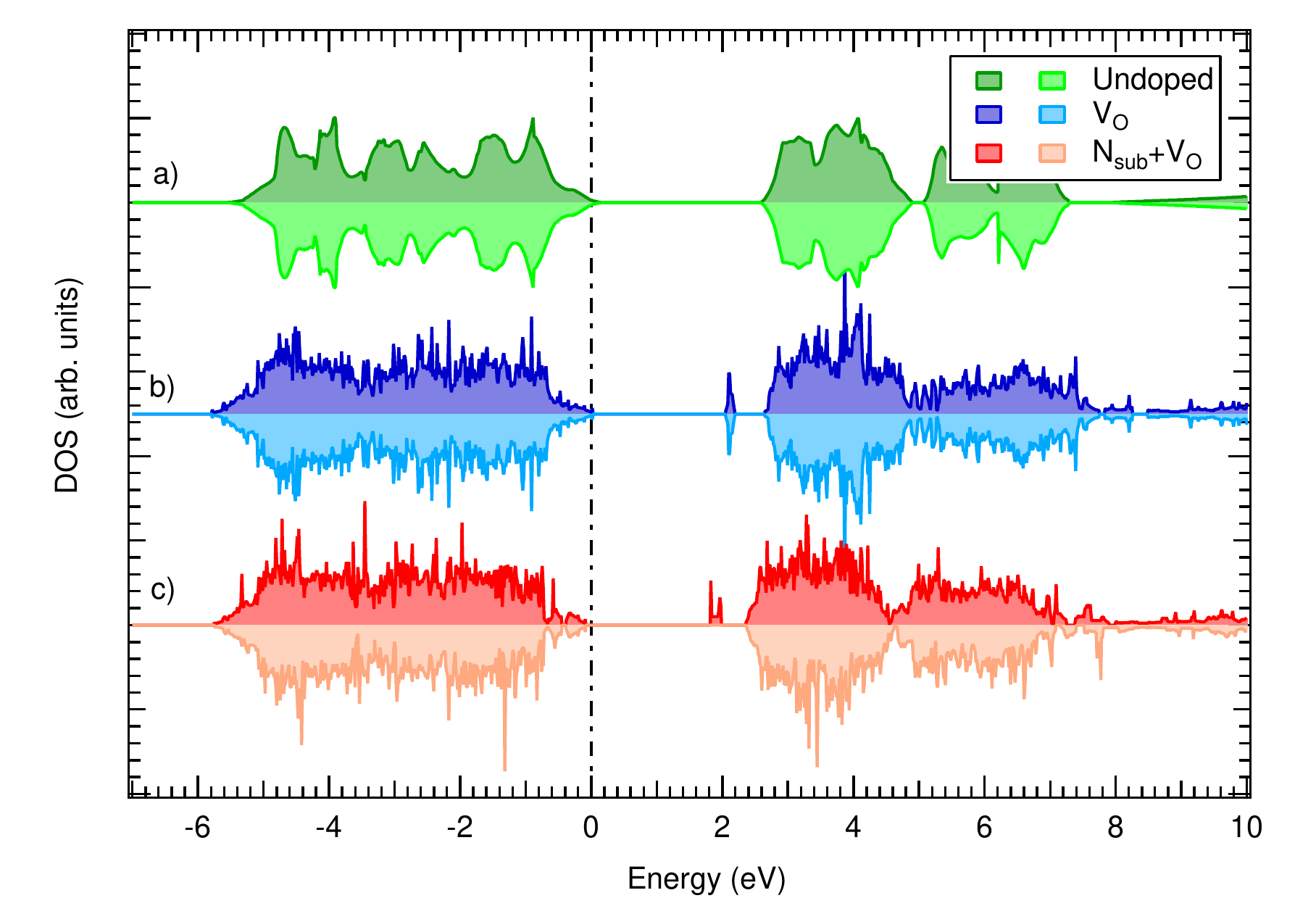}
\caption{Spin resolved DOS for the V$_O$ (b) N$_{sub}$+V$_O$ (c), compared to the single rutile cell (a). The zero of the energy scale is set to the top of the valence band, without considering the in-gap defect levels.\label{fig_dft_dos}}
\end{center}
\end{figure}

The consequence of this charge redistribution on magnetism can be seen by looking at the spin-resolved DOS, shown in Fig. \ref{fig_dft_dos}. The spin-up and spin-down DOS of V$_O$ supercell (see Fig. \ref{fig_dft_dos}(b)), like the undoped and the N$_{sub}$, do not show any difference, reflecting an antiferromagnetic alignment of the spin on Ti atoms. The total magnetic moment of these configurations is thus zero; this also explains why non-clustered V$_O$s have not been taken into consideration in theoretical calculations for the HfO$_2$ and CaO cases\cite{hfo2_theory,cao_theory}. The two V$_O^{2+}$ electrons magnetic moments are antialigned due to the exchange interaction. In fact, as suggested also by Coey and collaborators\cite{DMO_TiO2_d0}, only clustered V$_O$s can display magnetism. A non-zero magnetic moment is expected in neutral (V$_O^{0}$) or single-charged (V$_O^{1+}$) vacancies\cite{TIO2_ldau}.

\begin{figure}[htbp!]
\begin{center}
\includegraphics[width=0.9\textwidth]{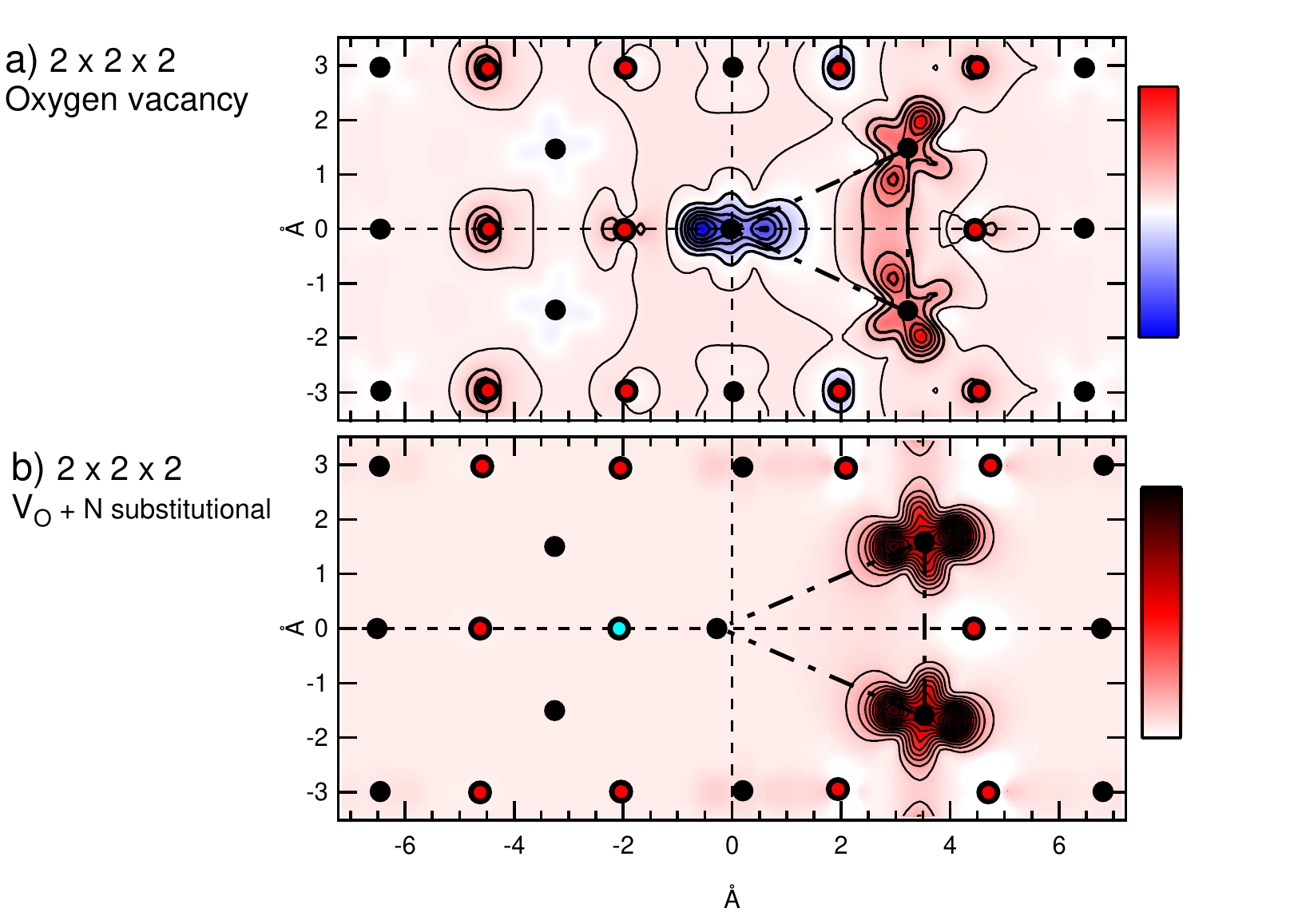}
\caption{Spin-difference electronic charge density along the (1\underline{1}0) plane (see Fig. \ref{fig_dft_cell}) for the V$_O$ (a) N$_{sub}$+V$_O$ (b) configurations. Blue color corresponds to negative values, red to positive one.\label{fig_dft_spin}}
\end{center}
\end{figure}

However, a nitrogen atom near the V$_O$ can attract one of the V$_O^{2+}$ electrons, leading to a net magnetic moment. The spin-resolved DOS on the N$_{sub}$+V$_O$ cell (Fig. \ref{fig_dft_dos}(c)) displays both a band-gap reduction (0.2 eV) and spin-polarized in-gap states. The spin-difference charge density plot shown in Fig. \ref{fig_dft_spin} shows how the presence of nitrogen can quench the spin-down density (blue color in Fig. \ref{fig_dft_spin}(a)) on the central Ti atom, thus leading to a non-zero total magnetic moment.

The nitrogen doping on rutile TiO$_2$ can increase the FM not only through the reduction of the band-gap, that may favor a long range interaction between V$_O$s. In fact, the substitutional N atoms can host part of the V$_O$ excess charge, increasing the number of magnetically active sites. Since every oxygen atom is bonded to three titanium atoms (and thus is shared by three oxygen octahedra), in a rutile lattice with a 4\% of uniform substitutional N-doping the probability for a V$_O$ to be near a dopant N is about 60\%. Therefore, in our films at least half of the V$_O$s, even if isolated, should display an enhanced magnetic moment. In a film with a 6.6 \% of uniform N-doping, the presence of a nitrogen atoms in the proximity of each V$_O$s is assured. This kind of mechanism is rather general and could be applied also to other N-doped magnetic oxides.

\section{Conclusions}
In conclusion, in this Chapter a review of magnetism in TiO$_{2-\delta}$ and N:TiO$_{2-\delta}$ grown with RF-sputtering is given. A weak room-temperature ferromagnetism has been detected in each sample, despite the absence of magnetic ions (TM or RE) doping. In-vacuum annealing systematically increases the saturation magnetization, while the annealing in oxygen results in the opposite effect. This can be regarded as a proof of the V$_O$ fundamental role in FM, also excluding the possibility of a contamination from the sample holder and the annealing stage.

The weak intensity of FM cannot be explained with a uniform distribution of magnetic spins; FM should be then explained by a clustering of defects in particular areas of the samples, as suggested from theoretical works in literature. The saturation magnetization per unit volume is higher in thin films than in the corresponding thick ones, suggesting the presence of plane of clustered defects, either close to the surface or to the interface. XPS measurements on Ti 2p core levels, a surface sensitive technique, did not unambiguously display Ti$^{3+}$ (or V$_O$) signatures, thus supporting the hypothesis of a buried FM interface defect layer in our films. ResPES at Ti L-edge, taken on a rutile thin film, has proved to be an effective tool to underline the presence of electronic 3d$^1$ states in the gap, even for a very small concentration of V$_O$s; however, the observed Ti$^{3+}$ in-gap states lie away from the Fermi energy and thus can be ascribed to isolated oxygen vacancies with a strongly localized electronic structure. DFT+U calculations on rutile supercells further support this description.

Upon nitrogen doping, the ferromagnetism in rutile TiO$_{2-\delta}$ is greatly enhanced. The main effect of the N-doping in the electronic structure is the band-gap reduction, which can be easily observed with optical transmission experiments. XPS and ResPES have been performed in order to underline the nitrogen related electronic structure in core-levels and in the valence band. The structural disorder induced by the N-doping can be seen through the broadening of XAS and XPS spectra. Finally, spin-resolved LDA+U calculations suggest that substitutional N-doping can induce a magnetic moment in otherwise non-magnetic isolated V$_O^{2+}$ through Ti oxidation.

In conclusion, even if the magnetism in undoped TiO$_{2-\delta}$ is still far from being exploited in future applications, it represents a challenging issue for solid state physicists. In order to understand the magnetism in TiO$_{2-\delta}$, further experimental investigations are required, in particular to map the bulk V$_O$ distribution. From the theoretical side, ab-initio calculations on large defects clusters are needed in order to give a realistic description of these systems.

\chapter{The LaAlO$_3$-SrTiO$_3$ heterostructure}
\section{Introduction}
\begin{figure}
\begin{center}
\includegraphics[width=0.9\textwidth]{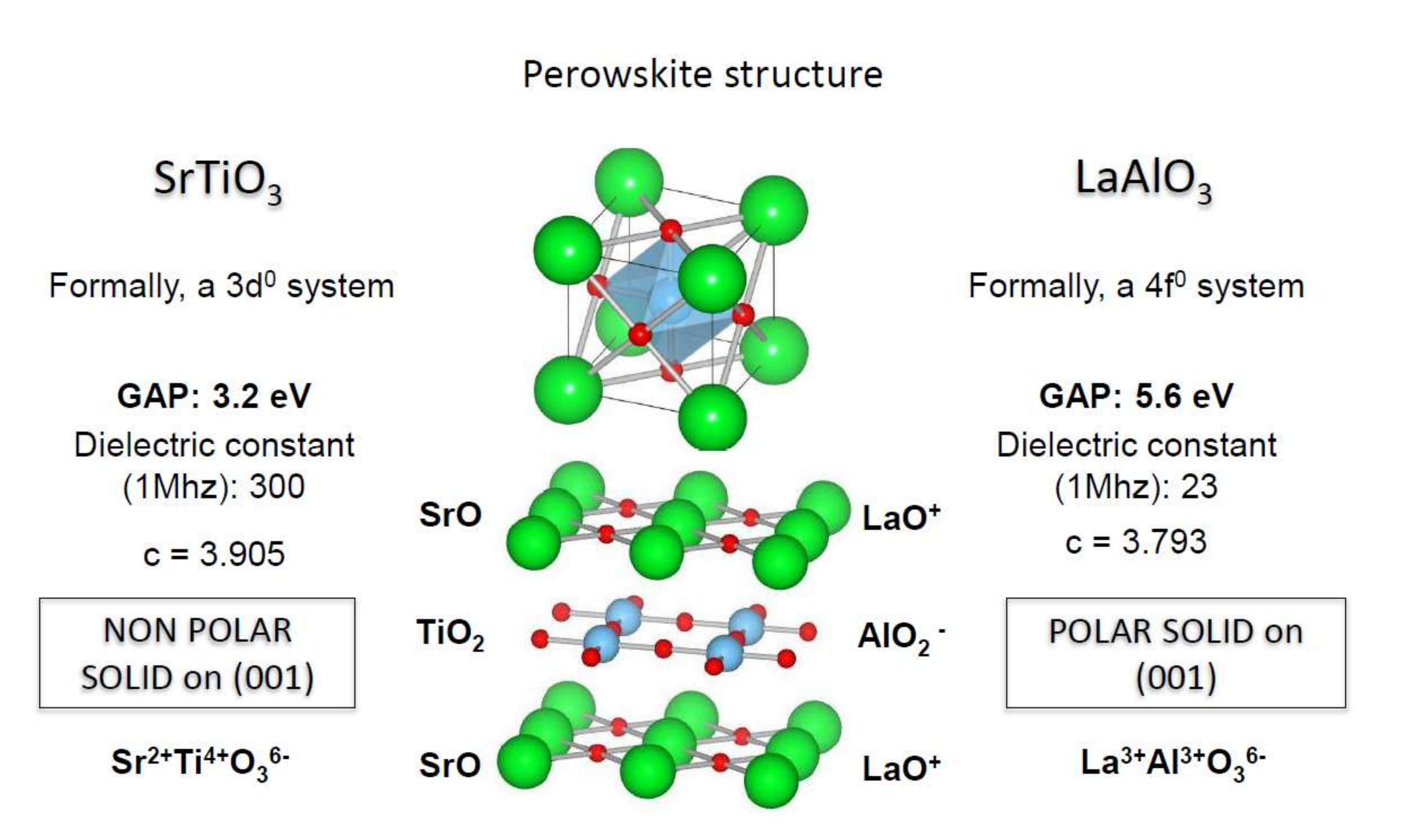}
\caption{A schematic illustration of LaAlO$_3$ and SrTiO$_3$
crystalline and electronic structure.\label{fig_LAOSTO_crystal}}
\end{center}
\end{figure}

Lanthanum aluminate (LaAlO$_3$, LAO in short) and strontium titanate (SrTiO$_{3}$, STO) are ionic insulators, with a band gap respectively of 5.4 and 3.2 eV. These materials belong to the perovskite group, which share the same chemical formula (ABO$_3$) and a similar cubic crystal structure (see Fig. \ref{fig_LAOSTO_crystal}). These compounds are very interesting due to their relative simple crystal structure and to the possibility to have rare earths or transition metals at the cationic A and B sites, displaying many different functional properties (metal-to insulator transitions, ferroelectricity, magneto-resistance and so on). Moreover, it is possible to use a perovskite as an epitaxial template for the growth of another one, creating a seamless amount of possible heterostructures. LAO and STO are formally band insulators, since they are closed-shell compounds (4f$^0$ for LAO and 3d$^0$ for STO) and are easily grown with chemical vapor deposition (CVD) or pulsed laser deposition (PLD); in fact, high quality STO samples are commonly used as a starting substrate for epitaxial growth. As shown in Fig. \ref{fig_LAOSTO_crystal}, the LAO-STO interface can be p-type when the bulk STO is terminated with a SrO plane (hole doping) and n-type with a TiO$_2$ plane (electron doping). In the latter case, an interesting phenomenon occurs: the interface of LAO-STO becomes conducting\cite{LAOSTO_ohtomo} and forms a quasi-2D electron gas (2DEG). The proof of the 2D character of the conductive layer was given later by means of the anomalous Shubnikov de Haas effect\cite{LAOSTO_sub_deHaas}.

Moreover, there's a thickness dependence of the metal transition: the 2DEG is observed only when the capping is at least 4 unit cell thick\cite{LAOSTO_science}. The main difference between LAO and STO resides in the layer charge polarity\cite{LAOSTO_rev1}: looking at the (001) planes (see Fig. \ref{fig_LAOSTO_crystal}), STO is a non-polar solid, since both Sr$^{2+}$ O$^{2-}$ and Ti$^{4+}$ O$_{2}^{2-}$ are charge-neutral, while LAO is a polar solid, as it is composed of La$^{3+}$ O$^{2-}$ and Al$^{3+}$ O$_{2}^{2-}$ charged layers. The p-type interface is thus formed by SrO - (AlO$_2$)$^{1-}$ planes, n-type by TiO$_2$ - (LaO)$^{1+}$ ones. The observed conductivity should be the response of the system to the diverging potential (the so called ``polar catastrophe'') created by LAO, both in the electronic and structural properties.

In non-oxide semiconductors, the relaxation of polar discontinuity in heterointerfaces is usually achieved by the so called ``atomic reconstruction'' process\cite{LAOSTO_rev1}: the interface stoichiometry is altered by interdiffusion, point defects, dislocations and in general by a structural roughening. In oxides, the possibility of multiple valence ions allow also an ``electronic reconstruction'' that, in this case, should move electrons from the surface to the empty Ti d-level, creating 3d$^1$ electronic states. A 2D lattice of electrons in a correlated material can raise phenomena\cite{LAOSTO_rev2} like MIT transitions, localized magnetic moments and even superconductivity; most of this phases have been observed, but not simultaneously in the same interface.

In principle, both atomic and electronic reconstruction could be present in the LAO-STO case. There are experimental hints that support either one theory or the other. For example, there are many experimental proofs of interdiffusion\cite{LAOSTO_chambers} (lanthanum ions moving inside STO, a form of atomic reconstruction), but this alone cannot be the solely responsible of the conductivity, since in principle should be present also in the n-type interface. Moreover, standard ``polar catastrophe'' (an electronic reconstruction picture) alone can't explain the lack of conductivity for thickness below 4 unit cell threshold. The thickness dependence could be explained by a band bending effect, induced by polarity discontinuity: the LAO VB DOS should be shifted to higher binding energies till, for a capping equal or major of 4 u.c., the VB maximum is superimposed to the buried empty levels of STO. The conduction should be caused by a tunneling effect from surface to the interface. However, a band bending has never been observed in therm of core-level shift, while a shift of 3.2 eV (needed to spam the electronic gap in STO) should be easily observed.

\begin{SCfigure}
\includegraphics[width=0.45\textwidth]{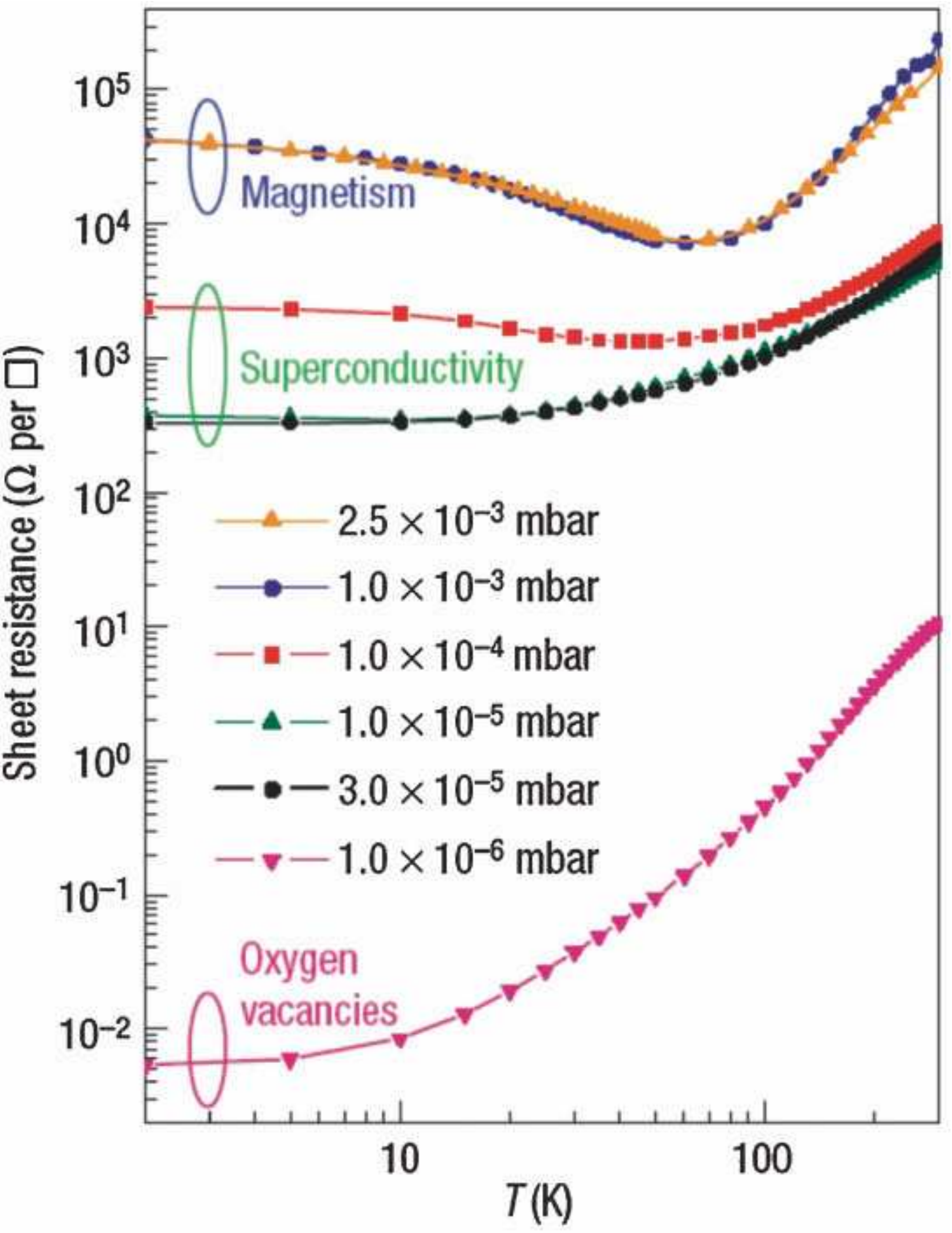}
\caption{Temperature dependence of the sheet resistance for SrTiO$_3$/LaAlO$_3$ conducting interfaces, grown at various partial oxygen pressures. Three regimes can be distinguished: low pressures lead to oxygen vacancies, samples grown at high pressures show magnetism, whereas samples grown in the intermediate regime show superconductivity. Adapted from Ref. \cite{LAOSTO_natmat}. \label{fig_LAOSTO_resistance}}
\end{SCfigure}

Finally, the sample quality deeply affects the transport measurement; an oxygen-poor growth atmosphere can induce oxygen vacancies and thus a 3D conductivity, while an excessively-rich one can even result in a 3D growth and thus in a different kind of heterostructure. In Nb-doped STO, the presence of oxygen-vacancies at the interface can induce superconductivity with T$_c$ below 400 mK, in a similar way to the one measured on LAO-STO\cite{LAOSTO_sc}. Three phases are usually identified, according to the O$_2$ partial pressure during growth (Fig. \ref{fig_LAOSTO_resistance}): one dominated by oxygen vacancies contribution (P$_{O_2}\simeq 10^{-6}$ mBar), one displaying superconductivity (P$_{O_2}\simeq 10^{-5}$ mBar) and one displaying a magnetic behavior (P$_{O_2}\simeq 10^{-3}$ mBar). It is quite a challenging task to find a unified description of all this phenomena.

Since the conduction is thought to occur at the interface, the filling of electronic states at Fermi level can't be observed with standard photoemission techniques: in conductive samples, even in a synchrotron facility, X-ray photoemission might not show the Fermi level, being more sensitive to the uppermost insulating oxide. The problem can be solved, as in the TiO$_2$ case, using the enhancement of selected spectral weight at resonance condition.

In this Chapter, an analysis of photoemission data of insulating and conductive LAO-STO film is given; samples have been grown with PLD in a P$_{O_2}$=1$\cdot$10$^{-3}$ mBar oxygen partial pressure at the University of Twente by A. Brinkmann research group. In the first section an XPS analysis of the stoichiometry of LAO-STO as compared to single-crystal LAO and STO is shown; the second section is devoted to ResPES, XAS and XLD analysis; the third section is devoted to intermixing analysis by means of AR-XPS. Finally, the last section is dedicated to conclusions and future improvement in research.

\section{XPS data}
The XPS measurements have been carried out on the 3 u.c. and 5 u.c. LAO-STO samples, plus two reference single-crystal LAO and STO samples terminated with the (001) surface. In addition, an \textit{insulating} 5 u.c. sample, grown at 10$^{-1}$ mBar O$_2$ partial pressure, has been analyzed. XPS has been used to measure the core-level electronic structure and to evaluate film stoichiometry. The STO and LAO samples have been treated with Argon ion sputtering in order to see the effect of oxygen loss. The XPS data shown in this section have been collected at Surface Science Lab of the Universit\'a Cattolica Brescia, with a non-monochromatized dual-anode PsP x-ray source; the Mg k$_\alpha$ line ($h\nu$=1256.6 eV) has been used to achieve a better resolution (about 0.7 eV), while the Al k$_\alpha$ line ($h\nu$=1486.6 eV, resolution of 0.8 eV) has been used when the maximum probing depth was needed (mainly for AR-XPS). Charging effects have not been detected during XPS, since this kind of X-ray source are rather unfocused. The analyzer was a SCIENTA R3000 XPS/UPS/ARPES, with a maximum energy resolution of 25 meV. For XPS, this analyzer has been used in the ``transmission mode'', which maximize the transmittance and work with a 30$^\circ$ acceptance angle; ``angular modes'' are also available, and have been used for ARXPS.
A survey view of the detectable core levels in 3 u.c. and 5 u.c. samples is shown in Fig. \ref{fig_XPS_Survey}:

\begin{figure}
\begin{center}
\includegraphics[width=1\textwidth]{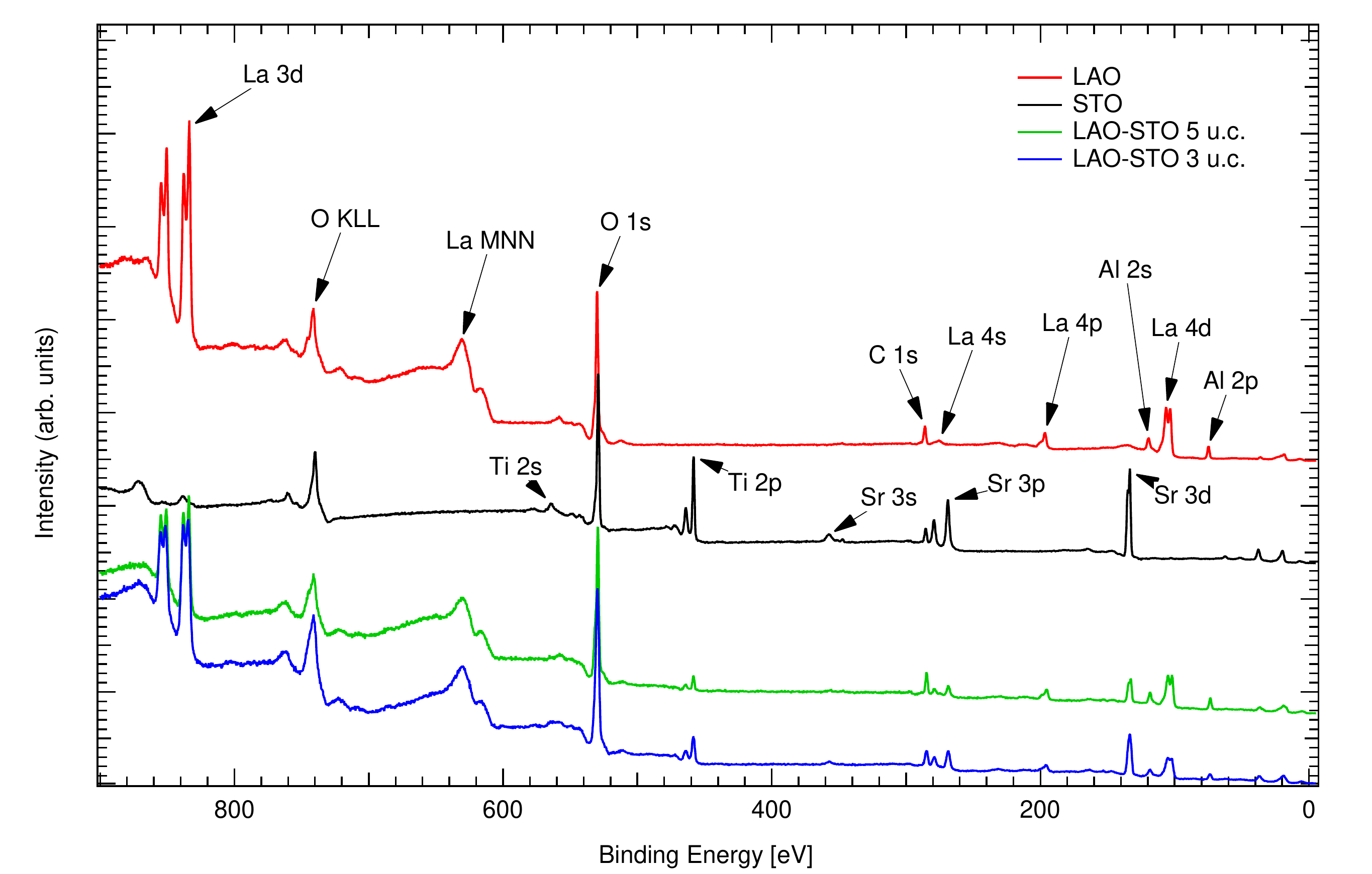}
\caption{XPS survey spectra taken in single-crystal LAO, STO and on
the 3 u.c and 5 u.c. LAO-STO samples. Peaks are labeled on the
figure. Spectra have been collected with a Mg k\textnormal{$_{\alpha}$}
x-ray source and have been normalized to the oxygen 1s
peak.\label{fig_XPS_Survey}}
\end{center}
\end{figure}

Carbon 1s peak is due to the usual atmospheric contamination: samples hav not been cleaned up since vacuum annealing or Ar$^+$ sputtering might have changed the oxygen stoichiometry. Moreover, C 1s peak can be used as reference for the binding energy scale (C 1s binding energy for carbon contamination is fixed at 284.8 eV).
Even in the survey spectra, La 3d and 4d peaks show already a complex structure which is not due to different oxidation state but to well-known multiplet structure\cite{DEGROOT}. The different intensity of LAO vs STO structures in the two samples is already visible in Fig. \ref{fig_XPS_Survey}; basically, in the 5 u.c. sample Ti and Sr peaks are more attenuated with respect to the Al and La ones due the larger LAO capping.

\begin{figure}
\begin{center}
\includegraphics[width=0.8\textwidth]{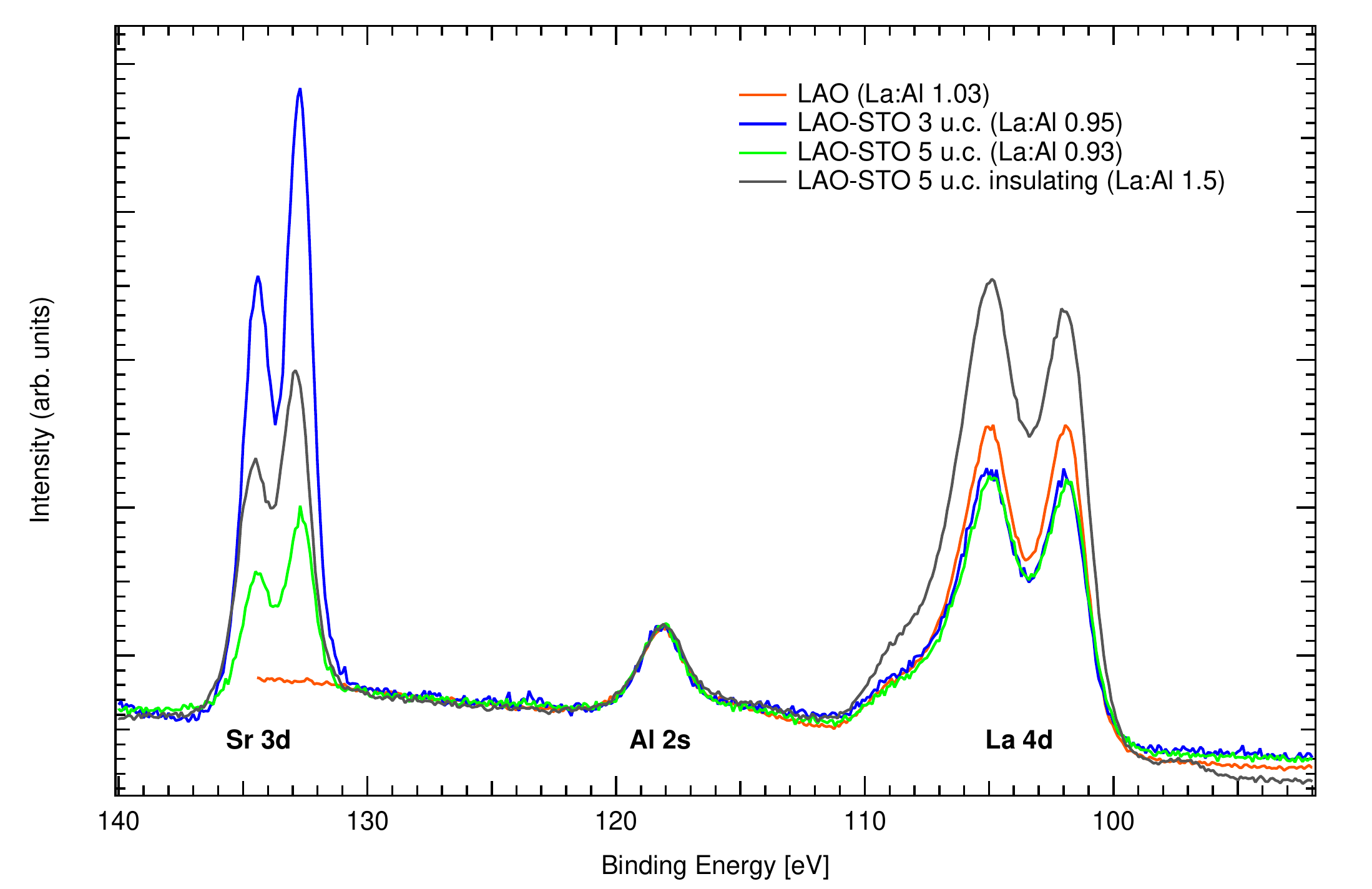}
\caption{XPS survey spectra taken on Sr 3d, Al 2s and La 4d on LAO
and heterostructures. Spectra have been  normalized to the Al 2s
area to allow a better comparison of lanthanum over aluminum ratio,
which are given in the legend. Calculated ratios are already
normalized with the photoionization cross section. Experimental
condition (i.e. the total resolution) has been kept the same for all
the samples.\label{fig_XPS_SrAlLa}}
\end{center}
\end{figure}

Since the 80-140 eV (in BE) region contains, in a relatively short range, most of the core level peaks required for a quantitative analysis, high resolution spectra (see Fig. \ref{fig_XPS_SrAlLa}) have been collected on this area to carry out the element quantification. The peak areas have been measured after a Shirley-type background subtraction and have been normalized according to the photoionization cross sections\cite{YehLindau}. First of all, Fig. \ref{fig_XPS_SrAlLa} spectra clearly show that the insulating 5 u.c. is very different from the other samples because, as already pointed out in the introduction, its growth regime is 3D instead of layer-by-layer. This sample still retains a crystalline order, as will be shown with AR-XPS data in the next section. The Al/La ratio in heterostructures is different from the LAO crystal one, showing a lanthanum deficiency that can suggest a La interdiffusion; anyway, LAO-STO samples are terminated with AlO$_2$ plane while single-crystal LAO is usually terminated with a mixture of AlO$_2$ and LaO phases, resulting an higher (closer to 1:1) ratio. The measured value for LAO-STO is compatible with the one that can be calculated with the AR-XPS model (see next section) and thus is not a sufficient proof of selective lanthanum diffusion. Moreover, data in Fig. \ref{fig_XPS_SrAlLa} have been collected in the ``transmission'' mode, thus the results are the average of the possible diffractive effect in the analyzer acceptance angle. In order to have precise theoretical results, one should rely on precise photoelectron-diffraction calculation to verify peak intensities enhancement at normal emission.

Probing the Ti/Sr ratio can be a more difficult task for two reasons: first, the most prominent peak (Sr 3d and Ti 2p) are 300 eV apart and the analyzer transmission function could not be constant; second, the total area of Ti 2p peaks has to be measured carefully accounting for the charge-transfer satellites. The results for this two peaks ratio are given in the AR-XPS section.

\begin{figure}
\begin{center}
\includegraphics[width=0.9\textwidth]{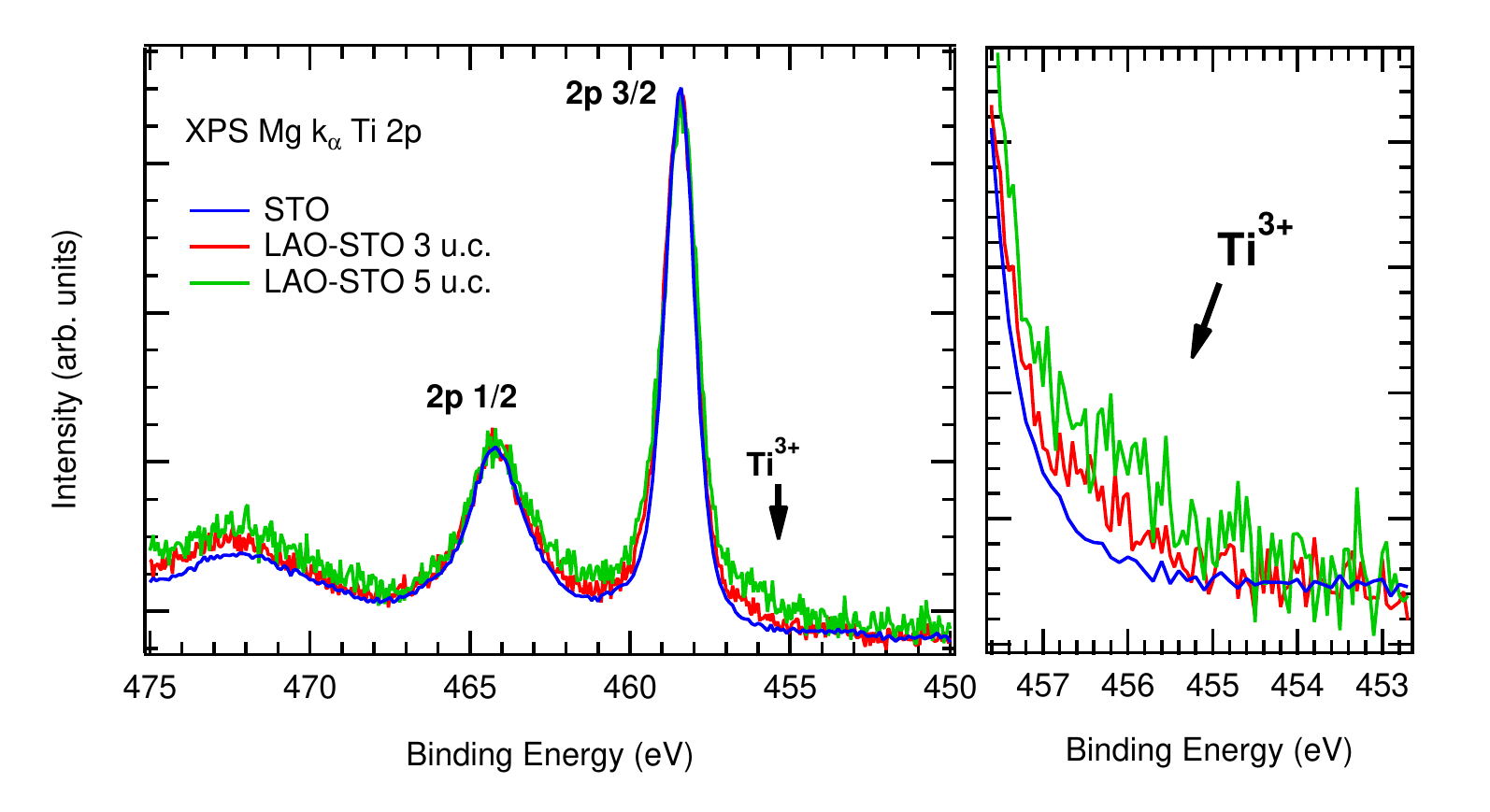}
\caption{High-resolution Ti 2p XPS taken on the 3 u.c LAO-STO,
conductive 5 u.c. LAO-STO and untreated single-crystal STO. The black arrows identify the position of Ti$^{3+}$ states, increasing from STO to the 5 u.c. sample. Similar data have been published in Ref.\cite{apl_LAOSTO}.\label{fig_XPS_Ti2p_a}}
\end{center}
\end{figure}

In Fig. \ref{fig_XPS_Ti2p_a} the Ti 2p XPS peaks of the 3 u.c., 5u.c. and pure STO samples are shown. The titanium 2p peaks are almost identical to that expected for a Ti$^{4+}$ ion. However, a weak spectral feature (indicated by the black arrow) is detectable on the lowest BE side of 5 u.c. spectrum in the position usually associated to Ti$^{3+}$ electronic states. These peaks can be detected only by comparison with different samples and can be easily confused as an additional experimental broadening of Ti 2p$_{3/2}$ peak. Recently, these electronic states have also been measured with hard X-rays (HAXPES) experiment to evaluate the Ti interface reconstruction in LaAlO$_3$-SrTiO$_3$\cite{LAOSTO_haxpes} with a large LAO capping.

\begin{figure}
\begin{center}
\includegraphics[width=0.9\textwidth]{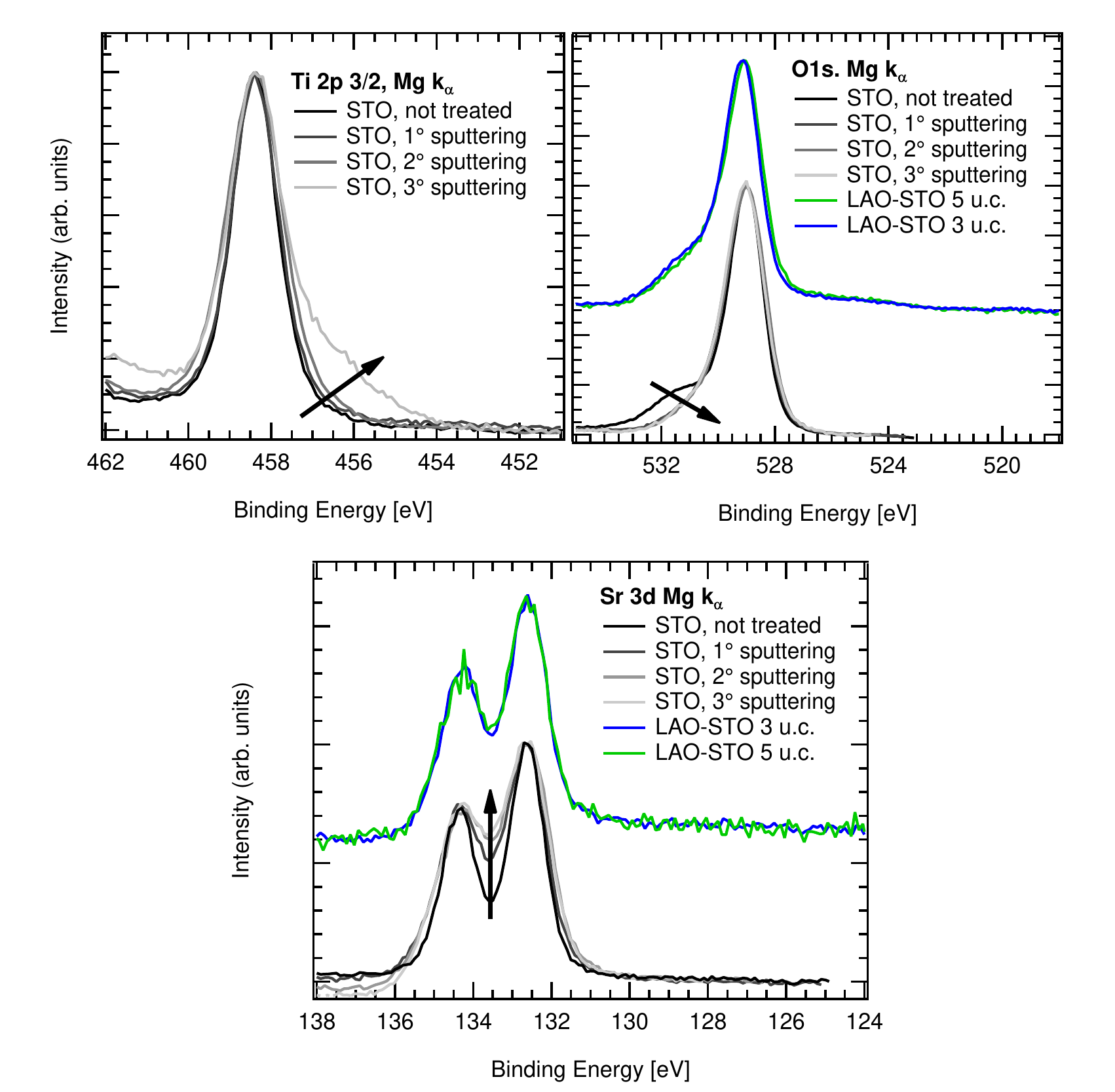}
\caption{Ti 2p$_{3/2}$ (a), Sr 3d (b) and O 1s (c) XPS spectra measured after subsequent Ar$^+$ sputtering treatments on the STO sample, as compared to LAO-STO data. The black arrows mark the changes due to sputtering effects. LAO-STO samples have not been sputtered in order to avoid structural damages or the creation of oxygen vacancies.\label{fig_STO_sp}}
\end{center}
\end{figure}

In order to underline the effect of an oxygen loss in XPS spectra, in Fig. \ref{fig_STO_sp} data taken on the clean STO after subsequent Ar$^+$ 1kV sputtering treatment are shown. The main effects on peak shapes are indicated by the black arrows: Ti 2p$_{3/2}$ gains the low BE component, Sr peaks become broader and O 1s peaks lose the component at higher BE due to surface contamination. The spectra (collected at the same resolution) of heterostructures are also shown; in particular, the Sr peaks are not as sharp as pristine STO and are comparable to the spectrum taken after one sputtering session. This can indicate a sort of disorder (cationic exchange or oxygen vacancies) around strontium atoms at the interface; other indications supporting this picture are given in the ARXPS section.

\begin{figure}
\begin{center}
\includegraphics[width=0.9\textwidth]{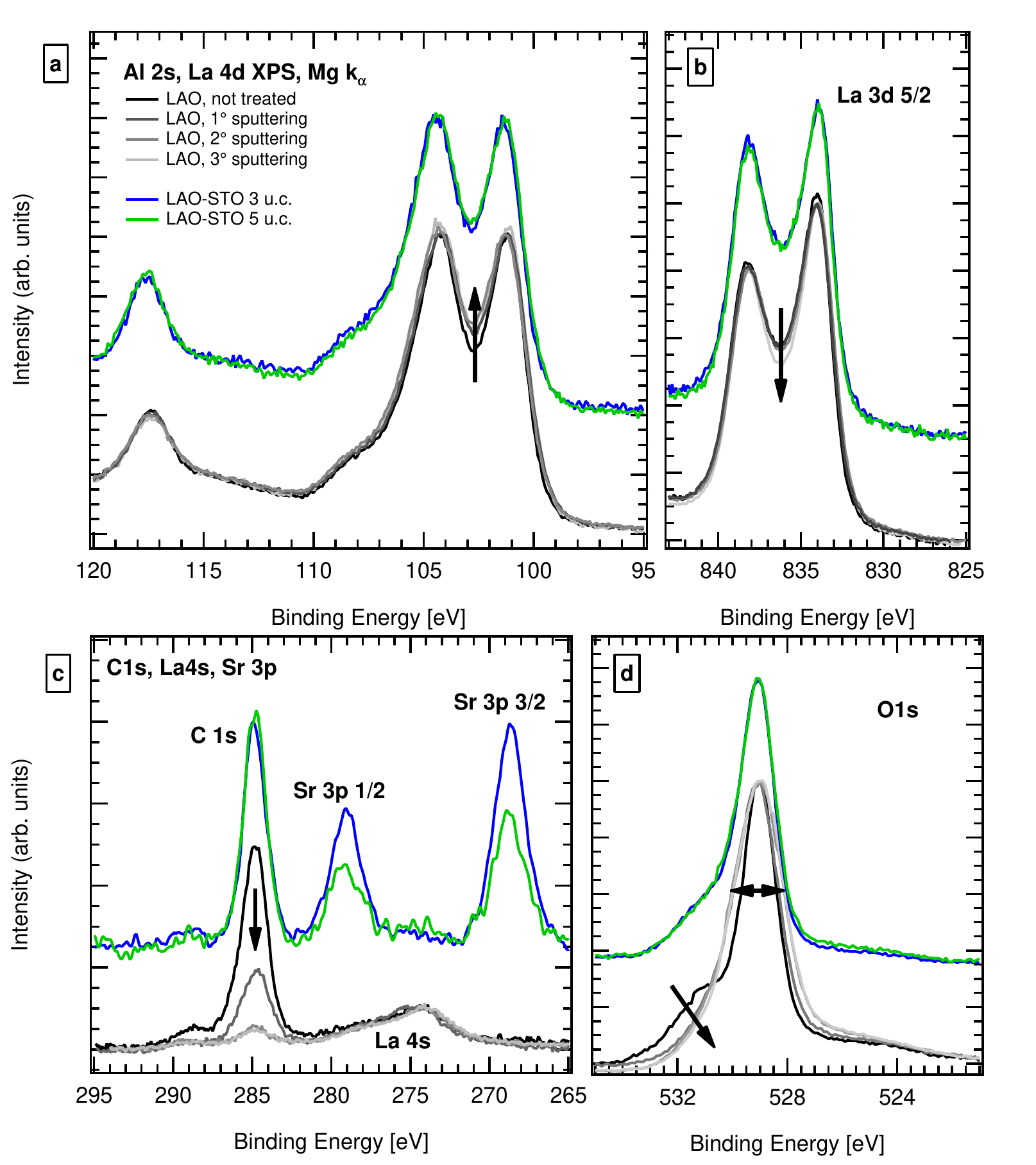}
\caption{XPS spectra from selected core level spectra taken during subsequent sputtering session on a single crystal LaAlO$_3$ sample. The colors of experimental data range from black (not treated) to grey (heavily sputtered). The black arrows underline the modification caused by sputtering.\label{fig_LAO_sp}}
\end{center}
\end{figure}

A similar analysis has been carried out on a clean LAO (001) sample; results are shown in Fig. \ref{fig_LAO_sp}. As expected the intensity of C 1s peak, which is associated to adventitious contamination, reduces upon sputtering; the shoulder of O 1s is quenched as in the STO, but the overall width of main peak also increases due to the increased disorder. The La 4d peak width (in a similar way of Sr 3d in STO) gets broader but surprisingly the La 3d$_{5/2}$ seems to behave in the opposite way. In fact, the splitting of La 3d in Fig. \ref{fig_LAO_sp} is not due to spin-orbit interaction but it is caused by multiplet structure, common in lanthanum (and in the other RE) oxides. The different broadening in this case can be explained by the presence of metallic lanthanum, whose La 3d$_{5/2}$ peak-shape is given only by a single peak that raise the low BE side of the overall XPS spectrum. Anyway, the shape difference in LaAlO$_3$ core level can give little information to the LAO-STO case.

\begin{figure}
\begin{center}
\includegraphics[width=0.9\textwidth]{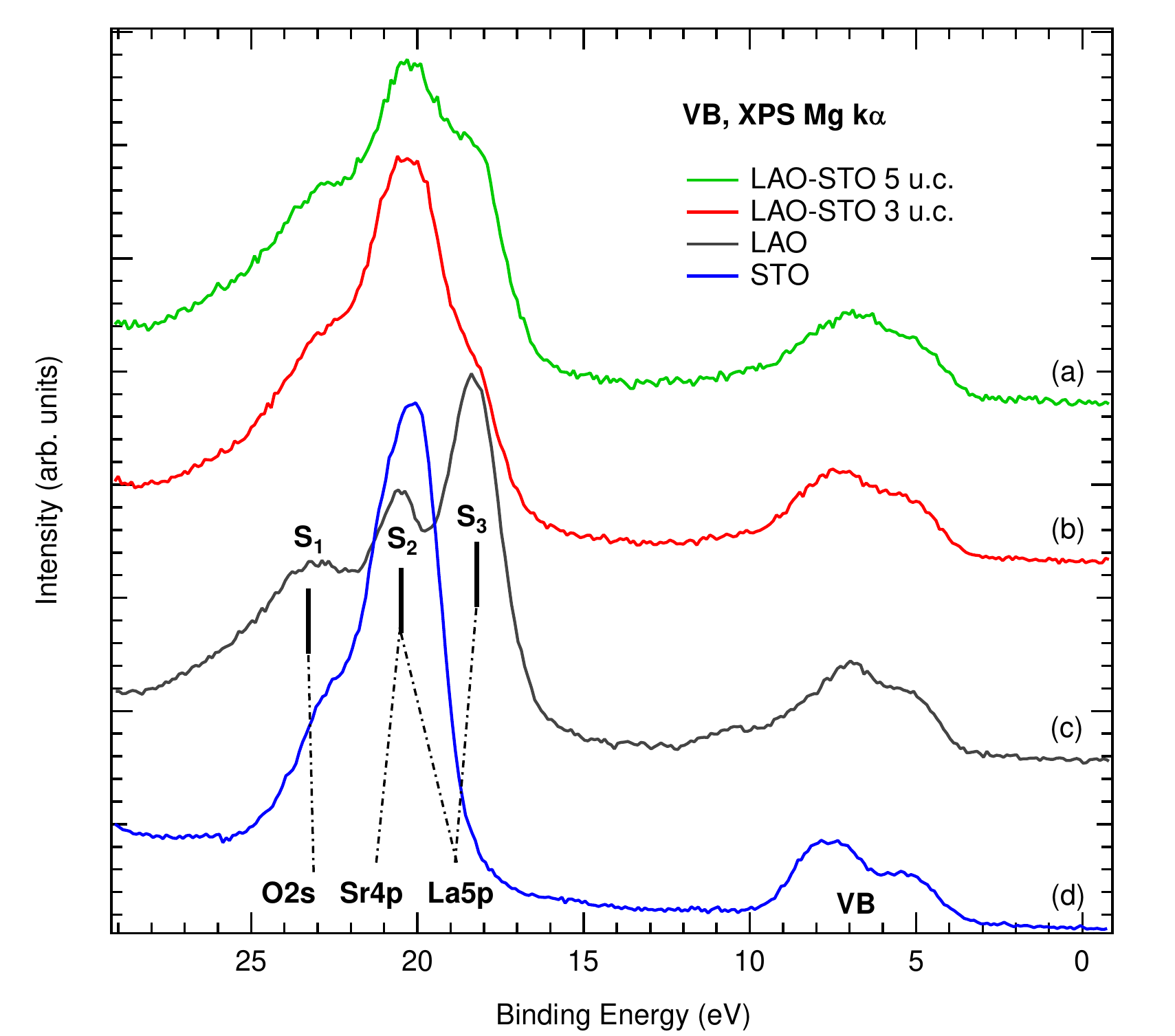}
\caption{XPS valence band spectra on single crystal (c and d) and heterostructures (a and b). Both the VB and the shallow core levels are shown.\label{fig_VB_1}}
\end{center}
\end{figure}

\begin{figure}
\begin{center}
\includegraphics[width=0.8\textwidth]{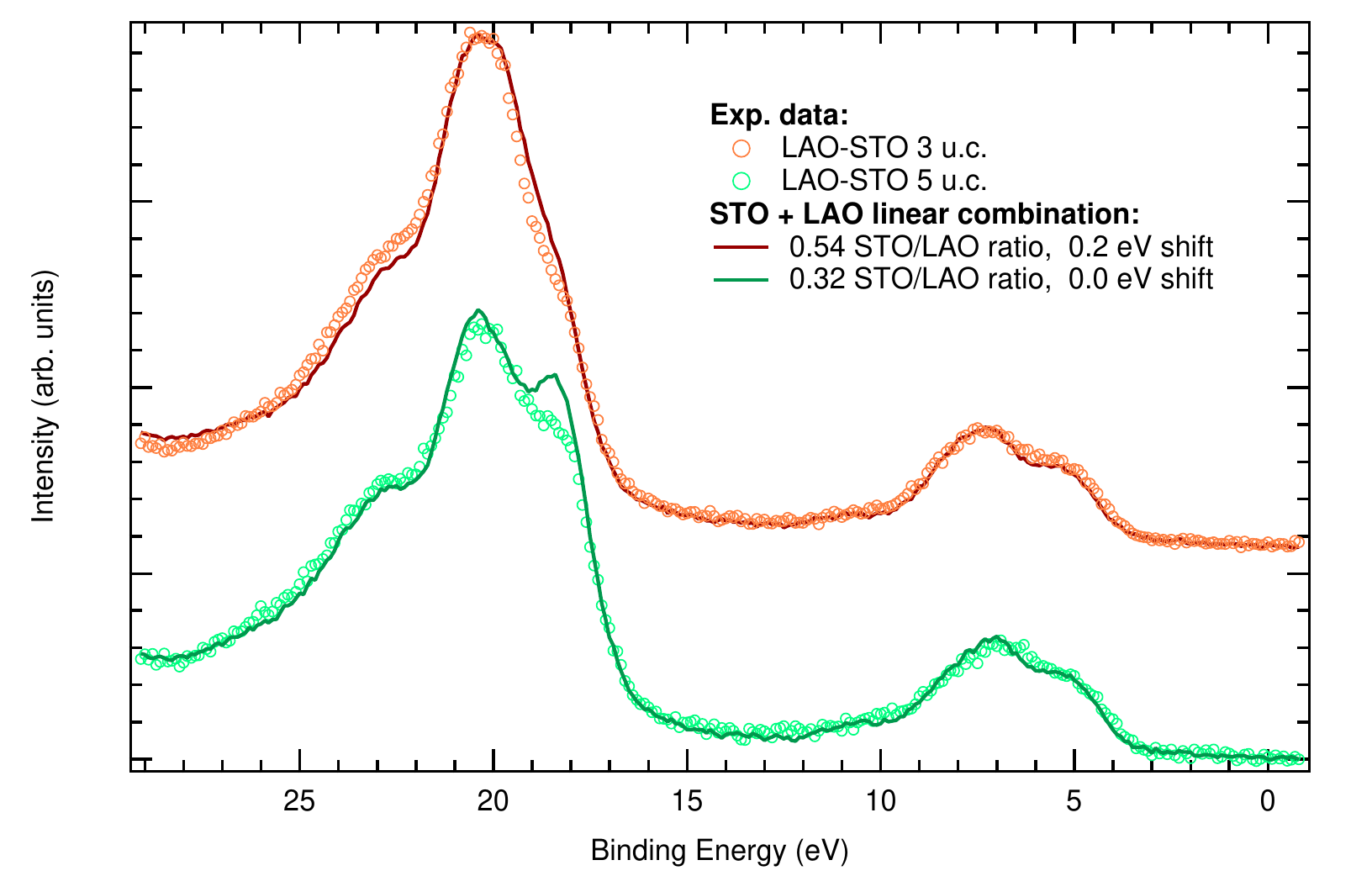}
\caption{3 u.c. and 5 u.c. LAO-STO valence band fitted with a linear combination of LAO and STO spectra. The starting intensity of the single components has been normalized to the total VB area and a maximum relative shift of $\pm$1 eV has been considered in the fitting procedure.\label{fig_VB_2}}
\end{center}
\end{figure}

Finally in Fig. \ref{fig_VB_1}, \ref{fig_VB_2}, \ref{fig_VB_2} the valence band (VB) spectra, taken with the Mg k$_{\alpha}$ X-ray source, are shown. The shallow core levels are labeled in Fig. \ref{fig_VB_1}; the O 2s spectral weight is located mostly at S$_1$ peak, Sr 4p is an unresolved doublet below S$_2$ peak and La 5p is splitted in S$_2$ and S$_3$ peaks. It is possible to describe the LAO-STO spectra as the linear combination of the single-crystal data; the spectra of LaAlO$_3$ and SrTiO$_3$ have been normalized in intensity, shifted and summed in order to minimize the difference with the multilayers. The results of this ``fitting'' procedure are given in Fig. \ref{fig_VB_2}; there is a minimal shift difference of 0.2 eV in the two linear combinations, which should reflect the difference in the band bending between the 3 u.c. and the 5 u.c. samples. Such a small value, not enough to justify the conductivity, is similar to the one that can be found with ResPES measurement, and will be discussed in the conclusion. The LAO/STO ratios indicated in Fig. \ref{fig_VB_2} are useless by themselves, since the absolute XPS intensity can vary in different spectra due to the different thickness of the LAO capping; aniway, it is possible to evaluate the difference in thickness by using the formula \ref{eq_exp3}. The STO signal attenuation of 5 u.c. respect to the 3 u.c. at normal emission can be written as:

\begin{equation}
\frac{I_{STO 5 u.c.}}{I_{STO 3 u.c.}} = e^{-\frac{z}{\lambda}}
\label{eq_LAO_STO}
\end{equation}
\\
The fitting procedure in this case gives a ratio of about 0.59. For electrons with a kinetic energy of 1250 eV, the TPP-2M formula in pure LAO gives an inelastic mean free path \textnormal{$\lambda=23.32$} \r{A}. By inverting Eq. \ref{eq_LAO_STO}, a difference in thickness of $z=12.30$ \r{A} is found, much higher than the expected 2 u.c. (7.6-7.8 \r{A}). This indicate that signal of the 5 u.c. sample is much more attenuated than expected respect to the 3 u.c. case.

In fact, the problem here is not related to the samples quality, but to the oversimplification of Eq. \ref{eq_LAO_STO} which doesn't take into account the elastic scattering. By replacing \textnormal{$\lambda$} with the calculated EAL, as described in Chapter 2, an extra-thickness of nearly 9 \r{A} is found, in good agreement with the expected value.

\begin{figure}
\begin{center}
\includegraphics[width=0.8\textwidth]{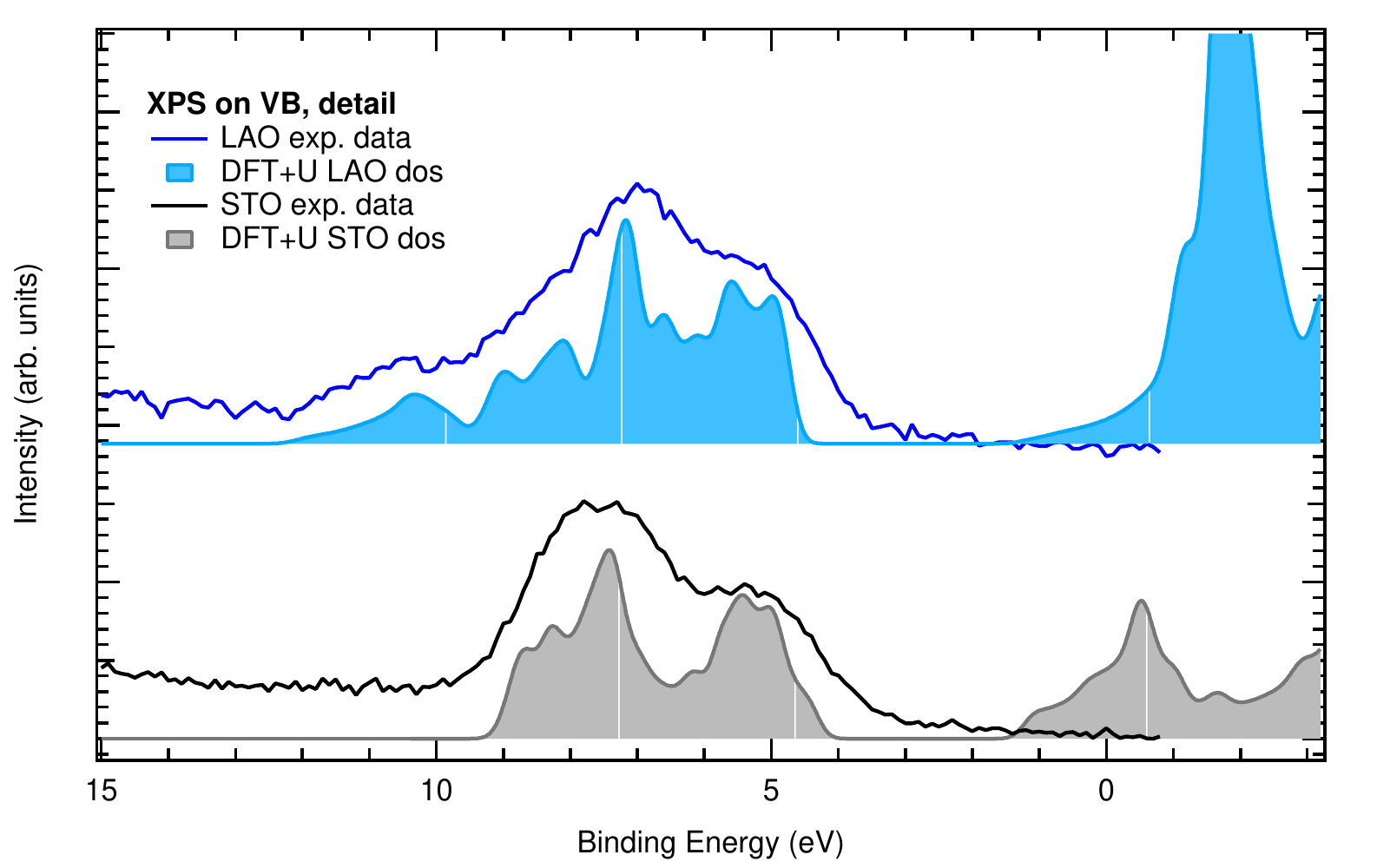}
\caption{Single crystal LaAlO$_3$ and SrTiO$_3$ valence band spectra, compared to DFT calculations of the density of states. The calculations have been done on a relaxed cubic perovskite single cell with the LDA+U approach.\label{fig_VB_3}}
\end{center}
\end{figure}

In Fig. \ref{fig_VB_3} the experimental LAO and STO valence band spectra, compared to the total density of state calculated with DFT on the cubic perovskite single cells, are shown. An LDA+U approach has been used in order to recover the correct band gap, as implemented in the \texttt{ABINIT}\cite{ABINIT} code. The parameters used in the calculation are\cite{DFT_U} U$_{STO}$=8.0 eV, J$_{STO}$=0.8 eV, U$_{LAO}$= 11.0 eV and J$_{LAO}$=0.68 eV. Since the VB is mostly composed by just one component (oxygen 2p states), the cross section of different electronic levels has not been considered. The agreement with calculation is fairly good; LAO states in particular are marked by the extra spectral weight at BE=11 eV, which is found also in DFT calculation. This good agreement can help to achieve a better energy alinement of the separate LAO and STO contributions in LAO-STO, especially in the ResPES measurement.

\section{XAS and XLD}
\begin{figure}
\begin{center}
\includegraphics[width=1\textwidth]{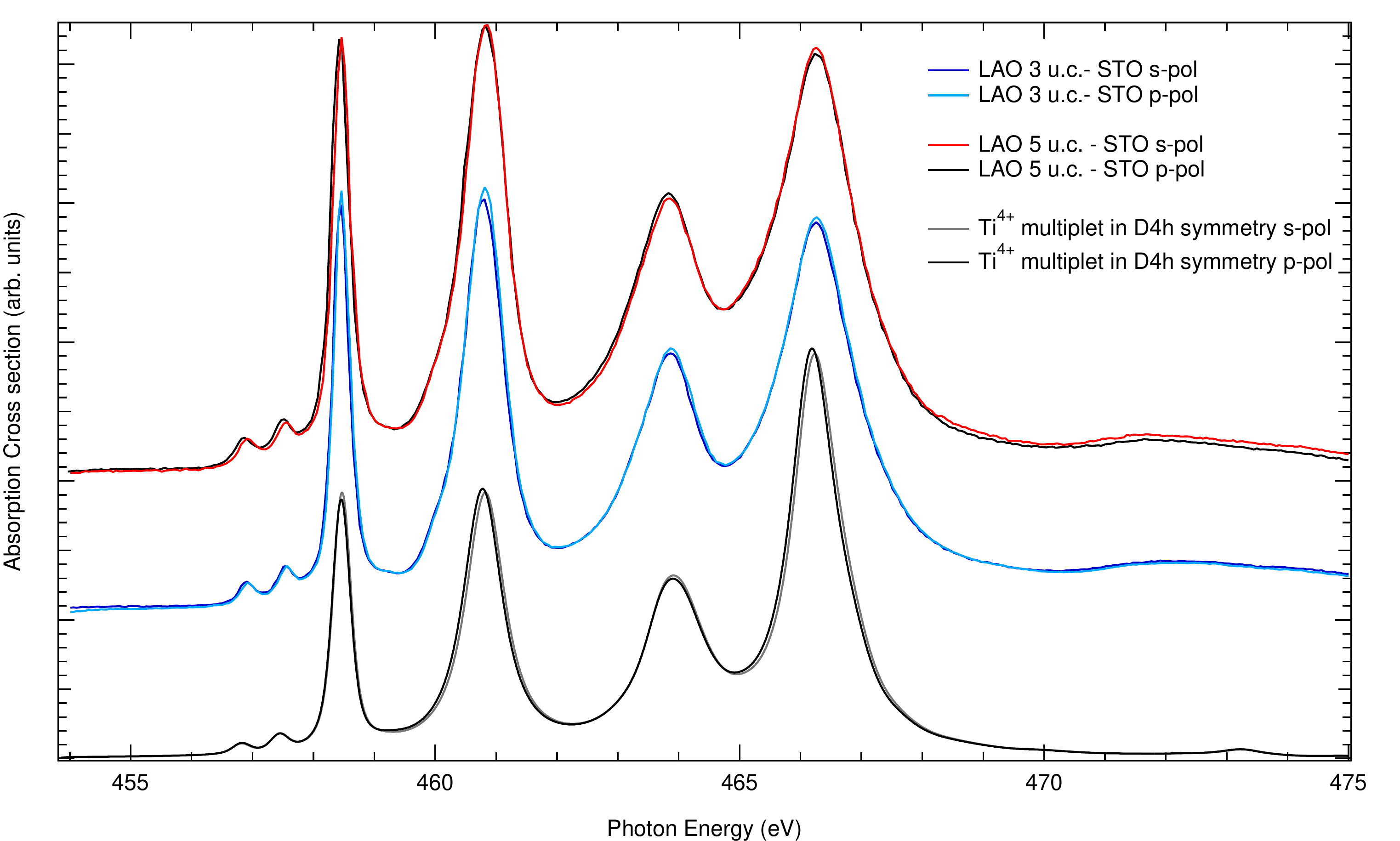}
\caption{3 u.c. and 5 u.c. LAO-STO Ti L$_{2,3}$ edge absorption spectra, compared with multiplet calculation. XAS spectra have been recorded at 30$^\circ$ incidence with linear polarized light (horizontal and vertical, see Fig. \ref{fig_BACH_setup}), and calculations have been done for an isolated Ti$^{4+}$ in a slightly distorted octahedral (D$_{4h}$) symmetry, following the work of Salluzzo et. al.\cite{LAOSTO_XLD}. X-ray linear dichroism results are given in Fig. \ref{fig_xld}.\label{fig_xas}}
\end{center}
\end{figure}

X-ray absorption spectra and X-ray linear dichroism spectra have been collected at the BACH beamline of the ELETTRA synchrotron. Measurements have been carried out in the same configuration of XPS (Fig. \ref{fig_BACH_setup}), with a photon resolution of 0.15 eV at $h\nu$=455.0 eV, at room-temperature. XAS spectra have been taken with a completely in-plane s polarization (vertical polarization) and with a partially out-of-plane p polarization (horizontal polarization). Total yield (i.e. drain current) has been the most reliable XAS detection mode, given the insulating nature of 3 u.c. samples. In Fig. \ref{fig_xas} XAS results are shown on Ti L$_{2,3}$ edge for 3 u.c. and 5 u.c. LAO-STO samples.

\begin{figure}
\begin{center}
\includegraphics[width=1\textwidth]{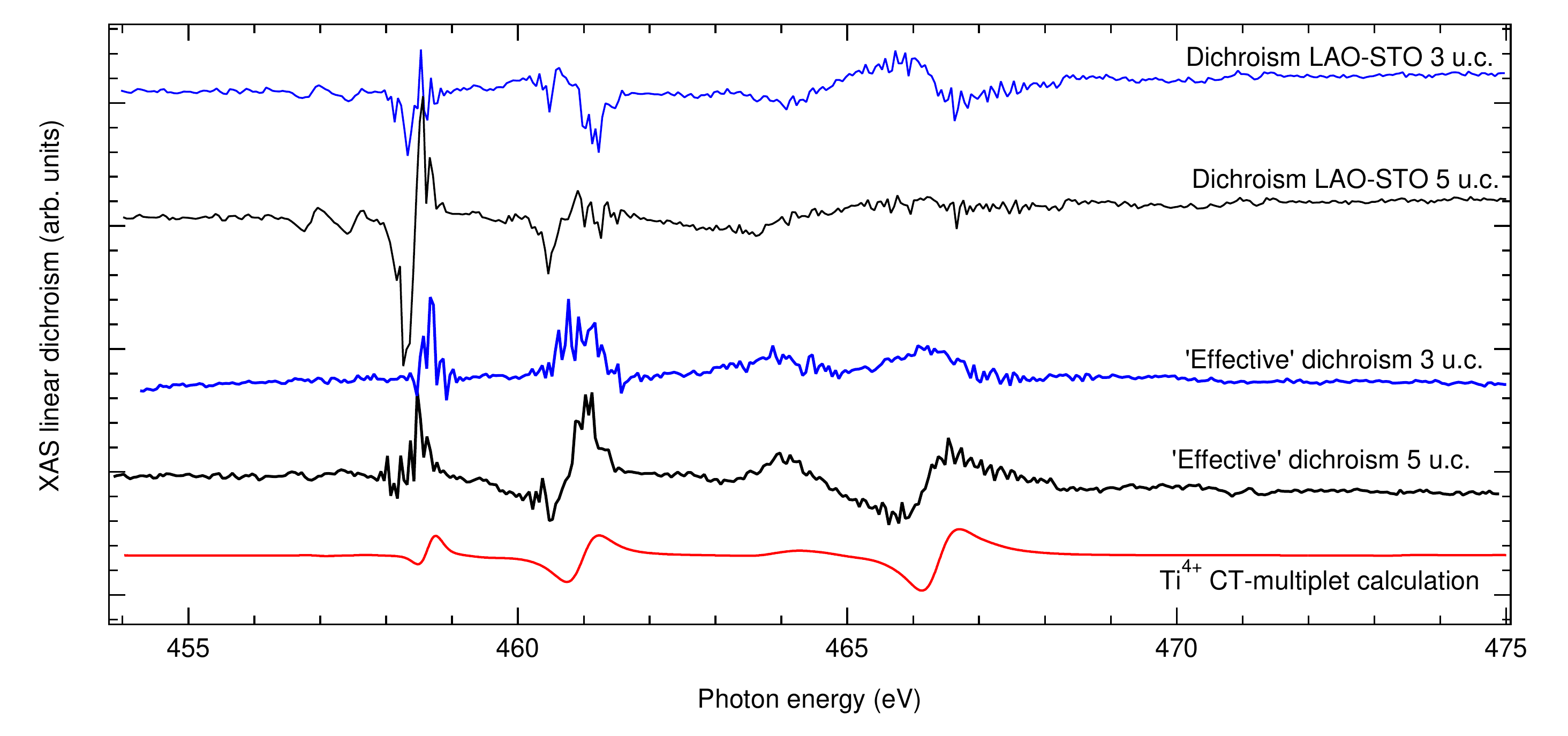}
\caption{3 u.c. and 5 u.c. LAO-STO Ti L$_{2,3}$ edge XLD spectra, compared to multiplet calculations. First, the spectra of Fig. \ref{fig_xas} have been subtracted from each other as measured (dichroism spectra); then a relative shift has been imposed in order to minimize the XLD signal (`effective' dichroism spectra). The conductive 5 u.c. sample shows a detectable dichroism, while minimization on 3 u.c. spectrum give a non dichroic signal (always positive) indicating that the two s and p original spectra are identical in shape. Multiplet calculation for a Ti$^{4+}$ ions in a D$_{4h}$ symmetry are shown in the bottom.\label{fig_xld}}
\end{center}
\end{figure}

As shown in Fig. \ref{fig_xas}, the shape of the Ti L$_{2,3}$ absorption edge can be well reproduced by a multiplet calculation for a Ti$^{4+}$ ion in an O$_h$ symmetry, with 10Dq=1.6 eV. The main difference from TiO$_2$ spectra is the absence of L$_3$ E$_g$ peak splitting (see Fig. \ref{Image_multiplet}), due to the higher symmetry of the [TiO$_6$] cluster in the heterostructure. Salluzzo et al.\cite{LAOSTO_XLD} have already reported a very weak dichroic signal at LAO-STO Ti L-edge, as a difference of circularly and linearly polarized XAS. Dichroism is not expected in an perfect O$_h$ environment, so that theoretical spectra (Fig. \ref{fig_xld}) have to be calculated by slightly lowering the symmetry group to a distorted octahedral (D$_{4h}$), with ds=dt=25 meV. However, from an experimental point of view, such differences are so small that could be generated by a minor shift (less than 0.1 eV) of photon energy during data acquisition.

In fact, the difference of two identical spectra shifted by a small quantity on the x-axis could give an effective dichroic signal, caused by the experimental set-up and not to by a real physics effect. In a genuine dichroic effect, the spectral shape has to be different (and not only translated) on in-plane and out-of-plane polarization. A check of the linear dichroism can thus be evaluated by considering the difference between the s-polarized and the shifted p-polarized spectra, till a minimum of total intensity is found as a function of displacement. The results of this procedure are given in Fig. \ref{fig_xld}.

The subtraction of Fig. \ref{fig_xas} XAS spectra yields a strong measurable XLD signal for both samples, but a careful check (the results of minimization are labeled as displaced spectra of Fig. \ref{fig_xld}) reveals that in the 5 u.c. the XLD signal persist even after the displacement, while in the 3 u.c. the minimum difference spectra is only positive, indicating the absence of a real shape differences in the p and s spectra.

These findings could indicate that the extra-energy caused by the discontinuity is relaxed through a ferroelectric-like distortion\cite{LAOSTO_XLD} of the Ti octahedron at the interface, and that this distortion is higher in conductive sample than in the insulating one. Moreover, both XAS and XLD data don't show any trace of Ti$^{3+}$ related features (theoretical spectra can be always calculated with a Ti$^{4+}$ configuration only). Since the XPS data provide evidence for Ti$^{3+}$ states, the differences in results can be ascribed to the different probing depth of the two technique, the XPS being the most sensitive to interface. The octahedron distortion can be a consequence of the strain effect due to the lattice mismatch\cite{LAOSTO_SXRD}.

\begin{figure}
\begin{center}
\includegraphics[width=0.8\textwidth]{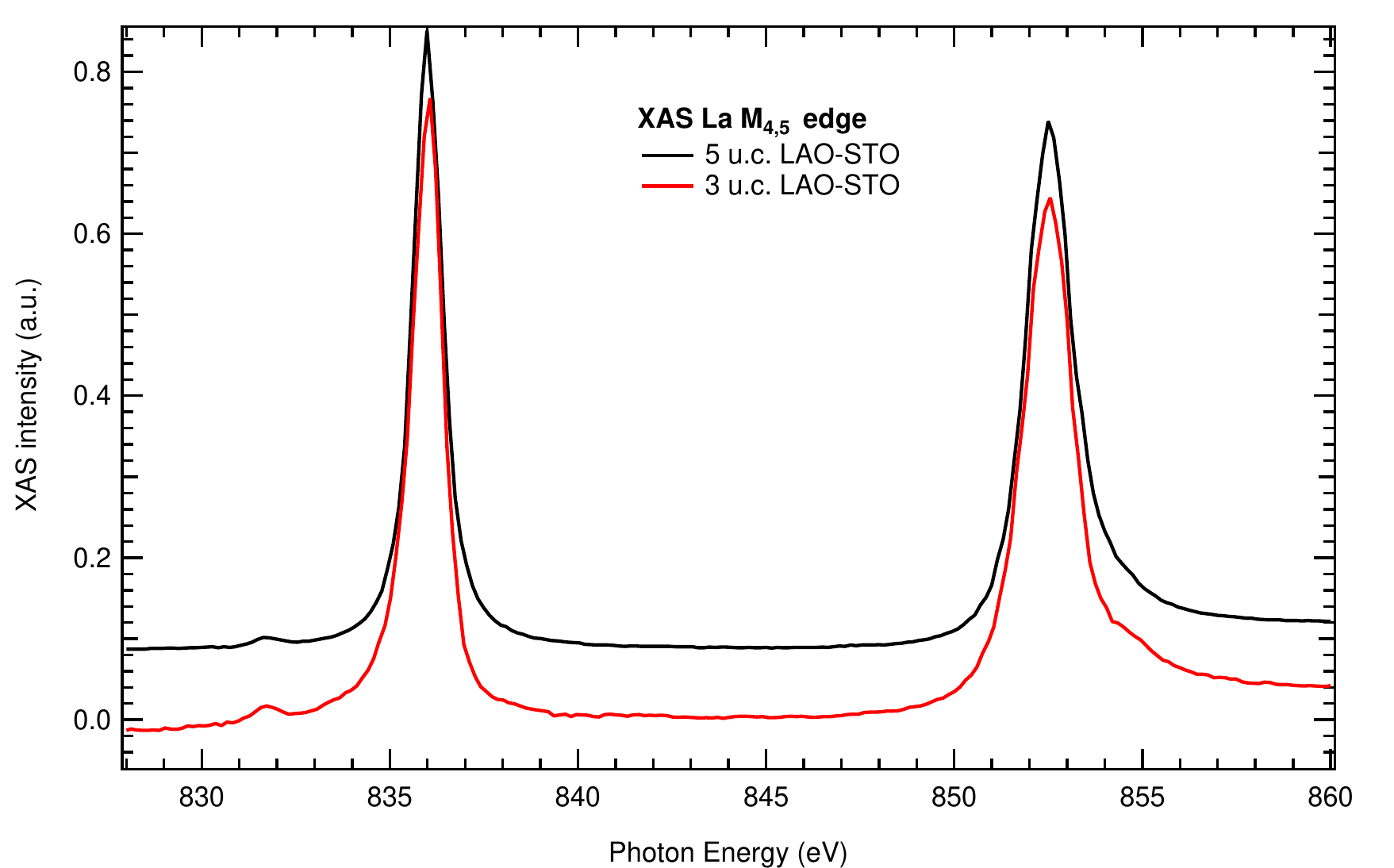}
\caption{3 u.c. and 5 u.c. LAO-STO lanthanum M$_{4,5}$ edge XAS
spectra.\label{fig_xas_la}}
\end{center}
\end{figure}

In Fig. \ref{fig_xas_la} the XAS spectrum at lanthanum M$_{4,5}$ edge of 3 and 5 u.c. samples is given. These spectra are composed by of a simple 3 peak structure, two main
peaks (3d$_{3/2}$ and 3d$_{5/2}$) and a weak small pre-peak. The shape can be easily calculated with multiplet calculation\cite{DEGROOT}, and doesn't correspond to the presence of metallic lanthanum. No dichroic signal has been detected. Oxygen k-edge XAS is not reported here since it contains contribution of STO, LAO and the surface contamination. Moreover, charging effect on insulating oxides are enhanced at oxygen absorption edge, preventing the collection of the 3 u.c. spectra.

\section{ResPES}
\begin{figure}[h!]
\begin{center}
\includegraphics[width=0.8\textwidth]{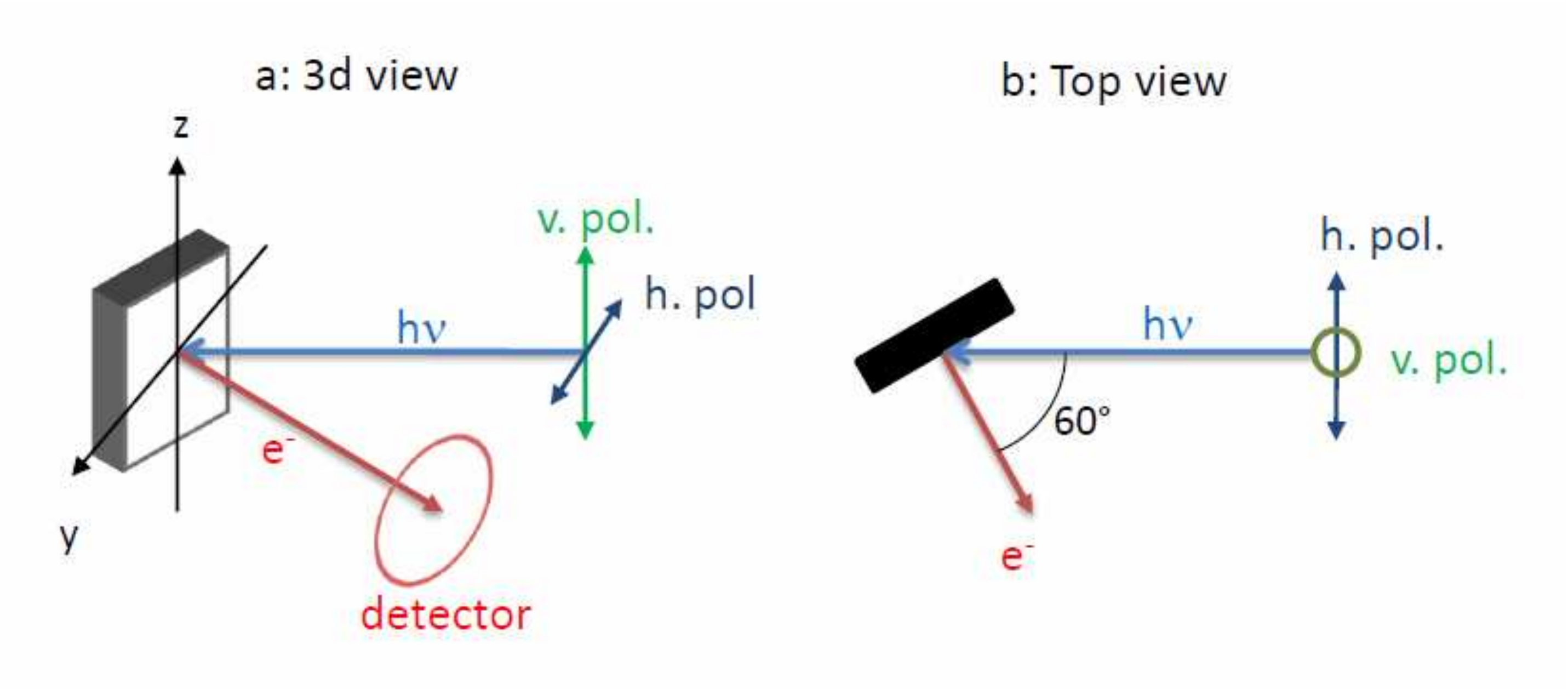}
\caption{Analyzer and sample geometry during photoemission and XLD
measurement; in A the 3D view, in B the top view. X-ray vertical and
horizontal polarization (v.pol. and h.pol.) are marked by green and
blue arrows.\label{fig_BACH_setup}}
\end{center}
\end{figure}

The ResPES data have been collected in two different sessions at BACH beam line with two sets of identical LAO-STO samples; results of first beam-time (already published in Ref.\cite{apl_LAOSTO}) have been used to better focus on the data collection of the second beam time. Experimental results of the two beam-times are given in two separated paragraphs. To achieve the maximum ``bulk'' sensitivity during measurement, photoelectron have been collected at normal emission angle (see Fig. \ref{fig_BACH_setup}), thus forming a 30$^\circ$ angle with the soft X-ray beam.

This set-up, which is the same used for XLD measurements, is suitable for measuring dichroic effects in this sample; in particular, the ``vertical'' polarization (Fig. \ref{fig_BACH_setup}(b)) has an electric field that oscillate entirely on sample plane, while the ``horizontal'' polarization has a strong out-of-plane component. One of the aim of this work has been to find possible dichroic effects in photoemission, due to the quasi-2D nature of the conductive layer.

\subsection{First data set}
\begin{figure}
\begin{center}
\includegraphics[width=0.9\textwidth]{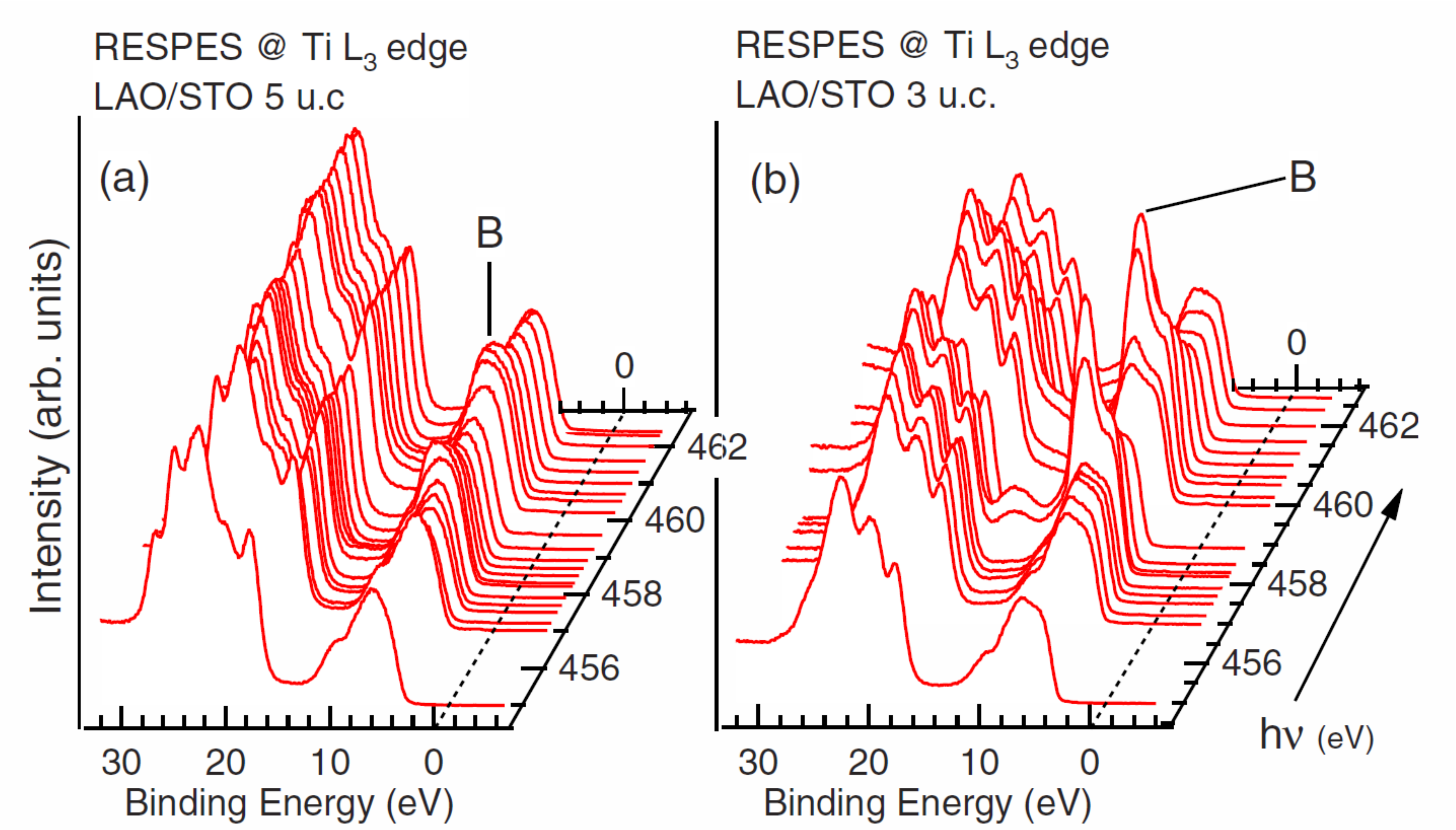}
\caption{3 u.c. (a) and 5 u.c. (b) LAO-STO resonant photoemission spectra taken at Ti L edge. Spectra contains shallow core levels (around BE=22 eV) and valence band (around BE=7 eV). Main (Ti$^{4+}$) resonances are marked with the letter B. Taken from Ref.\cite{apl_LAOSTO}.\label{fig_RESPES_1_sp}}
\end{center}
\end{figure}

As in the TiO$_2$ case, ResPES have been collected at the Ti
L$_{2,3}$ edge. The results for both the 5 u.c. and 3 u.c. LAO-STO are shown in
Fig. \ref{fig_RESPES_1_sp}.

\begin{figure}
\begin{center}
\includegraphics[width=0.8\textwidth]{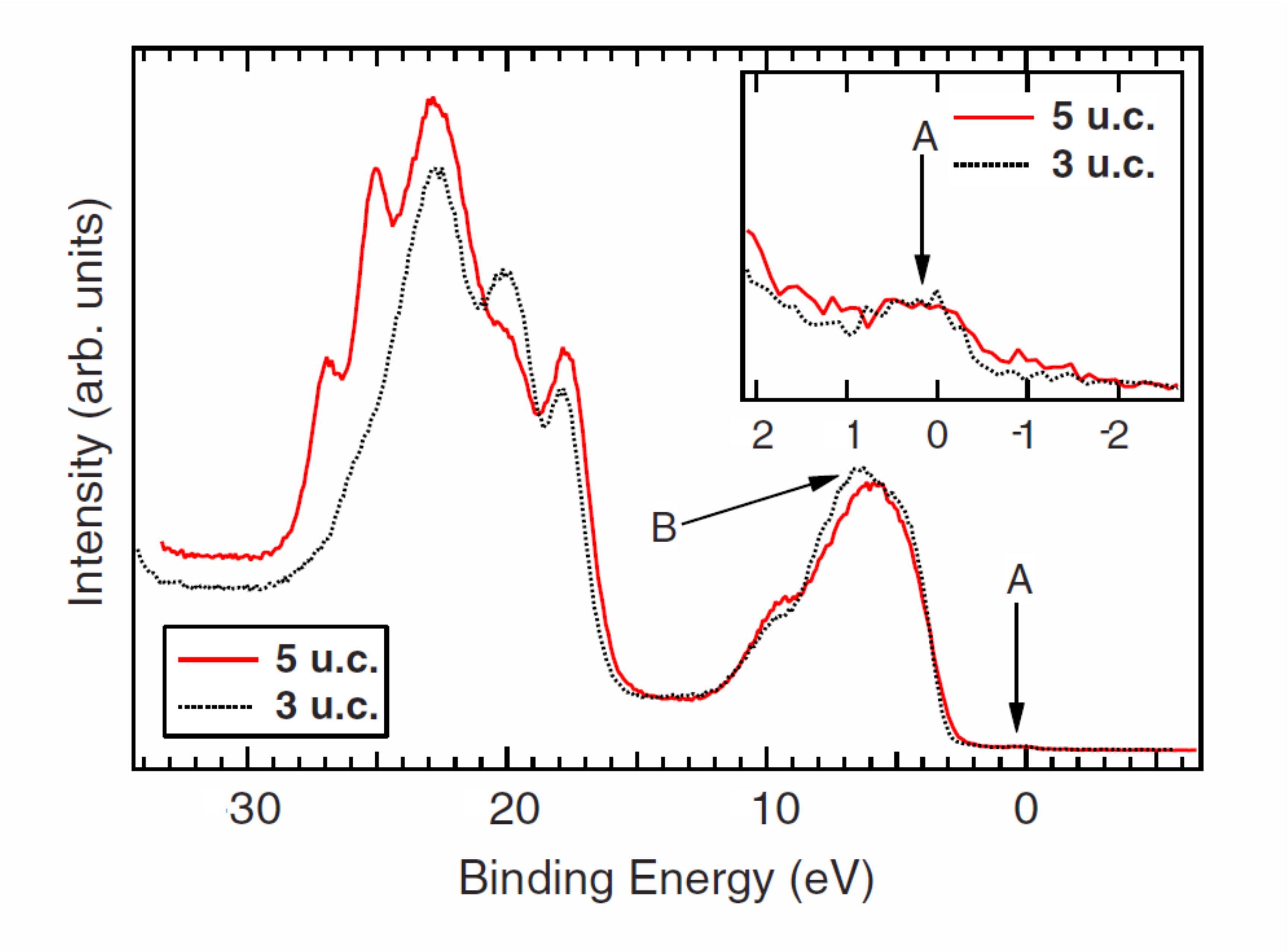}
\caption{Valence band spectra of the 3 and 5 u.c. samples collected at a photon energy of 457 eV. The inset shows the in-gap states for both samples. Taken from Ref.\cite{apl_LAOSTO}.\label{fig_RESPES_1_res}}
\end{center}
\end{figure}

The ResPES spectra Fig. \ref{fig_RESPES_1_sp} show a resonant behavior of the peak at BE=6 eV (marked as B in both ResPES set of data), which is ascribed to Ti states hybridizing with O-states in the valence band, as described in Chapter 3 for the TiO$_2$ case. The resonance is most intense for the 3 u.c. sample (Fig. \ref{fig_RESPES_1_sp}(b)), where the Ti is closer to the surface than for the 5 u.c. sample. The ResPES also reveals in-gap electronic states marked as A in Fig. \ref{fig_RESPES_1_res}. The analysis of these states is accomplished by examining two spectra collected with a photon energy of 457 eV, as shown in Fig. \ref{fig_RESPES_1_res}. At this excitation energy, only the Ti$^{3+}$ seems to resonate.

\begin{figure}
\begin{center}
\includegraphics[width=0.8\textwidth]{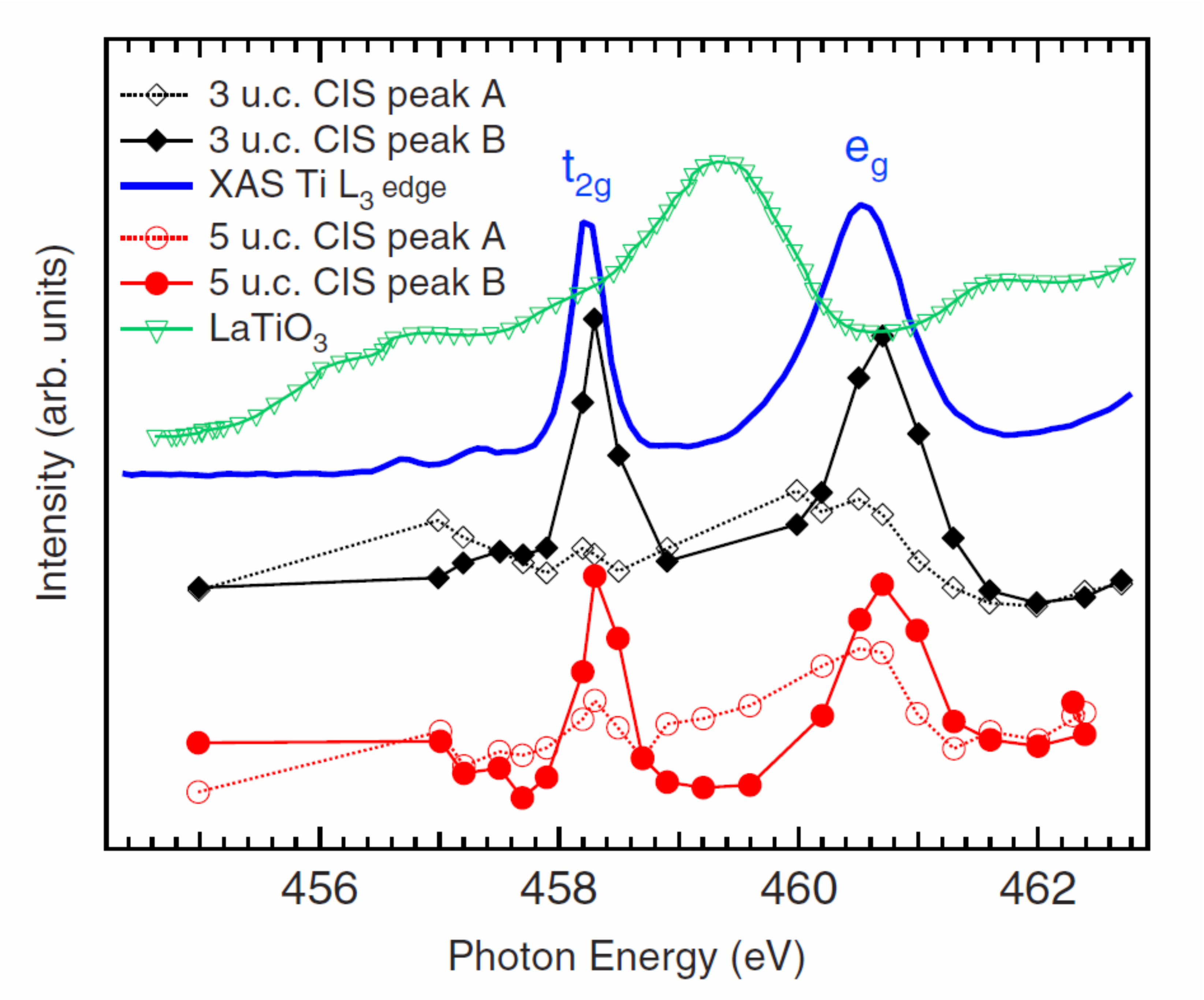}
\caption{XAS of the 5 u.c. sample (thick line) and of LaTiO$_3$ (adapted form Ref.\cite{LAOSTO_XLD}, triangles). CIS profile of the A and B features for the 3 u.c. sample (empty and filled diamonds, respectively). CIS profile of the A and B features for the 5 u.c. sample (empty and filled circles, respectively. Taken from Ref.\cite{apl_LAOSTO}.\label{fig_respes_1_cis}}
\end{center}
\end{figure}

Apparently, the in-gap states have the same weight in both samples. However, because of the thicker LAO overlayer in the 5 u.c. sample, the effective density of the in-gap states is larger for the conducting 5 u.c. sample. The excess of in-gap electronic states is consistent with the Ti$^{3+}$ states found in XPS Ti 2p core level data of Fig. \ref{fig_XPS_Ti2p_a}. If we assume that the only difference between the two samples is the thickness of the LAO overlayer, the attenuation of the signal based on the estimated for electrons with 457 eV kinetic energy would give an $I_{A}^{3 u.c.}/I_{A}^{5 u.c.}=1.97$ ratio between the intensities I of the A peaks of the two samples. Thus, as the measured intensities appear to be the same, the density of in-gap states for the 5 u.c. sample results to be 1.97 times that of the 3 u.c. sample. These estimates will be improved with the second data set results. Finally, CIS spectra are given in Fig. \ref{fig_respes_1_cis}.

The CIS spectra are shown for the in-gap states (Fig. \ref{fig_respes_1_cis}, A feature), as well as for the most resonating feature at BE=6 eV (Fig. \ref{fig_respes_1_cis}, B feature). The peak at 458.2 eV is ascribed to t$_{2g}$ states arising from crystal field splitting, while the broad peak at 460.6 eV is ascribed to e$_g$ states. As already explained in Chapter 2, the CIS spectra of the B feature follow the intensity of the Ti L$_3$-edge XAS (only the XAS for the 5 u.c. sample is shown, the other being quite similar) in agreement with their origin from Ti$^{4+}$ electronic states in the valence band. As in the case of defective TiO$_2$, the CIS spectrum of the A band shows a different behavior, with a quenching of the intensity below the t$_{2g}$ peak and a shift toward lower photon energies of the e$_g$ band. In spite of the weaker signal with respect to the CIS spectrum extracted form peak B, the CIS profiles of peak A show a higher intensity where the most intense XAS bands of a Ti$^{+3}$ ion are found (for comparison, the Ti L$_3$ edge of LaTiO$_3$ is given, Fig. \ref{fig_respes_1_cis}, thick line). This adds further evidence to the 3d$^1$ character of the in-gap states. However, the experimental noise on CIS spectra of Fig. \ref{fig_respes_1_cis} doesn't allow any prediction about the symmetry of Ti bonding at the interface.

\subsection{Second data set}
\begin{figure}[h!]
\begin{center}
\includegraphics[width=0.9\textwidth]{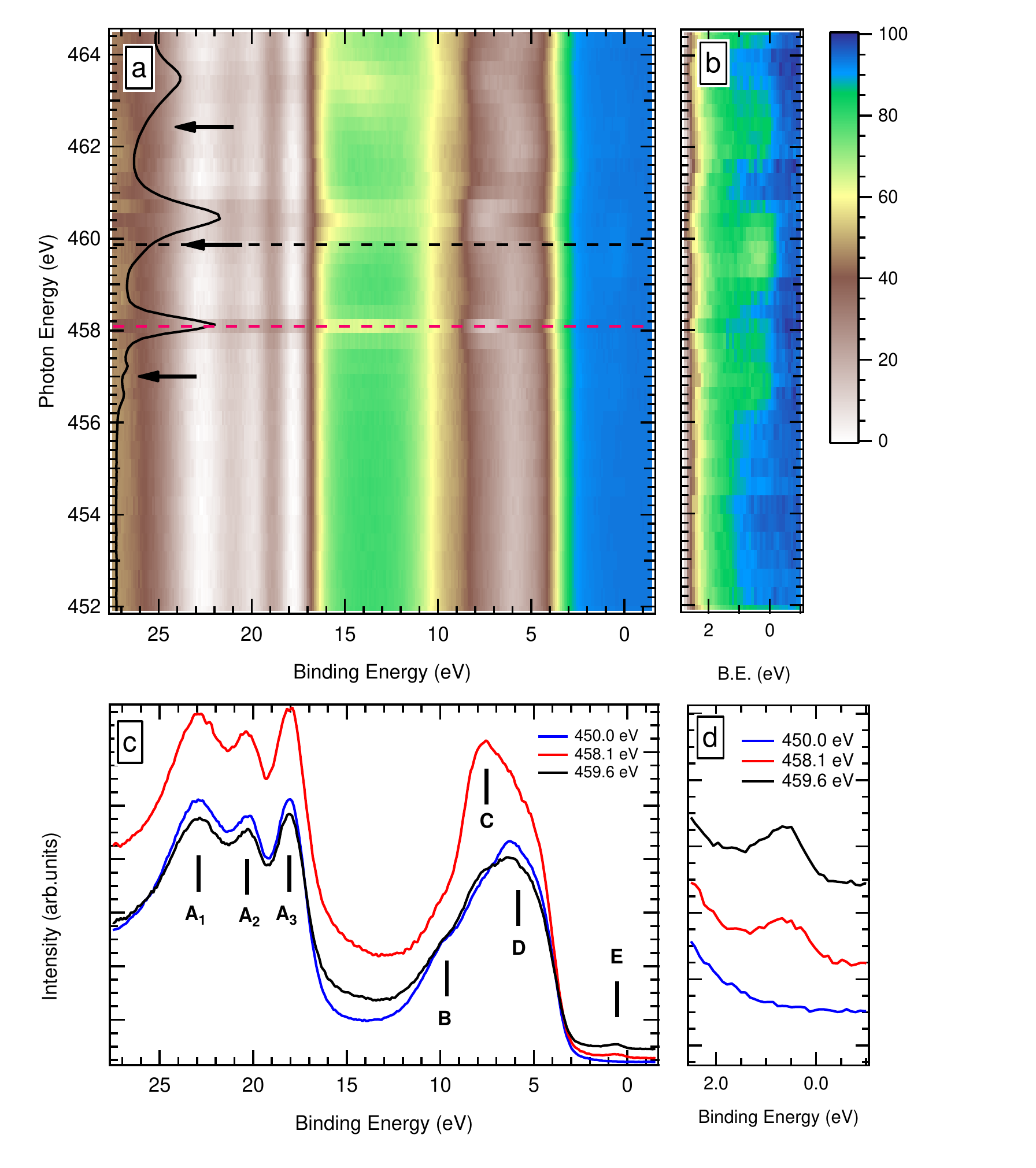}
\caption{Second ResPES data-set at L-edge on 5 u.c. LAO-STO sample. In panel (a) ResPES data in a 2D picture and in panel (b) a detailed view on Ti$^{3+}$ defects; on the left axis XAS (black line) is depicted. In panel (c), spectra taken at out-of resonance (blue line, blue dashed horizontal line of panel (a)), maximum of Ti$^{4+}$ spectra (red line) and maximum of Ti$^{3+}$ (black line) conditions. In panel (d) the detail of in-gap state spectra.\label{fig_respes_2_img5}}
\end{center}
\end{figure}

In order to improve the quality of CIS and in-gap state spectra another set of ResPES data has been collected; in this second beamtime also the polarization properties of VB photoemission have been proved. Most of the experimental data (CIS and ResPES spectra) shown in this paragraph have been taken with in-plane polarized X-rays, which give the maximum intensity on Ti resonances. Further comment on linear polarization effect will be given at the end of the paragraph. Following the case of TiO$_2$, the Ti L-edge ResPES more spectra have been collected in the pre-edge region where only the Ti$^{3+}$ states should be present. The X-ray flux intensity has been reduced for insulating samples in order to control the charging effects, resulting in an higher noise on experimental data. ResPES data for the 5 u.c. sample are shown in Fig. \ref{fig_respes_2_img5}.

The maxima of in-gap state intensity, marked with black arrow in Fig. \ref{fig_respes_2_img5}.(a), are at different photon energy than main resonance peak. In order to better visualize the in-gap states, in Fig. \ref{fig_respes_2_img5}.(b) an enlargement of ResPES image is depicted, with a different color scale. Panel (c) of Fig. \ref{fig_respes_2_img5} depicts selected photoemission spectra from panel (a) data, measured in out-of-resonance conditions (\textnormal{$h\nu$}=452 eV) and at the maximum intensity of Ti$^{4+}$ resonance (\textnormal{$h\nu$}=458.1 eV) and Ti$^{3+}$ (\textnormal{$h\nu$}=459.6 eV); the horizontal dashed lines mark the single spectra position on the main image. Peaks A$_1$,A$_2$ and A$_3$ are the shallow core levels, B and D are the most prominent VB features, C is the position of Ti$^{4+}$ resonance peak maximum and E is the Ti$^{3+}$ electronic state. Remarkably, the in-gap state appear to be at Fermi level, as expected from a conductive sample.

\begin{figure}[h!]
\begin{center}
\includegraphics[width=0.9\textwidth]{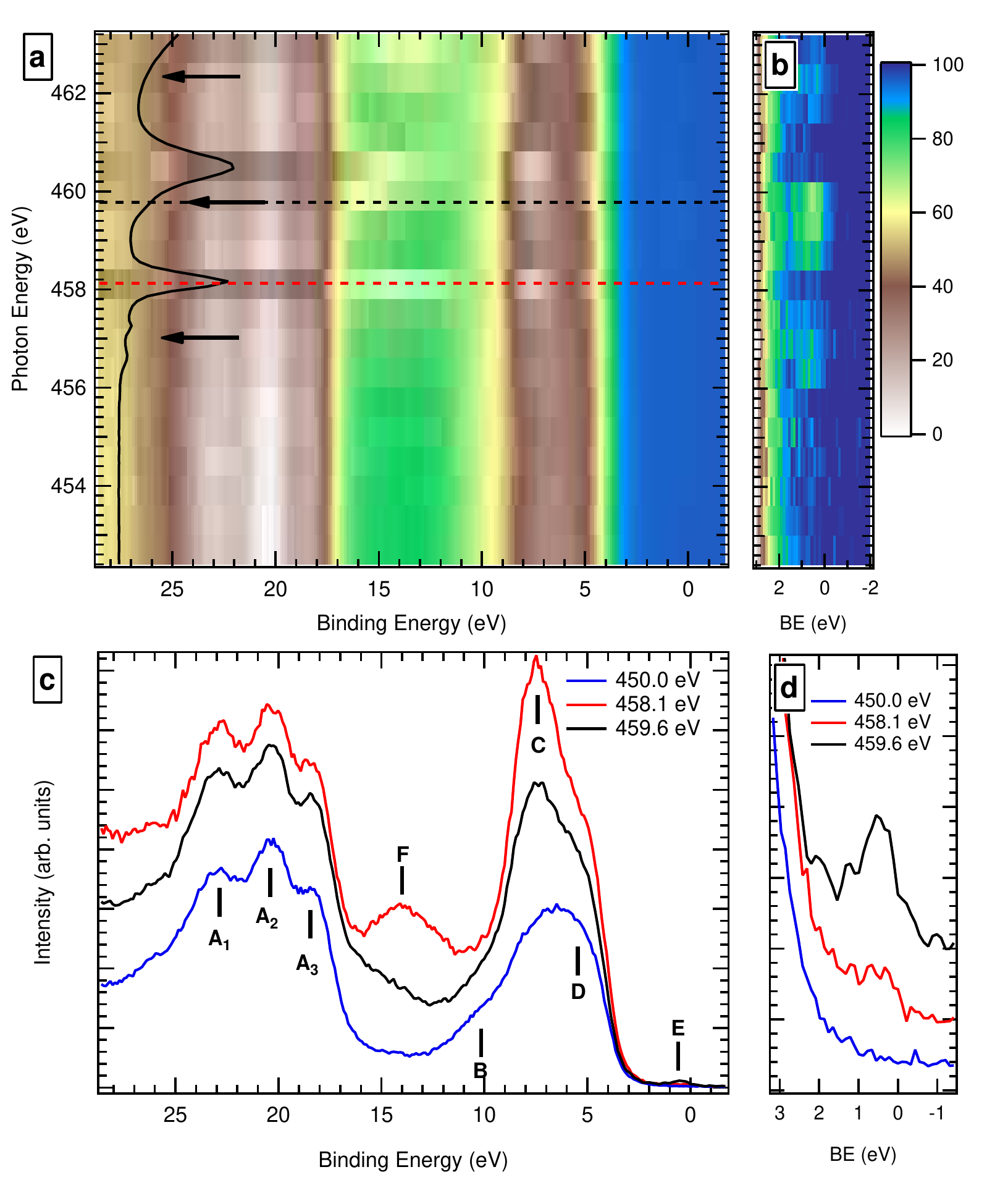}
\caption{Second ResPES data-set at L-edge on 3 u.c. LAO-STO sample.
In panel (a) ResPES data in a 2D picture and in panel (b) a detailed view on Ti$^{3+}$ defects; on the left axis XAS (black line) is depicted. In panel (c), spectra taken at out-of resonance (blue line, blue dashed horizontal line of panel (a)), maximum of Ti$^{4+}$ spectra (red line) and maximum of Ti$^{3+}$ (black line) conditions. In panel (d) the detail of in-gap state spectra.\label{fig_respes_2_img3}}
\end{center}
\end{figure}

As shown in Fig. \ref{fig_respes_2_img3} the ResPES of the 3 u.c. sample displays the same identical features of the 5 u.c., except for the relative intensity of main VB resonance and shallow core levels; the peak marked by F in Fig. \ref{fig_respes_2_img3} can be ascribed, as in the case of TiO$_2$, to the LVV Auger emission and is more pronounced due to the lower LAO capping. Again, the defect peak is detected near the Fermi level, its intensity being the only difference with respect to the 5 u.c. case.

\begin{figure}[h!]
\begin{center}
\includegraphics[width=0.9\textwidth]{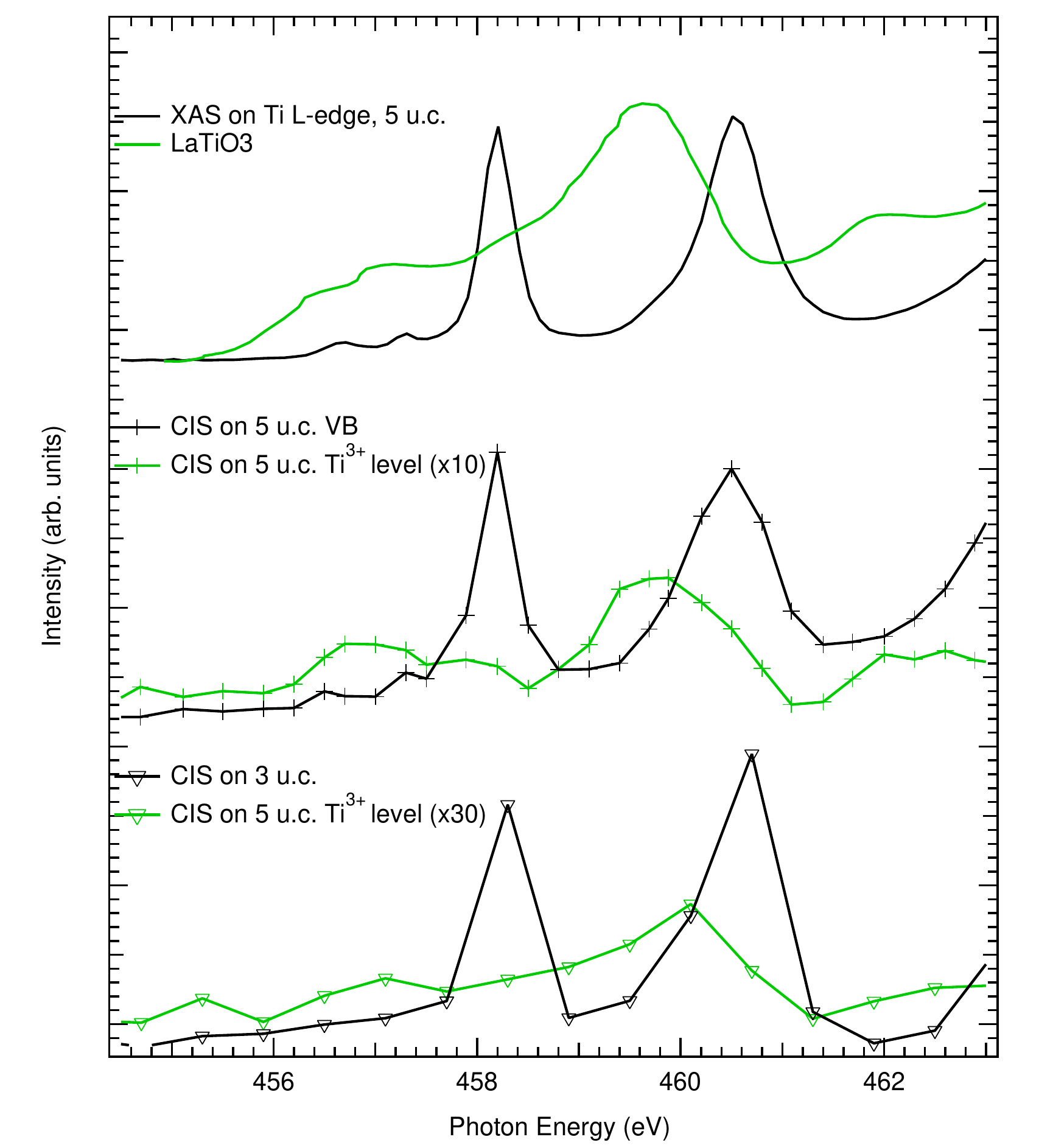}
\caption{CIS spectra of the second ResPES data-set; green and black solid lines are XAS taken at Ti L-edge on the 5 u.c. sample and on a LaTiO$_3$ crystal (adapted from Ref.\cite{LAOSTO_XLD}); cross markers are for the 5 u.c. CIS and triangles for the 3 u.c.; scaling factors have been used to visualize $Ti^{3+}$ CIS on the same scale of valence band resonances.\label{fig_respes_2_cis}}
\end{center}
\end{figure}

In Fig. \ref{fig_respes_2_cis} the CIS spectra for 3 u.c and 5 u.c. are shown and compared to Ti L-edge XAS spectra on samples and on a reference LaTiO$_3$ crystal. The Ti$^{4+}$ (Ti$^{3+}$) CIS has been collected at C (E) peak of Fig. \ref{fig_respes_2_img5}.(c) and
Fig. \ref{fig_respes_2_img3}.(c). The similarities between LaTiO$_3$ XAS and Ti$^{3+}$ CIS are evident, suggesting that of La-Sr intermixing may occur at the interface, yielding Ti$^{3+}$ states as in Sr doped LaTiO$_3$. Since the in-gap state intensity is lower in the 3 u.c. sample, different scaling factors have been applied to CIS spectra of Ti$^{3+}$.

\begin{figure}[h!]
\begin{center}
\includegraphics[width=0.9\textwidth]{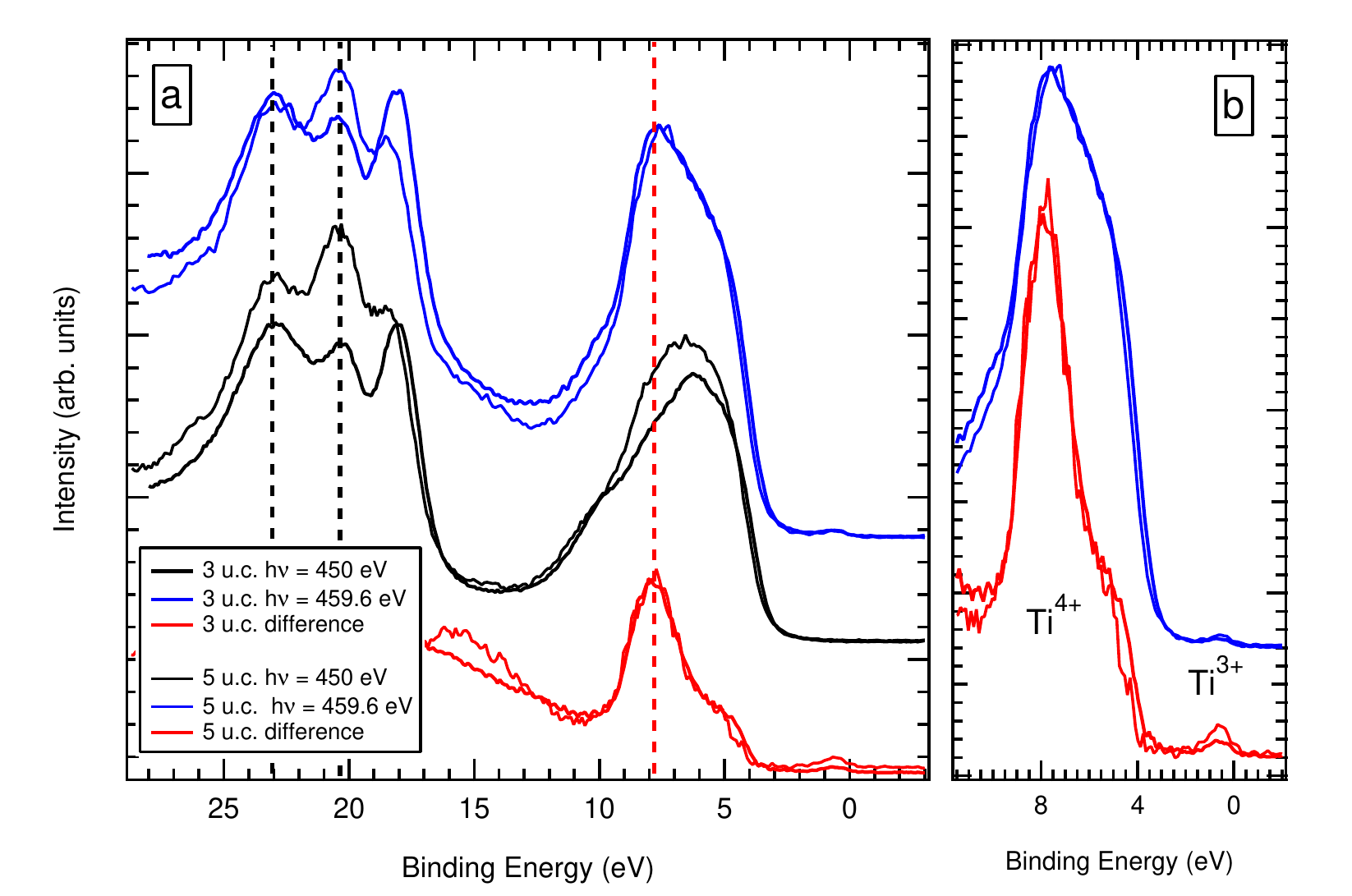}
\caption{Comparison of photoemission spectra taken in on-resonance (blue) and out-of-resonance (black) conditions on the 3 (thick line) and 5 (thin line) u.c. LAO-STO samples. The green spectra are the difference of the two and thus show the resonating contribution only. Vertical dashed lines mark the references that have been used for matching the relative energy scale. In panel B, a detail of Ti$^{3+}$ defects, normalized to the VB.\label{fig_respes_2_35}}
\end{center}
\end{figure}

A first comparison between the 3 u.c. and 5 u.c. LAO-STO samples is given in Fig. \ref{fig_respes_2_35}; black line spectra have been taken in out-of resonance conditions ($h\nu=450$ eV), blue line have been taken at the maximum of Ti$^{3+}$ resonance ($h\nu=459.6$ eV) and red spectra show only the resonating part given by the difference of the previous two. A normalization on the shallow core level areas has been done before the subtraction. By looking at the non-subtracted resonating spectra (black lines) of Fig. \ref{fig_respes_2_35} it seems that the intensity of defect states is similar in both samples; in fact the Ti$^{3+}$ states have an effective intensity which is much lower in the 3 u.c. LAO-STO, since the titanium photoemission signal should be more attenuated in the 5 u.c. sample because of the larger capping. The calculated ratio $I_{Ti 3+}^{5uc}/I_{Ti 3+}^{3uc}$ is nearly 2:1, considering the IMFP only.

This value can be affected by the relative normalization of the 5 u.c. and 3 u.c. ResPES data; a better comparison can be obtained by looking at the resonating part only (green spectra of Fig. \ref{fig_respes_2_35}) which is the difference of ResPES spectra with the corresponding out-of-resonance spectrum. Considering now the same intensity on Ti$^{4+}$ states (Fig. \ref{fig_respes_2_35}(b)), the relative ratio of 5 u.c. over 3 u.c. in-gap states becomes even larger (5:1). The intensity of Ti$^{3+}$ states marks thus the transition between the insulating and the conducting heterostructure. The resonating (red line) spectra of Fig. \ref{fig_respes_2_35} give also a fine way to adjust the relative BE scale, by pinpointing both the 3 u.c. and the 5 u.c. at the maximum of Ti$^{4+}$ states; this procedure is equivalent to fix the energy scale to the bulk STO Ti levels, which gives the highest spectral contribution to Ti$^{4+}$ states. The red vertical dashed line of Fig. \ref{fig_respes_2_35}(a) marks the reference used for the alignment, which results in the same BE also for the STO shallow core levels (O 2s and Sr 4p, vertical black dashed line). With this alignment procedure, the in-gap states of both samples are found at the same energy, i.e. crossing the Fermi edge.

\begin{figure}[ht!]
\begin{center}
\includegraphics[width=0.9\textwidth]{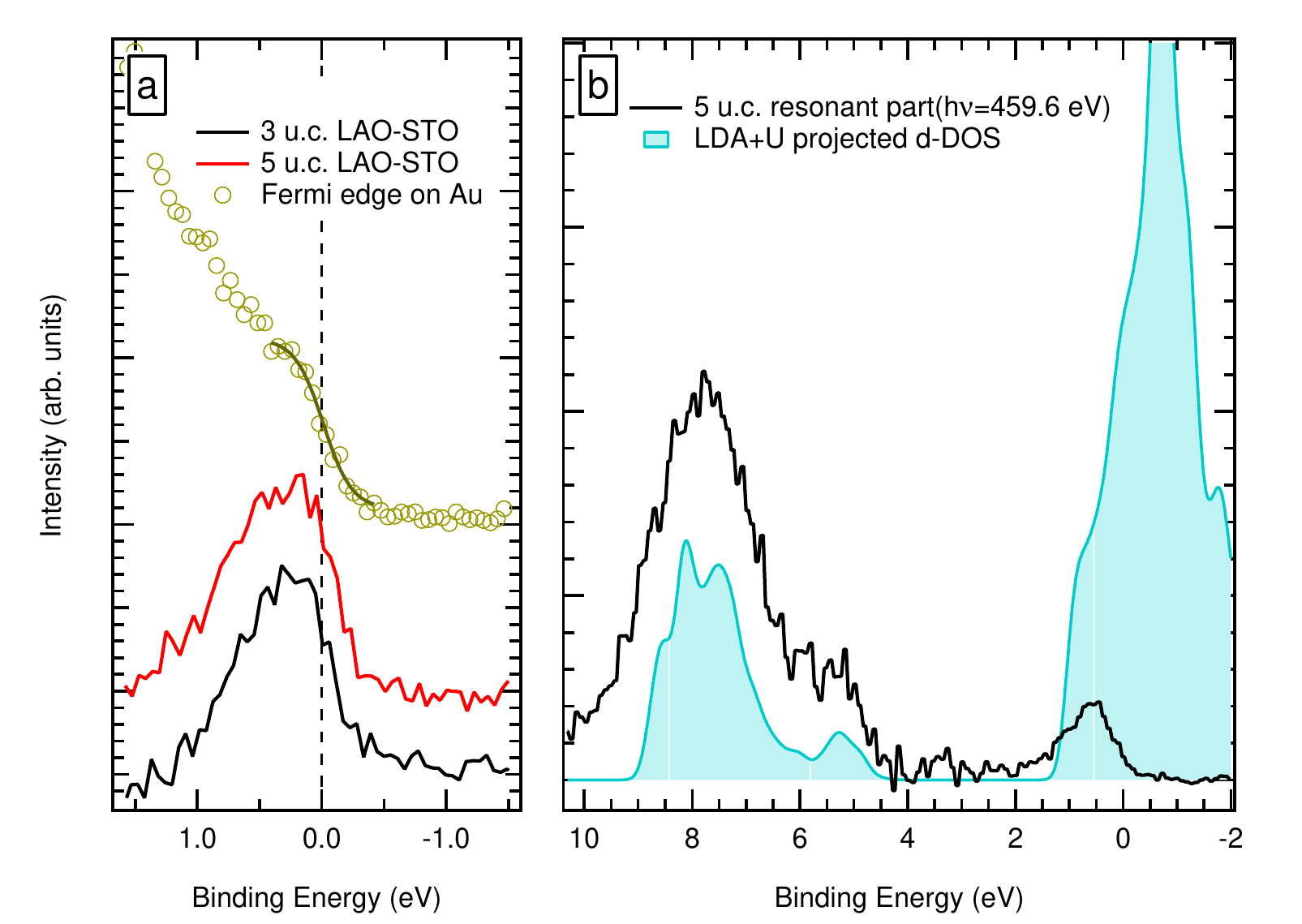}
\caption{High resolution spectra on in-gap state taken at $h\nu=459.6$ eV of 3 u.c. and 5 u.c. LAO-STO, compared with LDA+U calculations. The calculation parameters are the same of Fig. \ref{fig_VB_3}. In panel (A), the Fermi edge (yellow circle) of metallic Au deposited on the sample-holder, taken as an energy reference.\label{fig_respes_2_ingap}}
\end{center}
\end{figure}

High resolution spectra of Ti$^{3+}$ states are shown in Fig. \ref{fig_respes_2_ingap}, along with a comparison with DFT+U calculations. To better prove the energy scale, in
Fig. \ref{fig_respes_2_ingap}(a) the measured Fermi edge of metallic gold deposited on the sample-holder at the same photon energy is given. The spectral shapes of Ti$^{3+}$ states are rather similar and shows an asymmetrical profile that can be found in other 3d$^1$ (for example, the titanium oxychloride, TiClO\cite{LAOSTO_ticlo}) highly correlated compound. However, in this case the in-gap state shape is due to the filling of empty d-states, as can be seen aligning our data with DFT+U calculations, shown in Fig. \ref{fig_respes_2_ingap}(b). The DFT calculation has been carried out on a cubic single SrTiO$_3$ cell; the Hubbard U (U=8.0 eV) correction has been added to recover the correct 3.2 eV STO band-gap. The Ti$^{3+}$ state is found exactly at 3.2 eV with respect to the STO VB-band maximum. The non-correlated nature of these electronic states can be further proved by comparison with the correlated titanium 3d states measured with ResPES in Nb doped SrTiO$_3$\cite{STO_ResPES}.

\begin{figure}
\begin{center}
\includegraphics[width=0.9\textwidth]{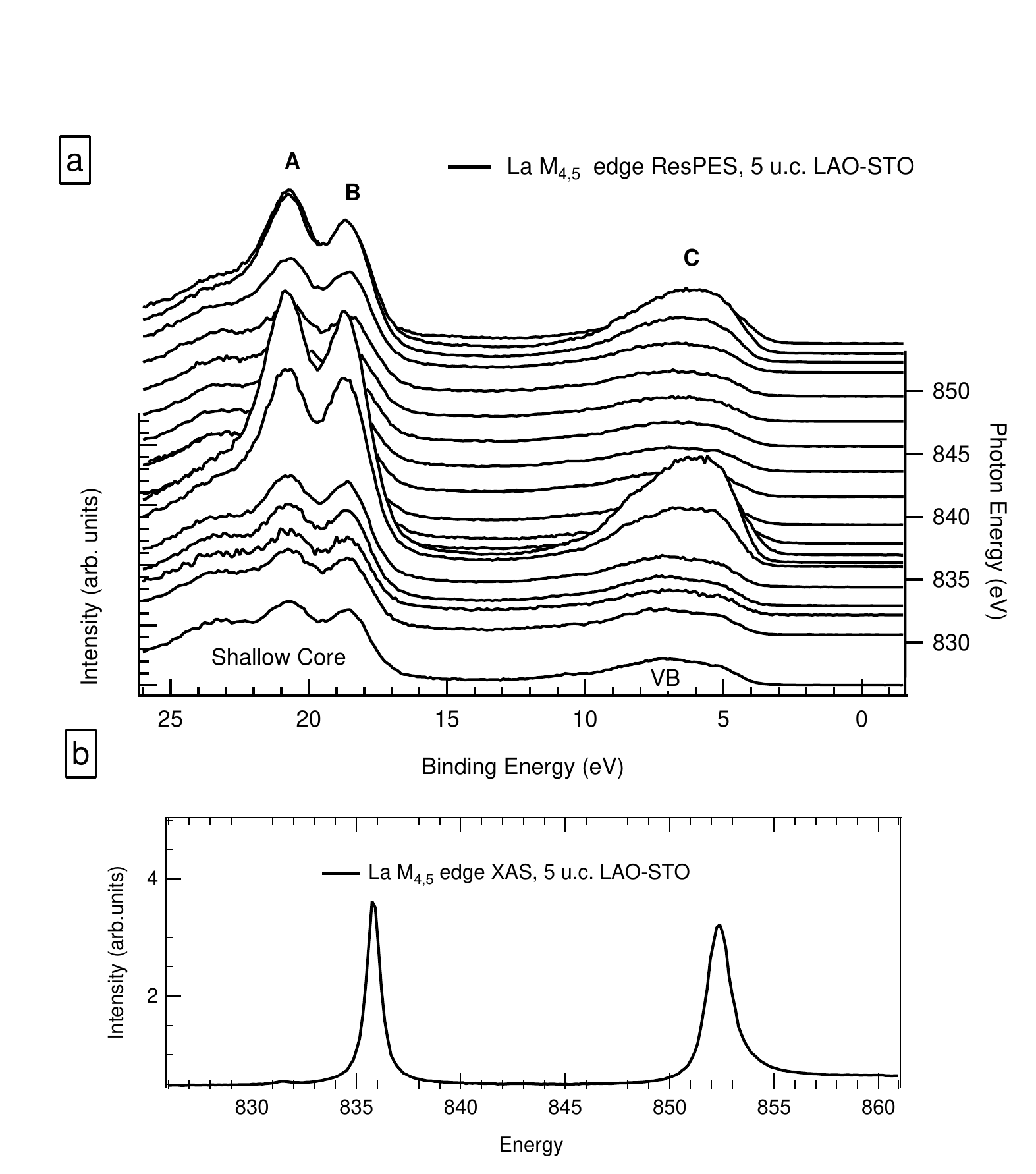}
\caption{In panel A, ResPES data and in panel B, XAS spectrum taken at La M$_{4,5}$ edge on the 5 u.c. sample.\label{fig_respes_2_la}}
\end{center}
\end{figure}

\begin{figure}
\begin{center}
\includegraphics[width=0.95\textwidth]{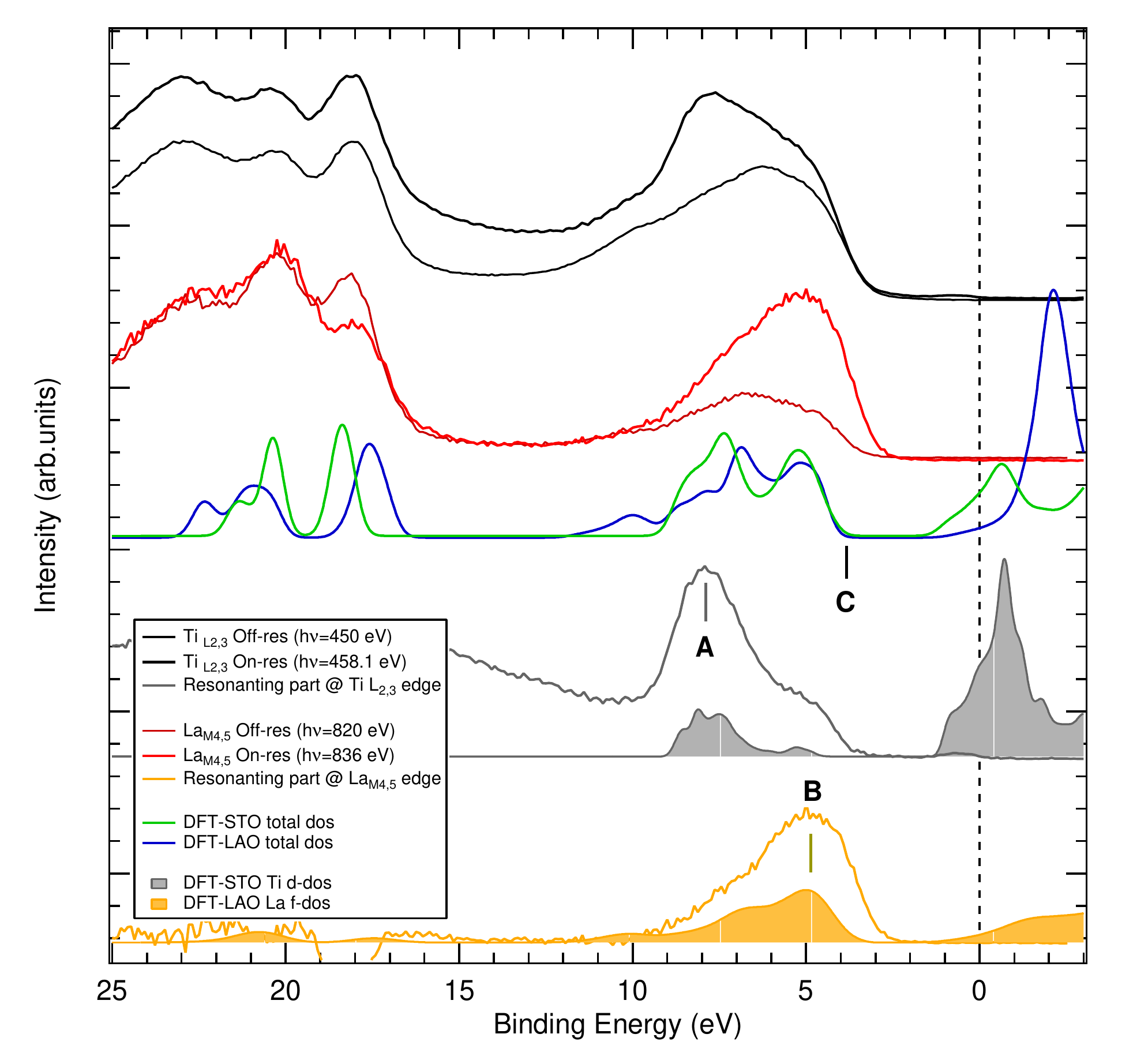}
\caption{Comparison of ResPES at Ti L and La M edges on the 5 u.c. sample. Top four spectra (black and red lines) are experimental data take at on and off-resonance condition; middle (blue and green lines) spectra are the total DOS calculated with DFT; lower spectra (yellow and grey) show the difference of ResPES spectra as compared to DFT projected DOS. The DFT calculation parameters are the same of Fig. \ref{fig_VB_3}.\label{fig_respes_2_con}}
\end{center}
\end{figure}

ResPES data at lanthanum M$_{4,5}$ edge are shown in Fig. \ref{fig_respes_2_la}. A strong resonant enhancement can be seen on the shallow core levels (La 5p); however, the in-gap state is not resonating, indicating that La states are not contributing to these electronic states and, ultimately to the transport properties. Finally, more information can be extracted from La ResPES data by considering the difference between resonant and off-resonant spectra. As in the case of Ti 3d states, the resonating spectral wight at lanthanum edge can be predicted with DFT+U La 4f projected DOS. By aligning the experimental data with calculation (in the 5 u.c. sample, peak A for titanium and peak B for lanthanum of Fig. \ref{fig_respes_2_con}), it is possible to find the mismatch of the top of valence bands (marked with C in Fig. \ref{fig_respes_2_con}) of STO and LAO separately, that is the band-bending. The \textit{relative} shift between the 3 u.c. and the 5 u.c. is again very low (0.15 eV) and it's comparable to the value extracted from XPS; this value is not enough to explain the difference in conductivity. However, the \textit{total} shift between LAO and STO, calculated from data of Fig. \ref{fig_respes_2_con}, can be quantified as 0.6 eV, in agreement with some theoretical model given in the literature\cite{bristowe}.

\begin{figure}[h!]
\begin{center}
\includegraphics[width=0.95\textwidth]{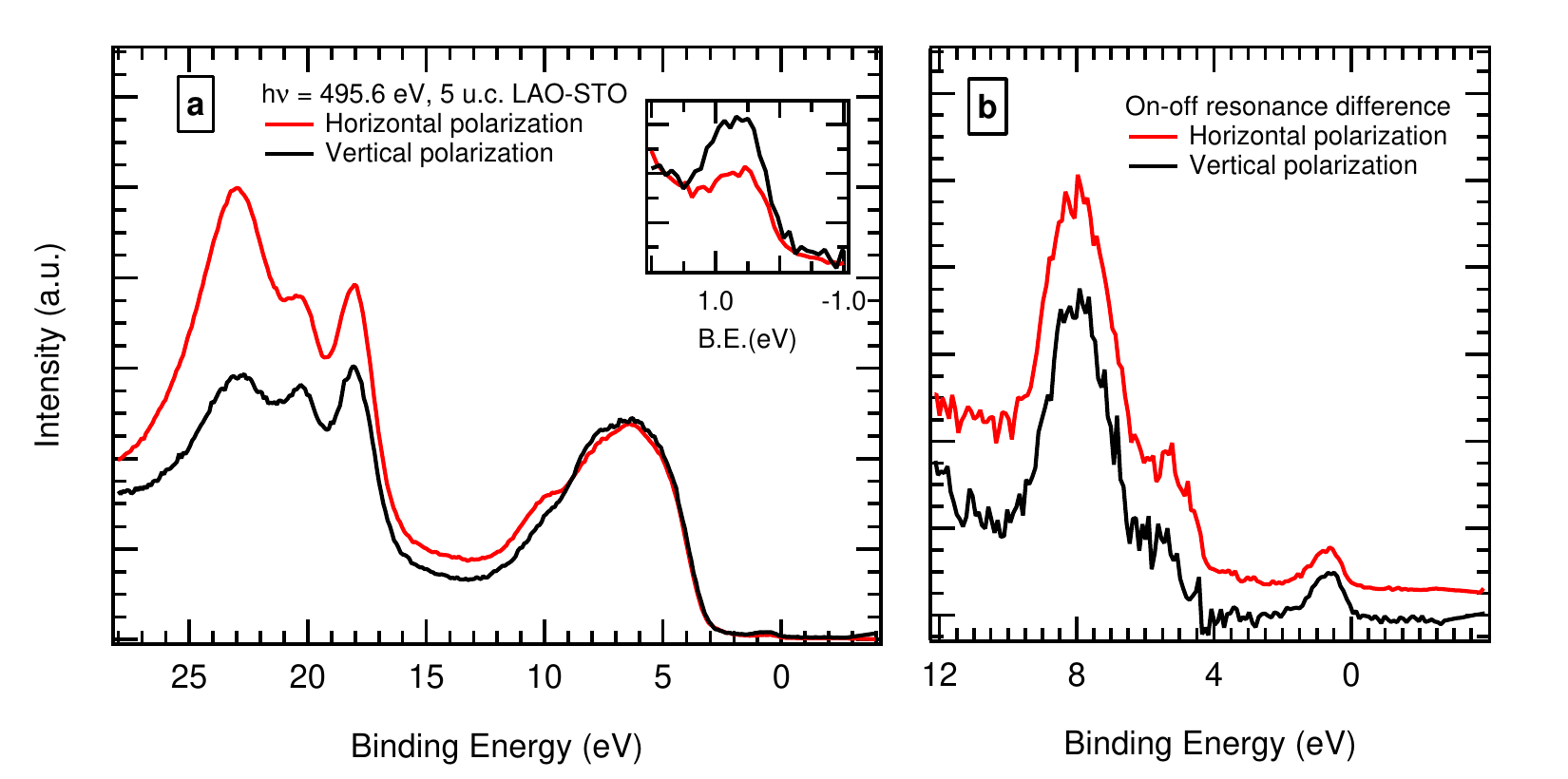}
\caption{A comparison of ResPES spectra taken with different polarization. In panel (a), spectra taken at maximum of Ti$^{3+}$ resonance in the 5 u.c. and 3 u.c. LAO-STO; in the inset, a detail on the in-gap level; in panel (b), the results of difference of on-resonant and off-resonant spectra.\label{fig_respes_2_pol}}
\end{center}
\end{figure}

Finally, ResPES at Ti edge have been collected either in vertical (completely in-plane) and horizontal (partly out-of-plane) linear polarized X-rays (see Fig. \ref{fig_Bach_Layout}); the intensity ratio of Ti$^{3+}$ resonances versus the total VB spectral weight is higher with the in-plane polarization geometry (see Fig. \ref{fig_respes_2_pol}), therefore the data shown on this section have been collected with that configuration. However, when subtracting the off-resonance contribution from the on-resonance spectra, the results are virtually identical in both polarizations (panel B of Fig. \ref{fig_respes_2_pol}). Apart from giving the optimal experimental condition for data collection, a polarization-dependent photoemission study can't give additional information concerning the physics of this system.

Summarizing, the ResPES and the XAS data suggest the following:
\begin{itemize}
    \item There is a small but detectable Ti$^{3+}$ spectral weight at Fermi level due to a filling of empty STO d-bands, both in the insulating and in the conductive samples. Those states are measurable only in resonance condition;
    \item The Ti$^{3+}$ peak has a similar shape in the two sample but differs in intensity;
    \item The CIS spectra agree with the Ti$^{3+}$ picture and suggest the presence of intermixing;
    \item The difference in band-bending effect is virtually absent;
    \item A small amount of lattice distortion has to be expected as measured by XLD.
\end{itemize}

AR-XPS measurements, presented in the next paragraph, have been carried out in order to better investigate the possibility of intermixing at the junction. Experimental results will be discussed properly in the final section of this Chapter.

\section{AR-XPS measurements}
\begin{figure}
\begin{center}
\includegraphics[width=0.6\textwidth]{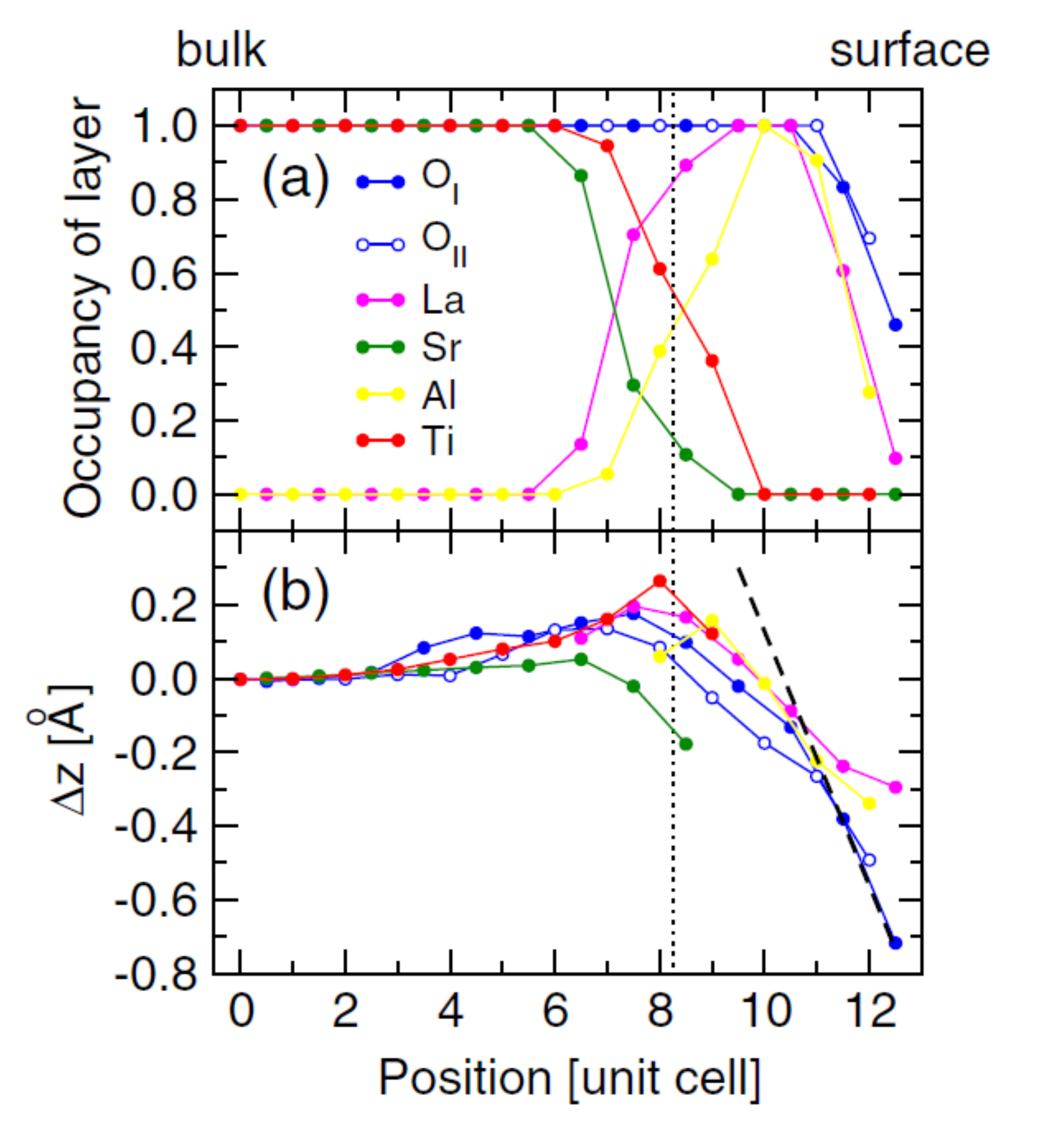}
\caption{Soft X-ray diffraction results on a 5 u.c. LAO-STO sample. In (a) the cationic occupancies and in (b) the cumulative displacement respect to STO layer spacing. The vertical dashed line mark the ideal interface. Taken from Ref.\cite{LAOSTO_SXRD}
\label{fig_LAOSTO_SXRD}}
\end{center}
\end{figure}

A cationic interdiffusion at LAO-STO junction has already been observed by means of X-ray diffraction, as shown in Ref.\cite{LAOSTO_SXRD} and \cite{LAOSTO_vonk}. In short, the experimental results usually display a reduction of the lanthanum content in the uppermost LAO layer, a strong preferential diffusion of La ions inside STO (Fig. \ref{fig_LAOSTO_SXRD}) and a fractional occupancy of the interface Ti sites, although a precise quantification of the latter cannot be done.

\begin{figure}
\begin{center}
\includegraphics[width=0.6\textwidth]{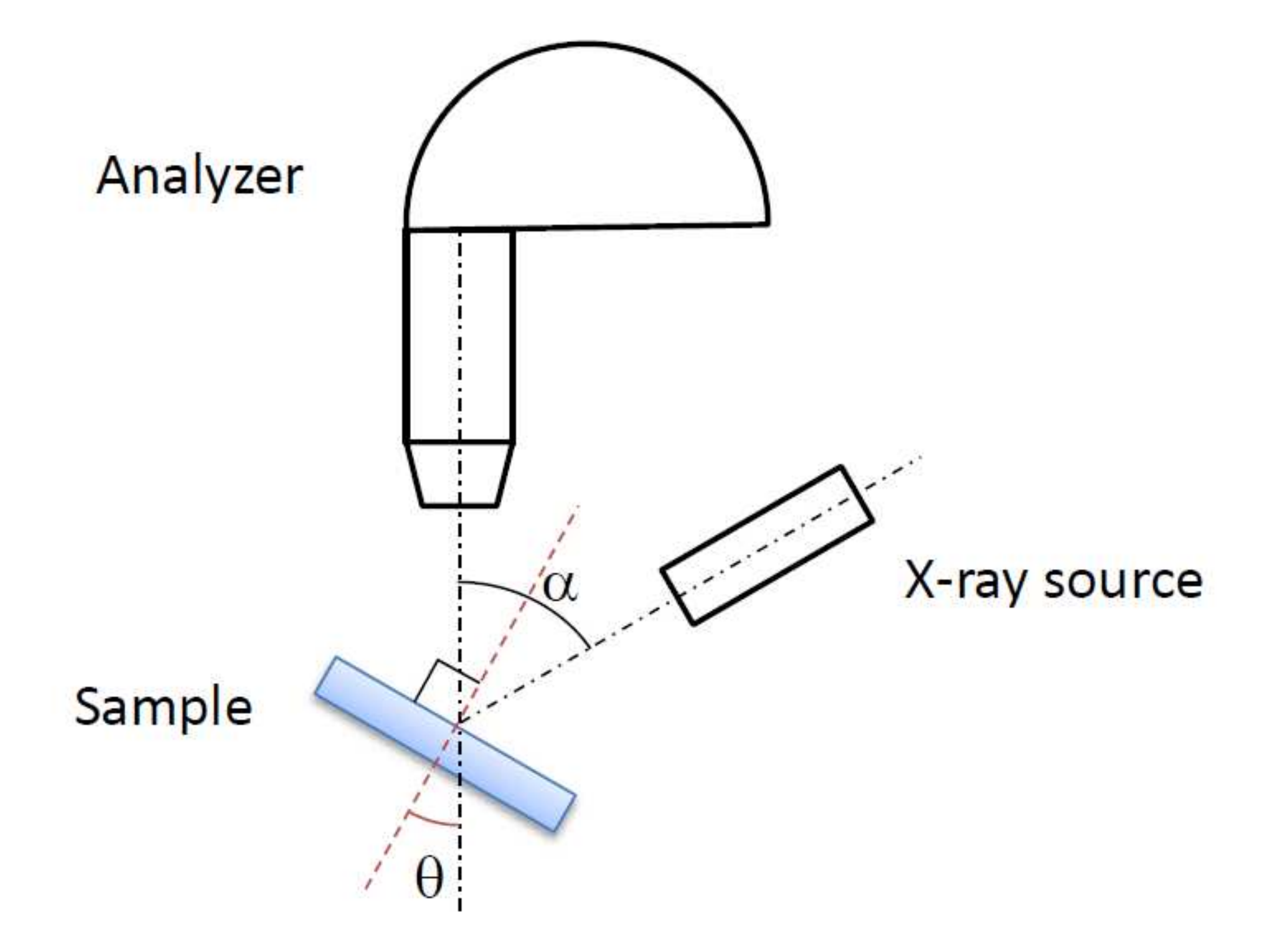}
\caption{Experimental geometry for AR-XPS measurement. The
analyzer-to-source angle $\alpha$ is fixed (55$^\circ$) , the sample
rotation is defined by the tilt angle
$\theta$.\label{fig_setup_arxps}}
\end{center}
\end{figure}

Given the strong angular dependence of XPS probing depth as a function of the emission angle, an angle-resolved XPS experiment has been carried out in order to reveal the chemical composition of each layer near the interface. The usual XPS experimental error in element quantification is usually large, around 10\%; using a set of angular-resolved measurement it is possible to reduce this error, although a certain knowledge of the structure has to be known in advance. The 3 u.c. and 5 u.c. LAO-STO samples have been analyzed by angle resolved XPS, with a non-monochromatized Al-K$_\alpha$ X-ray source; the Al anode has been chosen upon Mg because of the higher photon energy and thus because of the higher probing depth. This kind of source is not focussed, thus the probing area is determined by the analyzer focus dimension, which is around 1 mm; since the samples had a 5$\times$5 mm size, XPS signal from sample-holder hasn't been detected even at grazing emission. The electronic lenses of the analyzer (SCIENTA R3000 XPS/UPS/ARPES) allowed the selection of a small angular acceptance, that has been set to nearly $\pm$1$^\circ$. The experimental geometry is shown in Fig. \ref{fig_setup_arxps}. Both the analyzer and the source has been kept at fixed position, their relative angle $\alpha$ = 55$^\circ$ being chosen to eliminate the asymmetry correction. The alignment of the rotation stage has been checked by looking at the reflected spot of a laser beam collimated with the analyzer nose axis.

The electronic levels selected for this measurement are the Sr 3d, La 4d,
Al 2s core levels, because of the similar binding energy (see Fig. \ref{fig_XPS_SrAlLa}); the analyzer response should be similar in such a small energy range, and the spectra can be collected in a single acquisition. The photoemission at Ti 3s peaks, which have a similar BE, has a very low cross section, so Ti data have been collected by considering the Ti 2p$_{3/2}$ peak. When computing the total Ti 2p intensity, the satellite peaks have also been considered. The peak area has been evaluated after the removal of the satellite lines (due to the lack of monochromatization in the x-ray source) and after a Shirley-type background subtraction.

The LAO-STO interface has been modeled with a layered structure, which consider each cationic plane (so the half of a layer) as a different emitter, in order to recover the correct La/Al and Sr/Ti ratio; however, the local value of IMFP is considered constant inside each oxide (LAO or STO). In the first approximation, the XPS intensity of a peak at a given angle is thus a sum of the contribution of each layer:

\begin{equation}\label{eq_ARXPS_layer}
I_{m}(E_k,\theta) = \chi_m(E_k)\sum_{i=1}^{N} \alpha_i
\int_{z_i}^{z_{i+1}} P(\lambda,z,\theta)dz
\end{equation}
\\
where $m$ is the index of the XPS peaks, E$_k$ is the electron kinetic energy, $\alpha_i$ is the cationic occupancy of the half-layer $i$, $\chi_m(E_k)$ is the cross section and $P$ is the defined in Eq.\ref{eq_exp3}. The index $i$ is running on even (right) index for Sr and La (Ti and Al). The half-layer thickness $z_{i+1}-z_i$ is fixed to be equal to the lattice parameter of LAO or STO respectively, according to the dept $z$ with respect to the ideal interface. The integral of Eq.\ref{eq_ARXPS_layer} can be performed analytically when taking $\lambda$ as the IMFP; the summation has been done on up to 50 half-unit cell. An additional correction for the surface environmental contamination (given by a multiplicative factor equal to a $P(\lambda_{cont},d,\theta)$ function) is added at every peak intensity by considering a fix thickness of graphitic carbon. The values of the inelastic mean free path have been calculated with the TPP2M\cite{TPP2M} formula and are shown in the table \ref{tab_imfp}.

\begin{table}
\begin{center}
\begin{tabular}{lllll}
  \hline
  XPS peak & E$_k$(eV) & $\lambda_{LAO}$({\AA}) & $\lambda_{STO}$({\AA}) & $\lambda_{cont}$({\AA})\\ \hline
  La 4d & 1375 & 25.09 & 26.59 & 36.77 \\
  Sr 3d & 1345 & 24.67 & 26.14 & 36.14 \\
  Ti 2p & 1025 & 20.05 & 21.20 & 29.25 \\
  Al 2s & 1361 & 24.90 & 26.38 & 36.47 \\
  \hline
\end{tabular}
\caption{List of IMFP used in AR-XPS calculations, obtained by TPP-2M formula.\label{tab_imfp}}
\end{center}
\end{table}

\begin{figure}
\begin{center}
\includegraphics[width=0.9\textwidth]{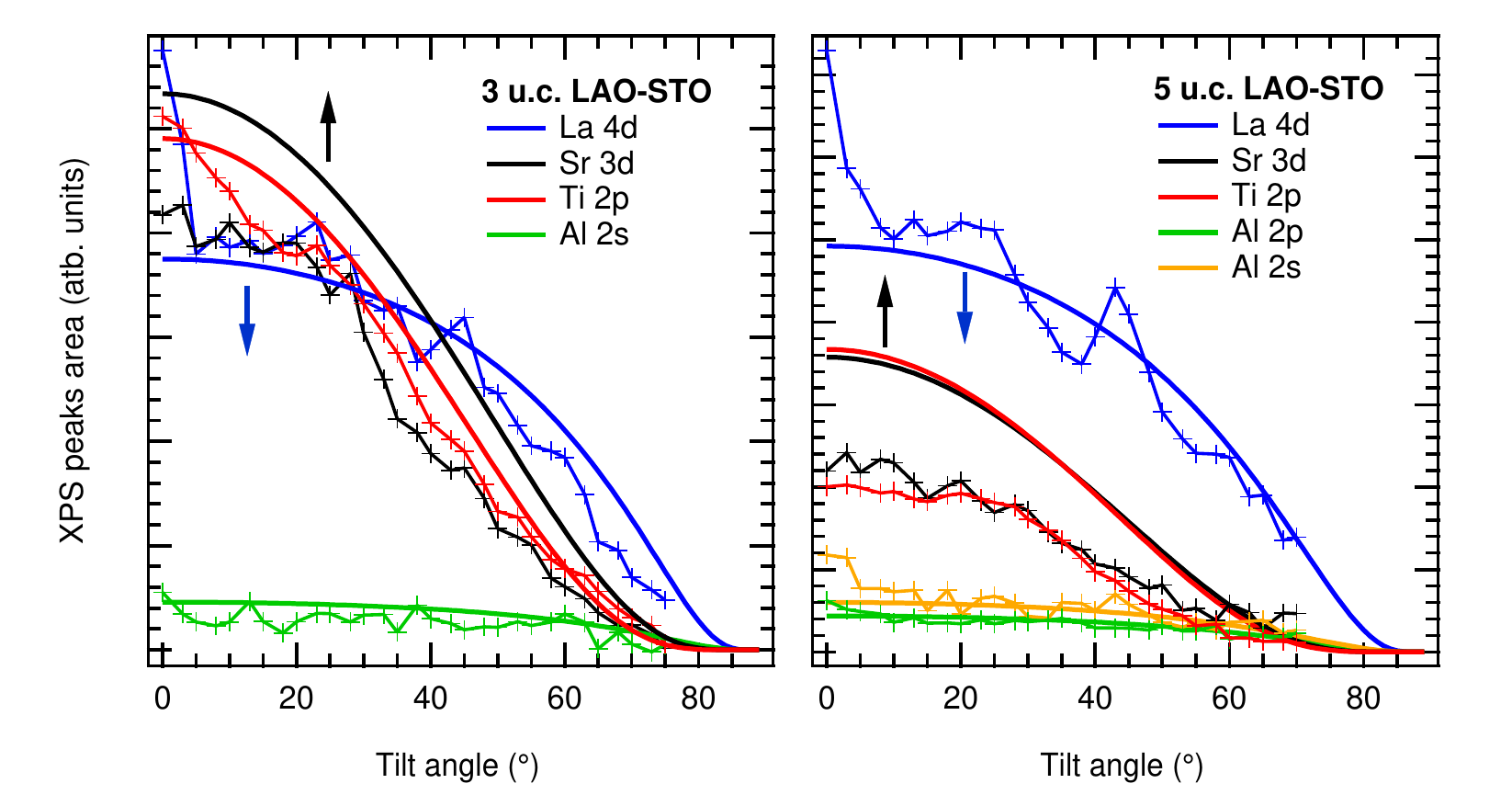}
\caption{Experimental AR-XPS measurement (filled circle) on the 3 u.c. and 5.u. LAO-STO, with the calculation results (thick line). The calculation has been done on a perfectly abrupt LAO-STO
configuration, with the implementation of IMFP only. The model has been normalized to La 4d data. The arrows show the effect of a possible La-Sr intermixing.\label{fig_arxps_lambda}}
\end{center}
\end{figure}

In a perfectly terminated LAO-STO interface the $\lambda_{STO}$ values for Al and La would not be needed, since photoelectrons from La and Al would travel only in the LAO layer; however, in the
intermixing case this value should to be considered (e.g. for electrons emitted from La ions that have diffused deeply into the STO). The contamination correction has been added to adjust the angular slope of AR-XPS data; it should affect La, Sr an Al peaks in the same way, because of the similar kinetic energy (and thus to the similar $\lambda_{cont}$). However, since the kinetic energy of Ti 2p
electrons is lower, a thick contamination layer can reduce the intensity ratio of Ti area versus the other elements. Finally, a common scaling factor is applied to match the experimental
data. The experimental results, depicted with the model results in the case of an abrupt interface, are given in Fig. \ref{fig_arxps_lambda}.

The peaks area is shown ``as it is'', without the cross-section normalization. The AR-XPS data clearly show the signature of X-ray photoelectron diffraction (XPD), consistent with the data reported in literature\cite{LAOSTO_XPD}; in particular, the XPD peaks at 0$^\circ$ and 45$^\circ$ can be detected, due to the forward-scattering events that occur in a cubic lattice with this kinetic energy. This signal modulation can increase considerably the amount of error in a fitting procedure. In any case, the result shown in Fig. \ref{fig_arxps_lambda} clearly doesn't match the experimental results. The calculations systematically overestimate the Ti and Sr contribution. Besides, an exchange between La and Sr ions at the interface would even increase the discrepancy with measured data (see the arrows in Fig. \ref{fig_arxps_lambda}). In order to improve our model, two things have to be done:

\begin{itemize}
  \item On the experimental data side, it is needed to remove the photoelectron diffraction contribution from the angular dependence of the peak area;
  \item On the theoretical side, it is needed to work with a better approximations for $\lambda$ than the simple IMFP.
\end{itemize}

The first issue can be solved by fitting the experimental data with a function that interpolates the angular dependence, averaging out the PED effects; following Eq.\ref{eq_exp3}, this function can be expressed in this form:

\begin{equation}\label{eq_arxps_fit}
f(\theta)=A \exp^{-\frac{B}{\cos{\theta}}}
\end{equation}
\\
with A and B being the parameters to be determined. The result of the data interpolation could in principle be fitted directly with the AR-XPS model; however, the estimated error on the A parameter is rather large as a results of the XPD modulation on the data. To better illustrate this concept, the experimental data on the next figures are replaced by the interpolation results with the Eq. \ref{eq_arxps_fit} with an error bar given by:

\begin{equation}\label{eq_arxps_fit_err}
f_{err}(\theta)=\pm \sigma_A \exp^{-\frac{B}{\cos{\theta}}}
\end{equation}
\\
where $\sigma_A$ is the estimated error of the A parameter in the fit.

\begin{figure}
\begin{center}
\includegraphics[width=0.75\textwidth]{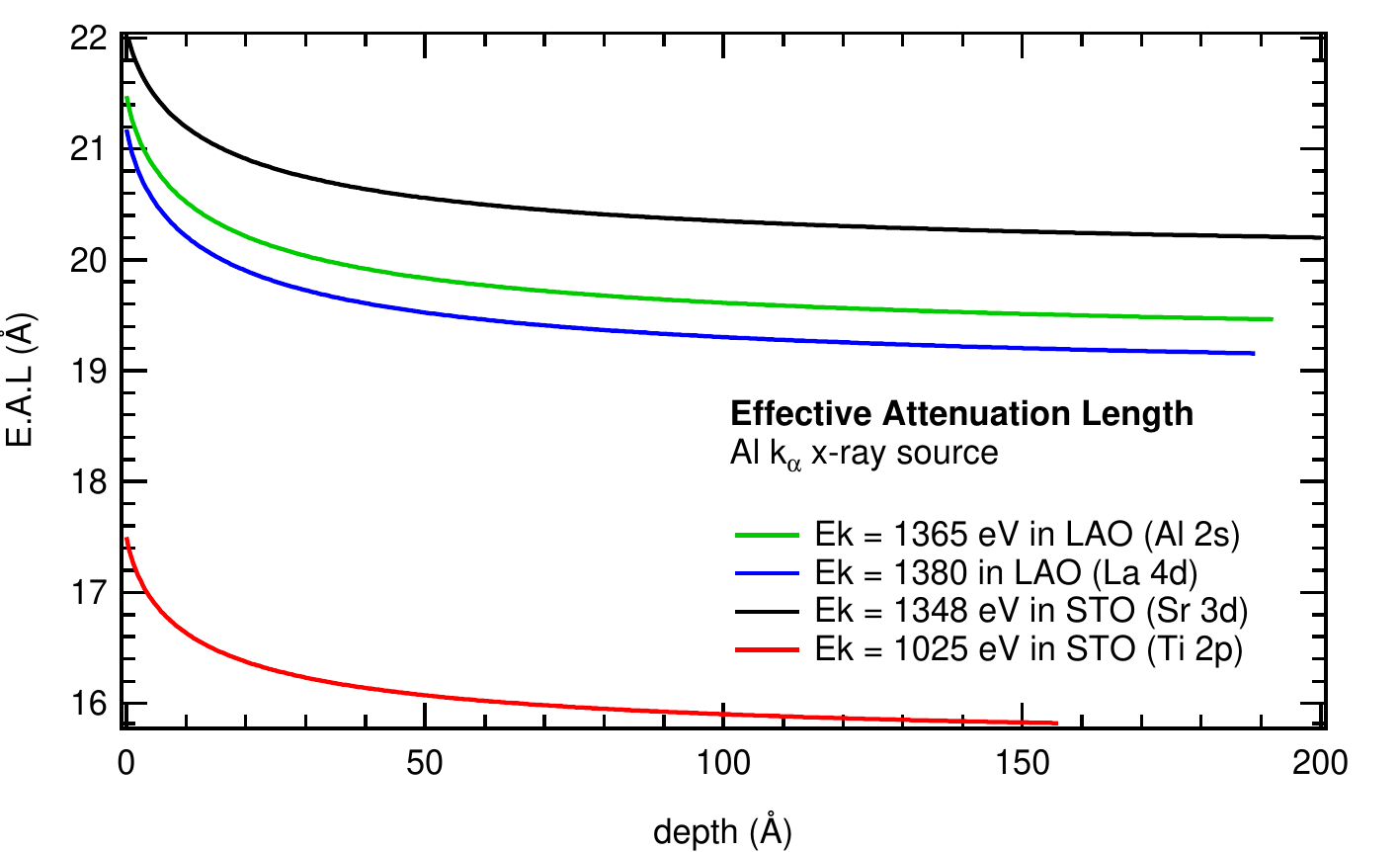}
\caption{Calculated Effective Attenuation Length (EAL) for the
measured XPS peaks in LAO and STO. From the NIST EAL database
software.\label{fig_arxps_eal}}
\end{center}
\end{figure}

As explained in Chapter 2, the theoretical model can be improved by replacing the IMFP with effective attenuation length (EAL), evaluated for a thin overlayer geometry. Calculated EAL's for the LAO-STO case are shown in Fig. \ref{fig_arxps_eal}. Since $\lambda_{EAL}$ is a function of the depth $z$, the integration of Eq. \ref{eq_ARXPS_layer} has to be done numerically. As shown in Fig. \ref{fig_arxps_eal}, in general EAL increases strongly at the surface, reflecting the fact that electrons from bulk STO should undergo to more elastic scattering events. The surface contamination has also to be evaluated trough a different approach than the simple graphitic layer; the carbon contamination can be estimated with an average EAL, that can be approximated with the phenomenological formula:

\begin{equation}\label{eq_arxps_carbon}
\lambda_{cont}(E_k)=0.0129 E_k^{0.7193} (nm)
\end{equation}
\\
The parameters of Eq. \ref{eq_arxps_carbon} have been obtained through a fitting procedure over many measurements on different samples\cite{carb_cont}.

\begin{figure}
\begin{center}
\includegraphics[width=0.9\textwidth]{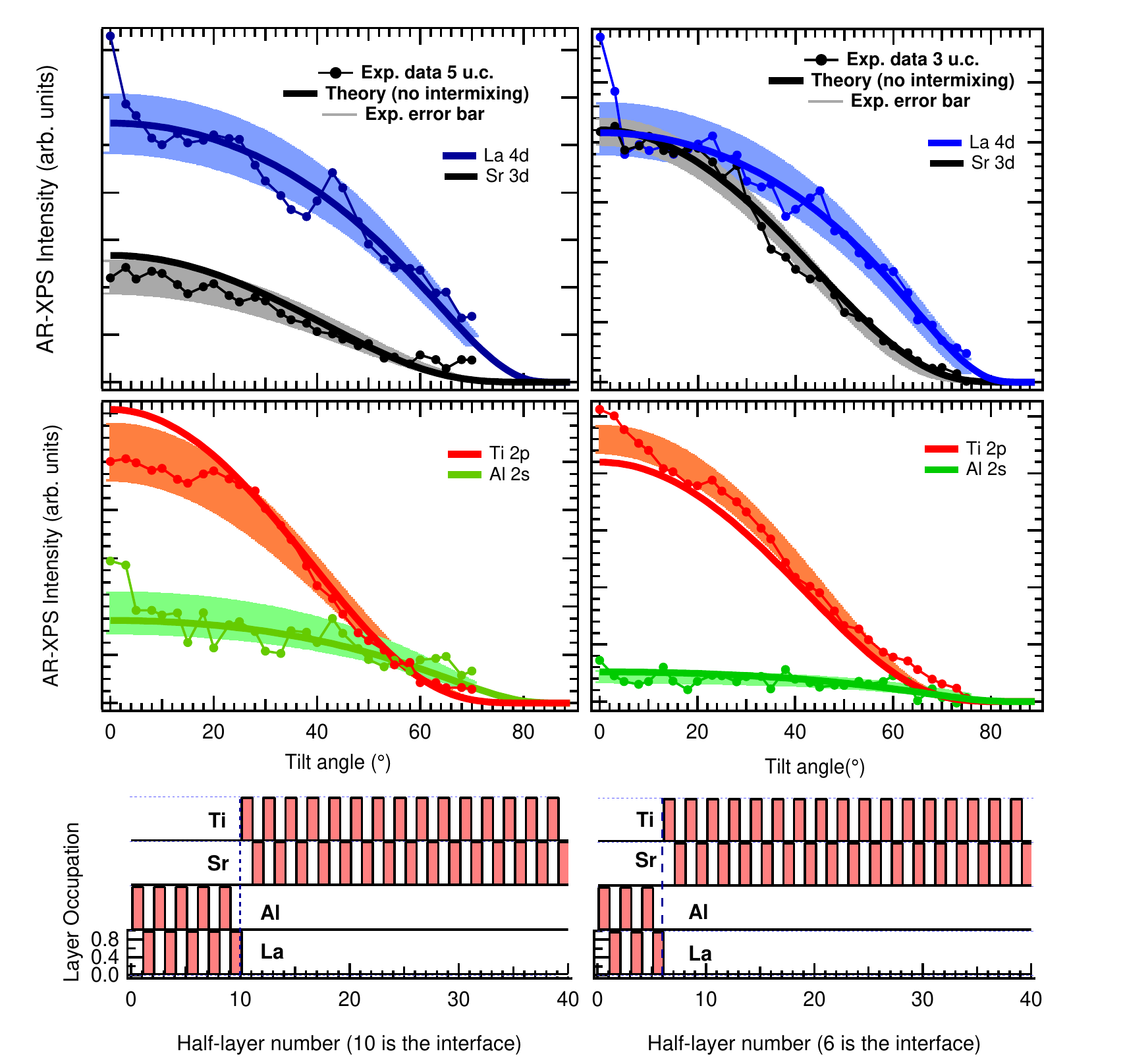}
\caption{Experimental AR-XPS measurement (filled circle) on the 3 u.c. and 5.u. LAO-STO, fitted with Eq. \ref{eq_arxps_fit}, with the calculation results (thick line). The confidence bands on the experimental data fitting are also shown. For the sake of clarity, the spectra for the perovskite A and B cations are displayed in different graphs. These calculations have been done on a perfectly abrupt LAO-STO configurations, with the introduction of the EAL. The layer occupation for each case is shown in the lower panel.\label{fig_arxps_ab_eal}}
\end{center}
\end{figure}

The results of the improved model and of the data fitting are shown in Fig. \ref{fig_arxps_ab_eal}. A 1.2 nm coverage of hydrocarbon contamination has been estimated by fitting AR-XPS data, for both the 3 u.c. and the 5 u.c. LAO-STO samples. A much better agreement with experimental results is found, even without intermixing effect. In particular, the Al 2s and La 4d areas are in good agreement with the theoretical model, while Sr and Ti peaks are sometimes a little off the experimental error bar. These results could be considered as a proof of the effectiveness of the EAL approach. It isn't possible to make a reliable structural predictions starting from the results of Fig. \ref{fig_arxps_ab_eal}. In fact, it should be noted that there's a common multiplicative factor; as an example of the possible mistakes, an excess of Sr can be fitted either by increasing the Sr occupancy or by decreasing the occupancies of \textit{all} the other elements. This, together with the XPD fluctuation, does not allow a precise fitting procedure in this case. In particular, in this case AR-XPS can't give a precise prediction of the \textit{depth} of the intermixing, but can anyway discriminate \textit{whether} an intermixing is present.

These AR-XPS data have been used to confirm or refine the results obtained with other techniques. In particular, the SXRD measurements show a reduction of the topmost LaAlO$_3$ layer into a reduced La$_{(1-x)}$AlO$_3$ and display a cationic intermixing occurring in the firs layer below the interface, with the formation of a La doped STO compound. The composition proposed for the mixed La-Sr layer in Ref.\cite{LAOSTO_vonk} can be expressed as La$_x$Sr$_{1-3/2 x}$; this formulation preserves the charge of a single plane, although it doesn't describe a complete filling of each layer. This kind of intermixing thus corresponds to a lattice rich of cationic vacancies.

\begin{figure}
\begin{center}
\includegraphics[width=0.85\textwidth]{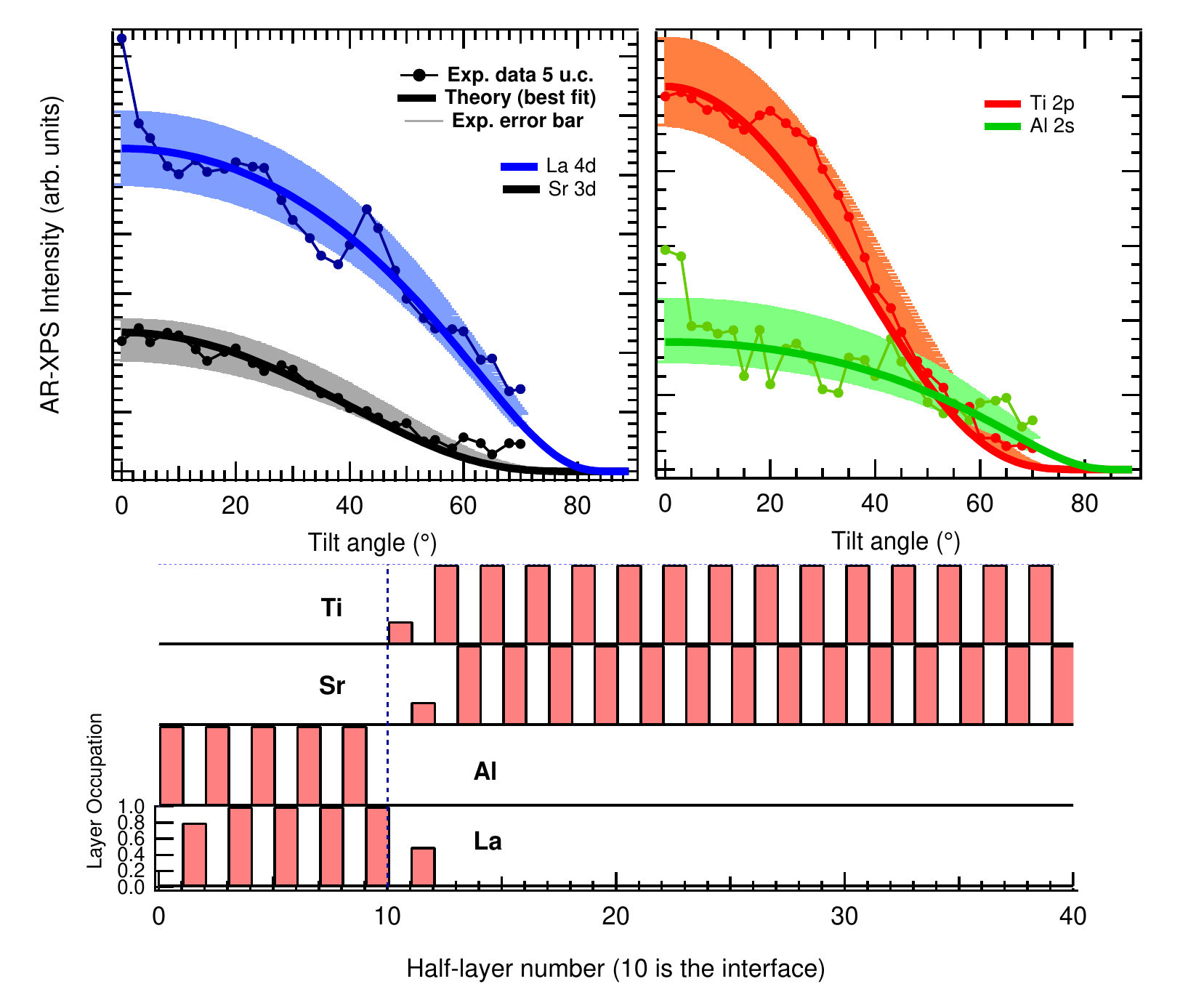}
\caption{Experimental AR-XPS measurement (filled circle) on the 5 u.c. LAO-STO, with the calculation results (thick line). The confidence bands on the experimental data are also shown. The cationic occupancy used in the calculation is shown in the lower panel.\label{fig_arxps_ab5_eal}}
\end{center}
\end{figure}

\begin{figure}
\begin{center}
\includegraphics[width=0.85\textwidth]{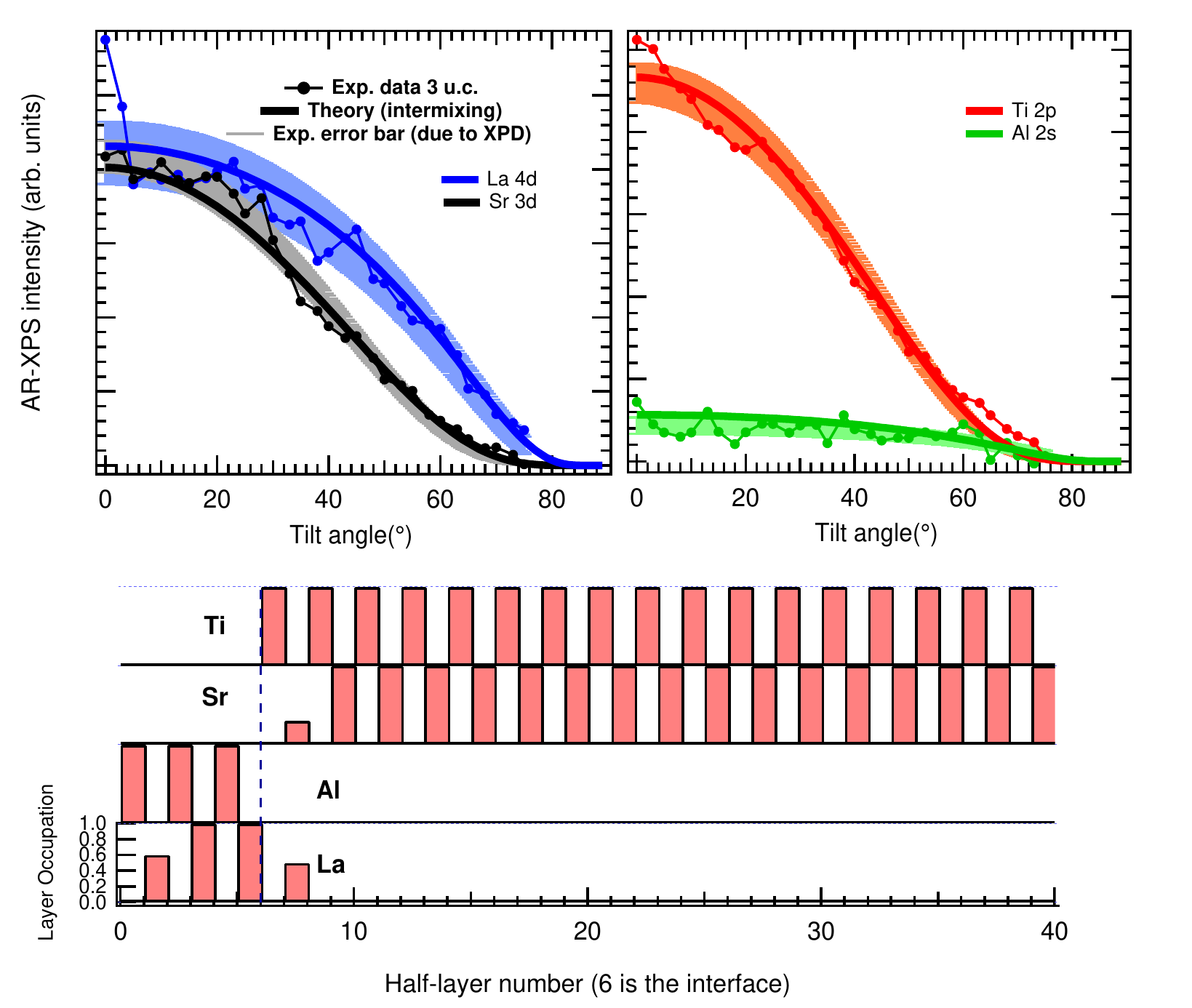}
\caption{Experimental AR-XPS measurement (filled circle) on the 3 u.c. LAO-STO, with the calculation results (thick line). The confidence bands on the experimental data are also shown. The cationic occupancy used in the calculation is shown in the lower panel. Note that the Ti occupancy at the interface layer is full.\label{fig_arxps_ab3_eal}}
\end{center}
\end{figure}

A refined solution for the 5 u.c. and 3 u.c. is given in Fig. \ref{fig_arxps_ab5_eal} and in Fig. \ref{fig_arxps_ab3_eal}. In both samples, the top layer has been set to a partial occupancy x=0.8 (5 u.c.) and x=0.5 (3 u.c.), the intermixing layer has been set to a La$_{0.5}$Sr$_{0.3}$. With these values, the total amount of La as compared to Al is nearly conserved, with a reduced content of Sr on the STO termination. In order to adjust the Ti 2p intensity, in the 5 u.c. the occupancy $\alpha_{Ti}$ at the interface has to be reduced; the best fit is with $\alpha_{Ti}$=0.3, but every value below 0.6 is within the experimental error. Remarkably, in the 3 u.c. case, a very good matching is found even without reducing the Ti occupancy at the interface. The different Ti stoichiometry in conducting and insulating LAO-STO can thus be reflected on the intensity of Ti$^{3+}$ states as measured by ResPES and XPS; however, further investigation would be needed to reduce the experimental uncertainty on Ti occupation.

\begin{figure}
\begin{center}
\includegraphics[width=0.75\textwidth]{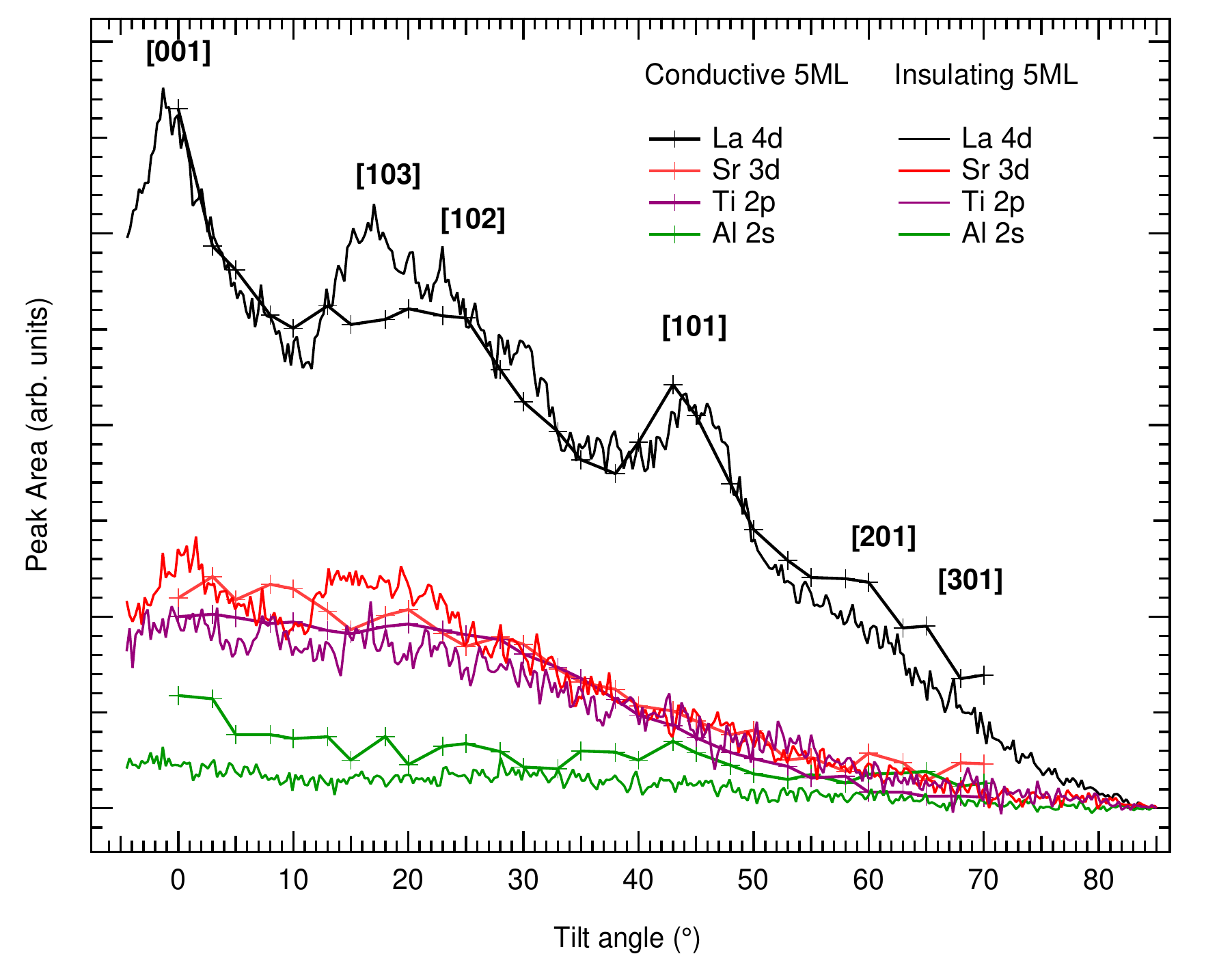}
\caption{AR-XPS data taken on conductive and insulating 5 u.c. samples. The lattice directions associated to the XPD forward scattering process are also shown. Data of the two samples are normalized to match the La 4d AR-XPS spectra. \label{fig_arxps_ins5}}
\end{center}
\end{figure}

Finally, in Fig. \ref{fig_arxps_ins5} AR-XPS data taken on the insulating 5 u.c. LAO-STO (grown with a 10$^{-1}$ mBar O$_2$ partial pressure), compared to the conductive 5 u.c., are shown. With this photon energy (Al k$_\alpha$ X-ray source), the photoelectron diffraction peaks can be interpreted with forward-scattering events and thus marked with the corresponding lattice direction, as shown in Fig. \ref{fig_arxps_ins5}. The possibility to observe the XPD modulation can be regarded as a proof of the crystalline order of these heterostructures. 
Consistently with results shown in Fig. \ref{fig_XPS_SrAlLa}, the Al 2s peak intensity is weaker in the insulating sample than in the conductive one, because of the different growth regime. By comparing the intensities of each cationic species, it seems that the insulating 5 u.c. sample is thus aluminium poor rather than lanthanum rich. The LAO-STO samples grown in 10$^{-1}$ mBar (or higher) O$_2$ pressure thus can't be compared with the ones grown at lower partial pressure, which also displays the physical properties of interest.

Summarizing the AR-XPS results, a clear signature of La intermixing in the bulk STO is found in both the 3 u.c. and 5 u.c. samples, in a similar way. A good data matching can be found following the structural model extracted from SXRD measurements, which should be accompanied by a reduction of Ti at the interface in the 5 u.c. case. In any case, since an abrupt model does not match exactly with the experimental data, every physical description of the conductivity in the LAO-STO heterostructures should account for an amount of disorder at the interface.

\section{Conclusions}
The most interesting signatures of the conductivity shown in this Chapter are the in-gap states detected with ResPES at the Ti L$_{2,3}$ edge. We provided the first experimental evidence\cite{apl_LAOSTO} of these states, probed with soft X-rays spectroscopies. Following our investigation, a similar ResPES analysis has been carried out by A. Koitzsch et al.\cite{LAOSTO_respes_copioni} on LAO-STO interfaces grown with different oxygen partial pressure. They reported the presence of in-gap states, especially in their ``oxygen poor'' sample. Also in this case the oxygen stoichiometry seems to play a crucial role to determine the presence of Ti 3d$^1$ electronic levels .

In our samples, the different spectral weight of in-gap states seem to be directly related to the interface conductivity. Surprisingly, the Ti$^{3+}$ states have also been clearly observed in the insulating sample, with the same peak shape and at the same energy of the conductive one. In a common metal, the rise of conductivity should be marked by an increasing DOS at Fermi level, as measured in LAO-STO, induced by a \emph{shift} of the Fermi-level, which has not be observed in this case. The XPS binding energies are almost the same in insulating and conductive LAO-STO. The presence of V$_O$ due to X-ray damages should be excluded, since the X-ray penetration depth is much larger than the LAO capping and thus a similar amount of X-ray defects should has been detected in each sample. Moreover, the photoemission spectra did not change upon the exposition to the synchrotron X-ray beam over the entire period of the data acquisition (nearly 5 days).

As a starting point we should consider that both samples are correlated systems, with a much higher effective mass in the ``insulating'' sample; in that case, however, a different qua\-si\-par\-ti\-cle peak-shape at Fermi level should be observed, as can be seen in 3d$^1$ correlated metals (for example, in V$_2$O$_3$\cite{v2o3_quasiparticle}) and in theoretical models (for example, the Dynamical Mean Field Theory\cite{dmft_review}). Moreover, as shown with DFT+U DOS calculations, the in-gap states can be described with a simple band-filling of the empty STO levels\cite{STO_ResPES}.

Since the difference in the Ti$^{3+}$ states of 3 and 5 u.c. samples is only a scaling factor without a detectable energy shift, the conductivity may be related to a phase separation mechanism which, in turn, is driven by the thickness of the LAO layer. In any case, the problem is to find a suitable ``doping'' mechanism that can justify the presence of these conductive Ti$^{3+}$ states at the LAO-STO interfaces.

\begin{figure}
\begin{center}
\includegraphics[width=0.6\textwidth]{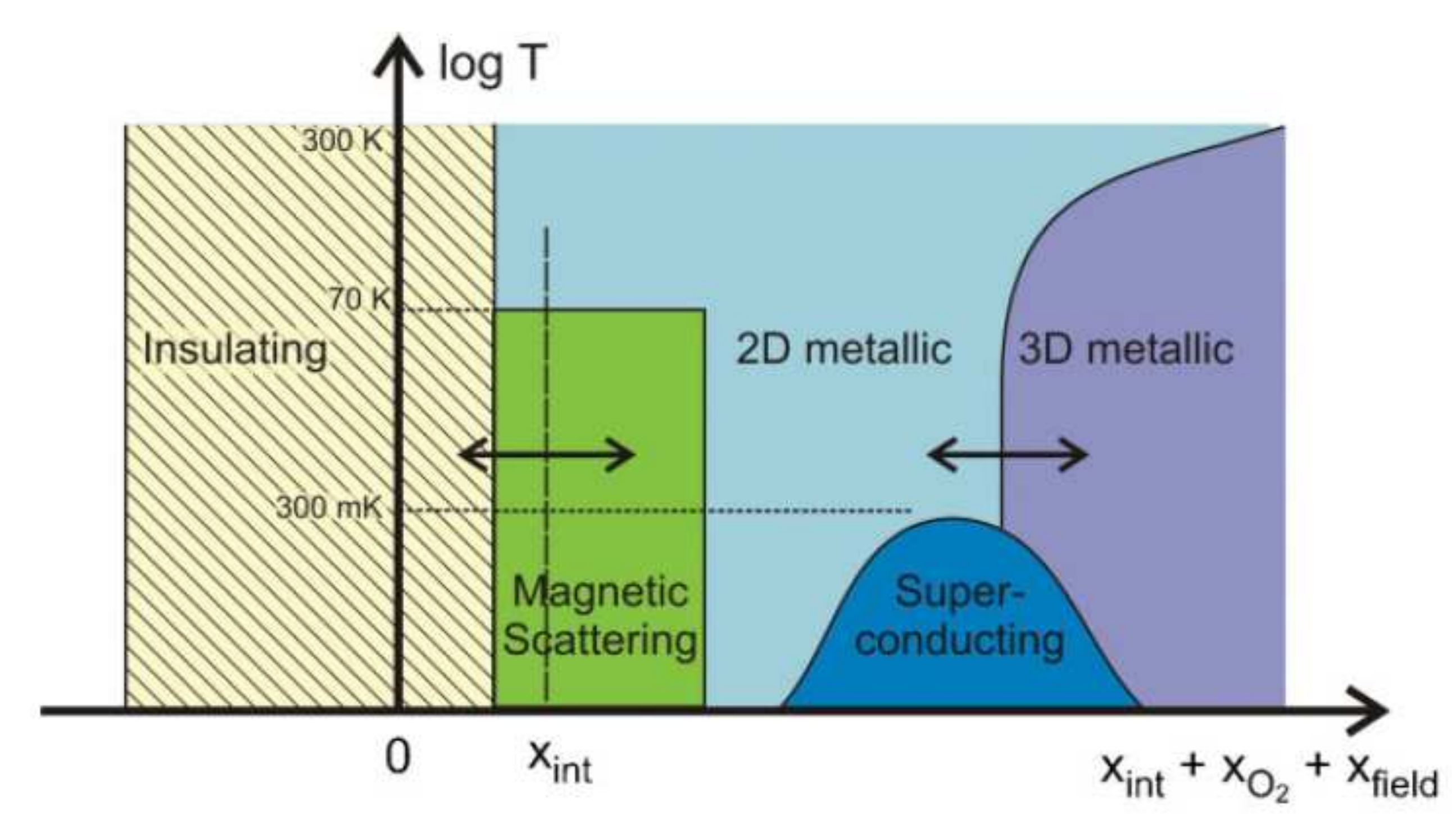}
\caption{Doping versus temperature phase diagram of SrTiO$_3$/LaAlO$_3$
interfaces. The doping scale consists of three possible contributions:
intrinsic carrier doping from the electronic reconstruction (x$_{int}$), carrier
doping from oxygen vacancies (x$_{O_2}$ ) and carrier doping by applied electric
fields (x$_{field}$). Observed transport effects are insulation at p-type interfaces\cite{LAOSTO_ohtomo}(x<0), 2D interface metallicity\cite{LAOSTO_ohtomo}, 3D bulk metallicity\cite{LAOSTO_ohtomo}, negative magneto resistance\cite{LAOSTO_magnet} below 70 K, magnetic hysteresis\cite{LAOSTO_magnet} below 300 mK, and superconductivity\cite{LAOSTO_sc} below T$_c$=200 mK. The exact position of x$_int$ in the phase diagram is not yet known and can vary due to other polarity discontinuity compensating mechanisms (such as x$_{O_2}$ ).Taken from Ref. \cite{LAOSTO_rev2}.\label{fig_laosto_phase}}
\end{center}
\end{figure}

The different conductivity mechanism found in the literature have been summarized\cite{LAOSTO_rev2} by M. Hujiben et al. in the phase diagram of Fig. \ref{fig_laosto_phase}. Three different type of doping are underlined: the intrinsic doping (x$_{int}$), related to the electronic reconstruction, the extrinsic doping (x$_{O_2}$, mostly due to V$_O$) and the external electric field (x$_{field}$). It should be noted that an electrical-field driven transition from the superconductive to the metallic state has been reported\cite{LAOSTO_el_field}, thus justifying the presence of the x$_{field}$ parameter in Fig. \ref{fig_laosto_phase}.
The 3D metallic phase correspond to a high density of V$_O$, due to the low P$_{O_2}$ during the growth. The superconductivity has been observed for an intermediate P$_{O_2}$ regime , while in the optimal regime the localized magnetic moments have been detected. The sample analyzed in this thesis work belong to the latter group. The value of x$_{int}$ in Fig. \ref{fig_laosto_phase} is still unknown; in particular, it cannot be put in the intermediate region because oxygen vacancies can have a non negligible role in the superconductivity.

The observed in-gap states could be ascribed for instance to a cationic disorder at the interface; in fact, STO can became conductive even for small La doping. In order to justify the 4 u.c. thickness threshold, the disorder should be dependent to the number of LAO capping layer; however, the cationic interdiffusion seems to be similar in conductive and insulating samples, as shown in section 5.5 and in the literature by N. Nakagawa et al.\cite{LAOSTO_tem_disorder}.

The metallic areas can also be generated by the presence of residual oxygen vacancies near the interface; as in the case of TiO$_{2-\delta}$, an oxygen vacancy could create localized defect states, that above certain concentrations can merge to yield a metallic band. In a recent article\cite{LAOSTO_oxy_pol}, Z. Zhong and co-workers report DFT calculations suggesting that the formation energy of oxygen vacancies in LAO-STO interfaces is thickness dependent. However, our samples should also display the presence of isolated magnetic moments that in turn cannot be provided by isolated oxygen vacancies alone, similarly to the TiO$_{2-\delta}$ case of Chapter 4. The magnetism could appear only in the case of clustered defects at the interface\cite{LAOSTO_rev2}. The possibility to reduce the influence of V$_O$ by mean of the growth conditions and the sharp thickness threshold suggest that interface V$_O$ can play a minor role in the conductivity, at least in our samples.

Finally, in the electronic reconstruction scenario the relaxation of the polar discontinuity is achieved by a 1/2 e$^-$ charge-transfer from the surface to the interface, which in fact should result in the presence of reduced Ti$^{3.5+}$ in the uppermost STO layer\cite{LAOSTO_ohtomo}. A practical implementation of this concept is the band-bending effect\cite{LAOSTO_disto}, which has never been observed in these systems by means of X-ray spectroscopies\cite{LAOSTO_chambers}. Moreover, ab-initio calculations carried out on a LAO-STO interface with different types of cationic disorder\cite{LAOSTO_chambers} show that the intermixing can be thermodynamically favorable and can reduce the LAO polarity effects.

Another kind of electronic reconstruction is a ferroelectric-like distortion of the first LAO layer at the interface. As expected, the average c cell parameter in the capping LAO (3.79 \r{A}) is usually lower than the STO planar distance\cite{LAOSTO_SXRD} (3.89 \r{A}); however, SXRD and HRTEM\cite{LAOSTO_tem_elongation} measurements show that the first unit cell at the interface is elongated by 4\%-9\% (up to 4.1 \r{A}). This elongation is explained by the authors with the presence of Ti$^{3+}$ ions at the interface (thus with a larger ionic radius with respect to Ti$^{4+}$), in a sort of Jahn-Teller effect. A structural deformation due to the strain in LAO capping that is spread also in STO has also been reported in a recent investigation with high-resolution XRD\cite{LAOSTO_sxrd_disto}.
As already pointed out in Section 5.3, the distortion of the oxygen octahedral cage around the interface Ti atoms has already been measured with XLD\cite{LAOSTO_XLD} and in a similar way in our samples. Surprisingly, Ti$^{3+}$ electronic levels cannot be observed by XAS, and XLD results can be explained completely with Ti$^{4+}$ CT-multiplet calculation only. However, CIS spectra clearly agrees with a Ti$^{3+}$ picture of the conductive Ti 3d states. In any case, in the electronic reconstruction frame the presence of the Ti$^{3+}$ states at Fermi edge in insulating LAO-STO cannot be explained, at least without assuming an inhomogeneous distribution of the distorted sublattice at the interface.

Another interesting theory that could fit our experimental results has been proposed by N. C. Bristowe and co-workers\cite{bristowe}; by recognizing that most of the measurement on this oxides have been done at atmospheric pressure (or, as in this case, with an environmental atmosphere surface contamination), the doping at the LAO-STO interface could be given by a surface contamination or surface defects.

\begin{figure}
\begin{center}
\includegraphics[width=0.6\textwidth]{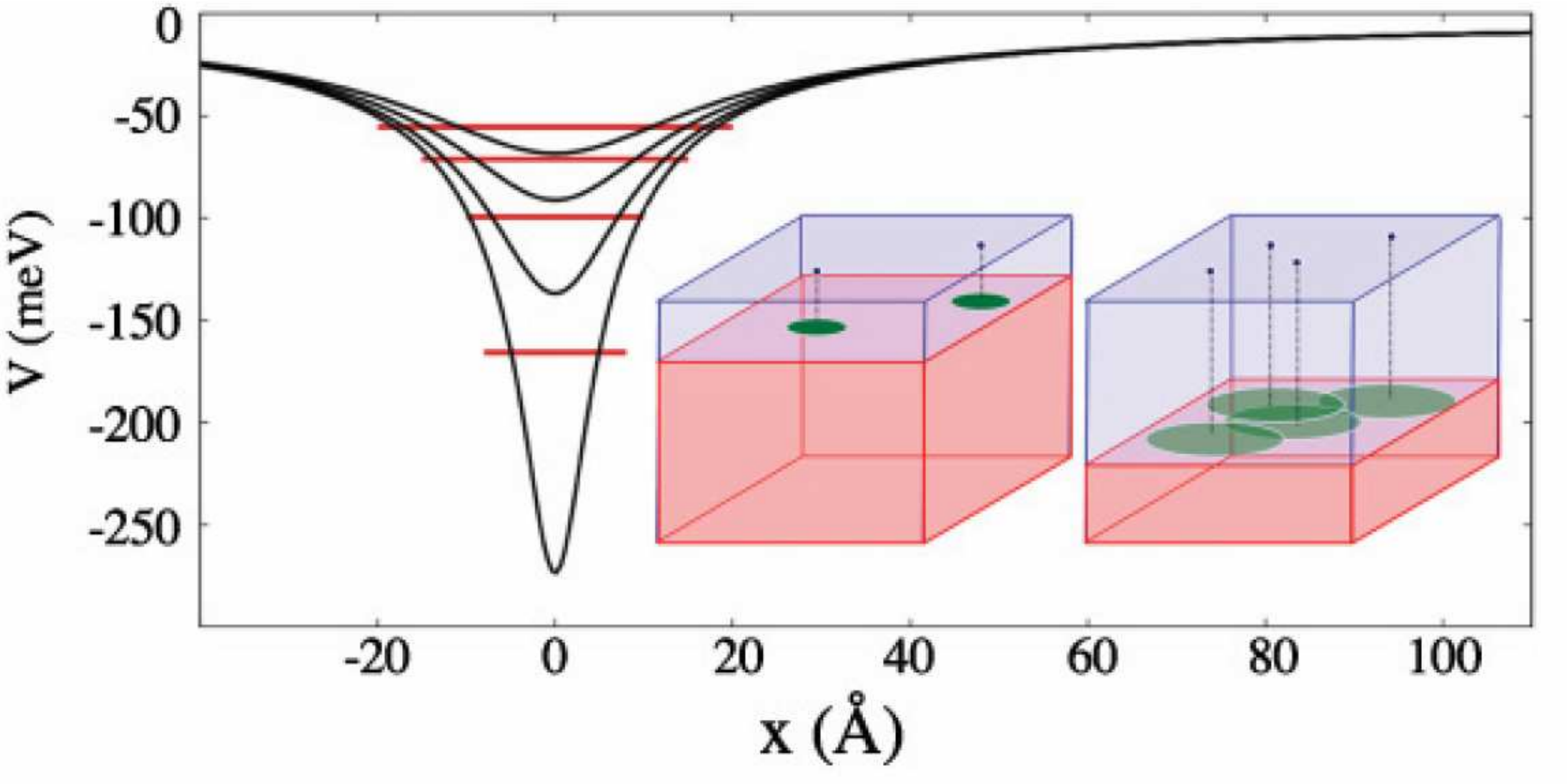}
\caption{Schematic view of the thickness dependence on interface electrons localization from the model of Ref.\cite{bristowe}. Trapping potential V created by a surface O vacancy as seen by interface electrons versus distance within the interface plane x (x=0 is directly below the vacancy) for film thickness d=a (deepest), 2a, 3a, 4a (shallowest), with harmonic estimates of corresponding donor ground states [taking m$_{eff}$ = 3m$_e$]. Inset: Sketch of range and density of trapped states.\label{fig_laosto_surf_model}}
\end{center}
\end{figure}

The relaxation of the polar discontinuity can thus be achieved by the same charge-transfer process of the electronic reconstruction model, but without the need of a band bending, which should also imply a conductive surface. In this model, the oxygen reduction at the surface should create a potential well for the induced localized charge in the 3d-level at the interface; in the authors view, the localization radius would be inversely proportional to the depth of the well, i.e. to the LAO thickness: hence the reason of the conductivity. A pictorial representation is given in Fig.\ref{fig_laosto_surf_model}.

This model can also explain a peculiar effect that has been reported on insulating LAO-STO interface. In fact, Y. Xie et al.\cite{LAOSTO_biased_afm} have shown that it is possible to draw (and erase) conductive lines on LAO-STO by moving a biased AFM tip on the LAO surface. The authors suggest that the driving mechanism of this process is the creation of defects at the surface; in the surface-reduction model, the biased tip just raises (or decreases, according to the bias sign) the vacancy stability. Recently published ab-initio DFT calculations of the V$_O$ formation energies in the LAO-STO further support this description\cite{laosto_surf_vo}. A possible way to check the influence of the surface reduction on the LAO-STO conductivity could be to use LAO-STO junction as a ``gas sensor''; it should be possible to measure a variable sample resistance upon the exposition to different gases or to directly measure the Ti-3d spectral weight with the aid of ResPES in a synchrotron facility.

In conclusion, the LAO-STO system displays a wealth of physical and chemical properties, which challenge both theoretical and experimental physicist. A single interface ``doping'' source is not enough to explain the various phenomena that occur at this interface. In fact, the theoretical models that involve perfectly abrupt interfaces and clean surfaces cannot completely explain the different experimental evidences. In this scenario, surface effects and electronic phase separation are thought to play a major role, so far often disregarded.

\addcontentsline{toc}{chapter}{Bibliography}
\bibliographystyle{unsrt}
\bibliography{Main_tesi}

\end{document}